\documentclass[a4paper,10pt,twoside,openright]{report}
\usepackage{fancyhdr}
\usepackage{amsmath}
\usepackage{amsfonts}
\usepackage{amssymb}
\usepackage{graphicx}
\usepackage{bbm}
\usepackage{algorithm}
\usepackage{algorithmic}
\usepackage{xspace}
\usepackage{color}
\usepackage[table,xcdraw]{xcolor}
\usepackage{multirow}
\usepackage{booktabs}
\usepackage{marginnote}
\usepackage[marginparsep=0.5cm, marginparwidth=2cm]{geometry}
\usepackage{url}
\usepackage{doi}
\usepackage{caption}
\usepackage{subcaption}
\usepackage{hyperref}
\AtBeginDocument{\usepackage{sidecap}}

\newcommand{\meqnote}[1]{\marginnote{\textit{\small{#1}}}[1cm]}
\newcommand{\mnote}[1]{\marginnote{\textit{\small{#1}}}}
\newcommand{\mynote}[2]{\marginnote{\textit{\small{#1}}}[#2cm]}

\newcommand{\iter}{{\rm t}}
\newcommand{\tv}{\mathbf{t}}

\newcommand{\inv}{\circleddash 1}

\newcommand{\ZZ}{\mathcal{Z}}
\newcommand{\state}{\{\sigma\}}	\newcommand{\sig}{\sigma} \newcommand{\sigv}{\boldsymbol{\sig}}
\newcommand{\OO}{\mathcal{O}}
\newcommand{\boltz}{\mu_{\beta}}
\newcommand{\freeEnt}{\phi} \newcommand{\FreeEnt}{\Phi}
\newcommand{\free}{f} \newcommand{\Free}{F}

\newcommand{\EE}{\mathcal{E}}
\newcommand{\GG}{\mathcal{G}}
\newcommand{\VV}{\mathcal{V}}
\newcommand{\FF}{\mathcal{F}}
\newcommand{\lh}{\hat{l}}
\newcommand{\er}{Erd\H{o}s-Renyi\xspace}
\newcommand{\pin}{p_{\mathrm{in}}}
\newcommand{\pout}{p_{\mathrm{out}}}
\newcommand{\ijlink}{\langle i j \rangle}
\newcommand{\Qmap}{Q^{\rm MAP}}
\newcommand{\qh}{\hat{q}}
\newcommand{\mapping}{\varphi}
\newcommand{\partition}{\{ l \}}

\renewcommand{\a}{a} \renewcommand{\b}{b}
\newcommand{\fac}{\mu}
\newcommand{\messv}{\psi}
\newcommand{\messf}{\tilde{\psi}}
\newcommand{\bfree}{\mathbb{F}}

\newcommand{\messvd}{\phi}
\newcommand{\messfd}{\tilde{\phi}}

\newcommand{\K}{K} \newcommand{\M}{\mathbf{M}}
\newcommand{\xdist}{\phi}
\newcommand{\alphacs}{\alpha_{\rm CS}}
\newcommand{\xiv}{\boldsymbol{\xi}}

\newcommand{\Tr}{{\rm Tr}}

\DeclareMathOperator{\saddle}{{\rm SP}}

\newcommand{\alphagmin}{\alpha_{{\rm min}}}
\newcommand{\alphacal}{\alpha^{{\rm cal}}}
\newcommand{\D}{D} \newcommand{\Dh}{\hat{\D}} \newcommand{\Db}{\bar{\D}}		
\newcommand{\dv}{\mathbf{d}} 
\renewcommand{\dh}{\hat{d}} \newcommand{\db}{\bar{d}}
\newcommand{\dhv}{\mathbf{\hat{d}}} 

\newcommand{\cc}{\mu}
\newcommand{\fsf}{\epsilon} 
\newcommand{\kv}{\mathbf{k}}

\newcommand{\dun}{d_1}	\newcommand{\ddeux}{d_2}	\newcommand{\dtrois}{d_3}

\newcommand{\Jacobian}{\mathbf{J}}
\newcommand{\one}{\mathbf{1}}

\newcommand{\xt}{\tilde{x}}	\newcommand{\xtv}{\tilde{\mathbf{x}}}

\newcommand{\rank}{R} \newcommand{\ri}{s}  \newcommand{\ribis}{s'}
\newcommand{\US}{M}  \newcommand{\ui}{\mu} \newcommand{\uibis}{\mu'}
\newcommand{\VS}{P}  \newcommand{\vi}{p}   \newcommand{\vibis}{p'}
\newcommand{\YS}{L}  \newcommand{\yil}{l}	\newcommand{\yibis}{l'}
\newcommand{\msex}{{\rm MSE_X}} \newcommand{\mseu}{{\rm MSE_U}} \newcommand{\msev}{{\rm MSE_V}}

\newcommand{\ppi}{\gamma}

\newcommand{\ie}{i.e.\xspace} \newcommand{\eg}{e.g.\xspace}
\newcommand{\etal}{et al.\xspace} \newcommand{\iid}{i.i.d.\xspace}
\renewcommand{\v}[1]{\mathbf{#1}}
  \newcommand{\awgn}{AWGN\xspace}
\newcommand{\inference}{inference\xspace} \newcommand{\statphys}{statistical physics\xspace}
\newcommand{\mythesis}{this thesis\xspace}
\newcommand{\cs}{CS\xspace} \newcommand{\amp}{AMP\xspace} \newcommand{\bp}{BP\xspace} \newcommand{\lasso}{LASSO\xspace}
\newcommand{\gamp}{GAMP\xspace}  \newcommand{\tap}{TAP\xspace}  \newcommand{\cgamp}{c-GAMP\xspace}	\newcommand{\calamp}{Cal-AMP\xspace}	\newcommand{\prgamp}{PR-GAMP\xspace}
\newcommand{\sbm}{SBM\xspace}
\newcommand{\glm}{GLM\xspace} \newcommand{\glms}{GLMs\xspace}
\newcommand{\gbm}{GBM\xspace}    \newcommand{\bigamp}{BiGAMP\xspace}	\newcommand{\dl}{DL\xspace}	\newcommand{\mf}{MF\xspace} \newcommand{\mcs}{MCS\xspace}
\newcommand{\pdf}{pdf\xspace}
\newcommand{\Dkl}{D_{{\rm KL}}}

\newcommand{\mse}{\mathrm{MSE}} \newcommand{\nmse}{\mathrm{nMSE}}
\newcommand{\ov}{O}
\newcommand{\modbp}{mod-bp\xspace}
\newcommand{\mri}{MRI\xspace}
\newcommand{\dmd}{DMD\xspace}

\newcommand{\pbig}{PBiGAMP\xspace}

\renewcommand{\eqref}[1]{eq.~(\ref{#1})\xspace}
\newcommand{\figref}[1]{Fig.~\ref{#1}\xspace}
\newcommand{\secref}[1]{sec.~\ref{#1}\xspace}
\newcommand{\exref}[1]{example~\ref{#1}\xspace}
\newcommand{\appliref}[1]{application~\ref{#1}\xspace}
\newcommand{\chapref}[1]{chapter~\ref{#1}\xspace}
\newcommand{\algoref}[1]{algorithm~\ref{#1}\xspace}
\newcommand{\appref}[1]{Appendix~\ref{#1}\xspace}

\newcommand{\RR}{\mathbb{R}}
\newcommand{\CC}{\mathbb{C}}
\newcommand{\Nintegers}{\mathbb{N}}
\newcommand{\XX}{\mathcal{X}}
\newcommand{\YY}{\mathcal{Y}}
\newcommand{\II}{\mathcal{I}}
\newcommand{\UU}{\mathcal{U}}
\newcommand{\cov}[1]{\mathcal{S}_+^{#1}}
\newcommand{\sym}[1]{\mathcal{S}^{#1}}

\DeclareMathOperator*{\argmin}{argmin}	\DeclareMathOperator*{\argmax}{argmax}

 \newcommand{\erfc}{{\rm erfc}}
\newcommand{\sign}{{\rm sign}}
\newcommand{\indic}{\mathbbm{1}}
\newcommand{\NN}{\mathcal{N}}
\newcommand{\CN}{\mathcal{CN}} 
\newcommand{\Cc}{\mathcal{CN}}
\newcommand{\f}{f} \newcommand{\fh}{\hat{f}} \newcommand{\fb}{\bar{f}}
 \newcommand{\gh}{\hat{g}} \newcommand{\gb}{\bar{g}}
\newcommand{\p}{p} \newcommand{\px}{p_X} \newcommand{\pf}{p_F} \newcommand{\pd}{p_D} \newcommand{\pyx}{p_{Y|X}} \newcommand{\pxy}{p_{X|Y}} \newcommand{\py}{p_{Y|Z}} \newcommand{\pyd}{p_{Y|Z,D}} \newcommand{\pu}{p_U} \newcommand{\pv}{p_V} 

\newcommand{\A}{\mathcal{A}}  \newcommand{\Au}{\mathcal{A}_U}  \newcommand{\Av}{\mathcal{A}_V}
\newcommand{\AM}{\mathbf{A}}

\newcommand{\DD}[1]{{\rm D} #1 \,}
\newcommand{\dd}[1]{{\rm d} #1}
\newcommand{\mypartial}[1]{\frac{\partial}{\partial #1 }}

\newcommand{\alphau}{\alpha_U} \newcommand{\alphav}{\alpha_V}

\newcommand{\xadd}{\tilde{x}}

\newcommand{\params}{$\{\theta\}$}

\newcommand{\F}{F}
\newcommand{\Fv}{\mathbf{F}}
\newcommand{\Y}{Y}
\newcommand{\y}{y} \newcommand{\yv}{\mathbf{y}} \newcommand{\ys}{M} \newcommand{\yi}{\mu}

\newcommand{\lr}{\mathbf{x}}
\newcommand{\x}{x} \newcommand{\xa}{\vec{\x}}
\newcommand{\xh}{\hat{x}} \newcommand{\xb}{\bar{x}}  \newcommand{\xs}{N}
\newcommand{\X}{X} \newcommand{\Xh}{\hat{X}} \newcommand{\Xb}{\bar{X}}
\newcommand{\xv}{\mathbf{x}} \newcommand{\xhv}{\hat{\mathbf{x}}} \newcommand{\xbv}{\bar{\mathbf{x}}}
 \newcommand{\Xhv}{\hat{\mathbf{X}}} \newcommand{\Xbv}{\bar{\mathbf{X}}}
 \newcommand{\ghv}{\hat{\mathbf{g}}} \newcommand{\gbv}{\bar{\mathbf{g}}}
\newcommand{\z}{z} \newcommand{\za}{\vec{\z}}
\newcommand{\Z}{z} \newcommand{\Zh}{\hat{Z}} \newcommand{\Zb}{\bar{Z}}
 \newcommand{\Zhv}{\hat{\mathbf{Z}}} \newcommand{\Zbv}{\bar{\mathbf{Z}}}
\newcommand{\zh}{\hat{z}} \newcommand{\zb}{\bar{z}}
\newcommand{\zv}{\mathbf{z}} \newcommand{\zhv}{\hat{\mathbf{z}}} \newcommand{\zbv}{\bar{\mathbf{z}}}

\renewcommand{\u}{u} \newcommand{\uh}{\hat{u}} \newcommand{\ub}{\bar{u}} \newcommand{\ua}{\vec{u}}
  
\newcommand{\uv}{\mathbf{u}} \newcommand{\uhv}{\hat{\mathbf{u}}} \newcommand{\ubv}{\bar{\mathbf{u}}}
 \newcommand{\Uhv}{\hat{\mathbf{U}}} \newcommand{\Ubv}{\bar{\mathbf{U}}}
\renewcommand{\v}{v} \newcommand{\vh}{\hat{v}} \newcommand{\vb}{\bar{v}} \newcommand{\va}{\vec{v}}
  
\newcommand{\vv}{\mathbf{v}} \newcommand{\vhv}{\hat{\mathbf{v}}} \newcommand{\vbv}{\bar{\mathbf{v}}}
 \newcommand{\Vhv}{\hat{\mathbf{V}}} \newcommand{\Vbv}{\bar{\mathbf{V}}}

\newcommand{\Qh}{\hat{Q}} 
\newcommand{\mh}{\hat{m}}

\newcommand{\Qmz}{\mathbf{Q}_Z} \newcommand{\Qmhz}{\mathbf{\hat{Q}}_Z}
\newcommand{\Qz}{Q_Z} \newcommand{\Qzz}{Q_Z^0} \newcommand{\qz}{q_Z} \newcommand{\mz}{m_Z}
\newcommand{\Qhz}{\hat{Q}_Z}  \newcommand{\qhz}{\hat{q}_Z} \newcommand{\mhz}{\hat{m}_Z}
\newcommand{\Qmx}{\mathbf{Q}_X}	\newcommand{\Qmhx}{\mathbf{\hat{Q}}_X}
\newcommand{\Qx}{Q_X} \newcommand{\Qzx}{Q_X^0} \newcommand{\qx}{q_X} \newcommand{\mx}{m_X}
\newcommand{\Qhx}{\hat{Q}_X} \newcommand{\Qhzx}{\hat{Q}_X^0} \newcommand{\qhx}{\hat{q}_X} \newcommand{\mhx}{\hat{m}_X}
	\newcommand{\Qmh}{\mathbf{\hat{Q}}}
\newcommand{\ix}{\mathcal{I}_X}
\newcommand{\iz}{\mathcal{I}_Z}
\newcommand{\ac}{S}

\newcommand{\Qmu}{\mathbf{Q}_U}	\newcommand{\Qmhu}{\mathbf{\hat{Q}}_U}
\newcommand{\Qu}{Q_U} \newcommand{\Qzu}{Q_U^0} \newcommand{\qu}{q_U} \newcommand{\mymu}{m_U}
\newcommand{\Qhu}{\hat{Q}_U}  \newcommand{\qhu}{\hat{q}_U} \newcommand{\mhu}{\hat{m}_U}
\newcommand{\iu}{\mathcal{I}_U}

\newcommand{\Qmv}{\mathbf{Q}_V}	\newcommand{\Qmhv}{\mathbf{\hat{Q}}_V}
\newcommand{\Qv}{Q_V} \newcommand{\Qzv}{Q_V^0} \newcommand{\qv}{q_V} \newcommand{\mv}{m_V}
\newcommand{\Qhv}{\hat{Q}_V}  \newcommand{\qhv}{\hat{q}_V} \newcommand{\mhv}{\hat{m}_V}
\newcommand{\iv}{\mathcal{I}_V}

\newcommand{\uhC}{a} \newcommand{\ubC}{A} \newcommand{\xhC}{b} \newcommand{\xbC}{B}

\newcounter{example}[chapter]
\renewcommand{\theexample}{\thechapter.\arabic{example}}
\newsavebox{\selvestebox}
\newenvironment{example}[1]
  {\newcommand\colboxcolor{E5E5E5}%
   \begin{lrbox}{\selvestebox}%
   \begin{minipage}{\dimexpr\columnwidth-2\fboxsep\relax}
   \refstepcounter{example}\par\medskip
   \textbf{Example~\theexample: #1} \rmfamily \\
    }
  {\end{minipage}
  \end{lrbox}%
   \begin{center}
   \colorbox[HTML]{\colboxcolor}{\usebox{\selvestebox}}
\end{center}}

\newenvironment{exampleSuite}
  {\newcommand\colboxcolor{E5E5E5}%
   \begin{lrbox}{\selvestebox}%
   \begin{minipage}{\dimexpr\columnwidth-2\fboxsep\relax}
    }
  {\end{minipage}
  \end{lrbox}%
   \begin{center}
   \colorbox[HTML]{\colboxcolor}{\usebox{\selvestebox}}
\end{center}}

\newcounter{application}[chapter]
\renewcommand{\theapplication}{\thechapter.\arabic{application}}
\newsavebox{\selvesteboxBis}
\newenvironment{application}[1]
  {\newcommand\colboxcolor{E5E5E5}%
   \begin{lrbox}{\selvesteboxBis}%
   \begin{minipage}{\dimexpr\columnwidth-2\fboxsep\relax}
   \refstepcounter{application}\par\medskip
   \textbf{Application~\theapplication: #1} \rmfamily \\
    }
  {\end{minipage}
  \end{lrbox}%
   \begin{center}
   \colorbox[HTML]{\colboxcolor}{\usebox{\selvesteboxBis}}
\end{center}}

\newenvironment{applicationSuite}
  {\newcommand\colboxcolor{E5E5E5}%
   \begin{lrbox}{\selvestebox}%
   \begin{minipage}{\dimexpr\columnwidth-2\fboxsep\relax}
    }
  {\end{minipage}
  \end{lrbox}%
   \begin{center}
   \colorbox[HTML]{\colboxcolor}{\usebox{\selvestebox}}
\end{center}}

\newcommand{\myempty}{\newpage
\thispagestyle{empty}
$\mbox{}$ }

\fancypagestyle{plain}{ %
  \fancyhf{} 
  \fancyfoot[CO]{ \thepage }
  \fancyfoot[CE]{ \thepage }
  
}

\let\origdoublepage\cleardoublepage
\newcommand{\clearemptydoublepage}{%
  \clearpage
  {\pagestyle{empty}\origdoublepage}%
}

\let\cleardoublepage\clearemptydoublepage

\begin{document}
 
 \begin{titlepage}
\begin{center}
 

\begin{minipage}{0.49\textwidth}
\includegraphics[height=2cm]{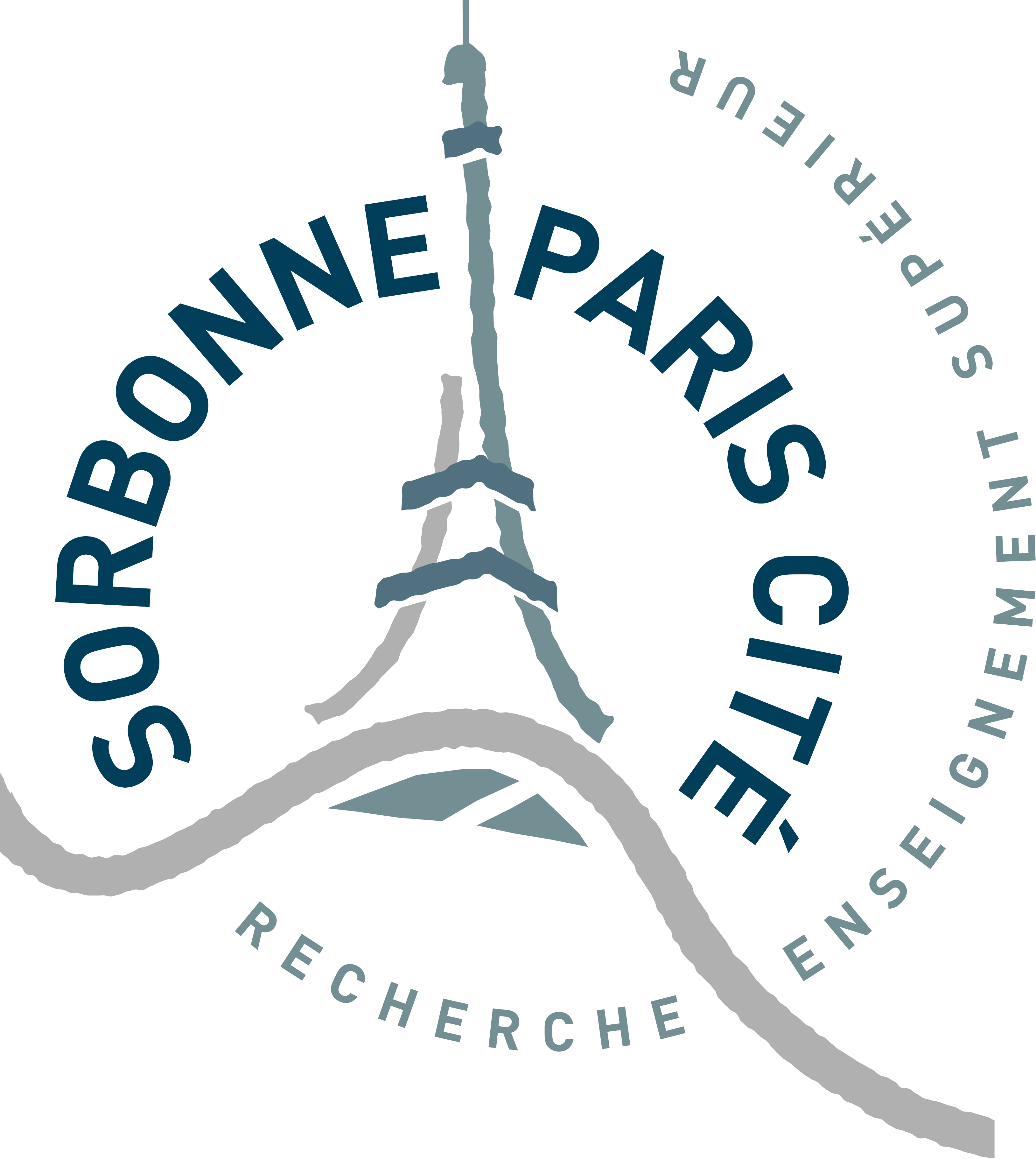} \hspace{0.5cm}
\includegraphics[height=1.5cm]{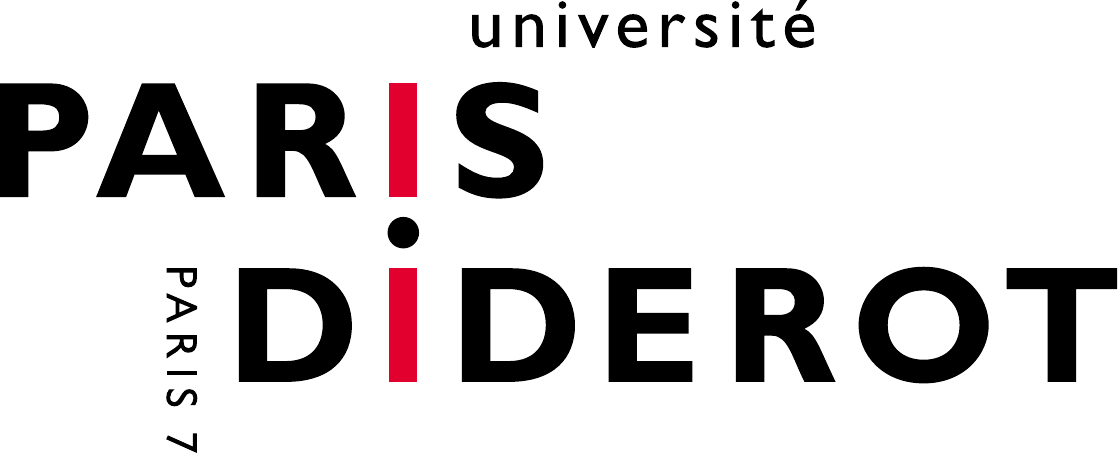} \\
 \Large Universit\'e Sorbonne Paris Cit\'e \\
 Universit\'e Paris Diderot
\end{minipage}
\begin{minipage}{0.49\textwidth}
\hspace{1.5cm}
\includegraphics[height=1.5cm]{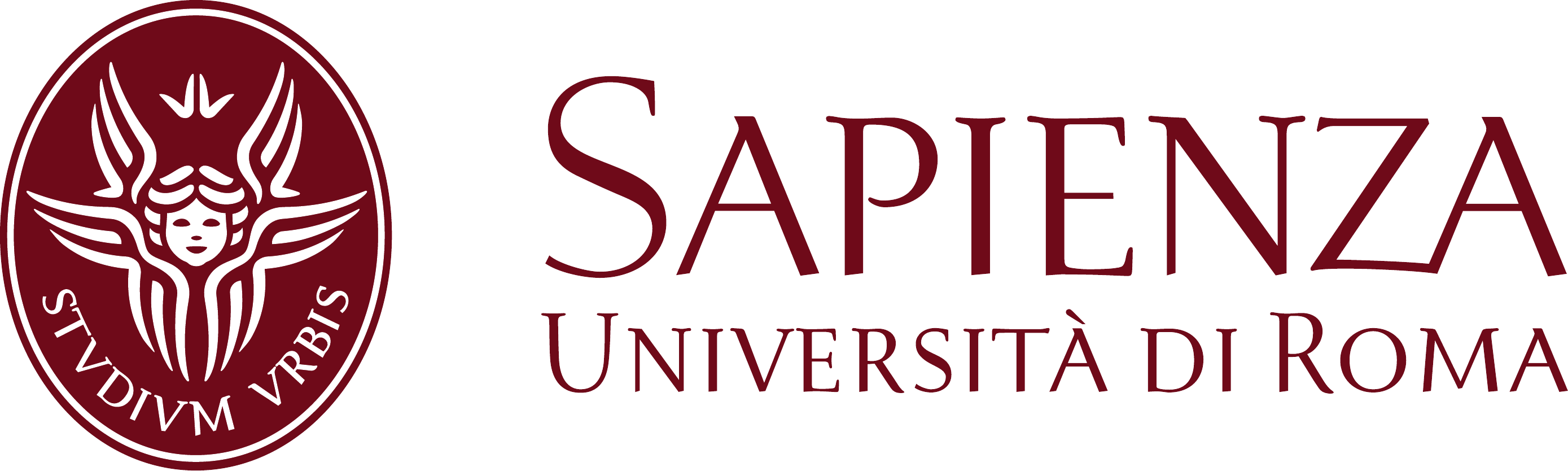}\\

\hspace{0.9cm} \Large Sapienza Universit\`a di Roma
\end{minipage}


\vspace{1.5cm}
\textsc{\Large PhD Thesis }\\[0.2cm]
in \textsc{\Large physics} \\[0.2cm]
\large presented by \\[0.2cm]
\Large \textsc{Christophe Sch\"ulke}\\[0.6cm]


\rule{\linewidth}{.1pt} \\[0.4cm]

\huge {\bfseries Statistical physics of linear and bilinear inference problems.}  \\[0.1cm]

\rule{\linewidth}{.1pt}  \\[1.5cm]


\begin{flushleft}
\large Defended on June 14, 2016, in front of the thesis committee: \\
\end{flushleft}
 
\begin{minipage}{0.8\textwidth}
 \begin{flushleft} \large
 \begin{tabular}{l l}
 Alfredo \textsc{Braunstein}, &  \textit{referee} \\
 Laurent \textsc{Daudet}, & \textit{president} \\
 Cyril \textsc{Furtlehner}, & \textit{member}   \\
 R\'emi \textsc{Gribonval},   &\textit{referee} \\
 Florent \textsc{Krzakala},  &\textit{thesis advisor (Paris), invited member} \\
 Federico \textsc{Ricci-Tersenghi},  &\textit{thesis advisor (Rome)} \\
 Lenka \textsc{Zdeborov\'{a}},  &\textit{thesis advisor (Paris)} \\
\end{tabular}

 \end{flushleft}
 \end{minipage}

\vfill

\includegraphics[width=0.5\textwidth]{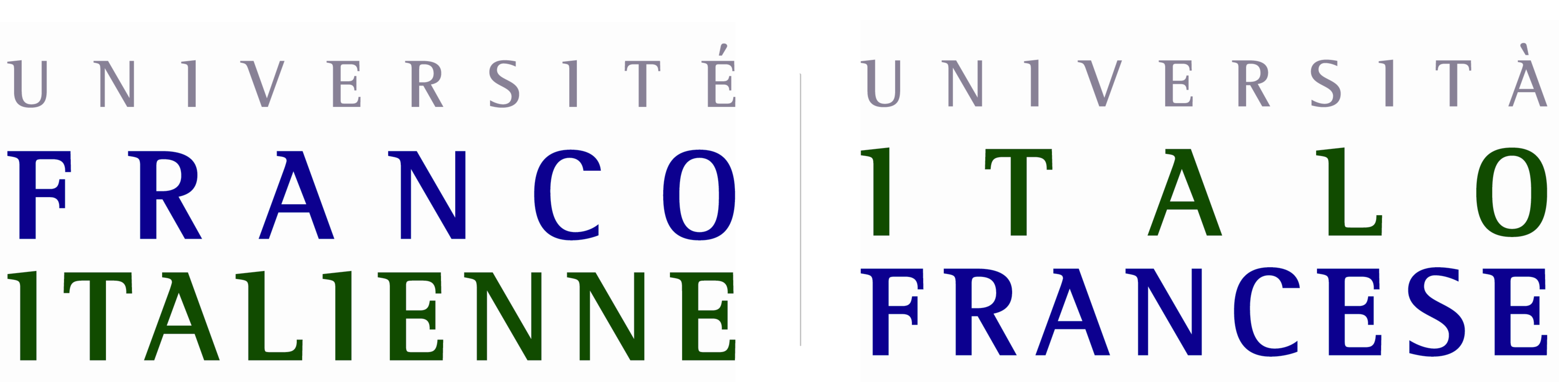}


\end{center}
\end{titlepage}
 
 \myempty

\thispagestyle{plain}
 \begin{minipage}{0.5\textwidth}
 $\mbox{}$
 \hfill
 \vfill
 \end{minipage}
\begin{minipage}{0.5\textwidth}
\vspace*{5cm}
\textit{Da steh ich nun, ich armer Tor! \\
Und bin so klug als wie zuvor.} \\

\vspace{0.3cm}
Goethe, Faust I
\vspace*{\fill}
\end{minipage}

  \myempty

\chapter*{Remerciements}
\addcontentsline{toc}{chapter}{\protect\numberline{}Remerciements}

Je tiens \`a remercier tous ceux qui m'ont soutenu sur le chemin qui a men\'e jusqu'\`a cette th\`ese 
et ceux qui m'ont aid\'e et accompagn\'e pendant ma th\`ese. Dans l'ordre d'apparition:
\begin{itemize}
 \item Mes parents, qui ont \'eveill\'e et cultiv\'e mon int\'er\^et pour la science en cherchant sans rel\^ache des r\'eponses \`a mes innombrables questions. 
  Ma m\`ere m'a \'egalement enseign\'e par l'exemple l'efficacit\'e et la ponctualit\'e; mon p\`ere, la patience et l'amour du d\'etail.
  Les deux pour m'avoir soutenu tout au long de mes \'etudes.
 \item Les professeurs qui m'ont fait aimer la science et la physique, en particulier Messieurs Larrue, M\'ezard, Montambaux et Sigrist.
 \item L'universit\'e franco-italienne pour la bourse doctorale de 3 ans, et Florent pour l'extension de 3 mois suppl\'ementaires.
 \item Mes directeurs de th\`ese Federico, Florent et Lenka pour m'avoir accept\'e comme th\'esard, 
 pour leurs conseils, leur accompagnement, leur motivation, pour m'avoir permis de participer \`a des conf\'erences et de nombreuses \'ecoles d'\'et\'e ou d'automne.
 \item Un merci tout particulier \`a Lenka pour son enthousiasme, son encadrement, ses conseils, pour m'avoir laiss\'e avancer \`a mon rythme, pour m'avoir laiss\'e me d\'ebattre quand 
 j'en avais envie et m'avoir aid\'e quand j'en avais besoin.
 \item Mes coll\`egues et collaborateurs en France\,: Jean, Thibault, Francesco, Alaa et l'\'equipe Sphinx.
  Mes coll\`egues \`a Rome: Bea, Chiara, Giulia, Yuliang, Jacopo et les \textit{brutti} Aur\'elien, Carlo et Paolo pour leur accueil chaleureux gr\^ace auquel je me suis
  rapidement senti chez moi en Italie.
 \item Arnaud, Jacopo, Marylou, Christian et Lenka pour leurs relectures et leurs corrections.
 \item Les membres du jury pour avoir accept\'e mon invitation et pour leurs commentaires et suggestions. 
\end{itemize}

 \chapter*{Abstract}
\addcontentsline{toc}{chapter}{\protect\numberline{}Abstract}

The recent development of compressed sensing has led to spectacular advances in the understanding of sparse linear estimation problems
as well as in algorithms to solve them.
It has also triggered a new wave of developments in the related fields of generalized linear and bilinear inference problems,
that have very diverse applications in signal processing and are furthermore a building block of deep neural networks.
These problems have in common that they combine a linear \textit{mixing} step and a nonlinear, probabilistic \textit{sensing} step, 
producing indirect measurements of a signal of interest.
Such a setting arises in problems as different as medical or astronomical imaging, clustering, matrix completion or blind source separation.

The aim of this thesis is to propose efficient algorithms for this class of problems and to perform their theoretical analysis.
To this end, it uses belief propagation, thanks to which high-dimensional distributions can be sampled efficiently, thus making
a Bayesian approach to inference tractable.
The resulting algorithms undergo phase transitions just as physical systems do. These phase transitions can be analyzed using 
the replica method, initially developed in statistical physics of disordered systems.
The analysis reveals phases in which inference is easy, hard or impossible.
These phases correspond to different energy landscapes of the problem.

The main contributions of this thesis can be divided into three categories.
First, the application of known algorithms to concrete problems: community detection, superposition codes and an innovative imaging system.
Second, a new, efficient message-passing algorithm for a class of problems called blind sensor calibration.
It could be used in signal processing for a large class of measurement systems that use arrays of physical sensors.
Third, a theoretical analysis of achievable performances in matrix compressed sensing and of instabilities in Bayesian bilinear inference algorithms.

\paragraph{Keywords:} Signal processing, compressed sensing, sparse estimation, community detection, generalized linear models,
generalized bilinear models, matrix factorization, low-rank matrix compressed sensing, phase retrieval, phase transitions,
Bayesian inference, belief propagation, message-passing algorithms, state evolution analysis, replica method.

\makeatletter
\@openrightfalse
\makeatother
\chapter*{R\'esum\'e}

Le d\'eveloppement r\'ecent de l'acquisition comprim\'ee a permis de spectaculaires avanc\'ees dans la compr\'ehension des probl\`emes d'estimation lin\'eaire parcimonieuse
ainsi que de leurs algorithmes de r\'esolution.
Ce d\'eveloppement a \'egalement suscit\'e un int\'er\^et renouvel\'e pour les probl\`emes d'inf\'erence lin\'eaire et bilin\'eaire g\'en\'eralis\'ee.
Ceux-ci trouvent diverses applications en traitement du signal et constituent de plus le composant de base des r\'eseaux neuronaux profonds.
Ces probl\`emes ont en commun de combiner un \'etape lin\'eaire avec une \'etape non lin\'eaire et probabiliste, \`a l'issue de laquelle des mesures sont effectu\'ees.
Ce type de situations se pr\'esente dans des probl\`emes aussi vari\'es que l'imagerie m\'edicale, l'astronomie, le clustering ou la s\'eparation de sources audio.

Cette th\`ese s'int\'eresse \`a des algorithmes pour la r\'esolution de ces probl\`emes ainsi qu'\`a leur analyse th\'eorique.
Pour cela, nous utilisons des algorithmes de passage de message, qui permettent d'\'echantillonner efficacement des distributions 
de haute dimension et rendent ainsi possible une approche d'inf\'erence bay\'esienne.  
Ces algorithmes connaissent des changements de phase tout comme de nombreux syst\`emes physiques.
Les diff\'erentes phases se laissent analyser \`a l'aide de la m\'ethode des r\'epliques, initialement d\'evelopp\'ee dans le cadre de la physique statistique des milieux d\'esordonn\'es.
L'analyse r\'ev\`ele qu'elles correspondent \`a des domaines dans l'espace des param\`etres dans lesquels l'inf\'erence est facile, difficile ou impossible, 
selon le paysage \'energ\'etique du probl\`eme.

Les principales contributions de cette th\`ese peuvent \^etre regroup\'ees en trois cat\'egories.
D'abord, l'application d'algorithmes connus \`a des probl\`emes concrets\,: d\'etection de communaut\'es, codes correcteurs d'erreurs ainsi qu'un syst\`eme d'imagerie innovant. 
Ensuite, un nouvel algorithme traitant une classe de probl\`emes d'inf\'erence appel\'ee calibration aveugle de capteurs, 
potentiellement applicable \`a de nombreux syst\`emes de mesure utilisant des r\'eseaux de capteurs physiques.
Enfin, une analyse th\'eorique des performances qui peuvent \^etre atteintes en inf\'erence bay\'esienne pour le probl\`eme de reconstruction
de matrices \`a petit rang \`a partir de projections lin\'eaires, ainsi qu'une analyse d'une instabilit\'e pr\'esente dans les algorithmes d'inf\'erence bilin\'eaire.

\paragraph{Mots-cl\'es:} Traitement du signal, acquisition comprim\'ee, estimation parcimonieuse,
d\'etection de communaut\'es, mod\`eles lin\'eaires g\'en\'eralis\'es, mod\`eles bilin\'eaires g\'en\'eralis\'es, 
d\'ecomposition de matrices, reconstruction de matrices de petit rang, reconstruction de phase, transitions de phase,
inf\'erence bay\'esienne, \textit{belief propagation}, algorithmes de passage de messages, \textit{state evolution}, m\'ethode des r\'epliques.

\newpage
\makeatletter
\@openrighttrue
\makeatother
\chapter*{Sommario}

Lo sviluppo degli algoritmi di \textit{compressed sensing} ha permesso grandi progressi nella comprensione dei problemi sparsi di teoria della stima
e degli algoritmi necessari per risolverli. Inoltre, ha contribuito a far crescere l'interesse per i problemi di inferenza lineare e bilineare generalizzati,
che hanno molte applicazioni nella teoria dei segnali e sono fondamentali per la descrizione dei modelli di reti neurali profonde.  
Questi problemi di inferenza hanno in comune il fatto che combinano un passo lineare ed uno non lineare, probabilistico, al termine del quale sono 
effettuate le misure del segnale di interesse. Inoltre, sono utili a risolvere problemi che emergono in contesti diversi come quello della 
ricostruzione di immagini in medicina e in astronomia, la determinazione di componenti diversi di un sistema (problema del \textit{clustering}) e 
quello del filtraggio di un segnale da un rumore in assenza di molte informazioni (\textit{blind sensor calibration}).

L'obiettivo di questa tesi \`e di proporre soluzioni efficienti a questo tipo di problemi e di farne un'analisi teorica. Dunque si fa uso della
\textit{belief propagation}, grazie a cui il \textit{sampling} di distribuzioni in alte dimensioni può essere fatto in modo efficiente. Questo rende trattabile un
approccio Bayesiano al problema. Gli algoritmi sviluppati hanno transizioni di fase proprio come i sistemi fisici e queste transizione di fase
possono essere studiate utilizzando il metodo delle repliche, sviluppato inizialmente nella meccanica statistica dei sistemi disordinati. 
Le analisi rivelano che ci sono fasi in cui l'inferenza \`e facile ed altre in cui \`e difficile o impossibile, a seconda del paesaggio di energia del problema.

I contributi principali di questa tesi si possono dividere in tre parti. Nella prima, l'applicazione di algoritmi noti a problemi concreti:
la ricostruzione di comunit\`a in un grafo (\textit{community detection}), lo studio degli \textit{sparse superposition codes} e un nuovo sistema di imaging. 
Nella seconda, un nuovo algoritmo per i problemi di inferenza della classe \textit{blind sensor calibration}, che potrebbe essere utile
a migliorare le stime dei sensori fisici. Nella terza, un'analisi teorica delle prestazioni raggiungibili nell'inferenza Bayesiana nel
problema di compressed sensing applicato alle matrici di basso rango, e di un'instabilit\`a presente negli algoritmi di inferenza bilineare.

\paragraph{Parole chiavi:} Elaborazione del segnale, \textit{compressed sensing}, segnali sparsi,
\textit{community detection}, \textit{generalized linear models},
\textit{generalized bilinear models}, \textit{matrix factorization}, \textit{low-rank matrix compressed sensing}, \textit{phase retrieval}, transizioni di fase,
\textit{Bayesian inference}, \textit{belief propagation}, \textit{message-passing algorithms}, \textit{state evolution analysis}, metodo delle repliche.

 
 \tableofcontents
 \listoffigures
 

 \chapter*{Introduction}
\markboth{Introduction}{Introduction}
\addcontentsline{toc}{chapter}{\protect\numberline{}Introduction}
 \pagenumbering{arabic}

\section*{Organization of the thesis}
\addcontentsline{toc}{section}{\protect\numberline{}Organization of the thesis}
The thesis is subdivided in 3 parts. 
The first one introduces key concepts in inference and statistical physics, and shows how the 
latter can be used to solve problems of the former.
The second part introduces a broad class of problems as well as related algorithms and analysis techniques.
The last part contains my main contributions to this class of problems.

\subsection*{Part I: Statistical physics of inference problems}
\subsubsection{Chapter \ref{chap:1}: Inference and statistical physics}
The first chapter separately introduces key concepts of inference and of statistical physics.
The goals of inference and the challenges commonly encountered are described, along with two general solving strategies.
The simple examples of denoising and linear estimation are given in order to illustrate the concepts, at the same time 
introducing two fundamental tools for all the problems encountered subsequently.
The fundamental tools of statistical physics \ie the partition function and the related free entropy and free energy are
introduced. Phase transitions are illustrated by the examples of the Ising model, the SK-model and their respective phase diagrams.
Strength and limitations of the statistical physics approach to inference problems are discussed, followed by an overview of 
the inference problems treated in the thesis.

\subsubsection{Chapter \ref{chap:communityDetection}: Community detection} 
The second chapter shows how insight gained from statistical physics can help solving an inference problem.
To this end, the problem of community detection is described along with its challenges. Belief propagation is 
introduced as an algorithm for estimating high-dimensional probability distributions. 
A study of a such an algorithm, published in~\cite{moiModularite} is made, showing the existence of algorithmic transitions between 
phases similar to the ones found in physical systems.

\subsection*{Part II: Linear and bilinear inference problems}
\subsubsection{Chapter \ref{chap:generalizedLinearModels}: Compressed sensing and generalizations} 
This chapter presents generalized linear models, focusing on compressed sensing.
The replica method---coming from physics of disordered systems---is used to perform a theoretical analysis of inference 
of generalized linear models. A Bayesian algorithm using belief propagation is introduced (\gamp).
The performances reached by \gamp are compared to the theoretical predictions previously obtained.
Limitations of \gamp are mentioned as well as possible remedies.

\subsubsection{Chapter \ref{chap:generalizedBilinearModels}: Generalized bilinear models} 
Chapter 4 presents generalized bilinear models, closely related to generalized linear models but more difficult to solve in practice.
Results of the theoretical analysis and a Bayesian message-passing algorithm for generalized matrix factorization are briefly presented. 
Generalized matrix compressed sensing is introduced.

\subsection*{Part III: Main contribution}
Part III contains my main contributions to inference of linear and bilinear models.
\subsubsection{Chapter \ref{chap:gampApplications}: Vectorial GAMP and applications }
\subsubsection{Chapter \ref{chap:blindSensorCal}: Blind sensor calibration }
\subsubsection{Chapter\ref{chap:matrixCS}: Analysis of matrix compressed sensing} 
Details about these chapters are given in the following section.

\section*{Main contributions}
\addcontentsline{toc}{section}{\protect\numberline{}Main contributions}
The main contributions of my thesis are published (or in preparation) in the following papers:
\begin{itemize}
 \item ``Blind calibration in compressed sensing using message passing algorithms''~\cite{moiNips},
 \item ``Reference-less measurements of the transmission matrix of a highly scattering material using a DMD and phase retrieval techniques''~\cite{moiOptique},
 \item ``Approximate message-passing with spatially coupled structured operators, with application to compressed sensing and sparse superposition codes''~\cite{moiCodes},
 \item ``Blind sensor calibration using approximate message passing''~\cite{moiCalibration},
 \item ``Multiple phases in modularity-based community detection''~\cite{moiModularite},
 \item ``Phase diagram of matrix compressed sensing''~\cite{moiMCS}.
\end{itemize}
They treat different inference problems using methods of statistical physics, relying on belief propagation and on the replica method.

\subsection*{Community detection}
Ref~\cite{moiModularite} analyzes a recently published community detection algorithm called \modbp, based on belief propagation
and on a physical intuition of the origin of the computational hardness in community detection.
In~\cite{moiModularite}, I reveal the existence of more algorithmic phases than previously known in community detection.
I introduce a new set of order parameters which allows to define an effective number of communities.
A study on synthetic and real-world networks is made, and
a simple multiresolution strategy for hierarchical networks is described and tested on a real-world network.

\subsection*{Applications of \gamp}
The \gamp algorithm can be applied to all problems of the class of generalized linear models and can therefore be used in many specific applications.
In~\chapref{chap:gampApplications}, I give a derivation of complex-valued \gamp (for which no derivation was published until recently) 
and present two applications of \gamp, treated in~\cite{moiCodes,moiOptique}.

In~\cite{moiCodes}, we perform a theoretical analysis of complex compressed sensing. 
Furthermore, we use structured operators (Fourier and Hadamard operators) as measurement matrices in compressed sensing,
which allows a drastic speed-up and allows to treat problems of bigger sizes. 
The second part of the paper focuses on superposition codes, that are capacity-achieving in a certain configuration using a message-passing decoder closely related to \gamp. 

In~\cite{moiOptique}, \gamp is used in an optics experiment. The goal of the experiment is to determine 
the transmission matrix of a highly scattering material, thus allowing imaging or focusing through the medium.
The use of phase retrieval---for which complex \gamp can be used---greatly simplifies the necessary experimental setup,
opening the way to further developments.

\subsection*{Blind sensor calibration}
In~\chapref{chap:blindSensorCal}, I present the work published in~\cite{moiNips,moiCalibration}.
Blind sensor calibration can be seen as a generalization of compressed sensing. 
While special cases of blind calibration (\eg blind \textit{gain} calibration) have been studied before using different types of algorithms,
I propose a Bayesian message-passing algorithm called \calamp that can handle much more general situations.
Additionally to real and complex gain calibration, two such situations are examined and \calamp tested on them.

\subsection*{Bilinear inference}
In~\chapref{chap:matrixCS}, I present two contributions to bilinear inference problems.
First, I provide an analysis that explains the convergence difficulties encountered by some algorithms in bilinear inference problems.
Secondly, I provide a theoretical analysis of low-rank matrix compressed sensing (in preparation in~\cite{moiMCS}).
I show that the theoretical analysis gives the same results as the one of the problem of matrix factorization.
I perform an analysis of a special case, which I compare with the results obtained by a recently published algorithm, \pbig.
Beside an excellent global agreement, interesting finite-size effects are observed that allow successful inference in a  hard phase.

\section*{Minor contributions}
Besides these main contributions that have been published, a few minor but possibly useful 
contributions are present in this thesis:
\begin{itemize}
 \item The use of a coherent set of notations for the estimators and variances updated in algorithms, 
 using hats, bars, upper and lower-case letters. This allows to keep the number of variables to a minimum 
 and to easily recognize the signification of each quantity.
 \item The use of the $f_k$ functions defined in~\appref{app:f}, that allow simplified expressions for the 
 state evolution equations, exclusively using functions used in the algorithms. Besides, the relation~(\ref{eq:fDerivateDifficult}) 
 allows to obtain the general state equations from the replica free entropy.
 \item The full derivation of the \gamp  state evolution equations starting from the replica analysis (which is nothing but a special case of the analysis made in~\cite{kabaMF}).
 \item The phase diagrams of noisy $1$-bit \cs and $1$-bit \cs of binary variables~(\figref{fig:3_oneBit_continuous}, \figref{fig:3_binaryOneBit}).
 \item Preliminary results on the state evolution for blind sensor calibration in~\appref{app:calampStateEvolution} and for phase retrieval in~\appref{app:PR_DE}.
\end{itemize}

 \part{Statistical physics of inference problems}
 \label{part:1}
 
 \chapter{Inference and statistical physics}
\label{chap:1}
In recent years, questions in fields such as signal processing, machine learning and information theory have increasingly drawn the attention of statistical physicists.
The differences in background, goals and spirit that separate mathematicians and physicists 
have turned out to be very fruitful and have lead to a new understanding of a number of problems. 
Therefore, the field is developing and several books already present information theory and statistical physics in a joint manner~\cite{mezardMontanari,nishimoriBook}, such that the way
between the two has become smoother.

The goal of this chapter is to introduce the concepts and notations of \inference (\secref{sec:inference}) and \statphys (\secref{sec:statphys}).
This concise introduction only aims at presenting what will be directly used in \mythesis, good reference books are~\cite{wassermanBook,mezardMontanari,nishimoriBook}.   
In \secref{sec:statphysinf}, I will explain how the \statphys approach can contribute to the understanding of inference problems, as well as its limitations.
Finally, in \secref{sec:myproblems}, I will briefly introduce the different inference problems on which I have worked and that will be treated in the rest of the thesis.

\section{Inference}
\label{sec:inference}
\subsection{General setting}
\begin{figure}
 \centering \meqnote{General inference setting}
 \includegraphics{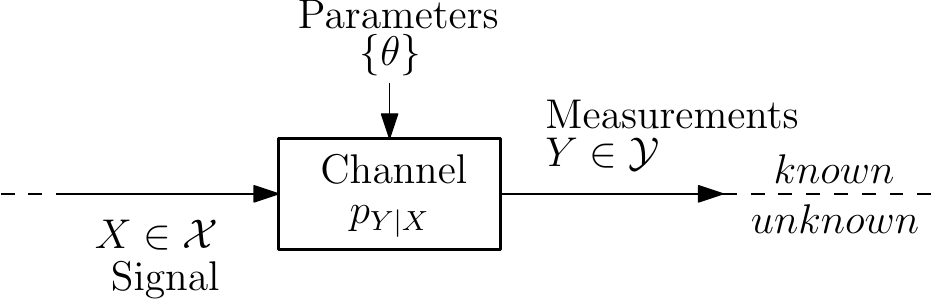}
 \caption[General setting of an inference problem]{General setting of an inference problem.
 An unknown signal $\X$ is observed through a channel $\pyx$ depending on a set of parameters \params,
 producing the measurements $\Y$. We will represent different inference problems following this scheme: 
 It shows the generative process and indicates which variables are known and unknown by the user.  }
 \label{fig:1_inference}
\end{figure}

Often, a signal of interest cannot be observed directly, but only through a channel that provides indirect measurements.
This channel is characterized by a probability distribution function (\pdf) that describes the statistical relation between signal and measurement.
In the most general setting, let us call
\begin{center}
\begin{tabular}{l l}
 $\X \in \XX$ & the signal, \\
 $\Y \in \YY$ & the measurements, and \\
 $\pyx: \YY \times \XX \to \RR^+$ & the channel.
\end{tabular}
\end{center}
The ensembles $\XX$ and $\YY$ can be discrete or continuous and of various dimensions, such as $\Nintegers$, $\RR$, $\CC$, $\RR^{10 \times 20}$, etc\dots
The channel often depends on a set of parameters \params, although we do not explicitly indicate this dependence for notational lightness.
This general inference setting, illustrated in~\figref{fig:1_inference}, is ubiquitous in a large number of fields, ranging from scientific experiments to telecommunications or internet advertising.

\begin{figure}[h]
 \centering
 \includegraphics{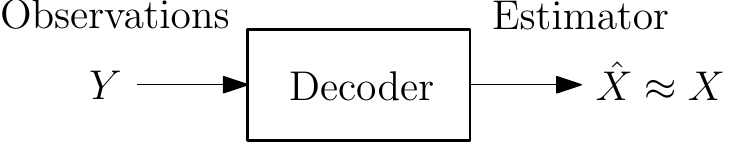}
 \caption[A decoder]{A decoder, producing an estimate $\Xh$ of $\X$ from the measurements $\Y$ produced in~\figref{fig:1_inference}.
 In order to give a good estimate, the decoder has to be designed from the knowledge we have about the channel $\pyx$.}
 \label{fig:1_decoder}
\end{figure}

The goal of inference is to obtain the best possible estimate $\Xh$ of $\X$ from the measurements $\Y$.
In order to reach this goal, three essential questions have to be answered:
\begin{enumerate}
 \item \textit{What is a ``good'' estimate of} $\X$?
 Obviously, the \textit{ideal} estimate is $\Xh=\X$. 
 But as nothing assures us that this estimate is possible to obtain, one needs to define a measure of success (or metric) that quantifies how good an estimate $\Xh$ is.
  \item \textit{What is the best possible performance achievable in this setting?}
 This depends both on the distribution of $\X$ and on the channel, and it is the most interesting question from an information-theoretical point of view.
 \item \textit{How do we produce a good estimate} $\Xh$? 
 In order to concretely obtain an estimate $\Xh$, one needs to design a function that returns an estimate $\Xh$ for every possible measurement $\Y$.
 This function, illustrated in \figref{fig:1_decoder}, is usually called the \textit{decoder} or the solver and has to be carefully designed. 
\end{enumerate}

Answering to the first question is part of a satisfying description of the problem and corresponds to choosing a distance (or metric) over the ensemble $\XX$.
For instance, for $\XX = \RR^N$, the  distance usually considered is the \textit{mean squared error} (MSE) \mynote{MSE}{1.1}
\begin{align}
 \mse(\Xh,X) = \frac{1}{N} \sum_i (\Xh_i - \X_i)^2 = \frac{|| \Xh - \X ||_2^2}{N}  	\label{eq:MSE}
\end{align}
while if $\XX$ is a discrete ensemble, a good metric is the \textit{overlap}, \ie the fraction of correctly guessed signal components  \mynote{Overlap}{0.8}
\begin{align}
 \ov(\Xh,\X) = \frac{1}{N} \sum_i \delta_{\Xh_i, \X_i}. 	\label{eq:overlap}
\end{align}

Answering to the second question is making the theoretical analysis of the problem, while answering to the third one is actually solving it.

\subsection{Achievable performances}
In order to make an information-theoretical analysis of a problem such as presented in~\figref{fig:1_inference}, two approaches coexist~\cite{mooreBook}.

The first one is the \textit{worst case} analysis. 
In it, we assume that it is not equally easy to obtain a good estimator $\Xh$ for all possible signals.
The worst case scenario focuses on the signals for which the achievable performance is the worst. 
Results obtained with this approach are strong in the sense that they give a strict lower bound, but are usually overly pessimistic and do not reflect the \textit{usually} achievable performances.

The second approach is the \textit{typical case} analysis, in which we focus on the performances \textit{usually} achievable.
In order to characterize this ``\textit{usually}'', we need to focus on a specific class of signals.
We therefore consider $\X$ to be a realization of a random variable $\mathsf{X}$, distributed according to a \pdf $\px$. 
The typical case analysis is therefore made in a statistical framework. Furthermore, $\px$ can be used to design a decoder.

\subsubsection{Elements of statistics}
\label{sec:statistics}
Let us briefly recall a few elements and notations of statistics. We write
\begin{align}
 \mathsf{X} \sim \px	\label{eq:def_sim}
\end{align}
to signify that a random variable $\mathsf{X}$ is distributed according to $\px$.
This means that the probability of the random variable $\mathsf{X}$ to take the value $\X$ is $\px(\X)$. 
For notational lightness, we will abusively use the same notation $\X$ both for the random variable and its realization.
The differential entropy of $\mathsf{X}$ is then defined by \mynote{Differential entropy}{0.7}
\begin{align}
 H_{\X} &= - \int \dd \X \px(\X) \log \px(\X)	  \label{eq:def_entropy}
\end{align}
and is a quantity that measures the uncertainty of $\mathsf{X}$: a random variable with zero differential entropy has no uncertainty, meaning it can only take a single value.
Another useful quantity is the Kullback-Leibler (KL) divergence between two probability distribution functions $\p$ and $q$, defined by \mynote{KL divergence}{0.7}
\begin{align}
 \Dkl(q||\p) &= \int \dd \X q(\X) \log \frac{q(\X)}{\p(\X)}	\label{eq:def_KL}
\end{align}
and that, although not symmetric, is a kind of measure of distance between pdfs.

Furthermore, we say that $\mathsf{Y}$ and $\mathsf{X}$ are independent random variables, if and only if their joint pdf $\p_{\X,\Y}$ 
can be written as a product in the following way: \mynote{Independence}{0.7}
\begin{align}
 \p_{\X,\Y}(\X,\Y) = \px(\X) \p_{\Y}(\Y).	\label{eq:def_independence}
\end{align}
If this is not the case, we say that $\mathsf{Y}$ and $\mathsf{X}$ are correlated, and 
the conditional probability of $\Y$ knowing $\X$ is given by \mynote{Conditional probability}{0.7}
\begin{align}
 \pyx(\Y|\X) &= \frac{\p_{\X,\Y}(\X,\Y)}{\px(\X)}.	\label{eq:def_bayes}
\end{align}
We can then define the mutual information \mynote{Mutual information}{0.75}
\begin{align}
 \II_{\X,\Y} &= \int \dd \X \dd \Y \p_{\X,\Y}(\X,\Y) \log \frac{\p_{\X,\Y}(\X,\Y)}{\px(\X) \p_{\Y}(\Y)}		\label{eq:def_information}
\end{align}
that measures the degree to which $\mathsf{Y}$ and $\mathsf{X}$ are correlated.


\subsubsection{Analysis of given decoders}
These quantities and formulas defined above are instrumental for analysing the achievable performances of an inference problem from an information-theoretical point of view.
Another interesting task is the analysis of the performances of a given decoding scheme.

\subsection{Decoding strategies}
Interesting inference problems are characterized by the fact that obtaining a good estimate $\Xh$ is not trivial, \ie there is no analytic formula allowing to obtain $\Xh$ from $\Y$.
In that case, a decoding strategy has to be designed and implemented in an algorithm. 
In the following, we describe two approaches to designing a decoder.
\subsubsection{Minimization approach}
\begin{figure}[h]
\centering \mynote{Convex minimization}{1.3}
 \includegraphics[width=0.45\textwidth]{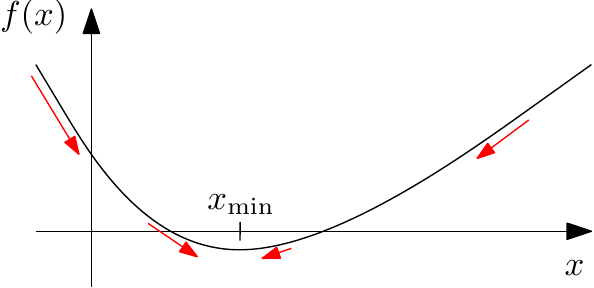}
 \caption[Convex minimization]{Convex minimization problems are well posed because convex functions have a unique minimum $x_{\rm min}$. 
 Furthermore, as the magnitude of the function's gradient (red arrows) increases with increasing distance to the minimum, gradient-descent type algorithms can converge quickly.
 In higher dimensions, the picture stays the same and gradient descent algorithms can efficiently find the minimum. }
 \label{fig:1_convex}
\end{figure}

In the minimization approach, the problem of decoding is reformulated as a minimization problem.
A cost function $C(\X,\Y)$ is chosen and the estimator $\Xh$ is taken as \mynote{Minimization problem}{0.6}
\begin{align}
 \Xh^{\rm MIN} &= \argmin_x C(x,\Y).	\label{eq:def_minimization}
\end{align}
The advantage of this approach is that minimization problems are very well studied and that fast and reliable methods exist if the function to minimize has the right properties.
In particular, as illustrated in~\figref{fig:1_convex}, minimization of convex functions is both well posed and efficiently solvable, and is furthermore very well documented~\cite{convexOpt}.
On the other hand, if the cost function is not convex, minimization can be a very difficult task~(\figref{fig:1_minima}).
In that case, solving~\eqref{eq:def_minimization} is not straightforward and a general strategy is to approach $C$ by a convex function $C_c$ \mnote{Convex relaxation} and solve the new minimization problem. 
However, the two problems are in general not equivalent, and minimizing the  \textit{convex relaxation} $C_c$ of $C$ leads to suboptimal results compared to minimizing $C$.

\subsubsection{Bayesian approach}
A second approach is probabilistic. Using~\eqref{eq:def_bayes} it is easy to show Bayes' theorem \mynote{Bayes' theorem}{0.7}
\begin{align}
 \p_{\X|\Y}(\X|\Y) &= \frac{\px(\X) \pyx(\Y|\X)}{\p_{\Y}(\Y)}, 	\label{eq:bayes}
\end{align}
and thus estimate the probability that the measurements $\Y$ were generated by a signal $\X$.
If $\p_{\X|\Y}$ is known, a sensible inference strategy is to use the \textit{maximum a posteriori} (MAP) estimator \mynote{MAP estimator}{0.45}
\begin{align}
 \Xh^{\rm MAP} &= \argmax_x \p_{\X|\Y}(x|\Y).		\label{eq:MAP}
\end{align}
Though intuitive, this estimator is not always the best, and other estimators can be constructed using the posterior distribution.
One of the most commonly used is the \textit{minimum mean square error} (MMSE) estimator \mynote{MMSE estimator}{0.65}
\begin{align}
 \Xh^{\rm MMSE} &= \int \dd \X \, \X \, \pxy(\X|\Y),	\label{eq:MMSE}
\end{align}
that minimizes the expected MSE between the signal and its estimator.

When \mnote{Bayes optimality} the functions $\px$ (called the \textit{prior}) and $\pyx$ (called the \textit{likelihood}) in~\eqref{eq:bayes} are known exactly, the probabilistic approach is said to be \textit{Bayes optimal}. 
It is also starting from~\eqref{eq:bayes} that the information-theoretical analysis of the problem is performed. 
An important advantage of the probabilistic approach to inference is therefore that a Bayes optimal decoder should be able to reach the performances predicted by the information-theoretic analysis of the problem.
If on the other hand the functions used in~\eqref{eq:bayes} are \textit{not} the right ones, for instance if wrong 
parameters \params \,are used, we say that there is a mismatch and the setting is not Bayes optimal.

The main drawback of the probabilistic approach is that in general, estimating $\pxy$ is hard, because the denominator in~\eqref{eq:bayes} has to be calculated by marginalization over $\X$, \mynote{Marginalization}{0.7}
\begin{align}
 \p_{\Y}(\Y) &= \int \dd \X \px(\X) \pyx(\Y|\X).	\label{eq:normalization}
\end{align}
In most cases, no analytical formula of this integral is known, and it thus has to be evaluated numerically, which is hard when $\X$ belongs to a high-dimensional space, such as $\RR^{N}$ with large integer $N$ for instance.
The same problem occurs for~\eqref{eq:MMSE}.

\paragraph{The two approaches} are not contradictory.
As a matter of fact, it is simple to see that the MAP problem~\eqref{eq:MAP} can be written as minimization problem using the cost function $-\pxy$.

\subsubsection{Challenges}
\label{subsec:challenges}
Two recurring difficulties in the implementation of both approaches are the \textit{curse of dimensionality} and the problem of \textit{finding a global minimum}.

The \textit{curse of dimensionality} is the fact that with increasing dimension of the signal $\X$, the number of points necessary for sampling a function defined on $\XX$ with a given precision increases exponentially.
To illustrate this, consider a function 
\begin{align}
 f:[0,1]^N \to \RR
\end{align}
with the regularity condition $\max || \nabla f ||_2^2 = C$.
We would like to be able to approximate the function's value at any point  by its value at the closest point of an $N$-dimensional grid of spacing $\Delta s$ 
(thus containing $M=(\frac{1}{\Delta s})^N$ sampling points).
The closest point $x_c$ is then at maximal distance of $\sqrt{ \frac{N}{4} \Delta s^2}$, such that the error of the estimate is bounded by
\begin{align}
 ||f(x) - f(x_c) ||_2^2 \leq C \frac{N}{4} \Delta s^2,
\end{align}
which can be rewritten as a function of the number of sampling points	\mynote{Curse of dimensionality}{0.8}
\begin{align}
 ||f(x) - f(x_c) ||_2^2 \leq  \frac{C}{4} N M^{-2/N}
\end{align}
from which one sees that the number of sampling points necessary to estimate a function with a given precision \textit{increases exponentially} 
with the dimension $N$.
This makes it rapidly impossible to solve extremization problems such as~\eqref{eq:def_minimization} or~\eqref{eq:MAP}
by sampling the function, and makes the numerical estimation of $N$-dimensional integrals such as~\eqref{eq:MMSE} or~\eqref{eq:normalization} 
very difficult as well.
In short, none of the two previously described decoding strategies are easy in high dimension.

\begin{figure}[h]
 \centering \meqnote{Nonconvex minimization}
 \includegraphics{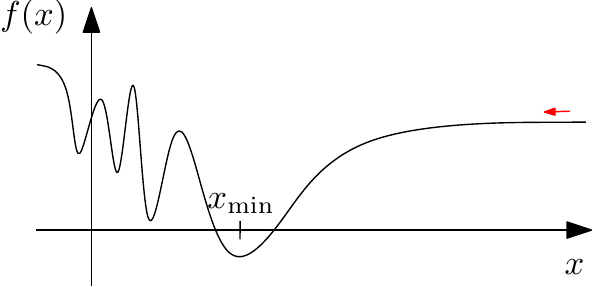}
 \caption[Non-convex minimization]{Finding the global minimum of a non convex function is difficult for two reasons. 
 First (left from $x_{\rm min}$), local minima can exist, in which algorithms can get stuck.
 Secondly (right from $\x_{\rm min}$), being far away from a minimum does not imply that the function's gradient (red arrow) has a big magnitude. 
 These two facts make gradient-descent algorithms useless for non-convex functions.
 }
 \label{fig:1_minima}
\end{figure}

The second difficulty is that of finding the global minimum of a function.
As mentioned before, in high dimension, performing this task by sampling would be very time consuming.
In~\figref{fig:1_convex} we have illustrated that finding the global minimum of a convex function is easy nonetheless.
For a non-convex function, however, there is no efficient method for finding the global minimum, as illustrated in~\figref{fig:1_minima}. 

\vfill

\pagebreak
\begin{example}{Denoising} 
\label{ex:awgn}
Let us start with the very simple example of a real  signal $\X\in \RR^N$ observed through a memoryless additive white Gaussian noise (AWGN) channel:
\begin{align}
 \pyx(\Y|\X)
             &= \prod_{i=1}^N \NN(\Y_i;\X_i,\Delta).
\end{align}
With no assumption on $\X$, a ``natural'' cost function to use in the minimization approach is the MSE \mynote{Least squares}{0.8}
\begin{align}
 C(\X,\Y) = N \times \mse(\X,\Y) = || \Y - \X ||_2^2.
\end{align}
In the probabilistic approach, making no assumption whatsoever on $\X$ is equivalent to 
taking the flat (and improper) ``prior'' $\px(\X) =1$, leading to 
\begin{align}
 \pxy(\X|\Y) = \prod_{i=1}^N \NN(\X_i;\Y_i,\Delta).
\end{align}
 It is then easy to see that
 \begin{align}
\Xh^{\rm MAP} = \Xh^{\rm MMSE} = \Xh^{\rm MIN} = \Y.
\end{align}
If we make the assumption that $\X$ has a Gaussian distribution, that is for all $i$
\begin{align}
  \px(\X_i) &= \NN(\X_i;0,1),
\end{align}
the probabilistic approach gives $\pxy \propto \NN(\X; \frac{\Y}{\Delta+1}, \frac{\Delta}{\Delta+1})$, yielding the estimators \mynote{Shrinkage}{1}
\begin{align}
 \Xh^{\rm MAP} = \Xh^{\rm MMSE} = \frac{\Y}{\Delta+1},
\end{align}
that \textit{shrink} the measurements by a factor $(\Delta+1)^{-1}$. 
Though not intuitive at first, this is the optimal way to estimate $\X$ knowing that it follows $\px$. 
The same result can be obtained with the minimization approach with a modified cost function containing a Tikhonov regularization term:
\begin{align}
 C(\X,\Y) &= || \X-\Y ||_2^2 + \Delta ||\X ||_2^2.
\end{align}
\end{example}

\begin{example}{Linear estimation}
\label{ex:linearMeasurements}
Let us take the other simple example of noiseless linear measurements of a $N$-dimensional signal $\X \in \RR^N$:
\begin{align}
 \Y = F \X	\label{eq:linearMeasurements}
\end{align}
where $F$ is a real $M \times N$ matrix.
One can study the (least-squares) minimization problem with the cost function
\begin{align}
 C(\X,\Y) = || \Y - F \X ||_2^2.
\end{align}
The solution to it has the simple expression
\begin{align}
 \Xh = F^+ \Y
\end{align}
in which the \textit{pseudo-inverse} $F^+$ is defined by
\begin{align}
 F^+ = (F^{\top} F)^{-1} F^{\top}.	\label{eq:pseudoinverse}
\end{align}
If $F$ is invertible, $F^+ = F^{-1}$ and $\Xh = \X$. 
However, if $F$ is not invertible, we know that~\eqref{eq:linearMeasurements} has either no solution 
or an entire subspace of solutions of dimension ${\rm dim}(Ker(F))$. 
In the latter case, $\Xh$ is in that space of solutions, but it is impossible to recover $X$ exactly without further information.
From an algorithmic point of view, the difficult part is the matrix inversion in~\eqref{eq:pseudoinverse}.
Note that with additional noise $\xi \sim \NN(0, \Delta)$,
\begin{align}
  \Y = F \X + \xi,	\label{eq:noisyLinearMeasurements}
\end{align}
the problem becomes more difficult. Not only is it always impossible to recover $\X$ exactly,
but it might as well be impossible to obtain a good estimate. 
In fact, even if $F$ is invertible, it can be \textit{ill conditioned}, which leads to  $\mse( F^{-1} \Y , \X) \gg \Delta$: the inference problem 
is not robust to noise.
Ill-conditioned matrices are characterized by a large \textit{condition number} $\kappa(F)$, defined by \mynote{Condition number}{0.8}
\begin{align}
 \kappa(F) \equiv \frac{\sigma_{\rm max}(F)}{\sigma_{\rm min}(F)},	\label{eq:conditionNumber}
\end{align}
in which the numerator and denominator are the biggest and smallest singular values of $F$.
$\kappa(F)$ is the maximal possible error enhancement ratio $\frac{|| F^{-1} \Y - \X ||_2^2}{|| \xi ||_2^2}$.
The finite numerical precision of computers is a noise-like perturbation of a signal, and for that reason
linear systems with very ill-conditioned matrices can be difficult to invert even in a noiseless setting.
\end{example}

\pagebreak
\section{Statistical physics}
\label{sec:statphys}
Statistical physics (or mechanics) has emerged as a field of physics with the work of Boltzmann at the end of the 19th century.
His work aimed at explaining the laws of thermodynamics from a microscopic approach, relying on the atomistic theory that was still an unproven conjecture at the time.
The probabilistic approach used in statistical physics is in apparent contradiction with the deterministic approach of classical point mechanics, but has proven to be correct and incredibly powerful.

In this section, I introduce some of the key concepts of statistical mechanics, independently of the previous section. 
The link between inference and statistical physics will be illustrated in~\chapref{chap:communityDetection}. 

\subsection{Equilibrium statistical physics in a nutshell}
Statistical mechanics typically studies a physical system at thermal equilibrium, composed of a large number $N$ of particles, described by $\{\sigma_i\}_{i \in [1,N]}$, 
whose energy $E$ is given by a Hamiltonian $H(\{\sigma\})$.
The system is closed but can exchange energy with a thermal bath at temperature $T$ (and inverse temperature $\beta=1/T$).
The system's probability to be in a microscopic state $\{ \sigma \}$ is then given by the \textit{Boltzmann distribution} \mynote{Boltzmann distribution}{0.7}
\begin{align}
 \boltz(\{\sigma\}) &= \frac{e^{-\beta H(\{\sigma\})}}{\ZZ}.	\label{eq:boltzmann}
\end{align}
The denominator $\ZZ$ is a normalization constant called the \textit{partition function}: \mynote{Partition function}{0.7}
\begin{align}
 \ZZ &= \int \dd \state e^{-\beta H(\{\sigma\})}	\label{eq:partitionFunction}
\end{align}
Statistical physics makes the assumption that microscopic states $\state$ cannot be observed or measured, 
but that it is possible to measure global quantities called observables. 
An observable $\OO$ is a real-valued function of the microscopic state $\state$, but cannot be measured instantaneously. 
Instead, measurement instruments always average observables over a period of time $\delta t$:
\begin{align}
 \langle \OO \rangle = \frac{1}{\delta t} \int_t^{t+\delta t} \dd t' \OO(\{ \sigma(t') \}).	\label{eq:averaged_observable}
\end{align} 
If the system is \textit{ergodic},\mnote{Ergodicity} it means that in the period $\delta t$, the system explores all possible microscopic states $\state$ and 
that the total fraction of time it spends in a state is equal to $\boltz(\{\sigma\})$.
Therefore, the expectation value of $\OO$ can be written as
\begin{align}
 \langle \OO \rangle = \int \dd \state \boltz(\state) \OO(\state).  \label{eq:av_observable}
\end{align}

This microscopic description of a system is closely linked to its macroscopic description through the \textit{thermodynamic potentials}.
The most used ones are the internal energy $U$, the entropy $S$ and the free energy $F$, linked by the formula \mynote{Free energy}{0.65}
\begin{align}
 F = U - TS  \label{eq:thermo_freeEnergy}.
\end{align}
The free energy is of particular importance for systems at thermal equilibrium, as it is the thermodynamic potential that these systems minimize.
The internal energy $U$ is the average of the observable $E$:
\begin{align}
 U = \langle E \rangle = \int \dd \state \boltz(\state) H(\state).
\end{align}
From this expression, we see that $U$ can be obtained from the partition function as follows:
\begin{align}
 U = - \mypartial{\beta} \log \ZZ.
\end{align}
Similarly, the free energy $F$ and the closely related free entropy $\Phi$ can be expressed as functions of $\ZZ$: \mynote{Free entropy and energy}{0.8}
\begin{align}
 \Phi = \log \ZZ \quad \mathrm{and} \quad F = -\frac{1}{\beta} \log{\ZZ}.
\end{align}

\subsubsection{The thermodynamic limit}
The power of statistical physics is its ability to deal with large system sizes $N$.
The thermodynamic potentials such as $U$, $\FreeEnt$ and $\Free$ are extensive (they increase with $N$), which leads to the definition of the intensive quantities
\begin{align}
  \freeEnt = \lim_{N \to \infty} \frac{\FreeEnt(N)}{N} && \free =  \lim_{N \to \infty} \frac{\Free(N)}{N} && 
\end{align}
called the \textit{free entropy density} and \textit{free energy density} respectively. 
The fact that these limits exist is the key point of the thermodynamic limit.
The study of $\freeEnt(\beta)$ and $\free(\beta)$ can reveal discontinuities (in value or slope) at some critical points $\beta_c$: 
at these temperatures, the system undergoes a \textit{phase transition} (of first or second order).

Let us sketch a proof that the free energy $\free$ is the thermodynamic potential that a system minimizes at thermal equilibrium.
We look at the distributions of energies per particle, $e_N=\frac{E}{N}$, of a system of size $N$ at thermal equilibrium:
\begin{align}
 \p_N \left( e_N \right) &= \int \dd \state \boltz(\state) \delta( H(\state) - E) = \Omega(E) \frac{e^{-\beta E}}{\ZZ} \\
	&= \frac{e^{-N \beta ( e_N - T s_N(e_N) )}}{\ZZ} = \frac{e^{-N \beta f_N(e_N)}}{\ZZ}	\label{eq:saddlePoint_into} 	
\end{align}
In the first line, we introduce the number  $\Omega(E)$ of states with energy $E$. 
In the second line, we introduce the entropy per particle, defined as $s_N(e_N) = \frac{\log \Omega(E)}{N}$, and 
 the free energy density $f_N(e_N) = e_N - T s_N(e_N)$.
In the thermodynamic limit (when $N \to \infty$), $e_N, f_N$ and $s_N$ converge to $e, f$ and $s$, and
\begin{align}
 p(e) &= \lim_{N \to \infty} p_N(e_N) = \delta \left( e - e^* \right)
\end{align}
where \mynote{Free energy minimization}{0.7}
\begin{align}
 e^* = \argmin_e f(e)
\end{align}

Therefore, in the thermodynamic limit, the system only explores the states of minimal free energy (or equivalently, of maximal free entropy).
However, in physical systems, $N$ is finite and the Boltzmann distribution allows fluctuations around the states of minimal free energy.

\begin{example}{Ising model}
\label{ex:Ising}
The Ising model is probably the most studied model in statistical physics.
In it, $N$ binary (Ising) spins $\{ \sigma_i = \pm1 \}_{i\in[1,N]}$ interact pairwise with the Hamiltonian
\begin{align}
 H(\state) = -\frac{1}{N} \sum_{\langle i j \rangle} \sigma_i \sigma_j,
\end{align}
where the sum is made over all pairs of spins $\langle i j \rangle$.
The system is characterized by its \textit{order parameter}, the magnetization \mynote{Magnetization}{1.05}
\begin{align}
 m = \frac{1}{N} \sum_{i=1}^N \sigma_i.
\end{align}
In this model, analytical calculations allow to obtain a simple expression of the free entropy $\freeEnt$ as a function of the magnetization $m$ for different temperatures, as shown below.

\begin{center}
\includegraphics[width=0.7\textwidth]{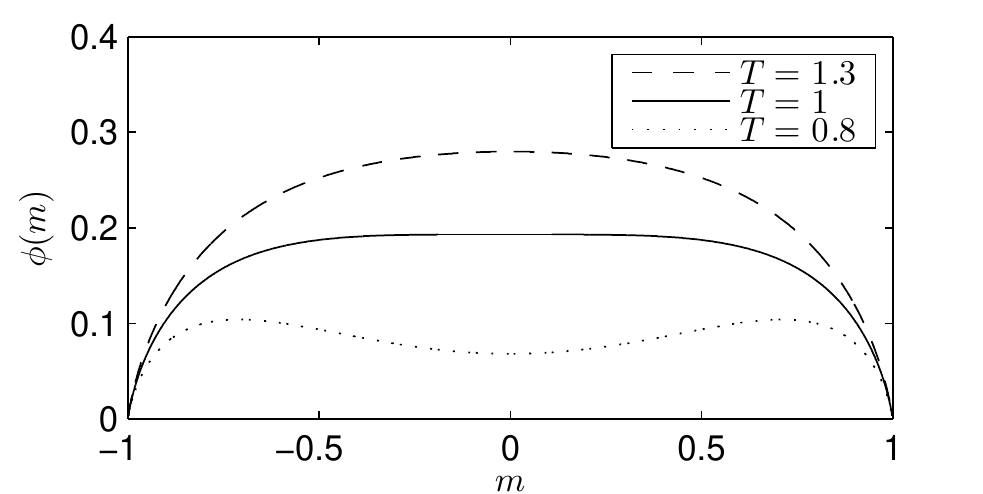}
\end{center}
\captionof{figure}[Free entropy of the Ising model as a function of the magnetization]{Free entropy of the Ising model as a function of the magnetization.}
\label{fig:Ising_free_entropy}

As the system maximizes its free entropy, its magnetization is given by
\begin{align}
 m = \argmax_m \freeEnt(m).
\end{align}
For $T>1$, this maximum is unique and $m=0$: the system is in the \textit{paramagnetic} phase. 
For  $T<1$, however, two maxima exist, separated by a free entropy barrier.

In the thermodynamic limit, this barrier is not crossable for the system, that remains in one of the two states as a consequence: 
it is the \textit{ferromagnetic} phase, that spontaneously exhibits a non-zero magnetization $m\neq 0$.
The ergodicity of the system is broken in that phase and~\eqref{eq:av_observable} cannot be used any more (it would incorrectly give $m=0$).
For finite $N$, the crossing of the free energy barrier is allowed by statistical fluctuations to states of higher free energy.
However, such transitions between the two states are exponentially rare as $N$ increases, and~\eqref{eq:av_observable} only holds for 
exponentially large integration times $\delta t$.
\end{example}

\subsection{Disordered systems}
\label{sec:1_disordered}
We speak of a disordered system when the Hamiltonian of the system is a function of the realization of a random variable $J$:
\begin{align}
 J \sim \p_J \quad \text{and} \quad E(\state)=H_J(\state).	\label{eq:disorderd_Hamiltonian}
\end{align}
All thermodynamic potentials and  observables of this system then explicitly depend on the realization $J$.
However, we can expect that systems with  different realizations of $J$ but large system size $N$ have the same characteristics.
We are therefore interested in computing the thermodynamic potentials averaged over the disorder $J$, starting with the average free energy
\begin{align}
 \langle F_J \rangle_J &= \int \dd J \, \p_J(J)  \log \ZZ_J  .  	\label{eq:quenched_average}
\end{align}
To calculate this average of a logarithm, the \textit{replica method} can be used, that starts from the identity \meqnote{Replica trick}
\begin{align}
 \langle \log \ZZ \rangle = \lim_{n \to 0} \mypartial{n} \langle \ZZ^n \rangle		\label{eq:replicaTrick}
\end{align}
and uses a series of non-rigorous mathematical tricks to calculate this quantity. \mnote{Glassy phases}
The characteristic of disordered systems is that they can have \textit{glassy phases} in which their energy landscape is chaotic, with exponentially many local minima.
In these glassy phases, the system gets stuck in metastable states, corresponding to such local minima. 
The exponential number of these metastable state make the dynamics of the system extremely slow, such that thermal equilibrium is never reached.


\begin{example}{SK model}
\label{ex:SK}
The Sherrington-Kirkpatrick model~\cite{SKModel} is a spin glass model in which $N$ Ising spins $\state$ interact pairwise and with couplings given by a coupling matrix $J$:
\begin{align}
 H(\state) = - \sum_{\langle ij \rangle} J_{ij} \sigma_i \sigma_j, 	 \label{eq:SKhamiltonian}
\end{align}
with $\langle i j \rangle$ runs over all pairs of spins.
The system is disordered because the coupling matrix $J$ is an \iid Gaussian random variable with mean $\frac{J_0}{N}$ and variance $\frac{1}{N}$.
Additionally to the magnetization $m$ as defined for the Ising model, we define a second order parameter, the Edwards-Anderson parameter \mynote{Edward Anderson parameter}{0.6}
\begin{align}
 q = \frac{1}{N} \sum_{i=1}^N \langle \sigma_i \rangle^2.	\label{eq:SKq}
\end{align}
As in the Ising model, the phase diagram (as a function of both $T$ and $J_0$) contains a para\-magnetic phase (P) in which $m=0$ and $q=0$,
a ferromagnetic phase (F) in which $m>0$ and $q>0$,
but also a so-called \textit{spin glass phase} (SG)  and a mixed ferromagnetic/spin glass phase (F-SG) in which $m=0$ but $q>0$.  
\end{example}
\vfill 

\begin{exampleSuite}
\begin{center}
\includegraphics[width=0.45\textwidth]{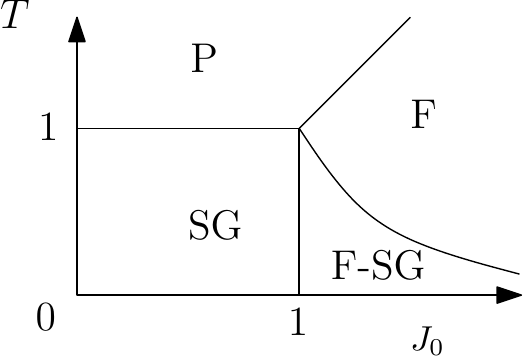}
\captionof{figure}[Phase diagram of the SK model]{Phase diagram of the SK model. Besides a paramagnetic and a ferromagnetic phase,
there are two additional, spin glass phases at low temperature. In these phases, the energy landcape is chaotic, leading to very slow dynamics.
The system is stuck in metastable states and thermal equilibrium is never reached.
}
\label{fig:SK_diagram}
\end{center}
\mynote{SK phase diagram}{-5}
\label{fig:sk_phase_diagram}
\end{exampleSuite}

\pagebreak

\section{Statistical physics for non-physical problems}
\label{sec:statphysinf}
In the probabilistic framework of statistical physics, many methods have been developed.
Some of them are analytical, such as the replica method, and were originally developed for the theoretic study of a certain class of physical systems, such as spin glasses~\cite{mezardParisiBook}.
Others, such as Monte Carlo algorithms, are numerical methods that have been developed to simulate physical systems.
Though developed for a given class of physical systems, these methods can be applied to any non-physical problem having a probabilistic formulation.
This is the case of inference problems, but also of many other problems in computer science, such as constraint satisfaction~\cite{constraintSatisfaction} or coding~\cite{ldpcKabashima}.

\subsection{Possibilities and limitations}
Let us mention a few advantages and limits of statistical physics methods when applied to non-physical problems.
The work presented in the following chapters of \mythesis is naturally as well concerned by all of those limitations.
\subsubsection{Rigour}
\paragraph{Limits} To the contrary of mathematics, physics use many non-rigorous methods---if they give useful results. 
Physicists do not shy away from using unproven identities, integrating a function without further verifications or inverting the order of limits.
While time-saving, this approach has the obvious disadvantage that no results found with methods from physics should be considered to be rigorous until made rigorous with methods from mathematics.
\paragraph{Advantages}
The bright side of the medal is that history has shown that non-rigorous physicist's methods can lead to accurate results.
As in physics of physical systems, theoretical results can be used to make predictions and design experiences to validate or disprove them.
In some of the problems of computer science examined by physicists, results obtained with non-rigorous methods could be verified by simulations, thus raising the interest 
of mathematician who, guided and inspired by the announced result, could prove them using rigorous methods.

\subsubsection{Finite sizes}
\paragraph{Limits}
In statistical physics, the thermodynamic limit allows great simplifications in many calculations. 
The corollary of this is that it is usually much more difficult to obtain results for finite-sized systems.
In physical systems, finite size effects are often minimal or unobservable because of the sheer number of particles (typically, $10^{23}$) that constitute macroscopic systems.
\paragraph{Advantages}
The advantage is that the behaviour of small-sized systems is often astonishingly close to the behaviour of their $N \to \infty$ counterpart.

\subsubsection{Typical vs. worst case}
\paragraph{Limits}
Adding to the fact that statistical physics methods are not rigorous, and as a side effect of the importance of the thermodynamic limit, results from statistical physics 
focus on the average case.
In information theoretical terms, this means that it is not possible to do worst case analyses with statistical physics.
When it comes to algorithms, worst case analyses can be very important as they give a lower bound on the algorithms' performances.
\paragraph{Advantages}
On the other hand, one can argue that worst case analyses often reveal little or nothing about the usual performances of an algorithm.
Furthermore, with increasing system sizes, the probability of the ``worst case'' actually happening decreases exponentially, and a typical instance of a problem is ever likelier to be close to the average instance.

\subsection{Examples}
In~\exref{ex:neuralNetworks}, we expose the problem of neural networks, that comes from biology and computer science but was studied extensively by physicists as well.
Other non-physical problems studied by the physicists include optimization, constraint satisfaction problems and error-correcting codes.
\begin{example}{Neural networks} 
\label{ex:neuralNetworks}
 Fascination for the human brain as well as the quest of artificial intelligence have triggered great interest for the study of neural networks.
 A neural network is an interacting assembly of real, artificial or simulated neurons.
 From the experimental insight into how biological neurons work, simplified neuron models have emerged.
 The simplest of them is the following:
 
 An idealized neuron performs a weighted sum of its inputs, $\{ \X_i\}$ with the weights $\{ J \}$.
 This sum gives the neuron's internal potential $h= \sum_i J_i X_i$. 
 If $h$ is bigger than a threshold $\tau$, the neuron ``fires'', outputting $\sigma=1$. Else, it remains ``silent'', $\sigma=0$. 
 With $\xi$ being a Gaussian noise and $\Theta(x>0)=1$, $\Theta(x<0)=0$, an idealized neuron can be represented as follows: \mynote{Idealized neuron}{1.5}
 \begin{center}
 \includegraphics{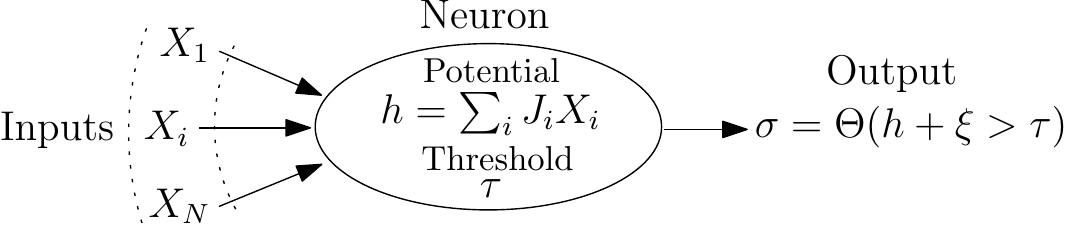}
  \captionof{figure}[An idealized neuron]{An idealized neuron.}
 \label{fig:idealized_neuron}
 \end{center}
 
 In a recurrent neural network, such as illustrated below, the neurons are interconnected and form a dynamical system. 
 The input of a neuron at time $t+1$ are the outputs of other neurons at time $t$: $\{ X_i(t+1) \} = \{ \sigma_i(t) \}$.
 The complex dynamics of such a network are characterized by attractor states, that can be seen as \textit{memories}. \mynote{Recurrent neural network}{1.5}
 \begin{center}
  \includegraphics[width=0.4\textwidth]{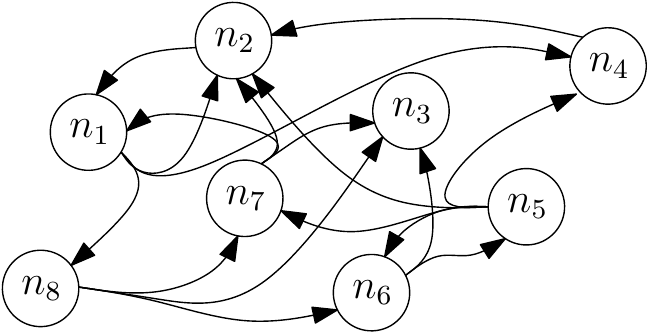}
   \captionof{figure}[Recurent neural network]{Recurrent neural network, in which each neuron is connected to a subset of the other neurons.}
 \label{fig:recurrent_network}
 \end{center}

 Simplified models of recurrent neural networks have been analysed using the replica method. 
 The system exhibits different phases that can be mapped to the paramagnetic, ferromagnetic and spin glass phases of the SK model.
 In particular, the analysis has allowed to determine the maximal number of memories that can be stored in such a network.~\cite{recurrentNetworks,neuralSystemsReview}
\end{example}

%
%
%
%
%

\section{Inference problems examined in this thesis}
\label{sec:myproblems}

\subsection{Community detection}
The goal of community detection is to detect communities in a network. 
For example, in a recurrent neural network as in~\exref{ex:neuralNetworks}, a community could be a sub-network of neurons 
that performs a specific task. If the neurons belonging to such a functional community are more connected with each other 
than with neurons of other communities, then finding these communities is \textit{a priori} an inference problem that could be possible to solve.

Some aspects of community detection as well as my contributions to it will be presented in~\chapref{chap:communityDetection}.

\subsection{Generalized linear and bilinear models}
\begin{figure}
 \begin{center}
  \includegraphics{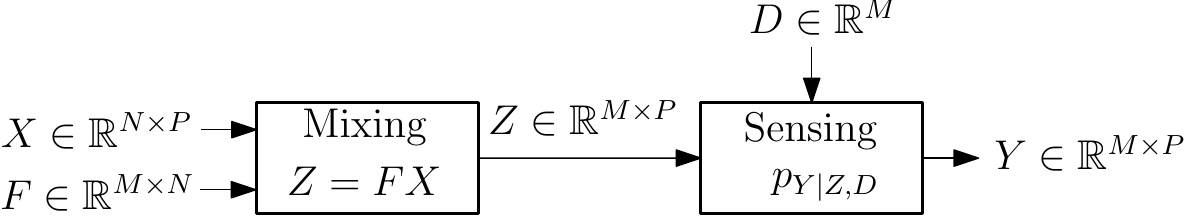}
 \end{center}
\caption[Inference problems with mixing and sensing]{General inference setting using a mixing step and a sensing step.
Depending on which variables are known and which have to be inferred, this general setting particularizes to different models examined in \mythesis.}
\label{fig:1_linBilin}
\end{figure}

We will present generalized linear models and generalized bilinear models in~\chapref{chap:generalizedLinearModels} and~\chapref{chap:generalizedBilinearModels} respectively, but briefly introduce them here.
Figure~\ref{fig:1_linBilin} presents a  general inference setting.
The table below explains how this general setting particularizes to the inference problems described in the following paragraphs.
\begin{table}[h]
\begin{center}
\begin{tabular}{@{}lccccc@{}} \toprule
 & \multicolumn{3}{c}{Variables}  &  & No. of signals \\ \cmidrule(r){2-4}
Problem class & $X$ & $F$ & $D$ & Measurements &  $P$ necessary\\ \midrule
Compressed sensing & ? & \checkmark &  \checkmark & $Z$  & $1$ \\
Generalized linear models & ? & \checkmark & \checkmark & $Y$ & $1$ \\
Blind calibration & ? & \checkmark & ? & $Y$ & $O(1)$  \\
Matrix factorization & ? & ? & \checkmark & $Z$ & $O(N)$ \\
Generalized bilinear models & ? & ? & \checkmark & $Y$ & $O(N)$   \\ \bottomrule
\end{tabular}
\end{center}
\caption{The general inference problem presented in~\figref{fig:1_linBilin} particularizes to different classes of problems
depending on which variables are known (\checkmark) and which have to be inferred (?).
With increasing number of variables to infer, the number of independently measured signals $P$ needs to increase for inference to be possible.}
\label{table:1_problems}
\end{table}

\subsubsection{Compressed sensing}
The compressed sensing (\cs) problem is closely related to the problem of linear measurements of~\exref{ex:linearMeasurements}.
The difference is that the measurement---or \textit{sensing}---matrix $F$ is taken to be random and have a \textit{compressive} measurement rate $\alpha = \frac{M}{N} < 1$. 
While in the general case, it is impossible to recover $X$, in \cs, we consider the case in which the signal $X$ is known to be \textit{sparse}:  
only a fraction $\rho$ of its components are non-zero.
In that case, information theoretical arguments show that the problem has a unique solution as soon as $\alpha > \rho$.
\cs has applications in fields such as medical imaging, which are motivations for developing ever more efficient algorithms.

\subsubsection{Generalized linear models}
Generalized linear models (\glm) are a class of problems generalizing the linear estimation problem.
In it, the variable $Z=FX$ is unobserved, but measured through a sensing channel $\py$.
The measurements $Y$ therefore contain in general less information about $X$ then they do in \cs.
As in \cs, one generally considers the setting in which the measurement matrix $F$ is random.
The compressive  regime $\alpha<1$ can still be studied if the signal is sparse, but depending on the 
measurement channel, oversampling regimes $\alpha>1$ can be necessary to compensate for the loss of information induced by the sensing channel 
and allow good estimates of $X$.

\subsubsection{Blind sensor calibration}
The blind sensor calibration problem is similar to the one of \glm, with the difference that the sensing channel depends on a variable $D$,
which is unknown and different for each sensor. This variable can for instance be a threshold, as $\tau$ is in~\exref{ex:singleLayer}.
The presence of these additional unknowns $D$ makes this inference problem harder than a \glm, in the sense that in general, more measurements 
are necessary for successful inference to be possible.
An alternative to increasing the measurement rate $\alpha$ is to measure a set of $P>1$ different unknown signals $\{ X \}$. 

\subsubsection{Generalized bilinear models}
As blind sensor calibration, the generalized bilinear model (\gbm) setting extends the \glm setting by introducing further variables.
The novelty is that the measurement matrix $F$ itself is unknown. 
To compensate for these additional unknowns, the number $P$ of measured signals generally has to increase with 
the signal size, \ie $\frac{P}{N} = O(1)$.
For an \awgn sensing channel, the setting particularizes to the well studied problems of dictionary learning or matrix factorization,
depending on the hypotheses made on $F$ and $X$.

\subsubsection{Illustration}
We can illustrate this class of generalized linear and bilinear models with the following toy example.

\begin{example}{Single layer feedforward neural network}
\label{ex:singleLayer}
 In~\exref{ex:neuralNetworks} we have introduced idealized neurons and recurrent neural networks.
 Here, we introduce feedforward neural networks as an illustrative example of inference problems.
 Unlike in recurrent networks, we consider a layer of $M$ neurons, that are not connected to each other, but take $N$ inputs from the preceding layer. \mynote{Feedforward neural network}{2}
 \begin{center}
 \includegraphics[width=0.5\textwidth]{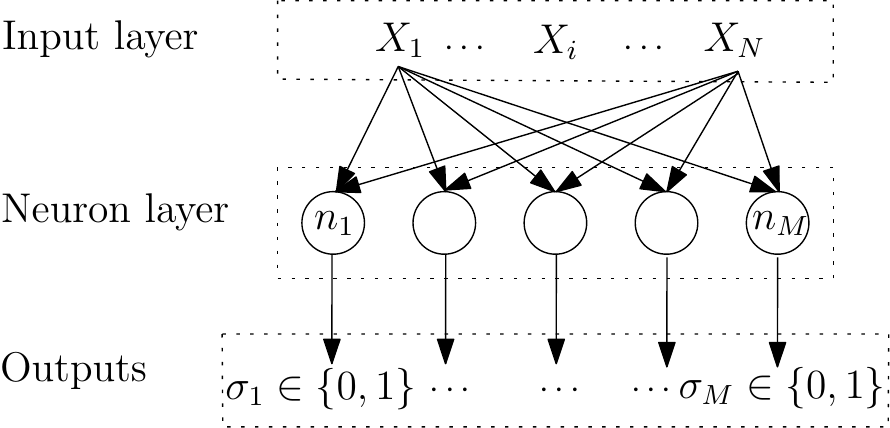}
 \captionof{figure}[A layer of neurons in a feedforward neural network]{A layer of neurons in a feedforward neural network.}
 \label{fig:feedforward_layer}
 \end{center}
 
 In computer science, such a layer is one of the basic building block of deep neural networks.
 The internal potential $h_{\mu}$ and output $\sigma_{\mu}$ of neuron $\mu$ are given by:
 \begin{align}
  h_{\mu} &= \sum_{i=1}^N J_{\mu i} X_{i} \\
  \sigma_{\mu} &= \Theta( (h_{\mu} + \xi)  - \tau_{\mu} )
 \end{align}
 where $\xi$ is white Gaussian noise and $\Theta(x>0)=1$, $\Theta(x<0)=0$.
 With the following correspondence of notations it can be seen that this setting is exactly of the type described in~\figref{fig:1_linBilin}:
 \begin{align}
  F \leftrightarrow J & & Z \leftrightarrow h & & D \leftrightarrow \tau & & Y \leftrightarrow \sigma.
 \end{align}

 The most studied problem with this setting is the \textit{perceptron}~\cite{perceptron}, in which the network is used as a classifier.
 A classifier attributes a class to each signal $X$, represented by the outputs $\{ \sigma \}$.
 The perceptron is able to classify signals into linearly separable classes after learning the set of weights $J$ and thresholds $\tau$.
 This learning takes place in a supervised fashion, by using known training signals and their classes $\{ X , \sigma\}_{\rm train}$.
 \newline
 Once the weights of the perceptron are fixed, many interesting problems can be considered. 
 For instance, obtaining an estimate of $X$ by measuring $\sigma$ is called $1$-bit compressed sensing, which
 is an example of the \glm setting.
 Suppose that one searches again to obtain an estimate of $X$, that the weights $J$ are known but the thresholds $\tau$ are not. 
 This inference problem  belongs to the class of blind sensor calibration.
 Finally, inferring jointly $X$ and $J$ from measurements $\sigma$ (supposing $\tau$ is known) is a non-linear matrix factorization problem,
 belonging to the \gbm class described above.
 
 \end{example}


 
 \chapter{Community detection}
\label{chap:communityDetection}

In~\chapref{chap:1}, I have presented important concepts of statistical physics, only mentioning that they can be used
for solving inference problems.
In the present chapter I show how, by treating one specific inference problem: community detection.

One motivation of community detection is the analysis of the subcommunities
a social group is divided into and how to detect these communities~\cite{karate}.
Initially limited by the difficulty of keeping track of social interactions in large social groups, 
community detection has experienced a revival of interest with the spectacular rise of online social networks 
such as facebook and twitter. Thanks to those, very large datasets are available, such that large-scale studies can be made, 
encouraging further and faster algorithms to be developed and studied.

After a formal presentation of community detection as an inference problem (\secref{sec:2_setting})
and of two solving approaches (\secref{sec:2_approaches}), I will focus on the \modbp algorithm, introduced in~\cite{modbp} 
by Zhang and Moore, that treats community detection as a statistical physics problem. 
As such, concepts like temperature and energy-entropy competition naturally appear, as well as phase transitions and glassy phases. 
With \modbp, we will present the belief propagation algorithm, that allows to sample from high-dimensional probability distributions.

My contribution to the field of community detection is a deeper study of the \modbp algorithm, published in~\cite{moiModularite} and 
presented in~\secref{sec:2_algoPT} and~\secref{sec:2_phases}. 
The main results are the definition of a new set of order parameters, the existence of multiple phase transitions and 
a study of several real networks.

\section{Setting}  
\label{sec:2_setting}
\begin{figure}
 \begin{center}
  \includegraphics{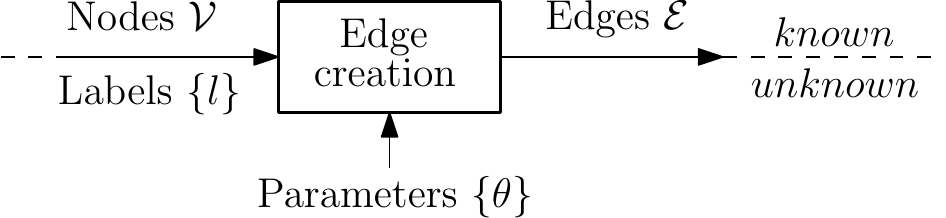}
 \end{center}
\caption[Community detection]{General inference scheme in community detection. Each node $i$ from the set $\VV$ has a label $l_i$ indicating a group.
Each pair of nodes can be linked by an edge, in a probabilitic process depending on the labels of the two nodes and on often unknown parameters. 
The goal of community detection is to infer $\partition$ from the observation of the graph $\GG=(\VV,\EE)$.}
\label{fig:2_setting}
\end{figure}

The general inference scheme in community detection is presented in~\figref{fig:2_setting}.
The setting is very general and applies to a great variety of domains in which networks appear.
Networks can be used as soon as a system of many interacting subsystems is studied~\cite{complexNetworks}. For example:
\begin{itemize}
 \item Social networks. Nodes of the network are people, edges of the network are a certain kind of social interactions.
 \item Transportation networks, in which nodes represent cities or airports, and edges represent roads or flights.
 \item Functional networks, such as networks of neurons or gene regulation networks.
 \item The internet, in which nodes are web pages, edges are hyperlinks.
 \item Citation networks, in which nodes are books or articles and edges are citations.
\end{itemize}
After these examples, let us define the community detection problem in mathematical terms.

\subsection{Graphs}

An \textit{graph} \mnote{Graphs} is a pair $\GG=(\VV,\EE)$ of a set of $N$ nodes (or vertices) $\VV$ and a set of $m$ edges (or links) $\EE \in (\VV \times \VV)^m$. 
We will consider only \textit{undirected} graphs, in which edges are bidirectional: if nodes $i$ and $j$ are linked, then so are $j$ and $i$ and they are said to be neighbours.
The number of neighbours of a node $i$ is called the \textit{degree} of the node, $d_i$.
A practical way of representing a graph is with its \textit{adjacency matrix} $A \in \RR^{N\times N}$, defined by
\begin{align}
 A_{ij} = \left\{ \begin{array}{cc}
           1 & \text{if } \ijlink \in \EE ,\\
           0 & \text{if not}.
          \end{array} \right.
\label{eq:adjacency}
\end{align}

In community detection,  \mnote{Groups} we consider networks that are subdivided (or \textit{partitioned}) into $q$ groups.
Thus, each node $i$ has a label $l_i \in \{1,\dots, q\}$ that indicates which of the $q$ groups it belongs to.
Furthermore, the edges are considered to be the result of a probabilistic edge creation process that depends on the labels of the nodes.
For many types of networks, such as social networks, this edge creation process depends on many other parameters that are in general unknown.
We call $n_a$ the fraction of nodes that are in group $a$.

The simplest assumption \mnote{Assortativity} that can be made is that the edge creation process is \textit{assortative},  meaning that two nodes belonging to the same 
group are more likely to be linked than two nodes belonging to different groups.

The goal of community detection \mnote{Partitions} is to find the nodes' hidden labels from the knowledge of the set of edges.
An estimated set of labels $\{ \lh \}$ is called a \textit{partition}. 
As the probabilistic edge creation process is in general unknown, Bayes optimal community detection is impossible in most settings.

The natural measure to compare the true partition $\partition$ (called \textit{ground truth}) to an estimated partition $\{ \lh \}$ is the overlap, defined in~\eqref{eq:overlap}.
Due to the permutation symmetry between group labels, a more adapted definition of the overlap in the case of community detection is~\cite{decelle}
\mynote{Overlap}{1.15}
\begin{align} 
 O(\partition, \{\lh\}) = \max_{\sigma} \left(  \frac{ \frac{1}{N} \sum_{i=1}^N \delta_{l_i, \sigma(\lh_i)}  - \max_a n_a}{1-\max_a n_a} \right) ,  \label{eq:defOverlap} 
\end{align}
where the maximum is over the set of all permutations $\sigma$ of $\{ 1, \dots, q \}$. An overlap equal to $1$ means that the nodes' labels were correctly inferred 
(up to a global renumbering of groups).

\subsection{Random graphs}
Let us introduce three simple models of graphs that are random, in the sense that the edge creation process is probabilistic. 
The first two models are models of random graphs \textit{without} underlying groups, while the third is a model of graphs with groups.

\subsubsection{The \er model}
In an \er graph~\cite{erdos}, the edge creation process is the simplest possible. Each pair of nodes $\langle i j \rangle$ is taken once, and added to the set of edges $\EE$ with a constant probability $p$.
This edge creation process does not take into account possible groups of the nodes. 
It produces graphs with a random number of edges and is characterized by a Poissonian distribution of degrees.

\subsubsection{The configurational model}
In the configurational model~\cite{configurationalModel}, edges are created from the set of nodes $\VV$ and the list of their degrees, $\{ d_i \}$.
The advantage of this model is that it allows to create random graphs with any desired degree distribution.
This is useful because the Poisson distribution obtained for \er graphs is unrealistic, in the sense that 
real networks usually do not have Poissonian degree distributions, but rather power-law, ``heavy-tailed'' distributions~\cite{complexNetworks}.
As in the \er model, the edge creation process is independent of possible node labels. 
Therefore, neither of these two models can be used for community detection. 
However, they can serve as \textit{null models}.

\subsubsection{The stochastic block model}
The stochastic block model (\sbm) is a simple model in which the edge creation process is linked to the labels of the nodes~\cite{sbm}.
Therefore, the structure of the resulting graph can be expected to contain information about the ground truth partition $\partition$,
and inference should be possible.

In the \sbm, edges are created by taking each pair of nodes $\langle i j \rangle$ once, and adding it to $\EE$ with a probability that depends \textit{only} on the labels $(l_i, l_j)$.
The \sbm is therefore fully characterized by a $q \times q$ matrix containing the probabilities $p_{ab}$ of two nodes of respective groups $a$ and $b$ to create an edge.

In its simplest version, this matrix is taken to have two distinct elements: one for the diagonal and one for the off-diagonal entries:
\begin{align}
 p(\langle i j \rangle \in \EE ) = \p_{l_i, l_j} = \left\{ \begin{array}{cc}
                                                    p_{\mathrm{in}} & \text{if } l_i=l_j \\
                                                    p_{\mathrm{out}} & \text{if } l_i \neq l_j \\
                                                   \end{array} \right. .  \label{eq:sbm}
\end{align}
In that version, we can define the parameter
\begin{align}
 \epsilon = \frac{p_{\mathrm{out}}}{p_{\mathrm{in}}},
\end{align}
and the network structure is assortative if $\epsilon<1$.

Two limiting cases are interesting: 
for $\epsilon=0$, only nodes of the same group can form edges, while for $\epsilon=1$, the model is equivalent to the \er model.
In the first case, the strategy for inference of groups is trivial, as nodes that are linked are known to belong to the same group.
Note, however that while $p_{\mathrm{in}}<1$, perfect recovery is in general impossible.
As $\epsilon$ increases, inference gets harder, and is obviously impossible for $\epsilon=1$.

In order to generate an instance of the \sbm, two other parameters have to be fixed: the number of groups $q$ and the fractions $\{n_a\}$ of nodes in each group, taken to be all equal to $1/q$ in the simplest version.

\begin{figure}
 \begin{center}
  \includegraphics[trim= 1 1cm 1 1, clip, width=0.3\textwidth]{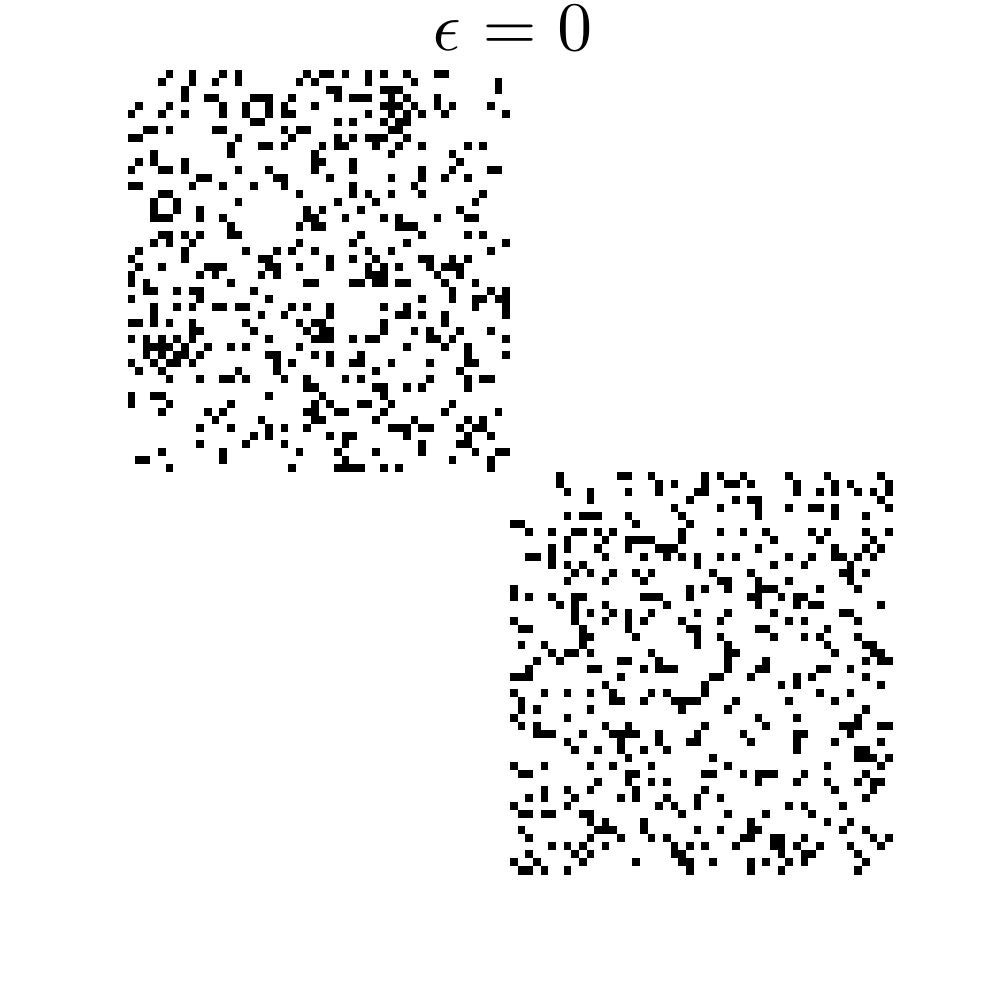}
  \includegraphics[trim= 1 1cm 1 1, clip,  width=0.3\textwidth]{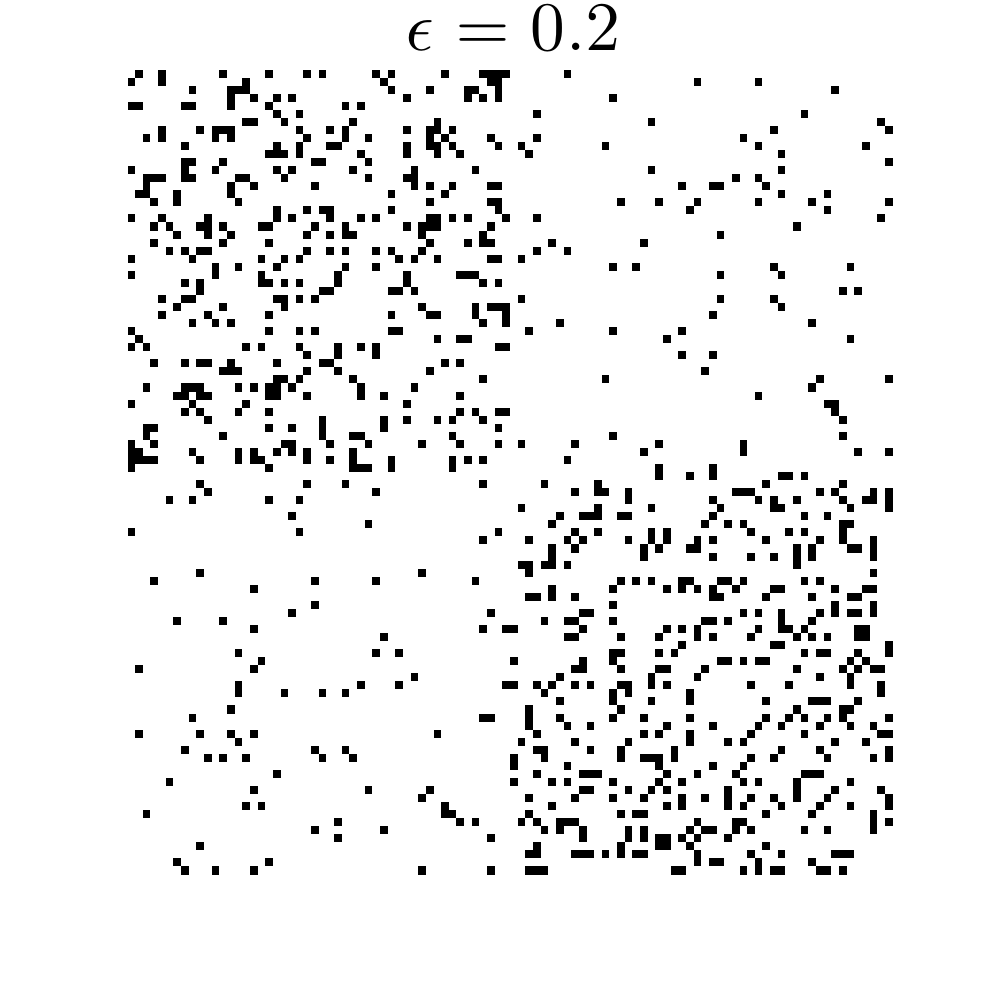}
  \includegraphics[trim= 1 1cm 1 1, clip, width=0.3\textwidth]{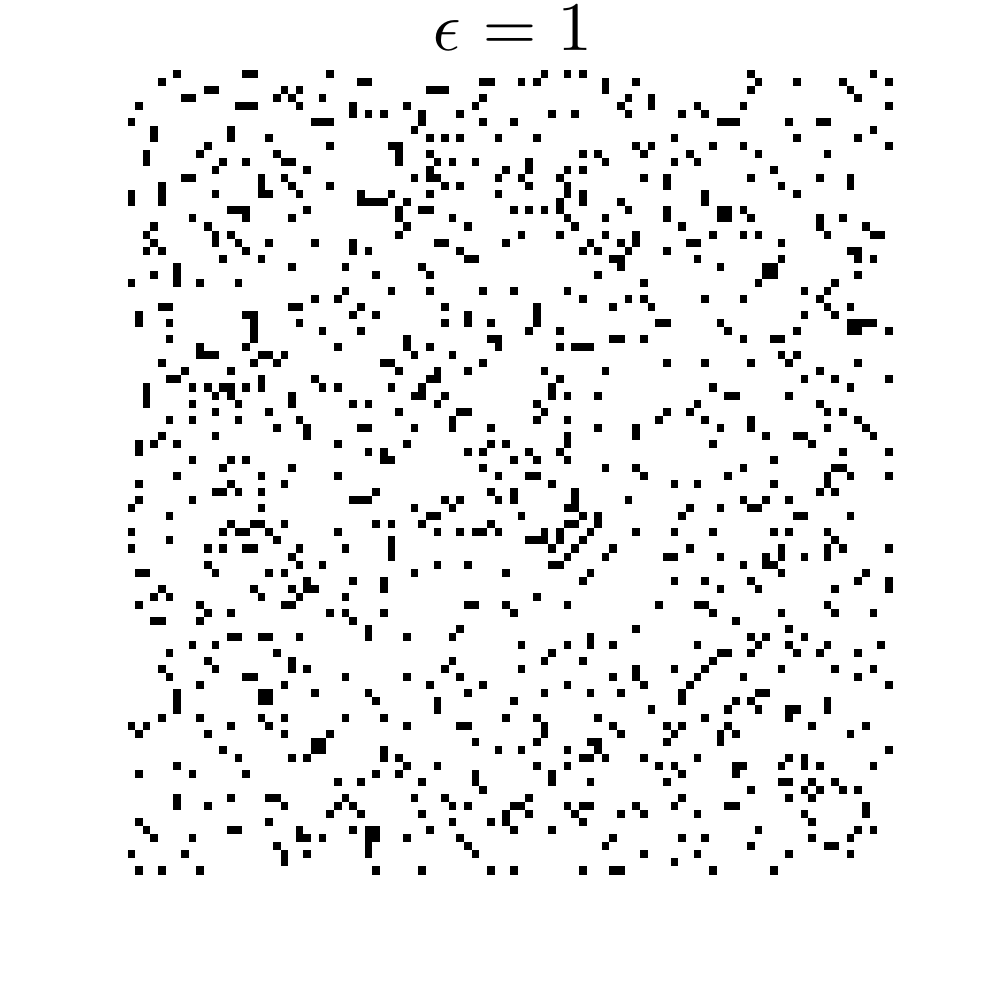}
 \end{center}
\caption[Adjacency matrices of SBM networks]{Adjacency matrices of graphs generated with the \sbm for $q=2$ and $\epsilon= \{ 0, 0.2, 1\}$ (left to right).
The group structure is visible in the adjacency matrices because the nodes are correctly ordered.
For the limiting case $\epsilon=0$, the communities are disjoint and inference is easy. 
For the limiting case $\epsilon=1$, the structure of the graph contains no information about the communities, as it is an \er graph, and inference is therefore impossible.}
\label{fig:2_sbms}
\end{figure}

The main advantage of the \sbm is its simplicity: it only requires to fix $N$, $q$, $p_{\mathrm{out}}$ and $p_{\mathrm{in}}$ in its simplest form.
Varying these parameters, one can interpolate between an easy and a hard inference problem, study the performance of algorithms and compare them.
Another advantage of the \sbm is that it can easily be generalized by leaving the ``diagonal'' scenario of~\eqref{eq:sbm}.
The simplicity of the model is also its main drawback. Just like the \er model, the \sbm produces unrealistic degree distributions, which indicates that the edge creation processes 
in real networks do not follow the \sbm.
Figure~\ref{fig:2_sbms} shows the adjacency matrices of graphs generated with different parameters of the  \sbm.

\section{Approaches and algorithms}
\label{sec:2_approaches}
In this section, we describe three of the many different approaches that exist in community detection.

\subsection{Spectral algorithms}
The first approach is a spectral approach, that is based on the computation of the eigenvalues and eigenvectors of a matrix. 
Several different choices of matrices can be made, reviewed in~\cite{spectralClustering}, but it is natural to use the adjacency matrix $A$ of the graph as a part of it, as it 
contains the entire structure of the network.
Another useful matrix is the diagonal, $N$ by $N$ matrix $D$ that contains the degrees of the networks nodes. 
From these two matrices, we can construct the graphs \textit{Laplacian} and normalized Laplacian matrices, defined as follows:
\begin{align}
 \text{Laplacian:} \quad L = D - A , \qquad \text{Normalized Laplacian:} \quad \mathcal{L} = D^{-\frac{1}{2}} L D^{-\frac{1}{2}}.
\end{align}
For one of these two matrices, or other related matrices, we then calculate the eigenvalues $\{ \lambda_1 , \dots , \lambda_N\}$, 
with $\lambda_1 \leq \cdots \leq \lambda_N$.
For a clustering of the networks into $q$ groups, the $q$ smallest eigenvalues are kept, as well as their corresponding $q$ eigenvectors $\{ y_{1}, \cdots, y_q\}$.
These eigenvectors are then clustered with a $k$-means clustering algorithm, after which each node can be assigned to a group.

Spectral algorithms remain popular for community detection, but have several limits. 
The most important of them is their bad performances when it comes to clustering \textit{sparse} networks, \ie networks for 
which the average degree $\langle d \rangle$ is much smaller than the number of nodes. 
This is often the case in real networks: for instance, each person is befriended with a small number of people,
that is not growing with the world population.

\subsection{Bayesian inference}
A second method is more principled and overcomes some of the inconvenients of spectral algorithms.
It follows the probabilistic approach to inference using Bayes' formula~(\eqref{eq:bayes}).
The present section presents the results obtained by Decelle \etal in reference~\cite{decelle}.

As mentioned previously, the edge creation process is usually unknown in real networks. 
Therefore, the ``channel'' $\p(A | \partition)$ is in general unknown, and Bayes' formula cannot be used.
For this reason, we focus on Bayesian inference of the \sbm. 
With~\eqref{eq:sbm}, one can write
\begin{align}
 \p(A_{ij}| l_i, l_j) = \p_{l_i,l_j}^{A_{ij}} (1-p_{l_i,l_j})^{1-A_{ij}}
\end{align}
and thus
\begin{align}
 \p(A | \partition ) = \prod_{i < j} \p_{l_i,l_j}^{A_{ij}} (1-p_{l_i,l_j})^{1-A_{ij}},
\end{align}
from which one can write the posterior probability
\begin{align}
\p(\partition | A) = \frac{\p(\partition)\prod_{i < j} \p_{l_i,l_j}^{A_{ij}} (1-p_{l_i,l_j})^{1-A_{ij}} }{ \p(A)},  	\label{eq:posteriorSBM}
\end{align}
where $\p(\partition)= \prod_i n_{l_i}$ is the prior distribution, with $n_a$ being the fraction of nodes in group $a$.
As explained in~\chapref{chap:1}, the problem of such posterior probability distributions is that they are hard to calculate.
Remember that the denominator $\p(A)$ has to be calculated by marginalization of the numerator, which implies a sum over all possible partitions $\partition$.
As there are $q^N$ of them, the sum is intractable even for reasonable network sizes.
The solution proposed in~\cite{decelle} is to use \textit{belief propagation} to calculate~\eqref{eq:posteriorSBM}.
Belief propagation (\bp), presented in more details in~\secref{sec:BP}, is an iterative algorithm that allows to estimate high-dimensional probability distributions as~\eqref{eq:posteriorSBM}.
Though not giving correct estimates under all circumstances, it is known for being exact in many cases. In the present case, \bp returns for each node $i$ and group $l$ the estimated probability
\begin{align}
 \messv_{i}^{l} = \hat{p} (l_i=l | A) \approx \p(l_i=l | A),
\end{align}
that verify $\sum_{l=1}^q \messv_{i}^l = 1$. Using these probabilities, the estimated partition $\{ \lh \}$ is obtained using the MAP estimator defined in~\eqref{eq:MAP}:
\begin{align}
 \forall i, \quad \lh_i = \argmax_l \messv_i^l.	\label{eq:MAPassignment}
\end{align}

\begin{figure}
  \begin{center}
   \includegraphics[width=0.6\textwidth]{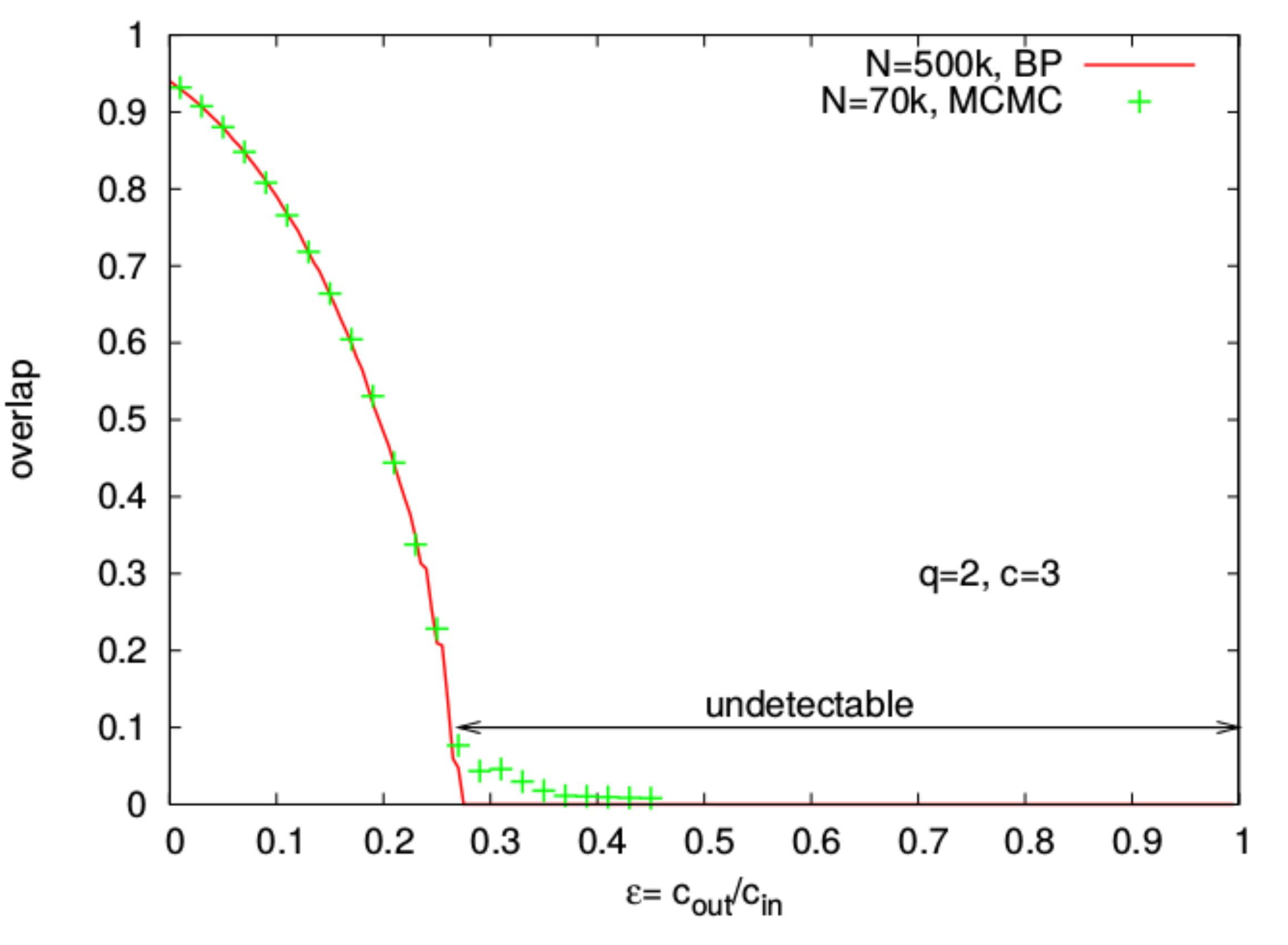}
  \end{center}
  \caption[Detectability transition in community detection]{(Figure taken from~\cite{decelle}). Bayesian inference of the \sbm with the algorithm presented in~\cite{decelle}.
  The networks are generated with the \sbm with 2 groups and average connectivity 3. 
  Besides the BP algorithm, a Monte Carlo Markov chain algorithm is used.
  A critical $\epsilon^*$ exists, above which the algorithm is in a paramagnetic phase and the community structure is undetectable.
  This $\epsilon^*$ is not even close to $1$, which disproves the wrong intuition that as long as $\epsilon<1$, inference of the groups \textit{should} be possible to a certain point.}
  \label{fig:2_bayesSBM}
\end{figure}

The main finding of~\cite{decelle} is that in Bayes optimal inference of community detection for the \sbm, there are different \textit{phases}, that are separated by phase transitions, 
depending on the value of the parameter $\epsilon$, just as in physical systems. 
Taking the vocabulary of the Ising model~(\exref{ex:Ising}), there is a ferromagnetic and paramagnetic phase. 
Just as the value of the \textit{magnetization} $m$ defines theses phases in the Ising model~(\exref{ex:Ising}), the overlap $O$ plays the role of an \textit{order parameter} in community detection:
In the paramagnetic phase, $O(\{ \lh \}, \partition) \approx 0$, while $O(\{ \lh \}, \partition) > 0$ in the ferromagnetic phase.
In the language of community detection, this means that below a critical $\epsilon^*$, the group structure is detectable, while above $\epsilon^*$, 
the group structure is undetectable. 
Figure~\ref{fig:2_bayesSBM}, taken from~\cite{decelle}, shows the phase transition.

\subsubsection{Non Bayes optimal case}
Probabilistic inference starting from~\eqref{eq:posteriorSBM} can also be made if the true edge creation process $\p^0(A | \partition)$ and the true prior $\p^0(\partition)$ are not known.
In that case, the inference setting is not Bayes optimal, but can still lead to good results if the supposed distributions $\p(A | \partition)$ and $\p(\partition)$ are close enough to their true counterparts.

One simple way to study such a setting is to generate a graph with the \sbm and a set of parameters $\{ q^0, \{n_a^0\}, \pin^0 , \pout^0 \}$ and to perform inference 
with a different set of parameters $ \{ q, \{n_a\}, \pin , \pout \} $.
Thanks to the simplicity of expression~\eqref{eq:posteriorSBM} as a function of these parameters, a step to optimality can be made by treating the parameters 
$\{ \theta \} = \{ \{n_a\}, \pin , \pout \}$ as 
variables that have to be inferred as well. This corresponds to performing inference starting from the posterior distribution
\begin{align}
 \p(\partition ,  \{\theta\} | A) = \frac{\p(\partition, \{ n_a \} ) \p(A| \partition, \pin, \pout) }{ \p(A)}, 	\label{eq:posteriorSBMlearning}
\end{align}
which can be done with \bp and an \textit{expectation-maximization} (EM) procedure.
Note that:
\begin{itemize}
 \item $q$ is not included in the set of learnable parameters $\{ \theta \}$. This comes from the fact that \bp uses the set of $q \times N$ variables $\psi_i^l$. 
 In EM, the values of the parameters to be learnt change from iteration to iteration. As $q$ is ``hard-coded'' in the \bp equations, it cannot vary as the other parameters.
 The number of groups $q$ is therefore a special parameter. In order to learn it with a \bp based algorithm, one has to run \bp several times with different values of $q$ 
 and compare the results: this is a \textit{model selection} procedure.
 \item Few additional variables have to be learned (there are $(q-1) +2$ of them) compared to the $N\times q$ probabilities $\psi_i^l$ infered by the algorithm without parameter learning.
 This allows EM to be successful. If the number of parameters to learn was comparable to $N \times q$, a whole different algorithm would be necessary.
\end{itemize}

\subsubsection{Advantages and limit} 
The findings of~\cite{decelle} have had an important impact on the theoretical understanding of community detection.
The existence of a phase transition in the Bayes optimal inference of the \sbm has been confirmed by theoretical results~\cite{proof1,proof2,proof3},
proving that for the \sbm with $q=2$ groups, it is \textit{impossible} for any algorithm to label nodes better than randomly when $\epsilon \in [ \epsilon^*, 1]$.
Unlike spectral methods based on the network's Laplacian, the \bp based algorithm proposed in~\cite{decelle} also works in the interesting regime of sparse networks.
Furthermore, this algorithm has inspired a novel spectral method for community detection, based on the so-called nonbacktracking matrix, that has the same phase transition 
as the \bp algorithm and nearly as good performances~\cite{spectralRedemption}.
Other advantages of the method are its speed (unlike spectral methods, \bp does not require to diagonalize matrices) and the fact it can be used with parameter learning.
However, its main limitation is that it is entirely constructed on a model that is not relevant for most real networks.

\subsection{Modularity maximization}
The main disadvantage of the Bayesian inference scheme presented above is that it heavily relies on the \sbm.
As mentioned already, the \sbm is not a good model for real networks, for which the edge creation process is in general complex and unknown.
For this reason, it is desirable to design an inference strategy that makes the least possible assumptions on how the network was created.
Another flaw of our presentation of community detection methods until now is the lack of an indicator that allows to estimate how good a proposed partition is.
In fact, the overlap can only be used for networks for which the true labels are known--and therefore do not require community detection.
The overlap is still useful as it allows to test the performances of algorithms on synthetic or labelled real-world networks. 
But in interesting cases, the overlap is not known and we thus have to introduce another indicator of success.

The \textit{modularity} is a quantity that measures the goodness of a partition based on the sole hypothesis that the network has an assortative structure and on the nodes' degrees.
The modularity is defined in~\cite{modularity} by \mynote{Modularity}{1.2}
\begin{align}
 Q(\partition) = \frac{1}{m} \left( \sum_{\langle ij \rangle \in \EE} \delta_{l_i,l_j} - \sum_{\langle ij \rangle} \frac{d_i d_j}{2 m} \delta_{l_i,l_j} \right). 	\label{eq:modularity} 
 \end{align}
The first term simply increases modularity each time two nodes connected by an edge are assigned to the same group.
The sum of the second term goes over all pairs of nodes $\langle ij \rangle$ and depends on the nodes' degrees. 
This term can be seen as choosing the configurational model as the null model for modularity: it makes sure that the modularity of a random partition stays small.
High values of the modularity indicate that there are more edges between nodes of the same group than between nodes of different groups: Thus, the higher the modularity, the better the partition.

With this quantity defined, a logical community detection strategy is \textit{modularity maximization}, for which several algorithms have been proposed~\cite{newman2004fast,duch2005community,aloise2010column,cafieri2011locally}.
One obvious handicap of modularity maximization is that finding the partition with highest modularity is a discrete combinatorial optimization problem~\cite{brandes2008modularity}. 
This is the discrete version of the curse of dimensionality presented in section~\ref{subsec:challenges}.
Effective heuristics thus have to be developed to perform modularity maximization.
Another drawback of modularity maximization is that it is prone to overfitting: Even in \er random graphs, high-modularity partitions exist and can be found~\cite{guimera2004modularity,reichardt2006statistical,lancichinetti2010statistical}.
This fact greatly weakens the claim that modularity is a good indicator of successful community detection. 
Finally, there is a fundamental resolution limit~\cite{resolutionLimit} that prevents the recovery of small-sized groups.

\subsubsection{Insights from statistical physics}
In~\cite{modbp}, Zhang and Moore introduce a community detection algorithm based on modularity that tackles the two first mentioned issues and propose a multiresolution strategy to overcome the third.
The algorithm, called \modbp, is of polynomial complexity with respect to $N$ (and thus fast), and is shown to not overfit, in the sense that it does not return high-modularity partitions for \er graphs.
This is achieved by treating modularity maximization as a statistical physics problem with an energy
\begin{align}
 H(\partition) = - m Q(\partition). 	\label{eq:modbpEnergy}
\end{align}
Finding the ground state of this system, \ie its state of minimal energy, is equivalent to maximizing the modularity of the network.
A further link to physics can be made by noting that the obtained model is a \textit{disordered Potts model}.
A Potts spin is a spin that can take $q$ different values. An Ising spin is a particular Potts spin with $q=2$. 
The Hamiltonian~(\ref{eq:modbpEnergy}) describes $N$ Potts spins interacting pairwise if they have the same value:
\begin{align}
  H(\partition) = -  \sum_{\langle ij \rangle} J_{ij} \delta_{l_i,l_j} 	\label{eq:modPotts}
\end{align}
with 
\begin{align}
 J_{ij} =  \indic(\langle i j \rangle \in \EE) - \frac{d_i d_j}{2 m}.	\label{eq:modPottsInteractions}
\end{align}
This system therefore presents great similarities with the SK model introduced in~\exref{ex:SK}.
The main differences are that Ising spins are replaced by Potts spins and that the couplings $J$ are not Gaussian random variables, but depend on the graph.
However, as the edge creation process is probabilistic, the couplings $J$ are random and therefore the model is disordered, like the SK model.

This similarity encourages us to consider that the phenomenology of both systems are similar, and in particular to predict the existence of paramagnetic, ferromagnetic and glassy phases.
With this analysis, modularity maximization corresponds to finding the state of lowest energy, \ie the equilibrium state at temperature $T=0$.
In the SK phase diagram of~\exref{ex:SK}, it is interesting to see that at $T=0$, the system is \textit{always in the spin glass phase}.
Let us remind that this glassy phase is characterized by a chaotic energy landscape with exponentially many local minima spread all over the space of configurations.
This picture perfectly agrees with the fact that modularity maximization of \er random graphs succeeds in finding many high-modularity partitions that are very different from one another.

This physical insight has led the authors of~\cite{modbp} to adopt an alternative strategy to modularity maximization,
which is to minimize the \textit{free energy} of the system with the Hamiltonian~(\ref{eq:modPotts}) at $T>0$.
In the SK phase diagram of~\exref{ex:SK}, we see that at high enough temperature, the system leaves the glassy phase, 
that is the cause of the problems of modularity maximization.
The starting point of the \modbp algorithm is therefore the Boltzmann distribution over partitions
\begin{align}
 \p(\partition ) = \frac{1}{\ZZ} e^{\frac{m Q(\partition)}{T}}. 	\label{eq:modbpBoltzmann}
\end{align}
As~\eqref{eq:posteriorSBM} in Bayesian inference of the \sbm, this high-dimensional probability distribution can be estimated using \bp.
The results of \modbp will be presented after an introduction to \bp in~\secref{sec:BP}.

\paragraph{Note:} As shown in~\cite{modbp}, performing community detection using~\eqref{eq:modbpBoltzmann} is 
equivalent to making Bayesian inference (though not Bayes optimal in general) of a generative model called \textit{degree-corrected} \sbm~\cite{PhysRevE.83.016107}.
This illustrates again the permeability between Bayesian inference and inference by cost function minimization and should be 
kept in mind: in that sense, using~\eqref{eq:modbpBoltzmann} is not completely model-free.  

\section{Belief propagation}
\label{sec:BP}
The belief propagation (\bp) algorithm~\cite{KschischangBP,PearlBP} was discovered independently in the fields of physics, coding and artificial intelligence for different purposes.
It allows to estimate (or sample from) a probability distribution over a high-dimensional space that takes the form
\begin{align}
 \p(\X) = \frac{1}{\ZZ} \prod_{\a=1}^M \fac_{\a} (\X_{\partial \a}),	\label{eq:bpProba}
\end{align}
where $\X = (\X_1, \cdots, \X_N)$ and each $\fac_{\a}$ represents a constraint involving a subset $\partial \a \subseteq \{1, \cdots, N\}$ of components of $\X$.
The \bp algorithm estimates a \pdf of this kind by passing messages $\messv$ from each variable $i$ to each of the constraints $\a \in \partial i$ it is involved in, and 
messages $\messf$ from each constraint $\a$ to each of the variables $i \in \partial \a$.

\begin{SCfigure}[1][h]
\includegraphics[width=0.5\textwidth]{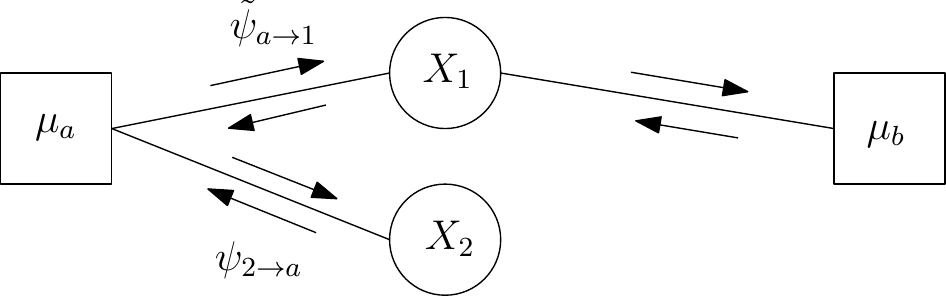}
 \caption[A factor graph]{A simple factor graph. Messages of type $\messv$ and $\messf$ are passed along each of the edges.
 This factor graph could for example represent the distribution \linebreak $P(\X_1,\X_2) = \delta(\X_1 - \X_2)\NN(\X_1; 0,1)$.}
 \label{fig:2_factorGraph}
\end{SCfigure}

The distribution~(\ref{eq:bpProba}) is usually visually represented by a so-called factor graph, composed of two types of nodes:
\begin{itemize}
 \item Variable nodes, represented as circles, that stand for individual components $\X_i$
 \item Factor nodes, represented as squares, that stand for the constraints $\fac_{\a}$.
\end{itemize}
An edge is present between the variable node $i$ and the factor node $\a$ when $i \in \partial \a$.
In this factor graph representation, a pair of messages $(\messv,\messf)$ is passed along each edge of the graph, as represented in~\figref{fig:2_factorGraph}.
More detailed introductions to \bp and interpretations leading to a better understanding of it can be found in~\cite{mezardMontanari,yedidia}.

\subsection{BP equations}
In the sum-product version of \bp, the messages are updated iteratively following the rule \mynote{BP equations}{1.3}
\begin{align}
 \messv_{i \to \a}^{t+1}(\X_i) &\propto \prod_{\b \in \partial i \backslash \a} \messf_{\b \to i}^{t} (\X_i) ,  	\label{eq:bp_v} \\
 \messf_{\a \to i}^{t+1}(\X_i) &\propto \int \prod_{j \in \partial \a \backslash i} \left( \dd \X_j \messv_{j \to \a}^{t}(\X_j) \right) \fac_{\a}(\X_{\partial \a}) . 	\label{eq:bp_f}
\end{align}
Note that the messages are probability distribution functions, hence the $\propto$ sign that indicates they have to be normalized.
The messages are often initialized at random, the only constraint being that they form a valid \pdf (\ie they are positive and sum to one).
After convergence of these equations (\ie when iterating them does not change their value anymore), the marginal distributions $\p(\X_i)$ is estimated by the \textit{belief} \mynote{Beliefs}{0.9}
\begin{align}
 \messv_{i}^t(\X_i) &\propto \prod_{\b \in \partial i} \messf_{\b \to i}^{t-1} (\X_i). 	\label{eq:bp_marginals}
\end{align}
Said differently, the marginals are \textit{fixed points} of the \bp iterations.
However, the \bp equations are only bound to converge to the correct marginals when the factor graph associated to the distribution is a \textit{tree} (by theorem).
Else, one speaks of \textit{loopy \bp}, as the factor graph contains loops.
There is no theorem guaranteeing the convergence of loopy \bp, nor that the fixed points of loopy \bp are the correct ones.
In many cases though, loopy \bp turns out to converge empirically and to give correct results, encouraging its use on factor graphs that are not trees.

An important quantity linked to \bp is the \textit{Bethe free entropy}. 
Given a set of marginals $\{ \{ \messv_i \},  \{ \messf_{\a} \} \}$, the Bethe free entropy is defined by \mynote{Bethe free entropy}{0.7}
\begin{align}
 \bfree &= \sum_{\a} \log \ZZ_{\a} + \sum_{i} \log \ZZ_{i} - \sum_{\a} \sum_{i \in \partial \a} \log \ZZ_{\a i}	,	\label{eq:bfree}	
\end{align}
where 
\begin{align}
 \ZZ_{\a} &=   \int \prod_{i \in \partial \a} \left( \dd \X_i  \messv_{i \to \a}(\X_i) \right) \fac_{\a}(\X_{\partial \a}),   \\
 \ZZ_{i} &=  \int \dd \X_i \prod_{\a \in \partial i} \messf_{\a \to i}(\X_i),  \\
 \ZZ_{\a i} &=  \int \dd \X_i \messv_{i \to \a}(\X_i) \messf_{\a \to i}(\X_i)  .
\end{align}

Derivating $\bfree$ with respect to the messages $\{ \{ \messv_{i \to \a} \},  \{ \messf_{\a \to i} \} \}$ gives back the \bp equations~(\ref{eq:bp_f}, \ref{eq:bp_v}) 
and is one way of deriving them. 
Therefore, the \textit{\bp fixed points are extrema of the Bethe free entropy}~\cite{mezardMontanari}. 

\subsection{Mod-bp}
Let us now write the \bp equations for~\eqref{eq:modbpBoltzmann} in order to obtain the \modbp algorithm.
Note that the calculations to obtain the Bayes optimal inference algorithm of the \sbm are very similar.

First of all, let us rewrite~\eqref{eq:modbpBoltzmann} under the generic form of~\eqref{eq:bpProba}:
\begin{align}
 \p(\partition) &= \frac{1}{\ZZ} \prod_{\ijlink} \underbrace{ e^{\beta \delta_{l_i, l_j} \left( \indic( \ijlink \in \EE) - \frac{d_i d_j}{2 m}\right) } }_{\fac_{\ijlink}(l_i,l_j)}	\label{eq:modbpProba} .
\end{align}
We see that the factor nodes of the corresponding factor graph are all the pairs $\ijlink$.
The factor graph is therefore very loopy, and \bp is not guaranteed to converge.
As each factor node $\ijlink$ has only two neighbouring variable nodes $i$ and $j$, we have $\partial \ijlink = \{ i , j\}$.
Furthermore, as the labels $\partition$ are discrete variables, the integral in~(\ref{eq:bp_f}) becomes a sum:
\begin{align}
 \messv_{i \to \ijlink}(l_i = l) &\propto \prod_{k \neq j} \messf_{\langle i k \rangle \to i}(l_i = l) , \label{eq:modbp_messv} \\
 \messf_{\ijlink \to i}(l_i = l) &\propto \sum_{l'=1}^q \messv_{j \to \ijlink}(l_j=l') \fac_{\ijlink}(l, l'), 	\label{eq:modbp_messf}
\end{align}
and we introduce the simplifying notation
\begin{align}
 \messv_{i \to \ijlink}^l &= \messv_{i \to \ijlink}(l_i=l).
\end{align}
Furthermore, as each factor node has only two neighbours, we can rewrite the \bp equations (\ref{eq:modbp_messv}, \ref{eq:modbp_messf})
in one equation of a single type of messages, $\messv_{i \to j}^l$:
\begin{align}
  \messv_{i \to j}^l &=  \frac{1}{\ZZ_{i \to j}} \prod_{k \in \partial i \setminus j} \left( 1 + \messv_{k \to i}^l (e^{\beta} - 1) \right) \prod_{k \neq i,j} \left( 1 + \messv_{k \to i}^l (e^{-\beta \frac{d_i d_k}{2m}} -1) \right) ,	\label{eq:modbp_bp}
\end{align}
which is the \bp equation for \modbp. The messages are initialized at random with the condition that they are positive and that for each pair $(i,j)$, $\sum_l \psi^l_{i \to j} =1$.
After convergence of the algorithm, the beliefs can be obtained by
\begin{align}
 \messv_i^l &= \frac{1}{\ZZ_i} \prod_{j \in \partial i} \left( 1 + \messv_{j \to i}^l (e^{\beta} -1) \right) \prod_{j \neq i} \left( 1 + \messv_{j \to i}^l (e^{-\beta \frac{d_i d_j}{2m}} -1 ) \right).
\end{align}
Note that the normalization constants $\ZZ_{i \to j}$ and $\ZZ_i$ are nothing but the sums over $l$ of the \textit{non-normalized} versions of $\messv_{i \to \ijlink}^l$ and $\messv_i^l$.

Just as in Bayesian inference of the \sbm, the estimated label of $i$ is obtained by the MAP estimate
\begin{align}
 \lh_i &= \argmax_l \messv_i^l,	\label{eq:MAPassignmentModbp}
\end{align}
from which one can calculate the retrieval modularity
\begin{align}
 \Qmap &= Q(\{ \lh \}). 	\label{eq:qmap}
\end{align}

As in the Bayesian \sbm inference, the authors of~\cite{modbp} detect the existence of two phases: a paramagnetic and a \textit{recovery} phase.
Additionally, they detect the existence of a spin glass phase. 
In~\cite{moiModularite}, I reveal the existence of further phases, as explained in the rest of the chapter. 

%
%
%
%

\section{Algorithmic phase transitions}
\label{sec:2_algoPT}
As in numerous statistical physics problems, the study of~\eqref{eq:modbpProba} leads to phase transitions at some given temperatures.
As previously noted, using the modularity as an energy function is similar to studying a Potts model in statistical mechanics~\cite{modPotts},
for which~\cite{qPottsTrans} has shown that a phase transition is always present.
\subsection{Paramagnetic, recovery and spin glass phase}
Zhang and Moore report the existence of three temperature ranges, in which the algorithm has a different qualitative behaviour.
\begin{itemize}
 \item At very low temperatures, the system is in a spin glass phase. 
In that phase, the problem is similar to modularity maximization: a chaotic energy landscape results in exponentially many local 
energy minima spread all over the space of partitions. In such a spin glass phase, \bp does not converge to a fixed point.
 \item At high temperature, the system is in a paramagnetic phase in which the \bp fixed point is the so-called \textit{factorized} or \textit{trivial} fixed point: 
$\forall (i,l), \messv_i^l = \frac{1}{q}$.
 \item In networks with communities, there can be an intermediate range of temperatures (the recovery phase), in which the algorithm converges 
to a nontrivial fixed point, from which group assignments can be obtained using~\eqref{eq:MAPassignmentModbp}.
\end{itemize}

\subsection{Model-based critical temperatures}
Modularity as a measure of goodness of a partition is appealing for real-world networks because it makes only the assumption 
of assortativity about the underlying edge creation process.
The drawback of this absence of model is that as a result, it is not clear how to choose the temperature $T$ at which to actually run \modbp.
Zhang and Moore analyze two generative models allowing to find useful characteristic temperatures.
\begin{itemize}
 \item For the configurational model, Zhang and Moore show that the phase transition between the spin-glass phase and the paramagnetic phase takes place at 
 \begin{align}
  T^* &= \left[ \log\left( \frac{q}{\sqrt{c}-1} +1 \right) \right]^{-1} ,	\label{eq:Tstar}
 \end{align}
where $c$ is the \textit{average excess degree}, calculated from the average degree $\langle d \rangle$ and 
the average square degree $\langle d^2 \rangle$, given by
\begin{align}
 c = \frac{\langle d^2 \rangle}{\langle d \rangle} -1	.	\label{eq:avExcessDegree}
\end{align}
\item In the \sbm with $q$ groups and $\epsilon = \pout / \pin$, Zhang and Moore 
show that \modbp is as successful as the Bayes-optimal algorithm, and that the phase transition between the paramagnetic 
and the recovery phase takes place at
\begin{align}
 T_R(\epsilon) &= \left[ \log \left( \frac{q \left[ 1 + (q-1) \epsilon \right]}{c(1-\epsilon) - \left[1 + (q-1) \epsilon\right]} +1  \right)  \right]^{-1} .	\label{eq:TR}
\end{align}
\end{itemize}
The recommendation of Zhang and Moore is to run \modbp at $T^*$, which seems to always lie inside the recovery phase.
On the other hand, $T_R$ cannot really be used, as it would require fitting the network to a stochastic block model and finding 
the parameters $q$ and $\epsilon$. 
As the \sbm is a bad model for real networks, this is not a good strategy.
However,~(\ref{eq:TR}) provides a useful upper bound for $T$:
\begin{align}
 T_0 = \left[ \log \left( \frac{q}{c-1} +1 \right) \right]^{-1}	,	\label{eq:Tzero}
\end{align}
which is the $\epsilon \to 0$ limit of~\eqref{eq:TR}. Indeed, above this temperature, even for \sbm networks with \textit{disconnected} components, 
\modbp will converge to the paramagnetic solution, and is therefore useless.

\subsection{Degenerate groups}
The rest of this chapter describes my contributions to the understanding of \modbp, as published in~\cite{moiModularite}.

In the paramagnetic phase, we said earlier that at the \bp fixed point, $\forall (i,l), \messv_i^l = \frac{1}{q}$. 
In practice, due to the numerical precision of the computer or incomplete convergence of the algorithm, 
there are small fluctuations around $\frac{1}{q}$. 
Due to these fluctuations, calculating a retrieval partition using~\eqref{eq:MAPassignmentModbp} is in general still possible
and would lead to random labels and thus a probably small, but non vanishing retrieval modularity $\Qmap$.

However, if the marginals were all strictly equal to $\frac{1}{q}$, then $\argmax_a \messv_i^a$ would be impossible to determine.
And in fact, the \textit{meaning} of the paramagnetic phase is that all groups are strictly equivalent, or \textit{degenerate},
which is to say that all nodes are in the same group and $\Qmap$ should therefore be exactly zero.
In order to obtain this, the algorithm has to check for degenerate groups before assigning 
a group to each node and assign the same ``effective'' group to nodes for which the maximization~\eqref{eq:MAPassignmentModbp} leads 
to different but degenerate groups.

This can be done by introducing a distance $d_{kl}$ between two groups $k$ and $l$:
\begin{align}
d_{kl} &= \frac{1}{N} \sum_{i=1}^N \left( \messv_i^k - \messv_i^l \right)^2.	\label{eq:dkls} 
\end{align}
If $d_{kl}$ is smaller than a chosen threshold $d_{\rm min}$, then we can consider that the groups $k$ and $l$ 
are degenerate and that they should not be distinguished. 
It generalizes the concept of degeneracy of all groups in the paramagnetic phase to any pair of groups.

\begin{SCfigure}[1][t]
 \centering
  \includegraphics[width=0.48\textwidth]{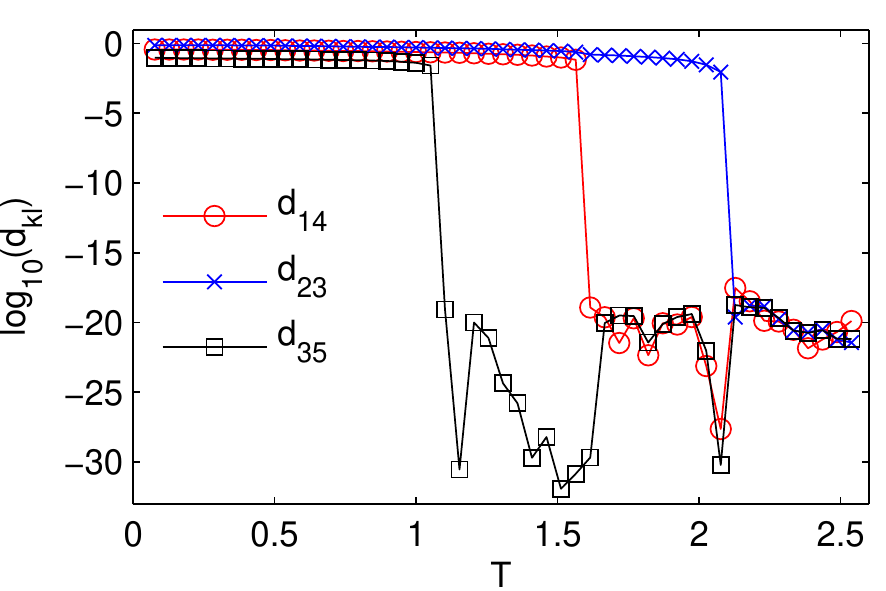}
\caption[Phase transitions of group degeneracies]{$d_{12}$, $d_{23}$ and $d_{35}$ as a function of temperature. In order to follow the groups at different temperatures, 
the temperature is increased step by step, and the messages are initialized with the final values they reached at the last temperature. The group distances $d_{kl}$ are like order parameters undergoing a phase 
transition at different temperatures, where they drop by more than ten orders of magnitude.
The dataset is ``political books'', run with $q=6$.}
\label{fig:2_distances}
\end{SCfigure}

An effective (or \textit{estimated}) \mnote{Effective groups $\hat{q}$} number of groups $\qh$ can then  be defined as the number of \textit{distinguishable} groups. 
We can define a mapping $\mapping$ between the $q$ groups used by the algorithm and the $\qh$ effective groups:
For each group $k$, $\mapping(k)$ is an integer between $1$ and $\qh$ representing one of the effective groups, and 
\begin{align}
 \forall(k,l), \mapping(k) = \mapping(l) \Leftrightarrow d_{kl} < d_{\rm min}. 	\label{eq:mapping}
\end{align}
With this mapping, we replace the group assignment procedure~\eqref{eq:MAPassignmentModbp} by
\begin{align}
 \lh_i = \mapping \left( \argmax_{l} \messv_i^{l} \right).	\label{eq:correctAssignment}
\end{align}
With this assignment procedure, $\Qmap$ is strictly zero in the paramagnetic phase.
Figure~\ref{fig:2_distances} shows that choosing a threshold $d_{\rm min}$ is meaningful because $d_{kl}$ undergoes a phase transition 
at which it sharply drops by several orders of magnitude.
The main finding is that group degeneracy is observed not only in the paramagnetic phase, but also inside the retrieval phase, 
in which case only a subset of groups are degenerate.
Figure~\ref{fig:2_distances} shows this for the popular network ``political books''~\cite{polbooks}, on which \modbp was run at different temperatures.

\section{Coexistence of phases}
\label{sec:2_phases}
\begin{figure}
\centering
   \includegraphics[width=0.24 \textwidth]{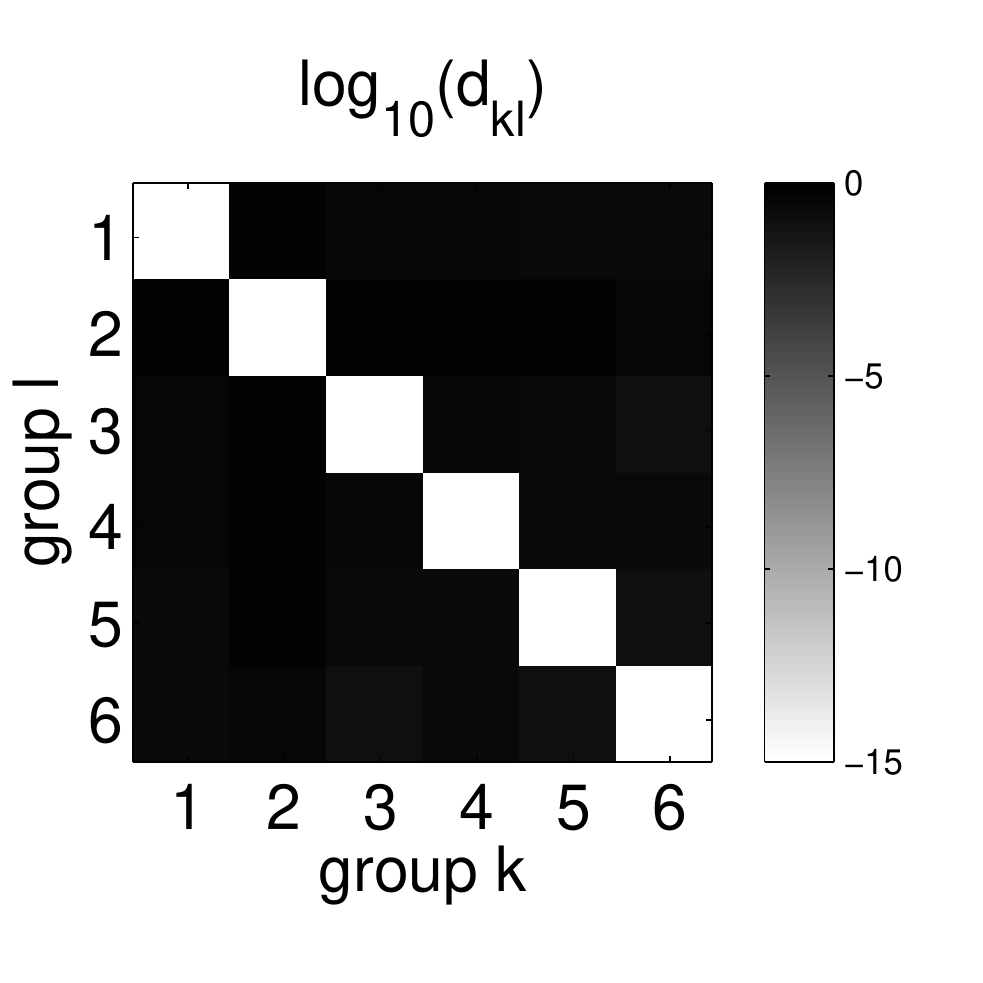}
   \includegraphics[width=0.24 \textwidth]{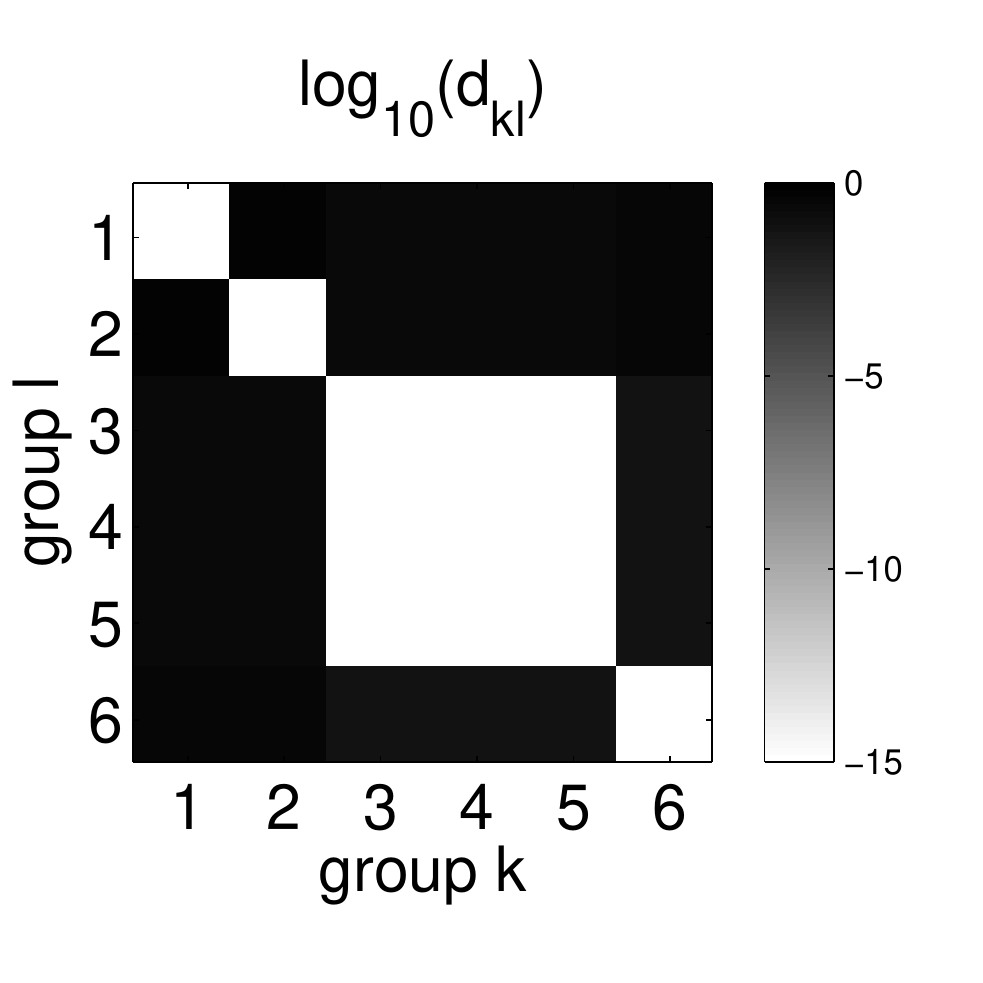}
   \includegraphics[width=0.24 \textwidth]{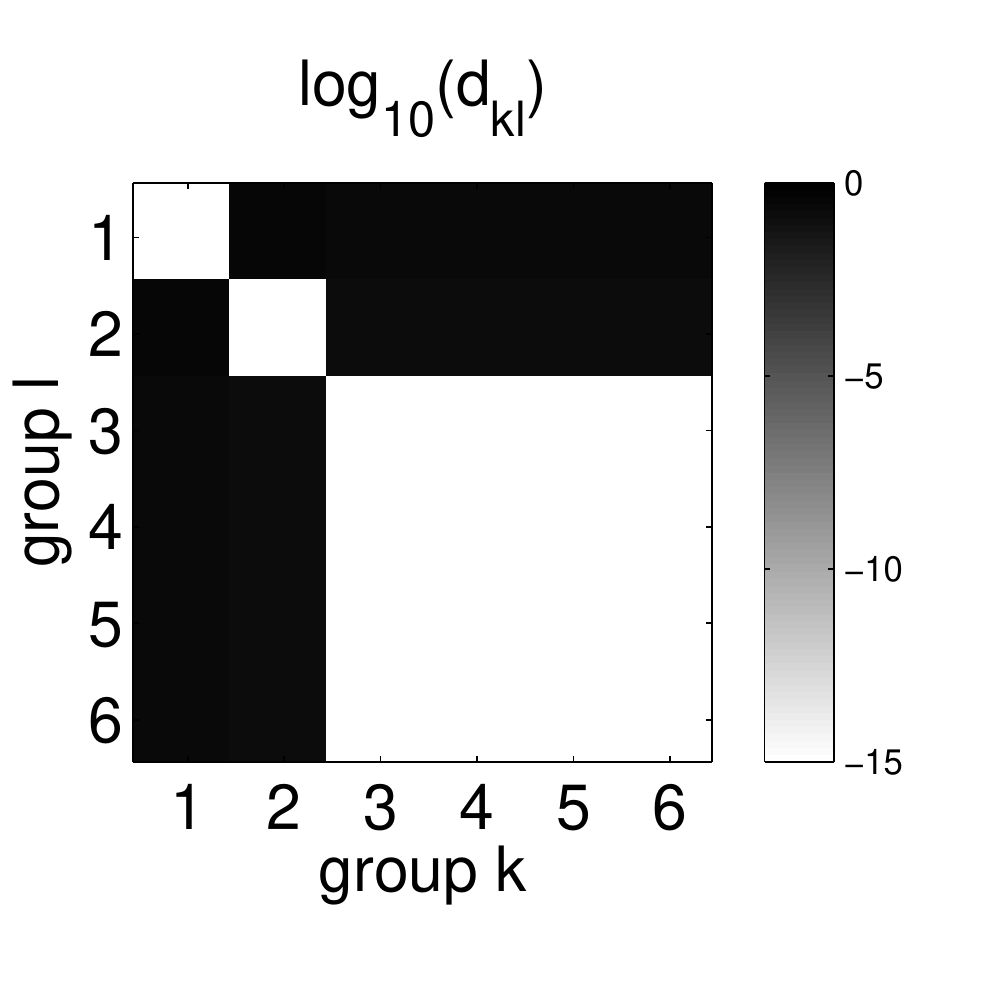}
   \includegraphics[width=0.24 \textwidth]{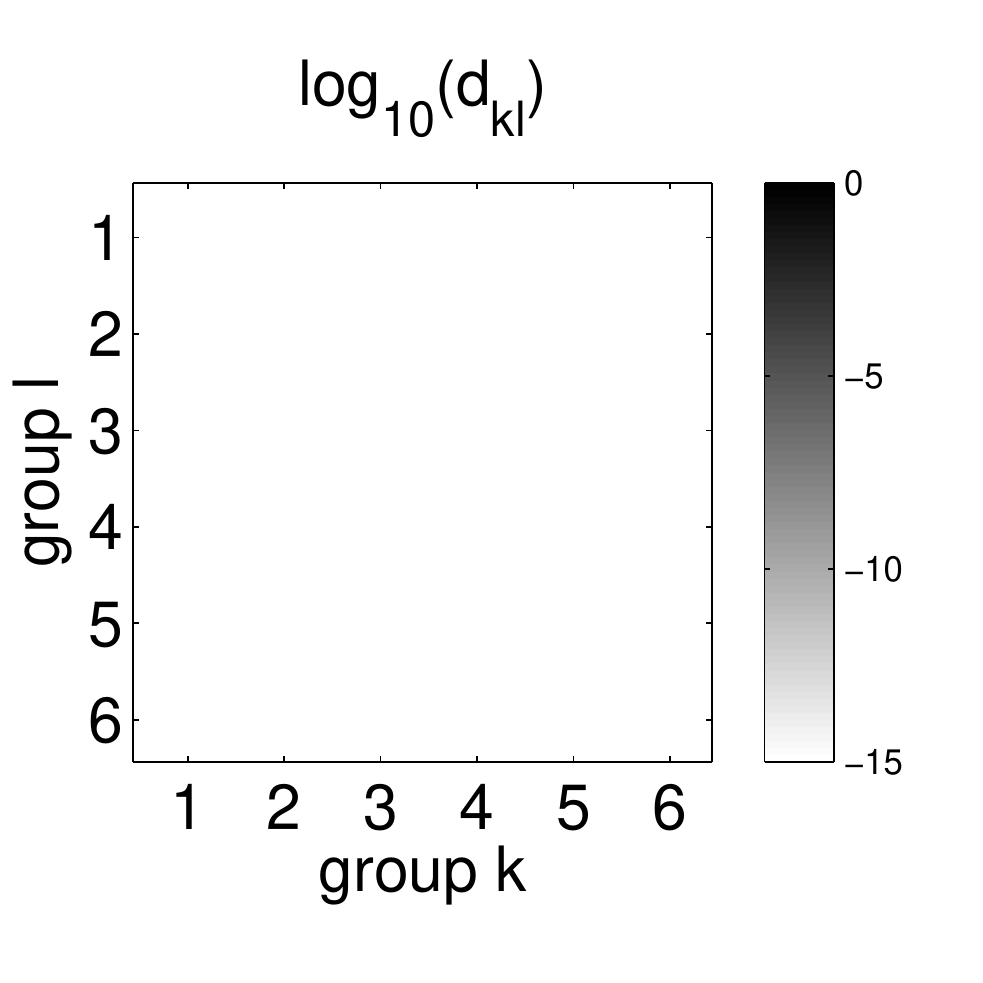}
\caption[Effective groups]{Matrices of distances between groups for different temperatures. The dataset is ``political books'', the algorithm is run with $q=6$ for $T= 0.26, 1.0, 1.28, 2.3 $ 
(from left to right). We observe the formation of a growing cluster of groups that are equivalent, allowing us to define a number of effective 
groups $\hat{q}$, that varies from $6$ at low temperature (left) to $1$ in the paramagnetic phase (right). 
Note that the area of the squares is not related to the number of nodes contained in the groups.}
\label{fig:2_groups}
\end{figure}

Thanks to the correct group assignment procedure in~\eqref{eq:correctAssignment}, one realizes that up to $q+1$ phases can exist 
for any network for which \modbp is run with $q$ groups: One for each $\qh \in [ 1 , q]$ plus a spin-glass phase.
Figure~\ref{fig:2_groups} shows this for the network ``political books''.

\begin{figure}
 \centering
 \begin{subfigure}{0.45\textwidth}
 \includegraphics[width=\textwidth]{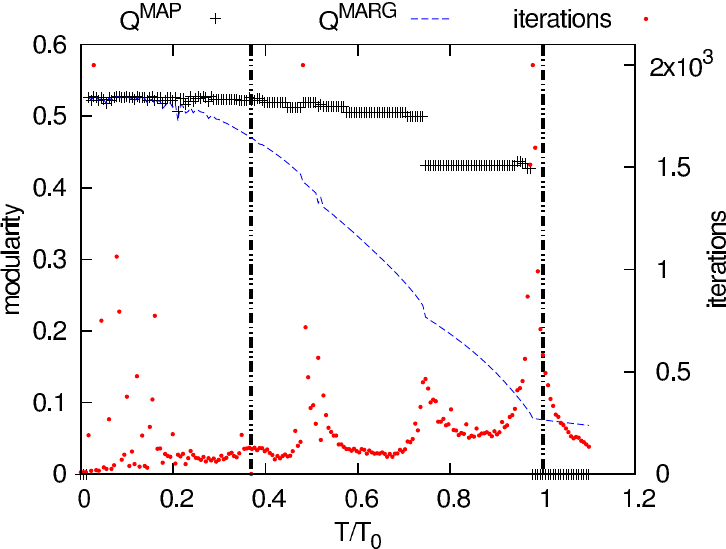}
 \caption{Modularity}
 \label{fig:2_polBooks_mod}
 \end{subfigure}
 \hspace{0.5cm}
 \begin{subfigure}{0.45\textwidth}
 \includegraphics[width=\textwidth]{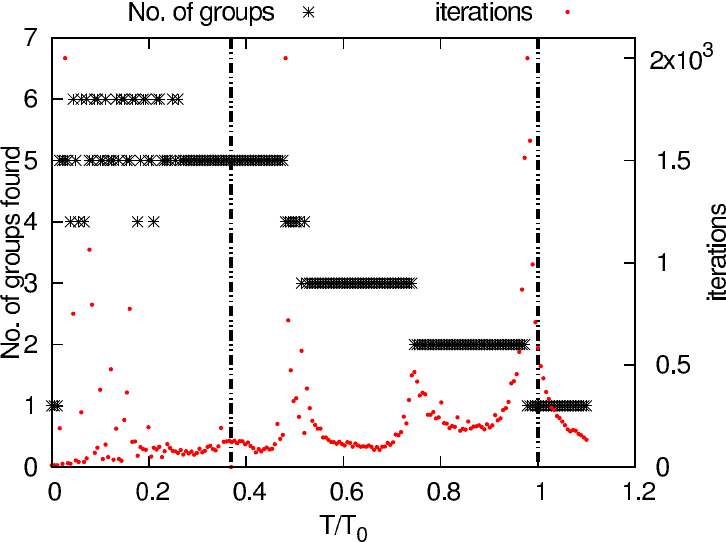}
  \caption{Number of groups}
 \label{fig:2_polBooks_groups}
 \end{subfigure}
 \caption[Phase transitions in ``political books'']{ Modularities and numbers of effective groups $\hat{q}$ obtained by sweeping a temperature range from $0$ to $1.2$  $T_0$ on the dataset ``political books'' with $q=6$.
The vertical lines indicate the positions of $T^*$ (left) and $T_0$ (right). 
Above $T^*$, the changes in $\hat{q}$ define quite homogeneous phases, separated by sharp transitions, where the number of iterations necessary to reach convergence increases greatly.
At low $T$, the phase is not homogeneous: depending on the starting conditions, $\hat{q}$ can be $4$, $5$ or $6$.
$Q^{\rm MAP}$ increases only minimally when $\hat{q}$ exceeds $3$, which agrees with the fact that $q^*=3$.
$Q^{\rm MARG}$ is an average modularity calculated from the BP marginals (\textit{without} the MAP estimate~(\ref{eq:MAPassignmentModbp}))~\cite{moiModularite}.}
\label{fig:2_polBooks}
\end{figure}

In this network, several phases coexist at lower temperature,
 whereas for higher temperatures, the phases exist in well-separated temperature intervals. 
 In the latter case, we can define ``critical'' temperatures $T_k$, separating a phase with $\qh = k$ from one with $\qh<k$.
 As can be seen on~\figref{fig:2_polBooks}, the number of iterations needed for \modbp to converge greatly increases around these critical 
 temperatures. 
 As noted previously, $T_0$ is a good reference temperature, and normalizing all temperatures by $T_0$ is a good way of 
 introducing a unified temperature scale that allows to compare critical temperatures of different networks and at different values of $q$.
 
 \subsection{Location of critical temperatures}
\begin{figure}
 \centering
 \begin{subfigure}{0.45\textwidth}
  \includegraphics[width=\textwidth]{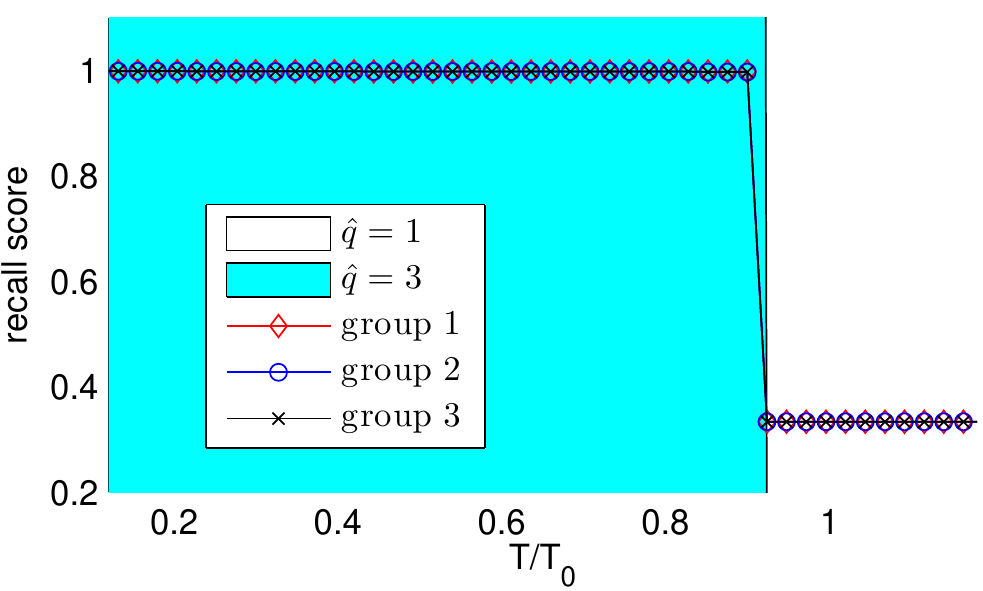}
  \caption{Degenerate $T_k$s.}
  \label{fig:2_degenerate}
 \end{subfigure}
 \begin{subfigure}{0.45\textwidth}
  \includegraphics[width=\textwidth]{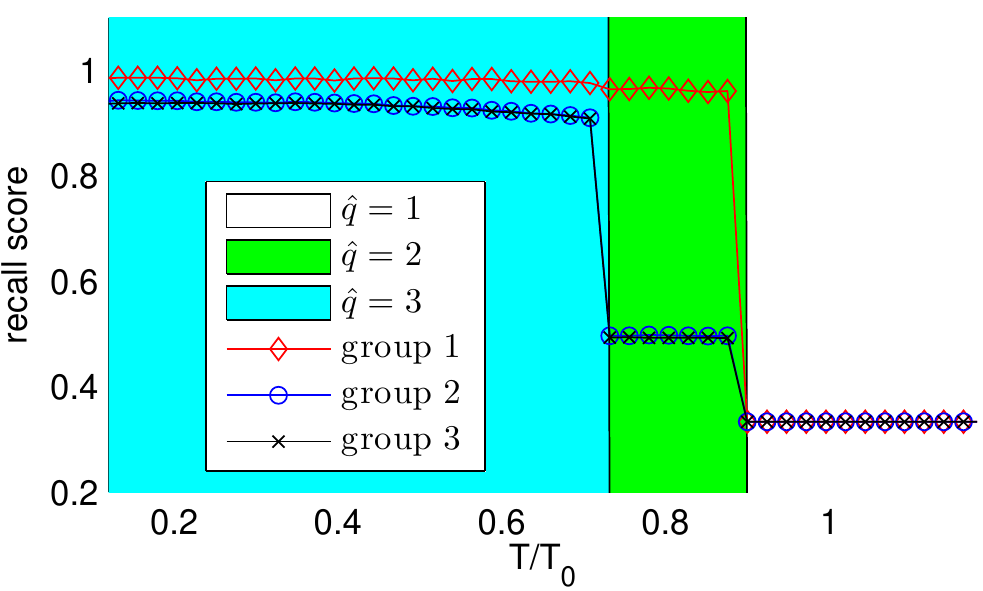}
  \caption{Distinct $T_k$s.}
  \label{fig:2_distinct}
 \end{subfigure}
 \caption[Degenerate phase transitions]{ Degeneracy of $T_k$'s on networks generated by the SBM with $N=5000$, $q^*=3$ groups and $c_{\rm out}=2$. \modbp was run with $q=3$.
  (a) All 3 groups have the same in-connectivity $c_{\rm in}=30$. There is no $\hat{q}=2$ phase because $T_1$ and $T_2$ are degenerate.
 (b) Group 1 has higher in-connectivity than the two others: $c_{11}=30$ whereas $c_{22}=c_{33}=15$. 
 $T_1$ and $T_2$ are distinct, and from the recall 
 scores we see that only group 1 is detected between $T_1$ and $T_2$, whereas groups 2 and 3 have an equally low recall score, 
 as in the partition given by the algorithm, they are merged to a single group. Below $T_2$, $\hat{q}=3$ and the algorithm 
 separates groups 2 and 3. The spin glass phase is not reached here.
 }
 \label{fig:2_degenerateTs}
\end{figure}

 In some cases, a subset of $n$ critical temperatures can be degenerate, in which case there is a phase transition betweel a phase with $\qh=k$ 
 and a phase with $\qh=k+n$.
 This is for instance the case for networks generated by the \sbm with~\eqref{eq:sbm}. 
 The picture (\figref{fig:2_degenerate}) then agrees with the description of three phases given in~\cite{modbp}.
 In contrast, the \sbm can be modified such that $\p_{rr} \neq \p_{tt}$ if $r \neq t$.
 The degeneracy of $T_k$'s is then lifted (\figref{fig:2_distinct}). 
 This figure also shows that, starting above $T_0$ and lowering the temperature, the groups are inferred in order of their strength.
 To show this, we use the \textit{recall score}, which allows us to see if one of the inferred groups corresponds well to a given real group.
 To quantify the similarity between a real group $G$ and an inferred group $\hat{G}_i$ that are not necessarily of the same size,
 we can use the Jaccard score (used in~\cite{hric2014community} for instance), defined by
 \begin{align}
  J(G, \hat{G}_i) = \frac{| G \cap \hat{G}_i |}{| G \cup \hat{G}_i |}.	\label{eq:Jaccard}
 \end{align}
 The recall score is the maximum of the Jaccard score:
 \begin{align}
  R(G) &= \max_i J(G, \hat{G}_i). 	\label{eq:recallScore}
 \end{align}
A recall score close to $1$ means that one of the inferred groups $\hat{G}_i$ is almost identical to group $G$.
Figure~\ref{fig:2_distinct} therefore shows that around $T/T_0 = 0.8$, the group with the biggest in-connectivity is nearly exactly returned by the algorithm,
whereas the two groups with lower in-connectivity are not. Only by further lowering the temperature does $\qh$ reach $3$, and all groups 
are correctly (though not perfectly) inferred.

\subsection{Running mod-bp with $q\neq q^*$}
In networks generated with the \sbm, the real number of groups $q^*$ is known and it is thus interesting to look at what happens when \modbp
is run with $q \neq q^*$.
The behaviour for $q=q^*$ is described in~\cite{modbp} and~\figref{fig:2_degenerate}.
If $q<q^*$ then \modbp cannot return $\qh=q$ and will merge some of the groups together to obtain $q$ groups.
The more interesting case is when $q$ is bigger than $q^*$.

First of all, it must be noted that as $q$ increases, the range of (normalized) temperatures of the spin-glass phase grows.
If $\epsilon$ is only slightly above the detectability threshold $\epsilon^*$~\cite{decelle,mossel2012stochastic}, then increasing $q$ can 
lead to a situation in which there is no recovery phase between the spin-glass phase and the paramagnetic phase.

However, we will focus on the case when $\epsilon$ is small enough for intermediate phases to be present.
As described previously, the phase transitions are degenerate if $\p_{rr}$ is the same for all groups.
Therefore, we generally observe only one intermediate phase, with $\hat{q}=q^*$. 
However, this is not always the case and \modbp can return partitions with different values of $\hat{q}$,
 depending on the initialization, similarly to what is observed in real networks in~\figref{fig:2_polBooks}.
 For the \sbm, two phenomena can often be observed, separately or simultaneously.
 \begin{enumerate}
  \item The first phenomenon is to have $\hat{q} = q^* +1$, with $q^*$ groups corresponding very well to real groups and a last 
 group containing a very small fraction of nodes.
 Depending on the initialization, this last group can even contain no node at all, in which case it can simply be discarded.
 This phenomenon is likely to come from the stochasticity of the \sbm and is also present for large networks with $10^5$ nodes.
 The modularity of such partitions with an additional group is usually equal or slightly higher than those found in 
 the $\qh=q^*$ phase of \modbp run with $q=q^*$, which explains why they are found. 
 On the other hand, I never observed more than one of these additional, almost empty groups, such that $\qh$ is always at most equal to $q^*+1$.
 
 \item The second phenomenon is that of distinct groups merging together in the retrieval partition, leading to $\hat{q}<q^*$.
 Such partitions have lower modularities than partitions with $\qh=q^*$ (found for different initializations),
 showing that the algorithm is not able to correctly maximize the modularity starting from any initialization.
 This is likely due to the existence of ``hard but detectable'' phases~\cite{decelle}, in which frozen variables
  cause algorithms to be stuck in suboptimal solutions.
  A simple way out of this problem is to run the algorithm several times with different initial conditions and to 
  select the configuration with the highest modularity.  
 \end{enumerate}
These two effects might coexist and produce retrieval partitions in which two of the groups are merged into a single one,
but an additional, almost empty group is present. In this case $\qh = q^*$, but the retrieval partition is incorrect. 
The existence of both of these phenomena should be considered as a warning on the reliability of \modbp for inference of the \sbm.
 
\subsection{Results on real networks}
\label{subsec:realNetworks}
\begin{figure}
\centering
\begin{subfigure}{0.49\textwidth}
\includegraphics[width=\textwidth]{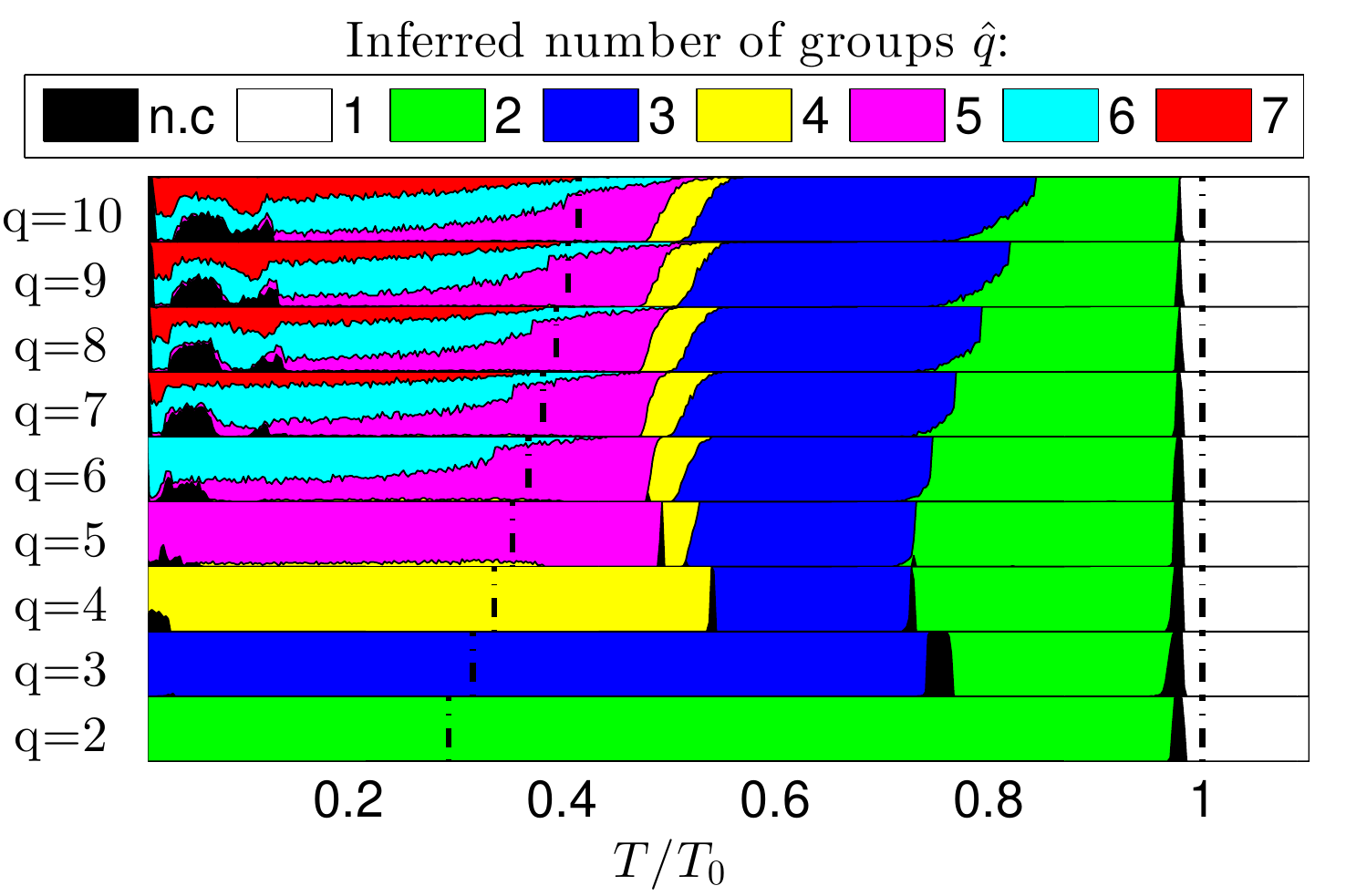}
\caption{Political books}
\label{fig:2_overview_books}
\end{subfigure}
\begin{subfigure}{0.49\textwidth}
\includegraphics[width=\textwidth]{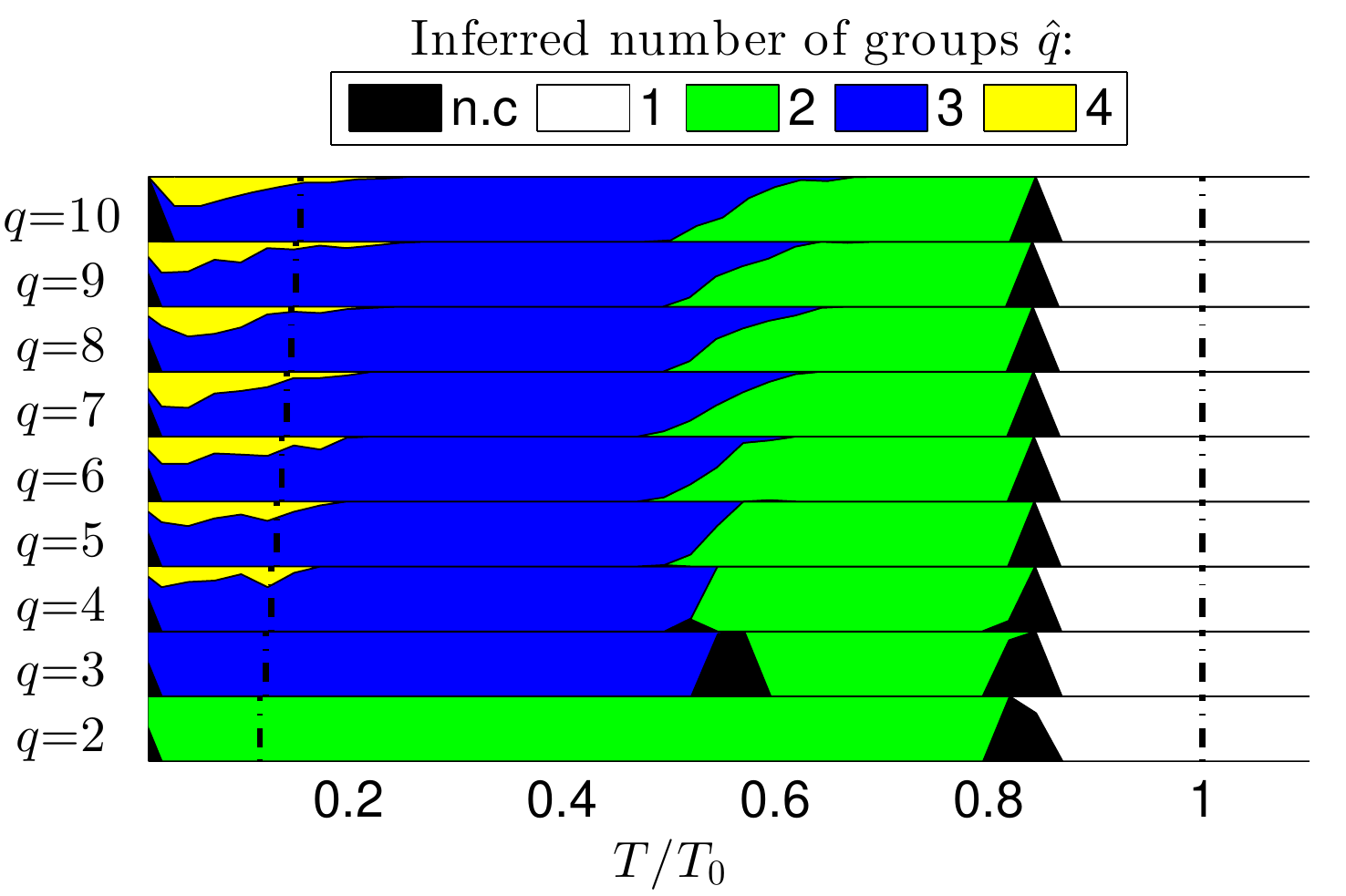}
\caption{Political blogs}
\label{fig:2_overview_blogs}
\end{subfigure}
\caption[Overview of political books and blogs]{These plots show the inferred number of groups $\hat{q}$ as a function of the normalized temperature $T/T_0$ and of $q$, for the ``political books'' and ``political blogs'' networks. 
The dotted lines mark $T=T^*$ (left line in each plot) and $T=T_0$ (right line in each plot). 
The ``n.c'' areas correspond to instances that did not reach the convergence criterion ($10^{-6}$) in 700 and 300 iterations respectively.
To take into account coexisting phases, the algorithm was run for $200$ (respectively $50$) different initializations at each temperature.
The position of $T_1$ is very stable across the different values of $q$, and is characterized by a diverging number of iterations.
The other critical temperatures $T_k$ are not always well defined due to overlaps between phases and to phase transitions becoming much less sharp; however, up to $q=4$, 
the phases stay well separated, with a clear divergence of the number of iterations at the phase boundaries.
Remarkably, the existence domains of each phase in terms of $T/T_0$ does not vary a lot with $q$.
 }
 \label{fig:2_overviews}
\end{figure}

For community detection on real networks, $q^*$ is in general unknown and there is no available ground truth.
From~\figref{fig:2_polBooks_groups} and the previous section, we know that \modbp can converge to partitions with different $\qh$  at the same temperature, depending on the initialization.
This motivates us to run \modbp several times at each temperature, which allows us to quantify the probability that a given $\qh$ is found for a given temperature $T$.
Fig~\ref{fig:2_overviews} shows the coexistence of phases in the ``political books''~\cite{polbooks} and ``political blogs''~\cite{blogs} datasets for different values of $q$.
The analysis made in these figures is similar to the one proposed in~\cite{multiresolution} for multiresolution community detection.

\begin{figure}
 \centering
 \begin{subfigure}{0.49\textwidth}
 \includegraphics[width=\textwidth]{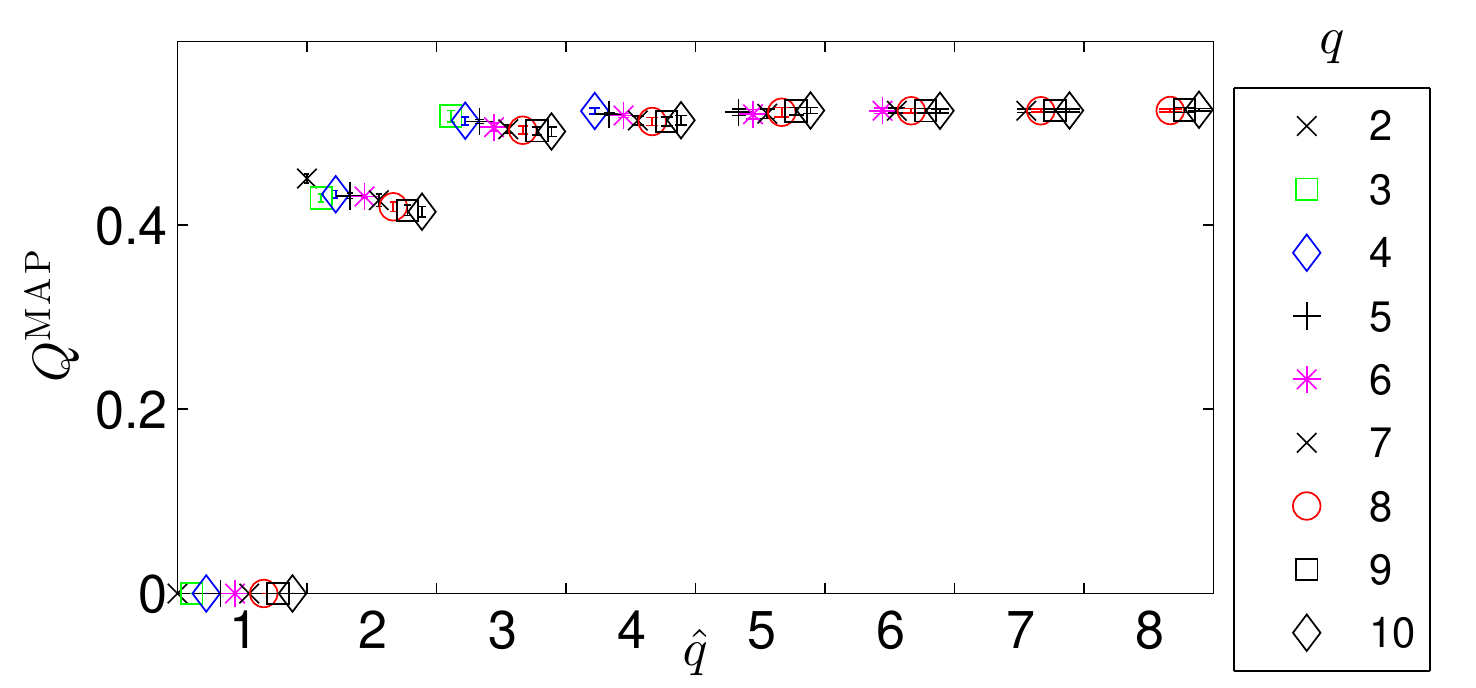}
 \caption{Political books}
\label{fig:2_overviewMod_books}
\end{subfigure}
\begin{subfigure}{0.49\textwidth}
 \includegraphics[width=\textwidth]{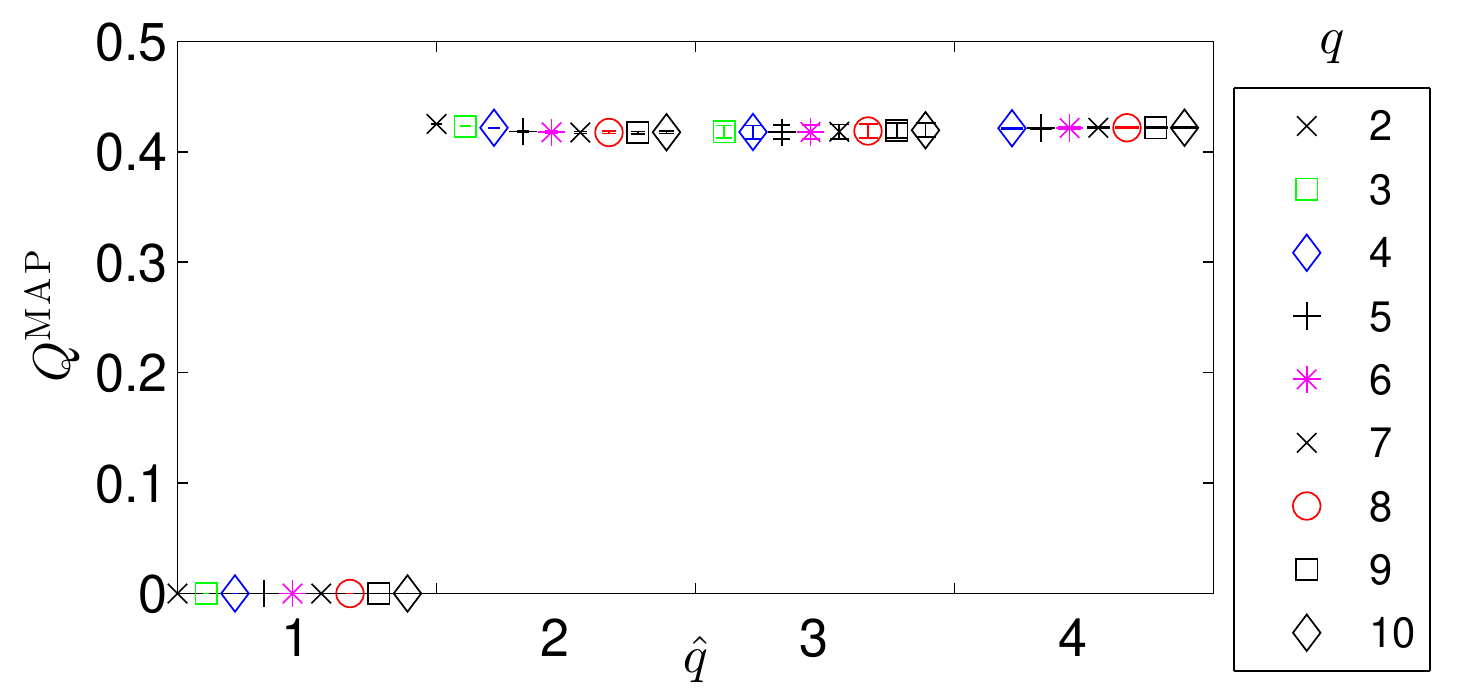}
 \caption{Political blogs}
\label{fig:2_overviewMod_blogs}
\end{subfigure}
 \caption[Retrieval modularities as a function of the effective number of groups]{$Q^{\rm MAP}$ as a function of $q$ and $\hat{q}$ for ``political books'' and ``political blogs'', using the same experimental results as in Fig.~\ref{fig:2_overviews}.
 Symbols represent the mean $Q^{\rm MAP}$ of all experiments with a given $q$ resulting in a given $\hat{q}$, along with 
 an error bar representing the standard deviation. 
 For ``political books'', $Q^{\rm MAP}$  only increases minimally for $\hat{q}>3$, and for ``political blogs'', it does not increase at all
 for $\hat{q}>2$, in concordance with the fact that $q^*=3$ and $q^*=2$ respectively for these networks. 
 }
 \label{fig:2_overviewsMod}
\end{figure}

These figures suggest that, at a given normalized temperature $T/T_0$, the results returned by \modbp only marginally depend on the chosen $q$ as long as $q>q^*$.
Moreover, we observe that within a phase with a given number $\qh$ of groups found, the partition $\partition$ only marginally depends on the temperature $T$.
Averaging over the several partitions found at different temperatures and with different initial conditions, we show in~\figref{fig:2_overviewsMod} that $\Qmap$ depends essentially on $\qh$ 
and only minimally on $q$. As in~\cite{modbp}, we consider that the largest $\qh$ leading to a significant increase of $\Qmap$ with respect to $\qh -1$ is a plausible estimate of $q^*$,
which agrees well with the commonly accepted ground truths of $q^*=3$ for ``political books'' and $q^*=2$ for ``political blogs''.

\begin{figure}
 \centering
  \begin{subfigure}{0.49\textwidth}
 \includegraphics[trim= 2cm 1.6cm 2cm 1.6cm, clip,width=\textwidth]{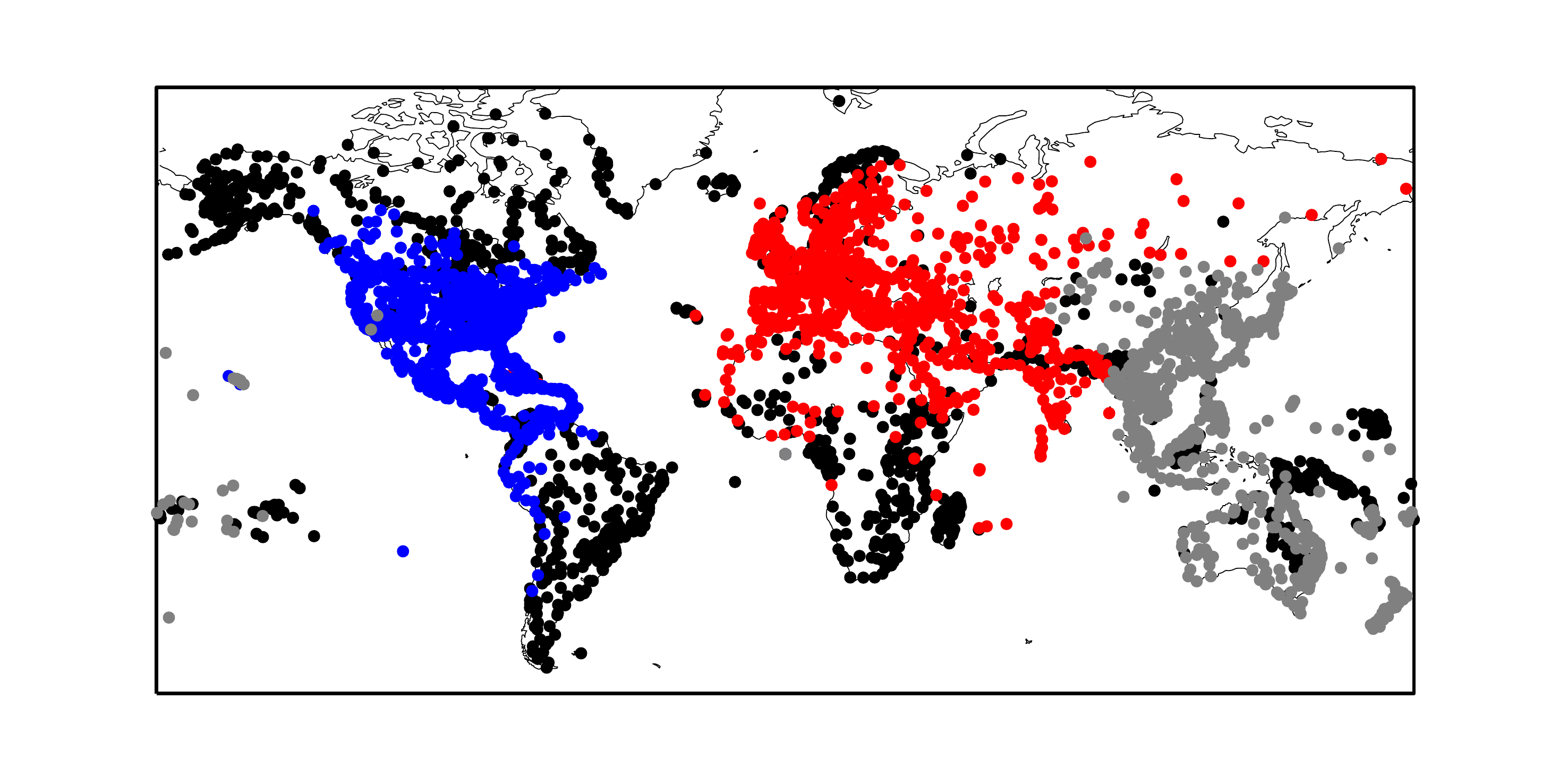}
 \caption{4 communities.}
 \label{fig:2_atn4}
  \vspace{0.5cm}
 \end{subfigure}
 \begin{subfigure}{0.49\textwidth}
 \includegraphics[trim= 2cm 1.6cm 2cm 1.6cm, clip,width=\textwidth]{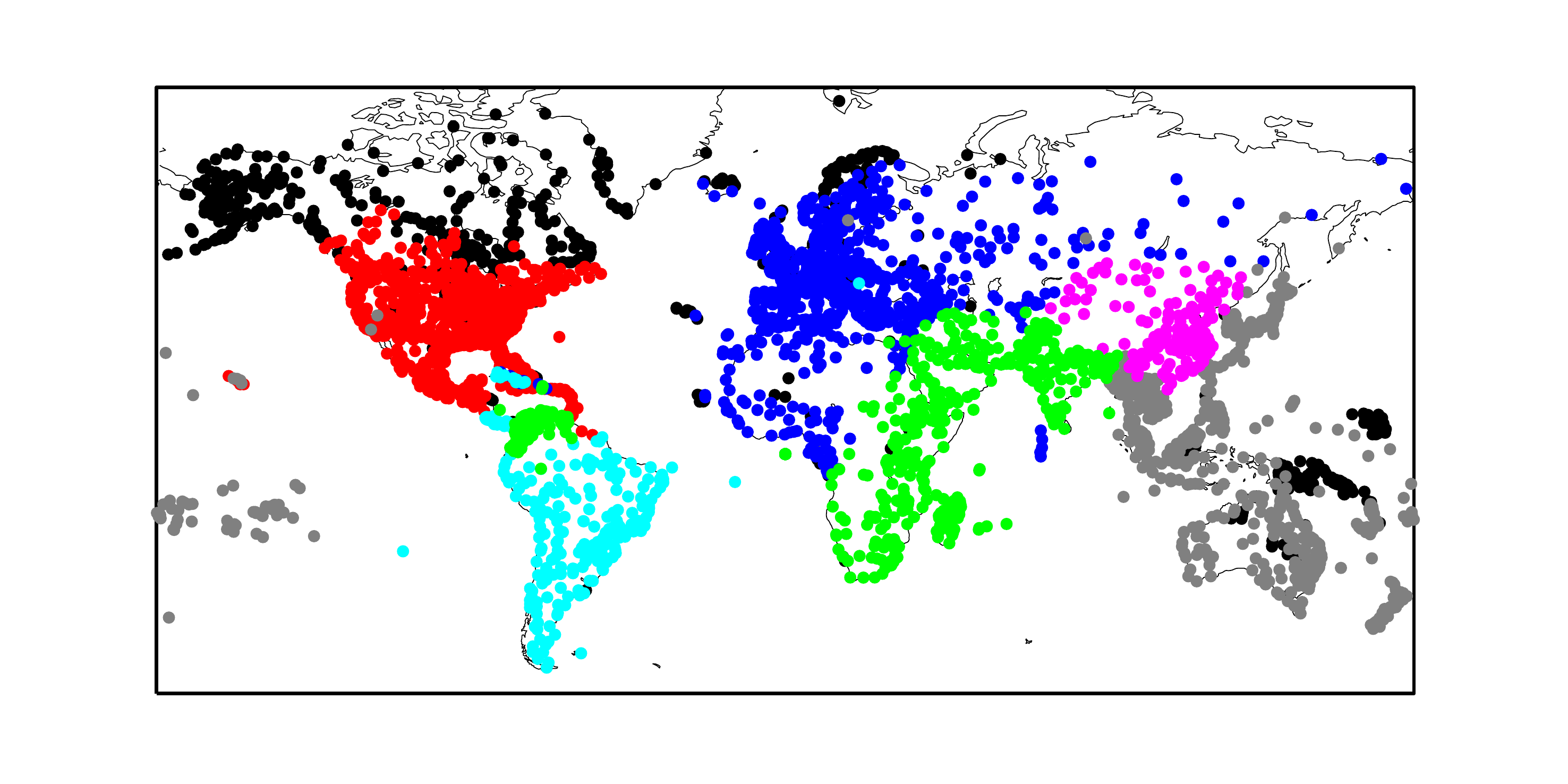}
  \caption{7 communities.}
 \label{fig:2_atn7}
 \vspace{0.5cm}
 \end{subfigure}
  \begin{subfigure}{0.49\textwidth}
 \includegraphics[width=\textwidth]{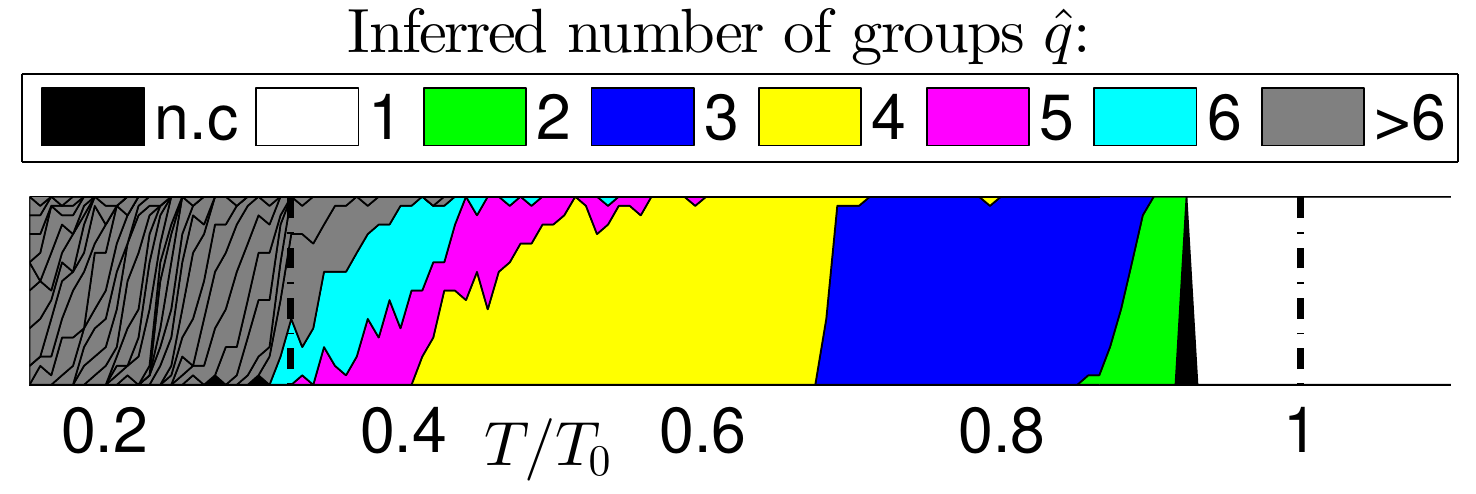}
  \caption{Coexistence of phases.}
 \label{fig:2_atnPhases}
 \end{subfigure}
  \begin{subfigure}{0.49\textwidth}
 \includegraphics[width=\textwidth]{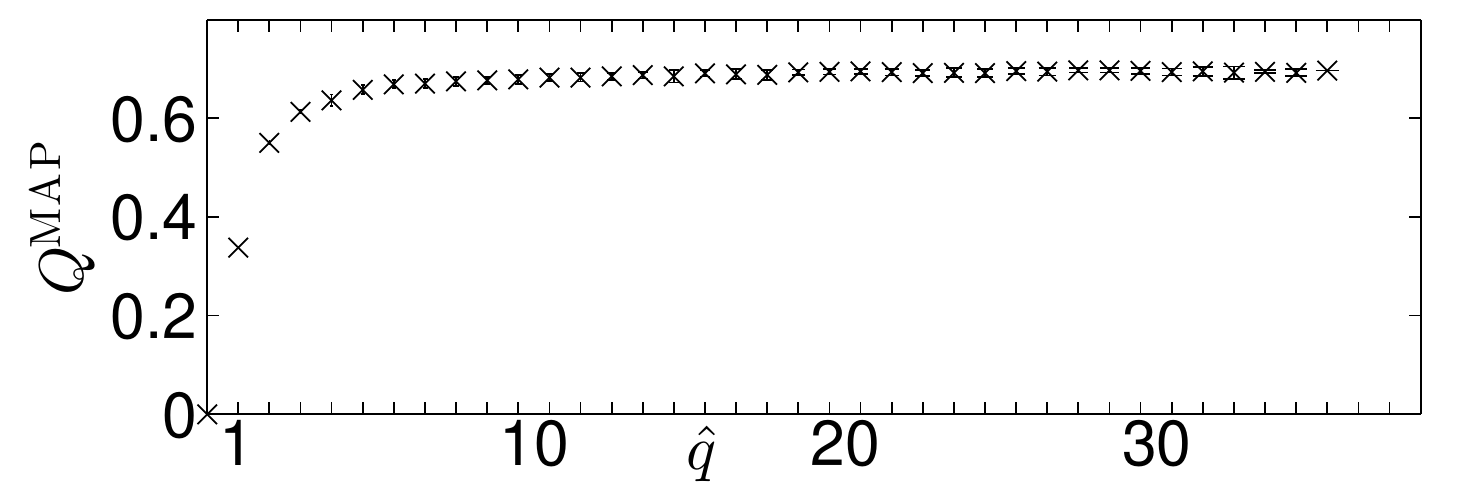}
  \caption{Increase of $\Qmap$.}
 \label{fig:2_atnMods}
 \end{subfigure}

 \caption[Overview of the air transportation network]{Clustering of cities in the ``air transportation network''~\cite{guimera2005worldwide} using mod-bp with $q=50$. 
 (a) and (b) correspond to $T/T_0=0.6$ and $T/T_0 = 0.35$ respectively. 
 The communities mainly correspond to geographical and geopolitical units. Increasing the number of communities, 
 substructures of bigger entities appear (\eg China separates from the east-asian cluster). 
 Lower-degree nodes are initially placed in the same group (\eg Alaska and Madagascar),
  but lowering $T$ lets new clusters appear (\eg South America).
  (c): As in Fig.~\ref{fig:2_overviews}, we show the frequency of retrieval configurations with different $\hat{q}$ as 
  a function of $T/T_0$ (on the base of 20 runs per temperature). While phases up to $\hat{q}\approx 7$ exist in broad ranges of temperatures, phases with higher $\hat{q}$
  exist on much narrower ranges and coexist with many different other phases, which makes it unclear which $\hat{q}$ is more 
  meaningful than others. (d):~$\Qmap$ increases only marginally with $\hat{q}$ for $\hat{q}\approx 7$.}
 \label{fig:2_airTransportation}
\end{figure}

To validate our results on a hierarchical network, we ran mod-bp on the ``air transportation network'', which is a network 
of cities in which an edge is present between each pair of cities connected by direct flights \cite{guimera2005worldwide,sales2007extracting}. 
A coarse-grained clustering results in a few communities of cities that are well 
 connected to each other. Each of these communities corresponds to geographical and geopolitical units that are clearly
  recognizable, which can be further subdivided in sub-communities. For example, the U.S and Mexico are two sub-communities of the 
  ``North America'' cluster.
  We ran mod-bp with $q=50$ for temperatures from $0$ to $1.2\times T_0$ and present the results in Fig.~\ref{fig:2_airTransportation}. As expected,
   the number of found communities increases with decreasing $T/T_{0}$, thus revealing substructures with increasing geographical 
   precision. Based on the modularity and the temperature range of the phases, $\hat{q}\approx 7$ seems to be a meaningful 
   number of communities. Further decreasing the temperature splits the communities into smaller ones, and individual countries 
   appear as single or even several communities.

\subsection{Discussion}
In addition to not requiring the knowledge of the generative model, a futher advantage of \modbp is that it has only two adjustable parameters, $T$ and $q$. 
However, for a given network, it is not clear how to choose them in order to obtain the optimal partition.
The recommendation of Zhang and Moore is to run \modbp at $T^*(q)$, defined in Eq.~(\ref{eq:Tstar}), for increasing values of $q$,
until it does not lead to any further significant increase in modularity.
Based on the experiments on synthetic and real networks presented in~\cite{moiModularite}, we conclude that an 
important additional step in this procedure is to calculate the effective number of groups $\hat{q}$ of each partition 
returned by the algorithm, which can be different from $q$. Furthermore, this phenomenon leads to a new rule for 
assigning a group to each node, given that some groups might be merged, which also affects the modularity.

Another possible way to proceed is to run \modbp with a large value of $q$, and sweep the temperature scale from $T_0(q)$ downwards.
As $T$ is lowered, the network is clustered into an increasing number of effective groups $\hat{q}$, and the found partitions have increasing modularities. 
Again, the procedure can be stopped once the modularity does not increase anymore in a significant way as $\hat{q}$ is increased.
 
For real networks, in which the generating process is in general not known and not as straightforward as in the SBM, the number of groups is in part let as a choice to the user.
In this case, running \modbp with a quite large value of $q$ and using $T$ as the parameter to search for the optimal partition seem both desirable and efficient.
To make the optimal choice, in addition to the value of the modularity of a partition with $\hat{q}$ groups, the range of temperatures where this $\hat{q}$ phase exists might indicate how relevant it is 
(as shown in Fig.~\ref{fig:2_overviews}).
In particular, if a $\hat{q}$ phase only exists on a narrow range of temperatures, then it is likely to be less important, because less stable with respect to changes in the model parameter ($T$ in the present case).
 
Furthermore, as seen on graphs generated by the SBM, it may occur that some group contains a very small number of nodes.
In this case, merging them with bigger groups will only slightly change the modularity and result into a more meaningful and stable partition.

\section{Conclusion}
This chapter treats the problem of community detection as a statistical physics problem.
Introducing a temperature, different phases are observed, that correspond to the ferromagnetic, paramagnetic and spin glass phases of a disordered physical system.
Understanding the characteristics of these phases is important in order to understand the phenomenology and the difficulty of community detection.
My contribution to the understanding of community detection is published in~\cite{moiModularite} and shows that the ferromagnetic phase is in fact subdivided into different phases. 
This is important to be aware of while performing modularity-based community detection and suggests a new multiresolution clustering strategy that was tested on the ``air transportation network''.

 
 \part{Linear and bilinear inference problems}
 \label{part:2}
 
 
 \makeatletter
\@openrightfalse
\makeatother
\chapter*{Notations}
From now on, I will use the following conventions.

\subsubsection{Acronyms and abbreviations}
\begin{tabular}{r l}
 \amp & Approximate message passing \\
 \awgn & Additive white Gaussian noise \\
 \bigamp & Bilinear generalized approximate message passing \\
 \bp & Belief propagation \\
 \calamp & Calibration approximate message passing \\
 \cs & Compressed sensing \\
 \dl & Dictionary learning \\
 \dmd & Digital micromirror device \\
 \eg & ``for example'' \\
 \iid & independent and identically distributed \\
 \ie & ``that is`` \\
 \glm & Generalized linear model \\
 \gamp & Generalized approximate message passing \\
 \gbm & Generalized bilinear model \\
 \lasso & Least absolute shrinkage and selection operator \\
 \mcs & Matrix compressed sensing \\
 MRI & Magnetic resonance imaging \\
 $\mse$ & Mean squared error \\
 $\nmse$ & Normalized mean squared error \\
 \pbig & Parametric bilinear generalized approximate message passing \\
 \pdf & Probability distribution function \\
 \prgamp & Phase retrieval \gamp \\
 RIP & Restricted isometry property \\
 SP & Saddle point \\
 \tap & Thouless Anderson Palmer
\end{tabular}

 \subsubsection{Ensembles}
 \begin{tabular}{ll}
 $\CC$ & set of complex numbers. \\
 $\CC^*$ & $\equiv \CC \setminus \{0\} $. \\
 $\Nintegers$ & set of natural numbers. \\
 $\RR$ & set of real numbers. \\
 $\RR^*$ & $\equiv \RR \setminus \{0 \}$\\
 $\cov{N}$ & set of symmetric, positive-definite $N \times N$ matrices with real coefficients.
\end{tabular}

\subsubsection{Operators}
\begin{tabular}{rl}
 $a^*$ & complex conjugate of $a$. \\
 $\mathbf{a}^{\top}$ & (complex) transpose of $\mathbf{a}$ .\\
 $\mathbf{a} \odot \mathbf{b}$ & elementwise product of $\mathbf{a}$ and $\mathbf{b}$ .\\
 $\mathbf{a}^2$ & unless stated otherwise, elementwise square of $\mathbf{a}$ .\\
 $\mathbf{a}^{-1}$ & inverse of $\mathbf{a}$ .\\
 $\mathbf{a}^{ \inv}$ & elementwise inverse of $\mathbf{a}$.
\end{tabular}

\subsubsection{Variables}
\begin{tabular}{rl}
 $a, \, A$ & scalar. \\
 $\mathbf{a}, \, \mathbf{A}$ & vector or matrix. \\
 $\one$ & vector or matrix of only ones. \\
 $\indic$ & identity matrix. \\
 $\vec{a}, \, \vec{A}$ & vector of unusual size.\\
 $\hat{a}, \, \hat{A}$ & mean/estimate of $a$. \\
 $\hat{\mathbf{a}}, \, \hat{\mathbf{A}}$ & mean/estimate of $\mathbf{a}$. \\
 $\bar{a}, \, \bar{A}$ & variance/uncertainty of $a$. \\
 $\bar{\mathbf{a}}, \, \bar{\mathbf{A}}$ & variance/uncertainty of $\mathbf{a}$. \\
 $a_i, \, [ \mathbf{a} ]_i$ & $i$-th component of $\mathbf{a}$ .\\
 $\mathbf{a}_i$ & vector or matrix indexed by $i$.
 \end{tabular}

\subsubsection{Functions}
\begin{align}
 \NN(x;\hat{x},\bar{x}) &= \frac{e^{-\frac{(x-\hat{x})^2}{2 \bar{x}}}}{\sqrt{2 \pi \bar{x}}}  & & \text{normalized Gaussian of mean $\hat{x}$ and variance $\bar{x}$.}  \nonumber \\
 \NN(\mathbf{x};\mathbf{\hat{x}},\mathbf{\bar{x}}) &= \frac{e^{-\frac{1}{2} (\mathbf{x}-\mathbf{\hat{x}})^{\top} \mathbf{\bar{x}}^{-1} (\mathbf{x}-\mathbf{\hat{x}}) }}{(2 \pi)^{N/2} \det(\mathbf{\bar{x}})^{\frac{1}{2}}}  & & \text{ multivariate Gaussian of mean $\mathbf{\hat{x}} \in \RR^N$ and covariance matrix $\mathbf{\bar{x}} \in \RR^{N \times N}$.} \nonumber\\
 \Cc(x;\hat{x},\bar{x}) &= \frac{e^{-\frac{|x - \hat{x}|}{\bar{x}}}}{\pi \bar{x}} & & \text{circular Gaussian of a complex variable $x$ with mean $\hat{x}$ and variance $\bar{x}$.} \nonumber \\
 \indic(x \in \XX) &= \left\{ \begin{array}{c c}
			    1 & \text{if } x \in \XX, \\
			    0 & \text{if } x \notin \XX. 
			    \end{array} \right. 		& & \text{indicator function.} \nonumber \\
&\delta(x) & & \text{Dirac $\delta$ function. \qquad \qquad \qquad \qquad \qquad \qquad} \nonumber \\
\Theta(x) &= \left\{ \begin{array}{c c}
                      1 & \text{if } x>0, \\
                      0 & \text{if } x\geq 0. \end{array} \right. & & \text{Heaviside step function} \nonumber.
\end{align}

\subsubsection{Norms}
\begin{align}
 || \mathbf{x} ||_2 &= \sqrt{ \sum_{ij} |x_{ij}|^2 } & & \text{$L_2$ norm of a real or complex vector or matrix.} \nonumber  \\
 || \mathbf{x} ||_1 &=  \sum_{ij} | x_{ij} |  & & \text{$L_1$ norm of a real or complex vector or matrix.} \nonumber \\
 || \mathbf{x} ||_0 &= \sum_{ij} \indic(x_{ij} \in \CC^*) & & \text{$L_0$ ``norm'' of a real or complex vector or matrix.} \nonumber \\
 || \mathbf{x} ||_* & & &\text{Nuclear norm of a matrix, equal to the sum of its singular values.} \nonumber
\end{align}

\subsubsection{Other}
\begin{align}
 f(x) &\propto g(x) & & \text{$f$ and $g$ are proportional up to a multiplicative constant that does not depend on $x$.} \nonumber \\
 a &= O(N) & & \text{$a$ scales as $N$, \ie there is a couple $(N^*,C)$ such that $\forall N>N^*, a \leq C N$.} \nonumber \\
 a &= O(1/N) & & \text{$a$ scales as $1/N$, \ie there is a couple $(N^*,C)$ such that $\forall N>N^*, a \leq C/N$.} \nonumber \\
 a &\equiv b & & \text{$a$ is equal to $b$ by definition.} \nonumber 
\end{align}

\newpage
\makeatletter
\@openrighttrue
\makeatother

 \chapter{Compressed sensing and generalizations} 
\label{chap:generalizedLinearModels}

In~\chapref{chap:communityDetection} I have shown, on the example of community detection, how tools and concepts from statistical physics could help 
in solving and understanding inference problems.
The \modbp algorithm, using belief propagation, undergoes a set of algorithmic phase transitions just as a physical system does.

This chapter introduces another broad class of inference problems called ``generalized linear models'', that can also be solved using \bp and 
for which different phases exist as well. This class of problems---along with their bilinear generalization (\chapref{chap:generalizedBilinearModels})---were
the main focus of my work.
In the context of compressed sensing, I introduce notations that are useful in all the inference problems I have studied.
I also show how to use the replica method to perform a theoretical analysis of an inference problem.
Finally, I show experimental and theoretical results for compressed sensing and quantized sensing.

My main contributions in the field of compressed sensing and generalized linear models are presented in~\chapref{chap:gampApplications}.

\section{Compressed sensing}
\label{sec:CS}
The idea behind compressed sensing (\cs) is the following:
Much of the digital data we acquire (pictures or music for example) can be reduced to a fraction of their initial size using compression algorithms.
The fact that compression is (nearly) lossless reveals that the uncompressed data contain no more information than their compressed version. 
In other words, the initial acquisition scheme of the picture is suboptimal, in the sense that much more data is acquired than what is necessary to store the picture in a compressed format. 
The idea of \cs is to change the acquisition process of signals, in order to acquire them in a ``compressed'' format in the first place.

\subsection{Setting}
\begin{figure}[h]
 \centering
 \includegraphics{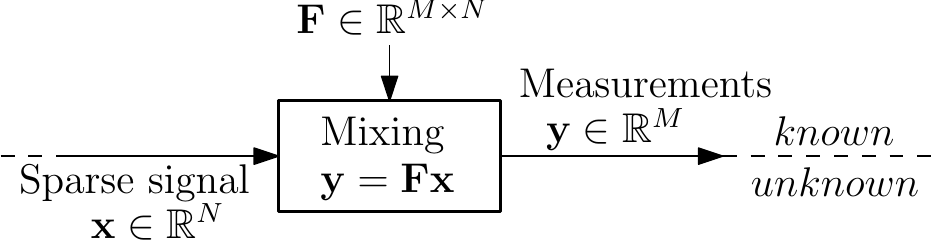}
 \caption[Noiseless compressed sensing]{Real-valued, noiseless compressed sensing. 
 Known variable are above the dashed line, variables to infer are below it. }
 \label{fig:3_CS}
\end{figure}
\subsubsection{The signal}
The fundamental concept in \cs is \textit{sparsity}.
The definition of a $\K$-sparse signal (with $\K \in \Nintegers$) is 
\begin{align}
 \xv \in \RR^N \text{ is } K\text{-sparse} &\Leftrightarrow ||\xv||_0 = \K \nonumber \\
 &\Leftrightarrow \xv \text{ has only } K \text{ non-zero components.}
\end{align}
We call \textit{sparsity rate} $\rho$ the ratio \mynote{Sparsity rate}{1}
\begin{align}
 \rho = \frac{\K}{N}.		\label{eq:rho_X}
 \end{align}
A signal $ \xv \in \RR^N$ can be compressed without loss if there is a basis of $\RR^N$ in which the signal is sparse.
\begin{align}
 \xv \text{ is compressible} \Leftrightarrow \exists \text{ a basis } \Psi \text{ such that } \xv = \Psi \xv' \text{ and } \xv' \text{ is sparse.}
\end{align}

\subsubsection{The matrix}
The setting of noiseless \cs is illustrated by~\figref{fig:3_CS}: linear measurements of a sparse signal are made with a measurement matrix $\Fv \in \RR^{M\times N}$.
The ratio of $M$ and $N$ is called the \textit{measurement rate}: \mynote{Measurement rate}{0.8}
\begin{align}
 \alpha = \frac{M}{N}.	\label{eq:alpha}
\end{align}
The goal of \cs is to recover the signal $\xv$ from the linear measurements 
\begin{align}
 \yv = \Fv \xv
\end{align}
in the regime where $\alpha<1$.
\cs is therefore the linear estimation problem of~\exref{ex:linearMeasurements} with the only difference that $\xv$ is sparse.
This sparsity assumption is, however, a very strong one, and therefore perfect recovery of $\xv$ can become possible despite having $M<N$.

Note that not all matrices are suitable for compressed sensing.
Many of the original papers focused on rigorous mathematical statements about the necessary properties of \cs matrices.
A good review of them can be found in the book~\cite{csBook}. 
Let us simply give an example to provide an intuition.
Consider for instance a signal that is sparse in the canonical basis.
If the rows of $\Fv$ are vectors of the canonical basis (and all different), each measurement component $y_{\mu}$ is one random component of $\xv$.
In order to be \textit{sure} to reconstruct the signal correctly \textit{no matter} which of its components are zeros, the only solution is to measure all components, as
each measurement carries information about one single signal component. 
Such a measurement matrix would therefore be completely unadapted to \cs, as it would require $M=N$.
On the contrary, a good measurement matrix carries information about \textit{all} of the signals components in \textit{each} of the measurements.

\textit{Random matrices} with independent identically distributed (\iid) entries are commonly used in \cs, as they satisfy the right conditions with high probability.
(In fact, nearly all matrices are suitable matrices for \cs~\cite{donohoCS}). 
Most of the time, we will therefore consider measurement matrices that are random matrices with \iid entries.

\subsubsection{Canonical setting}
Although \cs can be applied to any compressible signal, we will 
exclusively deal with signals that are sparse in the canonical base.
This simplifies the treatment of the problem, but note that one can always go to this setting: If $\xv$ is compressible in the base $\Psi$,
then $\yv = \Fv \xv$ is equivalent to $\yv = \Fv' \xv'$ with $\Fv'=\Fv \Psi$ and $\xv'$ sparse in the canonical base.

\begin{application}{MRI}
 \label{ex:mri}
 One application of \cs in the medical field is in magnetic resonance imaging (\mri)~\cite{lauterburMRI}.
 
 In \mri, the signal is a 2 or 3-dimensional image of a patient's body.
 Pixel intensities reflect the density of water in organs and can be weighted in different ways to reflect other properties, 
 as well as modulated by intravenous injection of contrast agents.
 As most natural images, the resulting images are approximately sparse in some domain. 
 In angiography, only blood vessels are visible, such that the image is sparse in the canonical basis.
 In other applications, the image can be sparse in some wavelet domain or in the gradient domain, using the total-variation transform.
 
 The measured physical quantity is the transverse magnetic field produced by the precession of proton spins in a strong 
 magnetic field $\vec{B}_0$.
 Additional, spatially varying magnetic fields allow to encode the position of each spin in its precession frequency.
 For a 1-dimensional object, the measurements $\yv$ are the Fourier transform $\FF(\xv)$ of the 1-d image $\xv$.
 
 An entire 2- or 3-dimensional image can be produced in an imaging sequence, in which parts of the Fourier space of the 
 image are acquired in successive measurements.		
 The more of them are available, the better the resolution of the final image.
 As the duration of each measurement is incompressible, high-resolution images require long imaging sequences.
 
 In \cs \mri~\cite{lustig}, random parts of the Fourier space are acquired and \cs is used to reconstruct the image using the sparsity assumption.
 This allows to obtain higher resolution images in applications such as cardiac imaging~\cite{otazo}, and to globally 
 reduce the length of exams, thus allowing more patients to be examined.
 
 \cs can also be used in x-ray tomography~\cite{ct-cs,gouillart}, allowing to reduce the patient's exposition to radiations.
\end{application}


\subsection{Geometric interpretation}
\begin{figure}[h]
 \centering
 \begin{subfigure}{0.4\textwidth}
 \includegraphics[width=\textwidth]{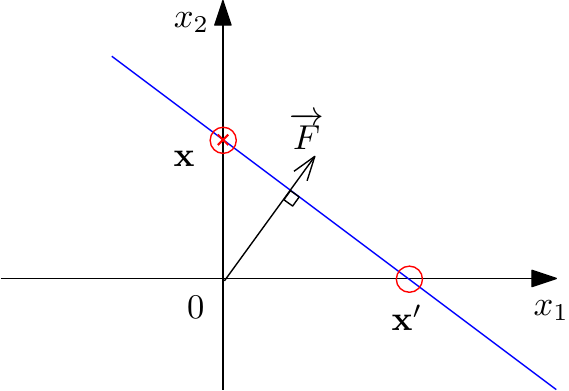}
 \caption{2 dimensions}
 \label{fig:3_geo1}
 \end{subfigure}
 \hspace{2cm}
 \begin{subfigure}{0.4\textwidth}
 \includegraphics[width=\textwidth]{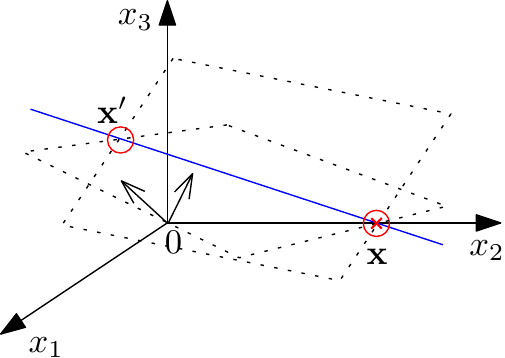}
  \caption{3 dimensions}
 \label{fig:3_geo2}
 \end{subfigure}
 \caption[Geometric insight into compressed sensing]{Geometric insight into noiseless \cs. 
 Each measurement $\y_{\mu}$ restricts the possible location of the solution to a hyperspace that is orthogonal to the measurement vector $\vec F_{\mu}$.
 Recovery is possible if the intersection of theses hyperplanes with the space of $K$-sparse vectors is unique.
In (a) and (b), these intersections are represented by red circles, while $\xv$ is represented by a red cross.
 (a) $N=2$, $M=1$. The measurement with $\vec F$ defines a space of possible solutions (blue line).
 The sparsity assumption reduces the number of possible solution to $2$. 
 Without the knowledge of which component of $\xv$ is zero, recovery is impossible: a second measurement is necessary.
 (b) $N=3$, and $M=2$ measurements restrict the space of possible solutions to the blue line, which is the intersection of the two (dotted) hyperplanes.
 The blue line contains two sparse solutions, but the signal, $\xv$, is sparser than $\xv'$.
 Note that if $\xv'$ were the signal, the probability that the blue line intersects the $x_2$-axis as it does would be vanishingly small, as its direction 
 is determined by the \textit{random} matrix $\Fv$.
 Therefore, $\xv$  is with high probability the sparsest of the possible sparse solutions . 
 }
 \label{fig:3_geo}
\end{figure}

In order to see how sparse signals can be recovered from an underdetermined linear system, one can rely on the geometric interpretation sketched in~\figref{fig:3_geo}.
The key point is that the ensemble of signals compatible with the measurements $\yv$ is a subspace of dimension $N-M$.
If $\xv$ is known to be $\K$-sparse, a sufficient condition for exact inference of $\xv$ to be possible is that this subspace contains 
only one $\K$-sparse element.

\subsection{Solving strategies}
\subsubsection{Oracle solver}
An \textit{oracle} \cs solver is a solver that knows which components of $\xv$ are non-zero.
Dropping the zero components of $\xv$, the linear system can be rewritten
\begin{align}
 \yv = \tilde{\Fv} \tilde{\xv},	\label{eq:oracleCS}
\end{align}
with $\tilde{\xv} \in \RR^{K}$ and  $\tilde{\Fv} \in \RR^{M \times \K}$. It is then obvious that $\tilde{\xv}$ and thus $\xv$ can uniquely be recovered as soon as $\tilde{\Fv}$ is invertible.

\subsubsection{Combinatorial optimization}
If the positions of the signal's zeros are unknown, a naive solving strategy would be to try out all the possible combinations of zero components and solve 
the corresponding reduced system~(\ref{eq:oracleCS}).
Because of the curse of dimensionality, this strategy becomes very rapidly unfeasable: for $N=100$ and $K=50$, there are 
$\binom{100}{50} \approx 10^{29}$ combinations to test.

This combinatorial optimization problem can be replaced by the following minimization problem:
\begin{align}
\xh = \argmin_{\xv} || \xv ||_0 \quad \text{ such that } \quad \Fv \xv = \yv.	\label{eq:l0min}
\end{align}
Under this form, no assumption is made on the exact sparsity of $\xv$. 
However, finding the sparsest possible $\xv$ verifying $\Fv \xv = \yv$ is the right strategy, as illustrated in~\figref{fig:3_geo2}.
The problem with~\eqref{eq:l0min} is that the cost function is hard to minimize, as the $L_0$ ``norm'' (which is not a norm properly speaking) is not a convex function.

\subsubsection{LASSO}
A natural way to deal with~\eqref{eq:l0min} is to use a convex relaxation of the cost function.
The $L_0$ ``norm''  can be replaced by the $L_1$ norm, which is convex. The condition $\Fv \xv = \yv$ can be enforced by a Lagrange multiplier,
which also allows to consider the case of noisy \cs,
\begin{align}
 \yv &= \Fv \xv + \boldsymbol{\xi}, \quad \text{with} \quad \boldsymbol{\xi} \sim \NN(\boldsymbol{\xi}; 0, \Delta \indic).	\label{eq:noisyCS}
\end{align}

This leads to the \textit{basis pursuit} or \textit{LASSO} (least absolute shrinkage and selection operator) problem~\cite{lasso}: \mynote{LASSO}{0.9}
\begin{align}
 \xhv = \argmin_{\xv} \frac{1}{2} || \yv - \Fv \xv||_2^2 + \lambda || \xv ||_1 ,	\label{eq:lasso}
\end{align}
in which the (inverse) Lagrange multiplier $\lambda >0$  can be adapted to the noise variance.
In principle, there is no guarantee that the problems~(\ref{eq:l0min}) and~(\ref{eq:lasso}) are equivalent.
The pioneering paper of \cs~\cite{tao} proves that under some conditions, they actually are in the limit $\lambda \to 0^+$.
This discovery has triggered a huge interest in \cs, as it revealed that the combinatorial problem~(\ref{eq:l0min}),
that was considered hopelessly untractable, could be solved by the simple convex minimization problem~(\ref{eq:lasso}). 

A sufficient condition for LASSO being equivalent to~\eqref{eq:l0min} is that $\Fv$ verifies the so-called 
\textit{restricted isometry property} (RIP)~\cite{CandesCS}.

\subsection{A \cs solver: Iterative thresholding}
Here we present a class of simple \cs solvers 
called iterative thresholding algorithms~\cite{iterativeThresholding1,iterativeThresholding2}.
Successive estimates $\xhv^{t}$ of $\xv$ are produced with the iterative rule \mynote{Iterative thresholding}{0.55}
\begin{align}
 \xhv^{t+1} = \fh \left( \xhv^{t+1} + \mu \Fv^{\top} \left( \yv - \Fv \xhv^{t} \right) \right)	\label{eq:ita}
\end{align}
in which $\mu$ is a step size  and $\fh$ is a nonlinear thresholding function that acts elementwise on its input.
Though usually written in this compact form, I present it in an expanded and annotated form in~\algoref{algo:iht}.
\begin{algorithm}
\caption{Iterative thresholding}
\label{algo:iht}
  \textbf{Main loop:} while $t<t_{\rm max}$, calculate following quantities:
  \begin{align*}
  \Zhv_{t+1} &= \Fv \xhv_{t}  	 & &\text{estimate of $\yv$ based on the current estimate $\xhv_t$}\\
  \ghv_{t+1} &= \yv - \Zhv_{t+1} & &\text{error (residual) of this estimate}\\
  \Xhv_{t+1} &= \xhv_{t} + \mu \Fv^{\top} \ghv_{t+1} & &\text{estimate of $\xv$ by gradient descent}  \\
  \xhv_{t+1} &= \fh (\Xhv_{t+1}) & &\text{new estimate of $\xv$ by sparsifying $\Xhv$}
  \end{align*}
\end{algorithm}

The two parameters that can be chosen in this algorithm are the step size $\mu$ of the gradient descent step
and the thresholding function $\fh$.
A natural choice for the latter is the hard thresholding function~\cite{iterativeThresholding1}: \mynote{Hard thresholding}{0.7}
\begin{align}
 H(\Xh_i) = \left\{ \begin{array}{cl}
	 \Xh_i & \text{if $\Xh_i$ is one of the $K$ biggest components of $\Xhv$,} \\
	  0 & \text{if not.}
	  \end{array}
          \right. \label{eq:hardThresholding} 
\end{align}
This thresholding function makes sure that the estimate $\xhv$ has always the right sparsity, by assuming that the smallest components of the estimate $\Xhv$ should be zeros.
Note that this choice of thresholding function also assumes that the exact sparsity $K$ of the signal is known, which is not the case in general.

Another choice of thresholding function is the \textit{soft thresholding function}, defined by \mynote{Soft thresholding}{1}
\begin{align}
 S_{\lambda} (\Xh_i) = \left\{ \begin{array}{cl}
                      \Xh_i + \lambda & \text{if $\Xh_i< - \lambda$,} \\
                      0 & \text{if $-\lambda \leq \Xh_i \leq \lambda$,} \\
                      \Xh_i - \lambda & \text{if $\lambda < \Xh_i$.}
                    \end{array} \right.		\label{eq:softThresh}
\end{align}
With $\fh = S_{\lambda}$, the resulting \textit{iterative soft thresholding} algorithm can be shown to correctly solve the LASSO problem~(\ref{eq:lasso})
in certain regimes, and can be made faster by good choices of step sizes $\mu$, varying across iterations~\cite{fista}.

In the rest of \mythesis, I will focus on probabilistic inference rather than minimization problems such as LASSO. 
However, the basic structure of all algorithms presented in the rest is the same as the structure of~\algoref{algo:iht}: 
iterative estimates of the different basic variables of the problem, obtained 
by linear combinations of previous estimators or by applying nonlinear thresholding functions to them.

\section{Generalized linear models}
\label{sec:GLM}
\begin{figure}[h]
 \centering
 \includegraphics[width=\textwidth]{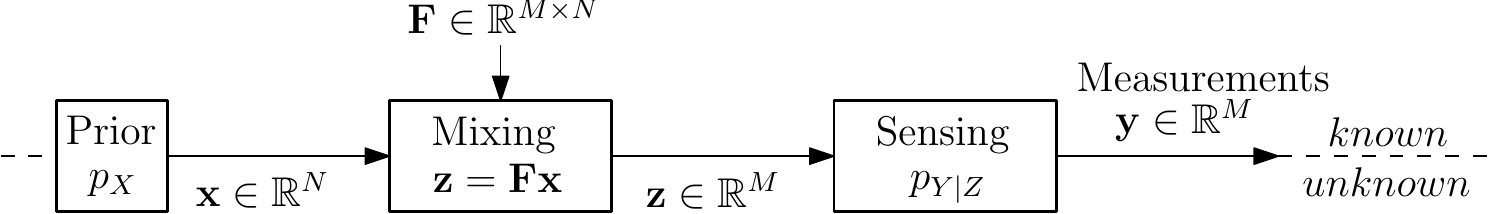}
 \caption[Generalized linear models]{A generalized linear model. The mixing step is followed by a sensing step, that acts componentwise on $\zv$ to give the measurements $\yv$.
 Noisy \cs is the simplest example of a generalized linear model.}
 \label{fig:3_GLM}
\end{figure}

The noiseless \cs problem presented in the previous section is a special case of the broader class of \textit{generalized linear models} (\glm), 
whose general setting 
is illustrated by~\figref{fig:3_GLM} 
\subsection{General setting}

In the most general setting, an unknown signal $\xv$ is multiplied by a known matrix $\Fv$ in a mixing step, producing an intermediary variable $\zv$.
In the subsequent sensing step, a \textit{sensing channel} leads to the measurements $\yv$.
The sensing channel always acts \textit{componentwise} on $\zv$: components of $\zv$ are never mixed in the sensing step.
In other words, the distribution $\py$ is separable:
\begin{align}
 \py(\yv | \zv) &= \prod_{\mu=1}^M \py^{\mu}(\y_{\mu} | \z_{\mu}).	\label{eq:separableChannelZERO}
\end{align}

\subsection{Additional conditions}
\label{sec:additionalConditions}
We consider a simplified setting with further conditions on the signal, the matrix and the sensing channel.
\subsubsection{The signal}
As we will follow a probabilistic approach to inference for \glms, we describe the signal by a prior distribution $\px$.
Unless stated otherwise, $\px$ has a variance of order $1$, such that typical components of $\xv$ have a magnitude of order one, and is separable, \ie
\begin{align}
 \px(\xv) &= \prod_{i=1}^{N} \px(x_i).	\label{eq:separablePx}
\end{align}
Note that the components are chosen to be \iid but that the generalization to the case in which they are only independent is easy.
However, independence of the components is an important assumption which is less straightforward to relax. 

\subsubsection{The matrix}
As in \cs, we will usually consider random measurement matrices with components that are \iid as matrices sampled this way 
satisfy the RIP with high probability. Unless stated otherwise, we consider that the matrix is drawn from a distribution
\begin{align}
 \pf(\Fv) = \prod_{i,\mu} \NN(\F_{\mu i}; 0 ,\frac{1}{N})	\label{eq:F_distrib}
\end{align}
and that it is exactly known.
Note that the variance of $\frac{1}{N}$ ensures that $\z_{\mu} = \sum_{i=1}^{N} F_{\mu i} x_i$ has a magnitude of order $1$.

\subsubsection{The sensing channel}
We will focus on the case in which the channel is the same for every measurement:
\begin{align}
 \py(\yv | \zv) &= \prod_{\mu=1}^M \py(\y_{\mu} | \z_{\mu}).	\label{eq:separableChannel}
\end{align}
As for the signal distribution, this condition can be easily relaxed to the more general case of~\eqref{eq:separableChannelZERO}.

\begin{example}{Noisy \cs}
\label{ex:noisyCS}
 The simplest example of a generalized linear model is noisy \cs, in which
 the sensing channel is an AWGN channel: \mynote{AWGN channel}{0.5}
\begin{align}
 \py(y|z) &= \NN(y;z,\Delta),	\label{eq:awgnChannel}
\end{align}
with noise variance $\Delta$. 

In a Bayesian approach, the sparsity of the signal is modeled by a sparse prior: \mynote{Sparse prior}{0.75}
 \begin{align}
  \px(x) &= (1-\rho) \delta(x) + \rho \xdist(x),		\label{eq:sparsePrior}
 \end{align}
where $\xdist$ is the distribution of the non-zero components of $\xv$.
We will often take $\xdist$ to be a Gaussian, which leads to the Gauss-Bernoulli distribution \mynote{Gauss-Bernoulli prior}{0.35}
\begin{align}
 \px(x) &= (1-\rho) \delta(x) + \rho \NN(x;0,1).	\label{eq:gaussBernouilli}
\end{align}
Taking $\Delta=0$, we recover the noiseless \cs setting.
\end{example}

\subsection{GAMP}
In~\cite{amp}, the authors proposed an inference algorithm for noisy compressed sensing based on belief propagation and named \textit{approximate message passing} (\amp).
This algorithm was generalized in~\cite{gamp} to \glms and called \textit{generalized approximate message passing} (\gamp).
A very similar algorithm was previously proposed in~\cite{kabaGAMP, kabaCDMA} along with a theoretical analysis, but had not drawn
attention at the time.
Being a probabilistic inference algorithm, \gamp's starting point is the posterior probability distribution obtained from Bayes' formula:
\begin{align}
 \p(\xv | \yv, \Fv) &= \frac{1}{\ZZ(\yv,\Fv)} \px(\xv) \int \dd \zv \py(\yv|\zv) \delta(\zv - \Fv\xv) 		\nonumber	\\
  &= \frac{1}{\ZZ(\yv,\Fv)} \prod_{i=1}^N \px(x_i) \prod_{\mu=1}^M \left( \int \dd \z_{\mu} \py(\y_{\mu}|\z_{\mu}) \delta(\z_{\mu} - \sum_{i=1}^N F_{\mu i} x_i) \right) . \label{eq:gampPosterior}
\end{align}
As \gamp treats real, continuous signals, it uses the MMSE estimator~\eqref{eq:MMSE}.
The probability distribution~(\ref{eq:gampPosterior}) is high-dimensional and \gamp estimates it using belief propagation. 


\subsubsection{From \bp to \tap}
\label{sec:bp_to_tap}
 \gamp is presented in~\algoref{algo:gamp}: Its full derivation can be found in~\cite{gamp} and in~\secref{sec:vectorial_gamp} (in a slightly more general setting). 
It is obtained in three steps:
\begin{enumerate}
 \item First, the \bp equations corresponding to~\eqref{eq:gampPosterior} are written.
 In  the case of \modbp for community detection (\chapref{chap:communityDetection}), $\xv$ was a discrete variable with $q$ values, and therefore a finite set of messages was introduced.
 In the present case, the messages $\{ \messv_{i \to \mu}(x_i) , \messf_{\mu \to i}(x_i) \}$ are distributions of continuous variables.
 Under this form, the \bp equations are untractable.
 
 \item In the $N \to \infty$ limit, the messages $\{ \messv_{i \to \mu}(x_i) , \messf_{\mu \to i}(x_i) \}$ can be expressed as functions of Gaussians, 
 parametrized by one mean and one variance per message. These means and variances are the messages passed and updated in approximate message passing.
 Means can be seen as estimators and are noted with hats ($\Xh, \xh$), whereas variances can be seen as uncertainties and are noted with bars ($\Xb,\xb$).
 Upper-case and lower-case quantities are estimators/variances of the same variable: as in~\algoref{algo:iht}, lower-case estimators take into account 
 the previous upper-case estimators \textit{and} the additional knowledge coming from the prior (for $\xv$) or the measurements (for $\zv$).
 The term ``approximate'' in \amp comes from the fact that this parametrization of the messages with Gaussians is exact only in the $N \to \infty$ limit.
 However, this parametrization relies on the central limit theorem, which is approximately verified even for finite values of $N$.
 Under this form, \amp can be implemented by iteratively updating $O( M \times N)$ variables.
 
 \item  The complexity of the algorithm can be greatly reduced using the \tap approximation, first introduced by Thouless, Anderson and Palmer in the context of spin glasses~\cite{tap}.
 It allows to only update local \textit{beliefs} instead of messages and relies on the fact that the factor graph is fully connected. 
 As a result, the number of variables to update at each iteration is $O(N)$ instead of  $O(M \times N)$, allowing a great simplification of the algorithm as well as a speedup. In particular, it 
 allows the use of fast transforms as in~\cite{moiCodes}. 
\end{enumerate}
These three steps are illustrated in table~\ref{table:3_bpToTap}. 

\begin{table}[h]
\begin{center}
\begin{tabular}{@{}lccr@{}} \toprule
 Step &  Messages & Updates & \# variables \\ \midrule
1:\bp & $\messv_{i \to \mu} (\x_i)$& $\messv_{i \to \mu}^{t+1} (\x_i) = \Phi \left( \{ \messf_{\mu \to i}^t (\x_i) \} \right) $ & $\infty$ \\
  & $\messf_{\mu \to i}(\x_i)$ & $\messf_{\mu \to i}^{t+1} (\x_i) = \tilde{\Phi} \left( \{ \messv_{i \to \mu}^{t+1} (\x_i) \} \right) $  & \\  \midrule
2:\amp &   $\messv_{i \to \mu} (\x_i) = \Phi\left( \NN(\x_i; \Xh_{i \to \mu}, \Xb_{i \to \mu}) \right)$ & $ (\Xh_{i \to \mu}^{t+1} , \Xb_{i \to \mu}^{t+1}) = \Phi \left( \{ \xh_{\mu \to i}^{t} , \xb_{\mu \to i}^{t} \} \right)  $ &  $O( M \times N)$ \\
  & $\messf_{\mu \to i} (\x_i) = \tilde{\Phi} \left( \NN(\x_i; \xh_{i \to \mu}, \xb_{i \to \mu}) \right)$&  $ (\xh_{\mu \to i}^{t+1} , \xb_{\mu \to i}^{t+1}) = \tilde{\Phi} \left( \{ \Xh_{i \to \mu}^{t+1} , \Xb_{i \to \mu}^{t+1} \} \right)  $ &  \\ \midrule
3:\tap & Beliefs & $ (\Xh_{i}^{t+1} , \Xb_{i}^{t+1}) = \Phi \left( \{ \xh_{i}^{t} , \xb_{i}^{t} \} \right)  $ & $O(N)$  \\
  & $\messv_{i} (\x_i) = \Phi\left( \NN(\x_i; \Xh_{i}, \Xb_{i}) \right) $ & $ (\xh_{i}^{t+1} , \xb_{i}^{t+1}) = \tilde{\Phi} \left( \{ \Xh_{i}^{t+1} , \Xb_{i}^{t+1} \} \right)  $  & \\ \bottomrule
\end{tabular}
\end{center}
\caption{The three steps for deriving \gamp from \bp, explained in the text.
$\Phi$ and $\tilde{\Phi}$ represent a general function, different in every occurrence.}
\label{table:3_bpToTap}
\end{table}

\subsubsection{The \gamp algorithm}
The final \gamp algorithm is the \tap version and is presented in~\algoref{algo:gamp}.
As we did for~\algoref{algo:iht}, we expand the algorithm, that could be written in a more compact form, for better legibility.
Note its similarity to~\algoref{algo:iht}: it relies on estimates of the same quantities as iterative thresholding does.
The difference is the way these estimates are produced and the fact that along with the estimators of each quantity, a corresponding uncertainty is calculated.
One of these uncertainties is used as the step size in the gradient descent step.

\begin{algorithm}[h!]
\caption{GAMP}
\label{algo:gamp}
  \textbf{Initialize} $\ghv_0=0$ and $(\xhv_0, \xbv_0)$ at random or according to $\px$. \\
  \textbf{Main loop:} while $t<t_{\rm max}$, calculate following quantities:
  \begin{align*}
  \Zbv_{t+1} &= \Fv^2 \bar{\xv}_{t} & & \text{Uncertainty on $\Zhv_{t+1}$. } \\
  \Zhv_{t+1} &= \Fv \xhv_{t} - \Zbv_{t+1} \odot \ghv_{t} & &\text{Estimate of $\zv$ based on the current estimate $\xhv_t$.} \\
  \zbv_{t+1} &=  \fb^Y(\yv|\Zhv_{t+1}, \Zhv_{t+1})& & \text{Uncertainty on $\zhv_{t+1}$. } \\
  \zhv_{t+1} &=  \fh^Y(\yv|\Zhv_{t+1}, \Zhv_{t+1})& &\text{New estimate of $\zv$ taking $\yv$ into account .} \\
  \gbv_{t+1} &= (\zbv_{t+1} - \Zbv_{t+1}) \odot (\Zbv_{t+1})^{\circleddash 2}& & \text{Auxiliary term} \\
  \ghv_{t+1} &=  (\zhv_{t+1} - \Zhv_{t+1}) \odot (\Zbv_{t+1})^{\inv} & & \text{Auxiliary term} \\
  \Xbv_{t+1} &= - \left(  (\Fv^2)^{\top} \gbv_{t+1} \right)^{\inv} & & \text{Uncertainty on $\Xhv_{t+1}$. }\\
  \Xhv_{t+1} &= \xhv_{t} + \Xbv_{t+1} \odot \left( \Fv^{\top} \ghv_{t+1} \right)  & &\text{Estimate of $\xv$ by gradient descent.} \\
  \xbv_{t+1} &= \fb^X(\Xhv_{t+1}, \Xbv_{t+1}) & & \text{Uncertainty on $\xhv_{t+1}$. }\\
  \xhv_{t+1} &= \fh^X(\Xhv_{t+1}, \Xbv_{t+1}) & & \text{New estimate of $\xv$ taking into account $\Xhv$ and $\px$.}
  \end{align*}
  \textbf{Stop} when $t=t_{\rm max}$, $|| \xbv_t||_2 < \epsilon$ or $|| \xhv_t - \xhv_{t-1} ||_2< \epsilon$.
\end{algorithm}

\subsubsection{Intialization}
\label{sec:initialization}
 The quantities $(\ghv_0, \xhv_0, \xbv_0)$ need to be initialized. 
 $\ghv_0$ can be fixed to $0$, $\xhv_0$ can be drawn at random from $\px$, $\xbv_0$ can be fixed to the variance of $\px$. 
 Other initialization schemes are possible.

\subsubsection{Update functions}
The update functions $\fh^X, \fb^X, \gh^Y$ and $\gb^Y$ that appear in \gamp are (in general) nonlinear functions of their arguments and act on them componentwise.

$\fh^X(\Xh,\Xb)$ and $\fb^X(\Xh,\Xb)$ are the mean and variance of the pdf $\propto \px(\X) \NN(\X;\Xh,\Xb)$ (as $\Xb$ is a variance it is positive).
They can therefore be written as 
\begin{align}
  \fh^X(\Xh,\Xb) &= \frac{f_1^X(\Xh,\Xb)}{f_0^X(\Xh,\Xb)}, 	\label{eq:def_fh}\\ 
  \fb^X(\Xh,\Xb) &= \frac{f_2^X(\Xh,\Xb)}{f_0^X(\Xh,\Xb)} -  \fh^X(\Xh,\Xb)^2, \label{eq:def_fb}
\end{align}
where we define, for all $k \in \Nintegers$,
\begin{align}
  f_k^X(\Xh,\Xb) = \int \dd x x^k \px(x) \NN(x;\Xh,\Xb)	\label{eq:def_f}.
\end{align}
Similarily, we can define $\fh^Y(y|\Zh,\Zb)$ and $\fb^Y(y|\Zh,\Zb)$ as being the mean and variance of the pdf $\propto \py(y|\z) \NN(\z;\Zh,\Zb)$ and 
\begin{align}
   f_k^Y(y|\Zh,\Zb) = \int \dd \z  \z^k \py(\y|\z) \NN(\z;\Zh,\Zb).	\label{eq:def_fy}
\end{align}
These functions are used to calculate new estimators and variances of $\z$,
\begin{align}
 \zh &= \fh^Y(y|\Zh,\Zb) & \zb &= \fb^Y(y|\Zh,\Zb), \label{eq:zhzb}
\end{align}
that are used to calculate the auxiliary ``gradient'' terms
\begin{align}
 \gh^Y(y|\Zh,\Zb) &= \frac{\zh - \Zh}{\Zb},  &  \gb^Y(y|\Zh,\Zb) &= \frac{\zb - \Zb}{\Zb^2}.  \label{eq:ghb}
\end{align}

In words, $\Xh$ and $\Zh$ are estimates of $\x$ and $\z$ obtained only from previous estimates.
$\xh$ is an estimate of $\x$ based on $\Xh$ but taking into account the prior $\px$, and $\zh$ is an estimate of $\z$ based on $\Zh$ but taking into account the measurement $\y$.

As these functions are means and variances of distributions, there is no general analytical formula for them, and evaluating them requires numerical integration~(\ref{eq:def_f},\ref{eq:def_fy}).
However, as they are integrals over a single variable, these integrals can often reliably be performed numerically, or in the best case be expressed as known functions.
Furthermore, note that in~\algoref{algo:gamp}, only matrix multiplications or elementwise operation are present. 
Last, note that $\Xbv$ can in principle be negative, because $\gbv$ can have positive or negative elements.
Looking at~\eqref{eq:ghb}, a sufficient condition for $\Xbv$ to be positive is that $\forall (\Zh,\Zb), \, \fb^Y(y|\Zh,\Zb)<\Zb$.
This condition is not respected for all sensing channels, which can lead to problematic negative ``variances'' $\Xbv$ that have to be handled carefully.

\subsubsection{Stopping conditions}
\label{sec:stopping}
Different stopping conditions can be implemented.
 Additionally to setting a maximum number of iterations, \gamp can be stopped if $|| \xhv_t - \xhv_{t-1} ||_2$ becomes smaller than a given threshold $\epsilon$, 
 at which the algorithm can be considered to have converged. 
 Another indicator of convergence is $|| \xbv_t ||_2$: when then algorithm converges to the right solution, the elements of $\xbv_t$, which are uncertainties, become smaller and smaller.

\begin{example}{Bayes optimal noisy \cs and \amp}
\label{ex:csGAMP} 
 Let us examine the case of noisy \cs in~\exref{ex:noisyCS}.
The update functions corresponding to the \awgn channel are:
\begin{align}
 \zh &= \fh^Y(y|\Zh,\Zb) = \frac{y \Zb + \Zh \Delta}{\Delta + \Zb}, & \zb &= \fb^Y(y|\Zh,\Zb) = \frac{\Delta \Zb}{\Delta + \Zb}.
\end{align}
Notice that $\zh$ is an average of the estimates $y$ and $\Zh$ weighted by the \textit{inverse} of their uncertainties $\Delta$ and $\Zb$.
In the case of a Gauss-Bernoulli prior, the update functions of $x$ have an analytical expression:
\begin{align}
 f_0^X(\Xh,\Xb) &= (1-\rho) \NN(\Xh;0,\Xb) + \rho \NN(\Xh;0,\Xb+1), \\
 f_1^X(\Xh,\Xb) &= \frac{\rho}{\Xb+1} \NN( \Xh;0,\Xb+1) \Xh , \\
 f_2^X(\Xh,\Xb) &= \rho \frac{\Xh^2 + \Xb (\Xb+1)}{(\Xb+1)^2}\NN(\Xh;0,\Xb+1).
\end{align}
These functions are the \textit{Bayes optimal} update functions when the signal really follows a Gauss-Bernoulli distribution.
If the real distribution of the signal is unknown, one can make an assumption about the prior and use the corresponding update functions.
The originally published \amp algorithm~\cite{amp} uses the soft thresholding function~(\ref{eq:softThresh})
\begin{align}
 \fh^X_{\rm \amp}(\Xh,\Xb) &=  S_{\Xb} (\Xh)
\end{align}
and 
\begin{align}
 \fb^X_{\rm \amp}(\Xh,\Xb) &= \left\{ \begin{array}{cc}
                                       \Xb & \text{if } |\Xh|>\Xb \\
                                       0 & \text{else.}
                                      \end{array}
                                      \right.
\end{align}
The resulting \amp algorithm solves the LASSO problem~(\ref{eq:lasso}) with $\lambda \to \infty$, and corresponds to taking the prior
\begin{align}
 \px(X) \propto e^{-\lambda |X|} .
\end{align}
This prior does not correspond to a distribution when $\lambda \to \infty$ and \amp is thus not Bayes optimal.
\end{example}


\section{Replica analysis}
\label{sec:replicaGLM}
The GLM setting can be analyzed with statistical physics methods in the thermodynamic limit,
\ie when the dimensions of the signal
 $\xs$ and of the measurements $\ys$ go to infinity, while the measurement ratio $\alpha$ remains fixed. 
 This analysis can be done with the replica method, which allows to calculate the free entropy linked to the pdf~(\ref{eq:gampPosterior}).
 In this section, we perform the replica analysis that results in a simple set of \textit{state evolution} equations.
 The analysis is very similar to the one of related inference problems~\cite{nishimoriBook,kabaDL,kabaSample}.
 For the \glm setting, the analysis was performed in~\cite{kabaGAMP} in a slightly more general setting, but only up to the general formula for the free entropy.
  The full analysis can be found in~\cite{krzakaCS} for the special case of noisy \cs, and in~\cite{yingying} for the special case of $1$-bit \cs.
  For a comprehensive introduction to the replica method, we refer the reader to~\cite{pedestrians,nishimoriBook,mezardMontanari}.

\subsection{Replica analysis: free entropy}
Treating an inference problem as a statistical physics problem consists in writing an energy function corresponding to the problem
and studying the partition function of the system.
Here, the relevant partition function is the normalization constant of the probability distribution~(\ref{eq:gampPosterior}):
\begin{equation}
\ZZ(\yv,\Fv) = \int \dd \xv \, \px(\xv) \int \dd \zv \p_{Y|Z}\left(\yv | \zv \right) \delta \left[ \zv - \Fv \xv \right], \label{eq:partition}
\end{equation}
from which the free entropy $\log \ZZ(\yv,\Fv)$ can be calculated.
Equation~(\ref{eq:partition}) is the partition function of a disordered system, as introduced in~\secref{sec:1_disordered}.
In order to determine the average properties of such a disordered system, one needs to average $\log \ZZ(\yv,\Fv)$ over all possible realizations of $\Fv$ and $\yv$, 
for which we use the replica method~\cite{mezardMontanari,nishimoriBook}.
It uses the  identity
\begin{align}
 \langle \log \ZZ \rangle &= \lim_{n \to 0} \frac{\partial}{\partial n} \langle \ZZ^n \rangle ,    \label{eq:replicaTrick2}
\end{align}
where $\langle \cdot \rangle$ denotes the average over $\Fv$ and $\yv$
and relies on the fact that an expression for $\ZZ^n$ can be found for integer $n$. 
This expression is then used for caculating the $n \to 0$ limit in~(\ref{eq:replicaTrick2}). 
Let us therefore start by calculating
\begin{align}
\ZZ(\yv,\Fv)^n = \int \prod_{a=1}^n \left\{ \dd \xv^a \, \px(\xv^a) \dd \zv^a \p_{Y|Z}\left(\yv|\zv^a \right) \delta\left[ \zv^a - \Fv \xv^a \right] \right\} \label{eq:partitionReplicated} 
\end{align}
and its average with respect to the realizations of $\yv$, generated by $\xv^0$ and $\Fv$:
\begin{align}
 \langle \ZZ^n \rangle  = \int &\dd \xv^0 \, \px^0(\xv^0) \dd \Fv \, \pf( \Fv) \dd \yv  \nonumber \\
    &\dd \zv^0 \py (\yv|\zv^0) \delta\left[ \zv^0 - \Fv \xv^0 \right] \ZZ(\yv, \Fv)^n . \label{eq:partitionReplicatedAveraged}
\end{align}
The indices $a$ represent so-called replicas of the system and are initially independent from each other.
Carrying on the calculation requires to couple them. 
To be more precise, each variable $\z_{\yi}^a = [\Fv \xv^a]_{\yi} $ is the sum of a large number of independent random variables and can therefore be approximated as a Gaussian random variable, with 
mean
\begin{align}
 \langle \z_{\yi}^a \rangle &= \sum_{i} \langle \F_{\yi i} \rangle x_{i}^a = 0	\label{eq:zeroMean}
\end{align}
because $\Fv$ has zero mean.
This allows to considerably reduce the number of integrals caused by the averaging over $\Fv$. 
However, $\z_{\yi}^a$ and $\z_{\yi}^b$ from different replicas $a$ and $b$ are not independent, as they are generated with the same matrix $\Fv$.
This can be seen by calculating the cross-correlation
\begin{align}
 \langle \z_{\yi}^a \z_{\yi}^b \rangle &= \sum_{i,j} \langle \F_{\yi i} \F_{\yi j} \rangle x_{i}^a x_{j}^b  \nonumber \\
				      &= \frac{1}{N} \sum_i x_{i}^a x_{i}^b \equiv  \Qx^{ab}	\label{eq:defQx}
\end{align}

The  multivariate random  variable $\za_{\yi} \equiv (\z_{\yi}^0, \dots, \z_{\yi}^n)$ is thus Gaussian with mean $0$ and covariance matrix $\Qmz = \Qmx$.
As in~(\ref{eq:partitionReplicatedAveraged}), $\Qmx$ can be anything, we have to integrate over it, such that
\begin{align}
  \langle \ZZ^n \rangle  &= \int \dd \Qmx \left[ \int \prod_{a=0}^n \dd \xv^a \, \px^a(\xv^a) \prod_{ a \leq b}  \delta\left( \xs \Qx^{ab} - \sum_{i} x_{i}^a x_{i}^b \right) \right]  \nonumber \\ 
  &\prod_{\yi=1}^{\ys} \left[ \int \dd \za_{\yi} \NN(\za_{\yi}; 0,\Qmz) \int \dd \y_{\yi} \py^0(\y_{\yi}|\z_{\yi}^0) \prod_{a=1}^n \py(\y_{\yi}|\z_{\yi}^a) \right].		\label{eq:ZwithDeltas}
\end{align}
Here, we use the convention that $\px^a=\px$ if $a\neq 0$.
We now see that the different replicas are coupled via $\Qmx$ in the first line. 
As we did with $\za_{\yi}$ , we now introduce the vector $\xa_{i} \equiv (\x_{i}^0, \dots, \x_{i}^n)$
and we use the integral representation of the $\delta$ function, introducing the conjugate variable $\Qmhx$ (details in appendix~\ref{app:replicaDetails}), which leads to
\begin{align}
  \langle \ZZ^n \rangle  &= \int \dd \Qmx \dd \Qmhx e^{-\frac{\xs}{2} \Tr(\Qmx \Qmhx)} \left[ \prod_{i} \dd \xa_{i} \px(\xa_{i}) e^{\frac{1}{2} \xa_{i}^{\top} \Qmhx \xa_{i}} \right] \nonumber \\
& \prod_{\yi=1}^{\ys} \left[ \int \dd \z_{\yi} \NN(\z_{\yi}; 0,\Qmz) \int \dd y_{\yi} \py^0(y_{\yi}|z_{\yi}^0) \prod_{a=1}^n \py(\y_{\yi}|z_{\yi}^a) \right].		\label{eq:ZwithQhat}
\end{align}
Finally, we assume the $x_{i}$'s and $\y_{\yi}$'s to be identically distributed. 
With the notations $\px(\xa)=\px^0(x^0)\prod_{a>0}\px(x^a)$
and $\py(\za)=\py^0(\y|\z^0) \prod_{a>0} \py(\y|\z^a)$, this leads to:
\begin{align}
  \langle \ZZ^n \rangle  &= \int \dd \Qmx \dd \Qmhx e^{- \frac{\xs}{2} \Tr(\Qmx \Qmhx)} \left[ \dd \xa \px(\xa) e^{\frac{1}{2} \xa^{\top} \Qmhx \xa} \right]^{\xs}  \nonumber \\
 & \left[ \int \dd \za \NN(\za; 0,\Qmz) \int \dd \y \py(\y|\za) \right]^{\ys}.
\end{align}
In the ``thermodynamic'' limit, we take $\xs$ and $\ys$ going to infinity with constant ratio $\alpha$.
This motivates us to rewrite the last equation as
\begin{align}
  \langle \ZZ^n \rangle  &= \int \dd \Qmx \Qmhx e^{-\xs \left[ \ac_n(\Qmx,\Qmhx) \right] }
\end{align}
and to use the saddle point method, according to which
\begin{align}
  \log \left( \langle \ZZ^n \rangle  \right) &= -\xs \min_{\Qmx,\Qmhx} \ac_n(\Qmx,\Qmhx) + O(1).  \label{eq:saddlePoint}
\end{align}
In the thermodynamic $N\to \infty$ limit, the $O(1)$ term has a vanishing contribution. 
We are therefore left with a minimization problem over the space of the matrices $\Qmx$ and $\Qmhx$, 
representing a total of $(n+1)(n+2)$ free paramters (as both matrices are symmetric).

\subsection{Replica symmetric assumption}
The idea of the replica symmetric assumption is that the $n$ replicas introduced in (\ref{eq:partitionReplicated}) are all equivalent, as they are purely a mathematical manipulation.
Based on this, we make the assumption that a sensible matrix $\Qmx$ does not make any distinction between the $n$ introduced replicas. 
We therefore parametrize $\Qmx$ and $\Qmhx$ in the following way: \mynote{RS assumption}{1.3}
\begin{align}
 \Qmx &= \left( \begin{tabular}{>{$}c<{$} | >{$}c<{$} >{$}c<{$} >{$}c<{$}}
               \Qzx & \mx & \cdots & \mx \\
               \hline
               \mx &  \Qx &  \cdots & \qx \\
               \vdots & \vdots & \ddots &\vdots \\
               \mx & \qx & \cdots & \Qx
              \end{tabular}
              \right) ,            
& \Qmhx &= \left( \begin{tabular}{>{$}c<{$} |  >{$}c<{$} >{$}c<{$} >{$}c<{$}}
               \Qhzx & \mhx & \cdots & \mhx \\
               \hline
               \mhx &  \Qhx &  \cdots & \qhx \\
               \vdots & \vdots  & \ddots &\vdots \\
               \mhx & \qhx &  \cdots & \Qhx
              \end{tabular}
              \right),		\label{eq:replicaSymmetry}              
\end{align}
allowing to be left with $8$ instead of $(n+1)(n+2)$ parameters over which to perform the extremization~(\ref{eq:saddlePoint}). 
Furthermore, $\Qzx$ is in fact known, as it is the second moment of the prior $\px^0$  and therefore we can set
\begin{align}
 \Qhzx &= 0
\end{align}
and thus the extremization is only over 6 variables: $(\mx, \mhx, \qx, \qhx, \Qx , \Qhx)$.

Let us now look in more details at the function $\ac_n$ to extremize:
\begin{align}
  \ac_n(\Qmx,\Qmhx) \equiv  \frac{1}{2} \Tr \Qmx \Qmhx - & \log \overbrace{\int \dd \xa \px(\xa) e^{\frac{1}{2} \xa^{\top} \Qmhx \xa}}^{\ix^n\left( \Qmhx \right) }  \nonumber \\
 		 - \alpha &\log \underbrace{\int \dd \za \NN(\za;0,\Qmz) \int \dd \y  \py(\y|\za)}_{\iz^n\left( \Qmz \right) }   		\label{eq:action}
 \end{align}
Thanks to the parametrization~(\ref{eq:replicaSymmetry}), the different terms have simple expressions. 
The trace can simply be written as
\begin{align}
 \Tr \Qmx \Qmhx &= 2 n \mx \mhx + n \Qx \Qhx + n(n-1) \qx \qhx ,
\end{align}
while we can use that
\begin{align}
 \xa^{\top} \Qmhx \xa  &= \Qhzx (x^0)^2  + (\Qhx-\qhx) \sum_{a>0} (x^a)^2 + \qhx ( \sum_{a>0} x^a )^2 + 2\mhx \x^0 \sum_{a> 0} x^a
\end{align}
and the Gaussian transformation $e^{\lambda \alpha^2} = \int {\rm D}t \, e^{\alpha \sqrt{2 \lambda} t}$,  
 where $\DD{t}$ is a Gaussian integration measure:
 \begin{align}
  \DD{t} &\equiv \dd t \, \NN(t;0,1),
 \end{align}
in order to write 
\begin{align}
\ix^n \left( \Qmhx \right)&= \int \DD{t}   \int \dd x^0 \, \px^0(x^0)  \left[ \int \dd x \, \px(x) e^{\frac{\Qhx-\qhx}{2} x^2 + (t \sqrt{\qhx} + \mhx x^0) x} \right]^n .
\end{align}
The second line in (\ref{eq:action}) can be simplified as well. The first step consists in writing the coupled Gaussian random variables $\z^0 \cdots \z^n$ 
as a function of $n$ independent, standard Gaussian random variables $u^a$ (for $a\in [1,n]$) and one additional standard Gaussian random variable $t$ that couples them all:
\begin{align}
 \z^0 &= \sqrt{\Qzz - \frac{\mz^2}{\qz}} \, u^0 + \frac{\mz}{\sqrt{\qz}} \, t  , & \z^a &= \sqrt{\Qz- \qz} \,  u^a + \sqrt{\qz} \, t.
\end{align}
Making the change of variables in the integral, we obtain the following expression for $\iz^n$:
 \begin{align}
 \iz^n \left( \Qmz \right)= \int \dd \y \, \int \DD{t} & \int \DD{u^0} \py^0\left(\y|\sqrt{\Qzz - \frac{\mz^2}{\qz}} \, u^0 + \frac{\mz}{\sqrt{\qz}} \, t \right) \nonumber  \\
 \times & \left[\int \DD{u} \py\left(\y| \sqrt{\Qz - \qz} \,  u + \sqrt{\qz} \, t \right) \right]^n,
 \end{align}
Looking back at the replica trick~(\ref{eq:replicaTrick2}), 
we have to study the quantity $\lim_{n \to 0} \frac{\partial }{\partial n} \ac_n$ and therefore the quantities 
 \begin{align}
 \ix(\Qmhx) &= \lim_{n \to 0} \frac{\partial}{\partial n} \log \ix^n \nonumber \\
    &= \int \DD{t}  \int \dd x^0 \, \px^0(x^0)  \log \left[ \int \dd x \, \px(x) e^{\frac{\Qhx-\qhx}{2} x^2 + (t \sqrt{\qhx} + \mhx x^0) x} \right] , \label{eq:ix} \\
    &= \frac{\sqrt{\qhx}}{\mhx} \int \dd t f_0^{X,0}\left(\frac{\sqrt{\qhx}}{\mhx}t, \frac{\qhx}{\mhx^2}\right) \log\left[ \frac{\sqrt{2 \pi} e^{\frac{\qhx t^2}{2(\qhx - \Qhx)}}}{\sqrt{\qhx-\Qhx}}  f_0^{X}\left(\frac{\sqrt{\qhx}}{\qhx - \Qhx}t, \frac{1}{\qhx - \Qhx}\right) \right], \nonumber
 \end{align}
as well as 
\begin{align}
\iz(\Qmz) &= \lim_{n \to 0} \frac{\partial}{\partial n} \log \iz^n  \nonumber \\
    &= \int \dd \y \, \int \DD{t} f_0^{Y,0}\left(\y | \frac{\mz}{\sqrt{\qz}} t , \Qzz - \frac{\mz^2}{\qz} \right) \log \left[ f_0^Y(\y |\sqrt{\qz}t, \Qz-\qz) \right]. \label{eq:iz}
\end{align} 
We use the shorter notations $f_i^Y, f_i^{Y,0}, f_i^X, f_i^{X,0}$ for $f_i^{\py}, f_i^{\py^0}, f_i^{\px^0}, f_i^{\py}$ as defined in~\appref{app:f}.
In the end, we obtain the free entropy $\phi$ as a saddle point \mynote{GLM free entropy}{0.9}
\begin{align}
 \phi = \saddle  & \left\{ - \mx \mhx - \frac{1}{2} \Qx \Qhx + \frac{1}{2} \qx \qhx +\ix(\Qmhx)  + \alpha \iz(\Qmz)	\right\} 	\label{eq:phi}
\end{align}
over a set of $6$ variables (because $\Qmz = \Qmx$). Note that the shift from a minimum in~(\ref{eq:saddlePoint}) to a saddle point in the equation 
above is a consequence to the hazardous $n\to 0$ limit in the replica method.

Note that $\ix$, $\iz$ and thus $\phi$ can be expressed using the information-theoretical quantities introduced in~\secref{sec:statistics}:
mutual information, entropy and Kullback-Leibler divergence. 

\subsection{State evolution equations}
\label{sec:stateEvol}
In the previous section, we have derived an expression for the free entropy as an extremum of a function over a set of parameters.
 In order to find the extremum in~(\ref{eq:phi}), we simply set all the partial derivatives of $\phi$ to $0$,
 which gives us \textit{saddle point equations}.
 This requires calculating the derivatives of the integrals $\ix$ and $\iz$:
 \begin{align}
  \frac{\partial}{\partial \Qhx} \ix(\Qmhx) &= \int \DD t \int \dd x^0 \px^0(x^0) \frac{ \int \dd x \, \px(x) x^2  e^{\frac{\Qhx - \qhx}{2} x^2 + (t\sqrt{\qhx} + \mhx x^0)x}}{\int \dd x \, \px(x) e^{\frac{\Qhx - \qhx}{2} x^2 + (t\sqrt{\qhx} + \mhx x^0)x}},  \nonumber \\
 \frac{\partial}{\partial \qhx} \ix(\Qmhx) &= \int \DD t \int \dd x^0 \px^0(x^0) \frac{ \int \dd x \, \px(x) \left( - \frac{x^2}{2} + \frac{t x}{2 \sqrt{\qhx}} \right)  e^{\frac{\Qhx - \qhx}{2} x^2 + (t\sqrt{\qhx} + \mhx x^0)\x}}{\int \dd x \, \px(x) e^{\frac{\Qhx - \qhx}{2} x^2 + (t\sqrt{\qhx} + \mhx x^0)\x}}, \nonumber \\
  \frac{\partial}{\partial \mhx} \ix(\Qmhx) &= \int \DD t \int \dd x^0 \px^0(x^0) \frac{ \int \dd x \, \px(x) \, x \, e^{\frac{\Qhx - \qhx}{2} x^2 + (t\sqrt{\qhx} + \mhx x^0)x}}{\int \dd x \, \px(x) e^{\frac{\Qhx - \qhx}{2} x^2 + (t\sqrt{\qhx} + \mhx x^0)x}} .
\end{align}
For $\iz$, we use the identity~(\ref{eq:fDerivateDifficult}), taking $s=q$ or $s=\frac{m^2}{q}$.
After an integration by parts, we obtain
\begin{align}
 \frac{\partial}{\partial \mz} \iz(\Qmz) &= \frac{1}{\mz} \int \dd \y \int \DD t \frac{ \left[ \frac{\partial}{\partial t} f_0^{Y,0}(\y | \frac{\mz}{\sqrt{\qz}} t , \Qzz  - \frac{\mz^2}{\qz} ) \right] \left[ \frac{\partial}{\partial t} f_0^{Y}( \y |\sqrt{\qz} t , \Qz -\qz ) \right]}{f_0^{Y}(\y | \sqrt{\qz} t , \Qz -\qz )}, \nonumber \\
 \frac{\partial}{\partial \qz} \iz(\Qmz) &= -\frac{1}{2 \qz} \int \dd \y \int \DD t \left( \frac{\frac{\partial}{\partial t} f_0^{Y}( \y |\sqrt{\qz} t , \Qz -\qz )}{f_0^{Y}( \sqrt{\qz} t , \Qz -\qz )} \right)^2 f_0^{Y,0}( \y |\frac{\mz}{\sqrt{\qz}} t , \Qzz  - \frac{\mz^2}{\qz} ),  \nonumber \\
 \frac{\partial}{\partial \Qz} \iz(\Qmz) &= \int \dd \y \int \DD t f_0^{Y,0}(\y | \frac{\mz}{\sqrt{\qz}} t , \Qzz  - \frac{\mz^2}{\qz} ) \left( \frac{\frac{\partial}{\partial \Qz} f_0^{Y}(\y | \sqrt{\qz} t , \Qz -\qz )}{f_0^{Y}(\y | \sqrt{\qz} t , \Qz -\qz )} \right).
\end{align}
 These expressions will be injected into the extremization equations of $\phi$ with respect to the elements of $\Qmx$ and $\Qmhx$ (remember that $\Qmx=\Qmz$ and $\Qmhx=\Qmhz$):
 \begin{align}
   \frac{\partial}{\partial \Qhx} \phi = 0 &\Leftrightarrow \Qx = 2  \frac{\partial}{\partial \Qhx} \ix(\Qmhx), &   \frac{\partial}{\partial \Qx} \phi = 0 &\Leftrightarrow \Qhx = 2 \alpha \frac{\partial}{\partial \Qx} \iz(\Qmz), \nonumber \\
   \frac{\partial}{\partial \qhx} \phi = 0 &\Leftrightarrow \qx = -2  \frac{\partial}{\partial \Qhx} \ix(\Qmhx), &   \frac{\partial}{\partial \qx} \phi = 0 &\Leftrightarrow \qhx = -2 \alpha \frac{\partial}{\partial \qx} \iz(\Qmz), \nonumber\\
   \frac{\partial}{\partial \mhx} \phi = 0 &\Leftrightarrow \mx =  \frac{\partial}{\partial \Qhx} \ix(\Qmhx), &   \frac{\partial}{\partial \mx} \phi = 0 &\Leftrightarrow \Qhx =  \alpha \frac{\partial}{\partial \mx} \iz(\Qmz). 
 \end{align}
 and using the update functions defined in~(\ref{eq:def_fh})--(\ref{eq:def_f}), we obtain
 \begin{align}
 \mx &= \int \DD t \int \dd x^0 \, x^0 \px^0(x^0) \fh^X \left( \frac{ \sqrt{\qhx} t + \mhx x^0 }{\qhx - \Qhx} , \frac{1}{\qhx - \Qhx}  \right)  ,  \\
 \Qx-\qx &= \frac{1}{\sqrt{\qhx}} \int \DD t \, t \int \dd x^0 \px^0(x^0) \fh^X \left( \frac{ \sqrt{\qhx} t + \mhx x^0 }{\qhx - \Qhx} , \frac{1}{\qhx - \Qhx}  \right)  , \label{eq:Q_DE_gen_inter} \\
   \Qx &= \int \DD t \, \int \dd x^0 \px^0(x^0) \left[ \frac{ f_2^X \left( \frac{ \sqrt{\qhx} t +  \mhx x^0 }{\qhx - \Qhx} , \frac{1}{\qhx - \Qhx}  \right)}{\f_0^X \left( \frac{ \sqrt{\qhx} t + \mhx x^0 }{\qhx - \Qhx} , \frac{1}{\qhx - \Qhx}  \right)} \right] .
 \end{align}
These equations can be further simplified by using the transformation $t \leftarrow t+\frac{\mh}{\sqrt{\qh}}x^0$ and integrating by parts \eqref{eq:Q_DE_gen_inter}: \mynote{GLM saddle point / state evolution equations}{2}
\begin{align}
 \mx &= \sqrt{\frac{\qhx}{\mhx ^2}} \int \dd t \, f_1^{X,0} \left(\frac{\sqrt{\qhx }}{\mhx }t , \frac{\qhx }{\mhx^2} \right)\fh^X \left( \frac{ \sqrt{\qhx } t }{\qhx  - \Qhx } , \frac{1}{\qhx  - \Qhx }  \right) ,   \label{eq:mx_DE_gen} \\
 \Qx-\qx &= \sqrt{\frac{\qhx}{\mhx^2}} \int \dd t \,  f_0^{X,0} \left(\frac{\sqrt{\qhx}}{\mhx}t , \frac{\qhx }{\mhx^2} \right)  \fb^X \left( \frac{ \sqrt{\qhx } t }{\qhx - \Qhx} , \frac{1}{\qhx - \Qhx} \right) ,  \label{eq:Qx_DE_gen}  \\
 \qx &= \sqrt{\frac{\qhx}{\mhx^2}} \int \dd t \,  f_0^{X,0} \left(\frac{\sqrt{\qhx}}{\mhx}t , \frac{\qhx}{\mhx^2} \right)  \left[ \fh^X \left( \frac{ \sqrt{\qhx} t }{\qhx - \Qhx} , \frac{1}{\qhx - \Qhx} \right) \right]^2 .  \label{eq:qx_DE_gen}
\end{align}
and for the conjugate variables, we obtain \mynote{GLM saddle point / state evolution equations}{2}
\begin{align}
  \mhx &= \frac{\alpha}{\mx}  \int \dd \y \int \DD t \frac{ \left[ \frac{\partial}{\partial t} f_0^{Y,0}\left(\y | \frac{\mx}{\sqrt{\qx}} t , \Qzx - \frac{\mx^2}{\qx} \right) \right] \left[ \frac{\partial}{\partial t} f_0^{Y}\left( \y |\sqrt{\qx} t , \Qx -\qx \right) \right]}{f_0^{Y}\left( \y |\sqrt{\qx} t , \Qx -\qx \right)},  \label{eq:mh_DE_gen} \\
  \qhx &= \frac{\alpha}{\qx}  \int \dd \y \int \DD t \left[ \frac{\frac{\partial}{\partial t} f_0^{Y}\left( \y |\sqrt{\qx} t , \Qx -\qx \right)}{f_0^{Y}\left( \y |\sqrt{\qx} t , \Qx -\qx \right)} \right]^2 f_0^{Y,0}\left(\y | \frac{\mx}{\sqrt{\qx}} t , \Qzx  - \frac{\mx^2}{\qx} \right) ,  \label{eq:qh_DE_gen}\\
  \Qhx &= 2 \alpha \int \dd \y \int \DD t f_0^{Y,0}\left(\y | \frac{\mx}{\sqrt{\qx}} t , \Qzx  - \frac{\mx^2}{\qx} \right) \left( \frac{\frac{\partial}{\partial \Qx} f_0^{Y}\left( \y |\sqrt{\qx} t , \Qx -\qx \right)}{f_0^{Y}\left( \y | \sqrt{\qx} t , \Qx -\qx \right)} \right) . \label{eq:Qh_DE_gen}
\end{align}
 The equations~(\ref{eq:mx_DE_gen}, \ref{eq:Qx_DE_gen}, \ref{eq:qx_DE_gen}) along with  the equations~(\ref{eq:mh_DE_gen}, \ref{eq:qh_DE_gen}, \ref{eq:Qh_DE_gen})
 constitute a closed set of equations that hold at the saddle points of $\phi$ in equation~(\ref{eq:phi}). (Note that~\eqref{eq:fDerivate} can be used in order to calculate the derivatives.)
 
 When they are iterated, they constitute the so-called state evolution equations. 
 These can also be obtained by the analysis of the \bp algorithm (\cite{amp,krzakaCS} for the special case of \cs, \cite{gamp,kabaMF} for a generic sensing channel) and are known to accurately describe the 
 algorithm's behaviour when the replica symmetric hypothesis is verified.
 
 Looking at the definition~(\ref{eq:defQx}), the ``physical'' meaning of $\mx$ is the overlap between $\xv$ and the estimate $\xhv$ obtained by sampling from~(\ref{eq:gampPosterior}).
In the same way, $\Qx$ is the squared $L_2$ norm of $\xhv$, and $\qx$ is the overlap between two estimates $\xhv$ and $\xhv'$ obtained by sampling from~(\ref{eq:gampPosterior}):
\begin{align}
 \mx &= \frac{1}{N} \sum_i x_i \xh_i, & \Qx &= \frac{1}{N} \sum_i \xh_i^2, & \qx &= \frac{1}{N} \sum_i \xh_i \xh_i'.
\end{align}
 From this, one can simply deduce the predicted mean squared error achieved by \gamp:
\begin{align}
  \mse &=   \Qx + \Qzx - 2\mx.  \label{eq:mse_general}
\end{align}

\subsection{Bayes optimal analysis}
Until now, we have not assumed exact knowledge of the true signal distributions and of the true measurement channel. 
When this is the case, the state evolution equations greatly simplify because of the so-called Nishimori conditions~\cite{zdeboReview}.
In our case, these ensure that the following equalities hold:
\begin{align}
 \Qx = \Qzx , \quad \Qhx = 0, \quad \mx = \qx, \quad \mhx &= \qhx.	\label{eq:nishConds}
\end{align}
Then, we only need to keep track of the variables $(\mx, \mhx)$, \renewcommand{\iter}{i}
and  the state evolution is obtained by choosing an initial value for $\mx^0$ and iterating for $\iter \geq 0$ the equations \mynote{Bayes optimal GLM state evolution}{1.8}
\begin{align}
 \mhx^{\iter+1} &= \frac{\alpha}{\mx^{\iter}} \int \dd \y \int \DD t  \frac{\left[ \frac{\partial}{\partial t} f^Y_{0}\left(\y |\sqrt{\mx^{\iter}}t,\Qzx  -\mx^{\iter}\right)\right]^2}{f^Y_{0}\left(\y |\sqrt{\mx^{\iter}}t, \Qzx -\mx^{\iter}\right)} ,   \label{eq:mh_DE_BO} \\
 \mx^{\iter+1} &= \frac{1}{\sqrt{\mhx^{\iter+1}}} \int \dd t \frac{\left[f^X_1\left(\frac{t}{\sqrt{\mhx^{\iter+1}}},\frac{1}{\mhx^{\iter+1}}\right) \right]^2}{f^X_0\left(\frac{t}{\sqrt{\mhx^{\iter+1}}},\frac{1}{\mhx^{\iter+1}}\right)} , \label{eq:m_DE_BO} 
 \end{align}
until convergence. \renewcommand{\iter}{t}
The expression of the predicted mean squared error~(\ref{eq:mse_general}) simplifies to
\begin{align}
  \mse &=   2( \Qzx - \mx).  \label{eq:mse_BO}
\end{align}

The initialial value $\mx^0$ indicates how close to the solution the algorithm is at initialization. 
In case of a random initialization of the algorithm, the expected initial overlap $\mx^0$  is of order $1/\xs$ ,
and should therefore be set to that value (or less) in the state evolution equations.

Note that the state evolution run with matching priors without imposing the Nishimori conditions~(\ref{eq:nishConds}) should in principle 
give the exact same results as the Bayes optimal state evolution analysis presented above, and thus be naturally attracted and follow the so-called ``Nishimori line'' defined by~(\ref{eq:nishConds}), 
as shown in~\cite{krzakaCS}.

\subsection{Partial information on $\xv$}
\label{sec:partialInf}
In some cases, it can be useful to consider that partial information on $\xv$ is available through additional measurements $\mathbf{\xadd}$ taken through a separable channel $\p_{\tilde{X}|X}$.
In that case, the $f_k^X(\Xh,\Xb)$ functions are replaced by $f_k^X(\xadd|\Xh,\Xb)$ and in the analysis, integration has to be performed over the realizations of $\mathbf{\xadd}$ as well.
The state evolution equations~(\ref{eq:mh_DE_gen}--\ref{eq:Qh_DE_gen}) are unchanged, but equations~(\ref{eq:mx_DE_gen}--\ref{eq:qx_DE_gen}) are replaced by
\begin{align}
 \mx &= \sqrt{\frac{\qhx}{\mhx ^2}} \int \dd \xadd \int \dd t \, f_1^{X,0} \left(\xadd | \frac{\sqrt{\qhx }}{\mhx }t , \frac{\qhx }{\mhx^2} \right)\fh^X \left(\xadd | \frac{ \sqrt{\qhx } t }{\qhx  - \Qhx } , \frac{1}{\qhx  - \Qhx }  \right) ,   \label{eq:mx_DE_gen_add} \\
 \Qx-\qx &= \sqrt{\frac{\qhx}{\mhx^2}} \int \dd \xadd \int \dd t \,  f_0^{X,0} \left(\xadd |\frac{\sqrt{\qhx}}{\mhx}t , \frac{\qhx }{\mhx^2} \right)  \fb^X \left( \xadd |\frac{ \sqrt{\qhx } t }{\qhx - \Qhx} , \frac{1}{\qhx - \Qhx} \right) ,  \label{eq:Qx_DE_gen_add}  \\
 \qx &= \sqrt{\frac{\qhx}{\mhx^2}} \int \dd \xadd \int \dd t \,  f_0^{X,0} \left(\xadd |\frac{\sqrt{\qhx}}{\mhx}t , \frac{\qhx}{\mhx^2} \right)  \left[ \fh^X \left( \xadd |\frac{ \sqrt{\qhx} t }{\qhx - \Qhx} , \frac{1}{\qhx - \Qhx} \right) \right]^2 .  \label{eq:qx_DE_gen_add}
\end{align}
Written in this more general form, the DE equations for $(\mx,\qx,\Qx)$ are even more similar to their counterparts for $(\mhx,\qhx,\Qhx)$ and require integration over 2 variables as well.

\section{Compressed sensing analysis}
\label{sec:CSanalysis}
\gamp particularizes to a \cs solver using the update functions presented in~\exref{ex:csGAMP}.
These same update functions can be injected into the state evolution equations in order to perform a theoretical analysis of compressed sensing.
As in~\exref{ex:csGAMP}, we look at the case of Bayes optimal \cs through an \awgn channel and with Gauss-Bernoulli priors.

\subsubsection{Algorithm dynamics and state evolution}
\begin{figure}[h]
 \centering
 \includegraphics[width=0.5\textwidth]{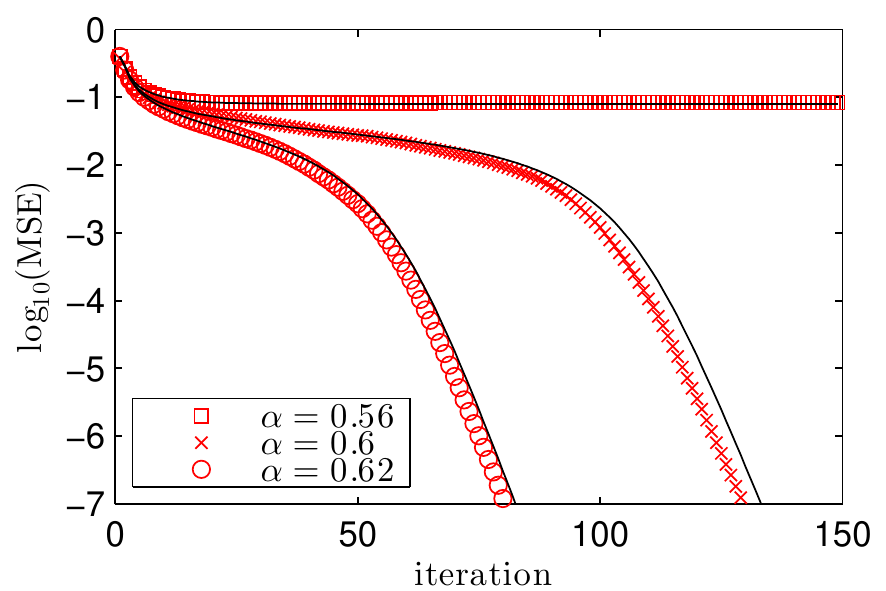}
 \caption[Algorithm dynamics and state evolution]{MSE between $\xv$ and $\xhv$ at each iteration, observed by running the algorithm with $N=10000$ (markers), and predicted by the state evolution (full lines),
 for a sparsity rate $\rho=0.4$ and three different measurement rates $\alpha$.
 The algorithm converges to the solution with a speed that depends on $(\rho,\alpha)$. 
 Close to the phase transition, the finite size of the signal induces a bigger deviation to the theory than further away from the phase transition.
 Same figure as in~\cite{krzakaCS}.}
 \label{fig:3_dynamics}
\end{figure}

Figure~\ref{fig:3_dynamics} illustrates the crucial point that \textit{the state evolution equations accurately describe the dynamics of \gamp}.
This can come as a surprise considering that the state evolution equations have been obtained independently of the algorithm.
However, they can be obtained as well starting from the \gamp algorithm and analyzing the distributions of the different updated quantities. 
Such an analysis can be found in~\cite{amp} or in~\secref{sec:complexCsStateEvolution} (for complex compressed sensing).
For real-valued compressed sensing, it has been made rigorous~\cite{rigorousState1,rigorousState2}.
This correspondence between the replica analysis and the belief propagation equations is linked to the hypothesis of replica symmetry, which is always verified in Bayes optimal inference~\cite{zdeboReview}. 

\subsubsection{Phase transitions}
\begin{figure}[h]
 \centering
 \begin{subfigure}{0.40\textwidth}
  \includegraphics[width=\textwidth]{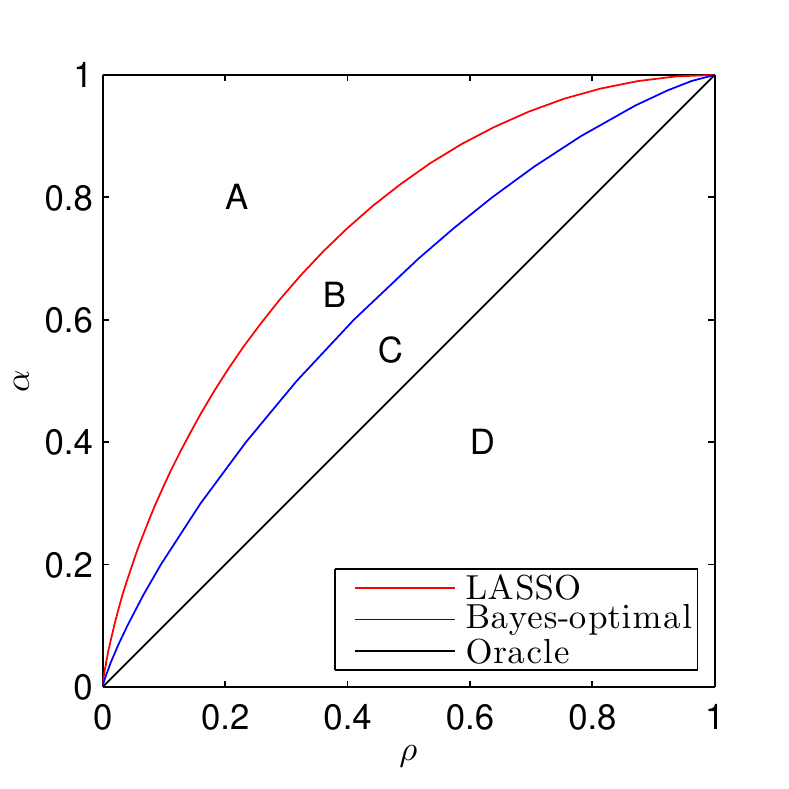}
  \caption{Phase diagram of \cs.}
  \label{fig:3_csPD_a}
 \end{subfigure}
 \begin{subfigure}{0.40\textwidth}
 \includegraphics[width=0.95\textwidth]{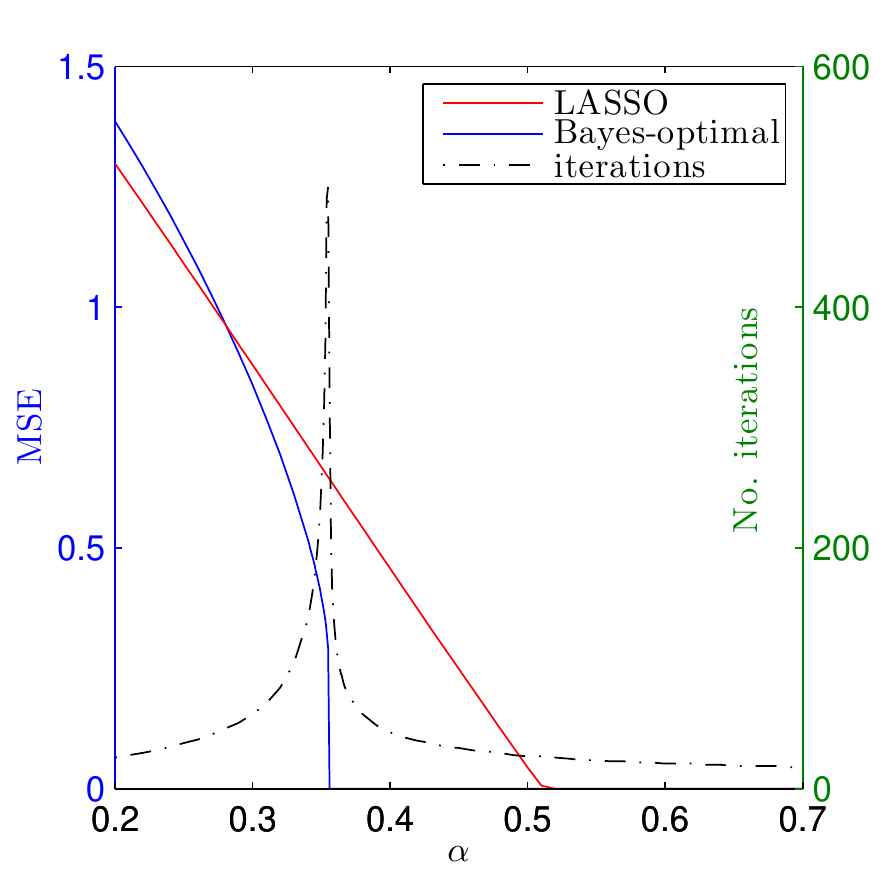}
 \caption{First and second order phase transitions.}
 \label{fig:3_csPD_b}
\end{subfigure}
\caption[Phases and phase transitions in \cs]{(a) Phase diagram of \cs in the $(\rho, \alpha)$ plane. Bayes optimal \gamp and \amp have different phase transitions, 
which allows us to distinguish between 4 different phases, described in the main text.
 (b) Phase transitions for $\rho=0.2$. The Bayes optimal phase transition is first-order (there is a jump in the MSE), whereas the \lasso phase transition is second order.
 At the Bayes optimal phase transition, the number of iterations of the algorithm diverges. Same figures as in~\cite{krzakaCS}.}
\label{fig:3_csPD}
\end{figure}

As seen in~\figref{fig:3_dynamics}, \gamp converges to the right solution or not depending on the values of $\rho$ and $\alpha$.
The MSE is therefore an order parameter of the problem that defines different phases depending on whether it is zero.
The corresponding \textit{phase diagram} is shown on~\figref{fig:3_csPD_a}.
The phases A-D have the following characteristics:
    \renewcommand*\theenumi{\Alph{enumi}}
    \renewcommand*\labelenumi{\theenumi)}
\begin{enumerate}
 \item  Easy phase: Both \amp and Bayes optimal \gamp converge to the solution. The \lasso phase transition 
is known as the \textit{Donoho-Tanner} phase transition~\cite{donohoTanner}. 
\item Easy phase: The convex relaxation \lasso ceases to be equivalent to \cs, but Bayes optimal \gamp
 converges to the solution. 
 \item Hard phase: Neither of the two algorithms converge to the solution. However, as an oracle algorithm would find it, recovery is possible.
 Specially designed, so-called \textit{spatially coupled} matrices allow recovery using Bayes optimal \gamp in this phase~\cite{krzakaCS}.
 The line separating B and C is known as the \textit{spinodal}, while the line separating C and D is the \textit{static} phase transition. 
 \item Impossible phase: The problem is hopelessly underdetermined as even oracle algorithms would fail. Recovery is impossible for any algorithm.
\end{enumerate}
    \renewcommand*\theenumi{\arabic{enumi}}
    \renewcommand*\labelenumi{\theenumi)}
Figure~\ref{fig:3_csPD_b} shows that the phase transition can be 
of first order (for Bayes optimal \cs, there is a discontinuity in the MSE) or of second order (for \lasso, the MSE goes to zero continuously).

\subsubsection{Energy landscapes and state evolution fixed points}
\begin{figure}[h]
 \centering
 \begin{subfigure}{0.48\textwidth}
  \includegraphics[width=\textwidth]{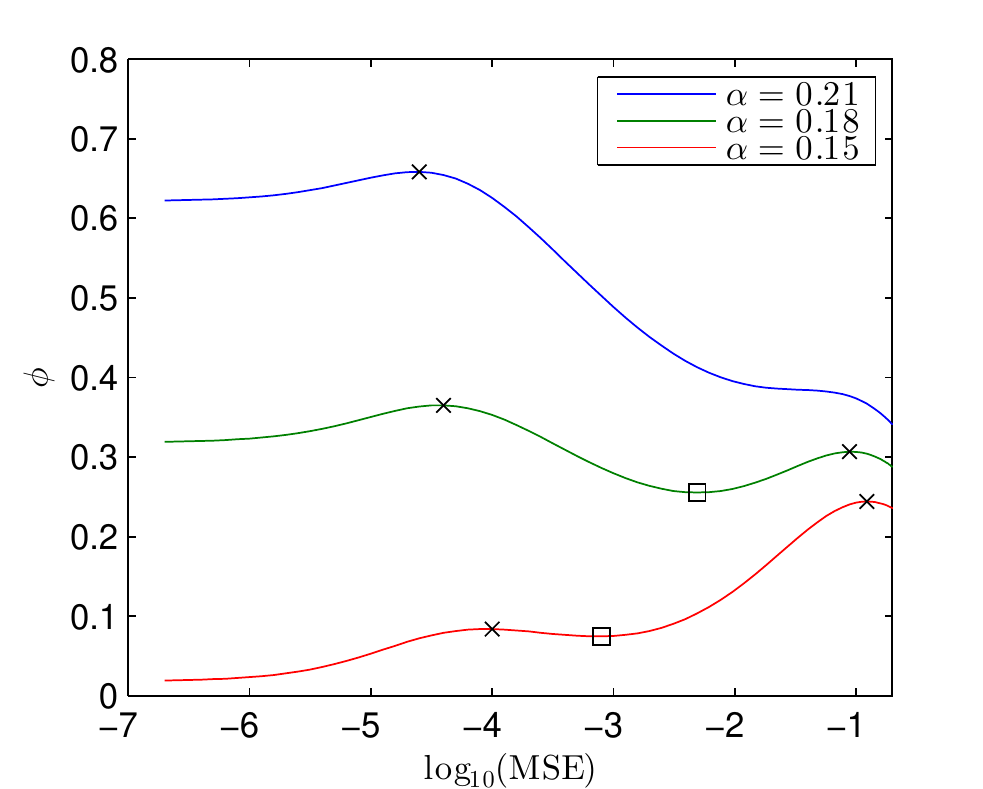}
  \caption{Free entropy.}
  \label{fig:3_cs_NRJ}
 \end{subfigure}
 \begin{subfigure}{0.48\textwidth}
 \includegraphics[width=\textwidth]{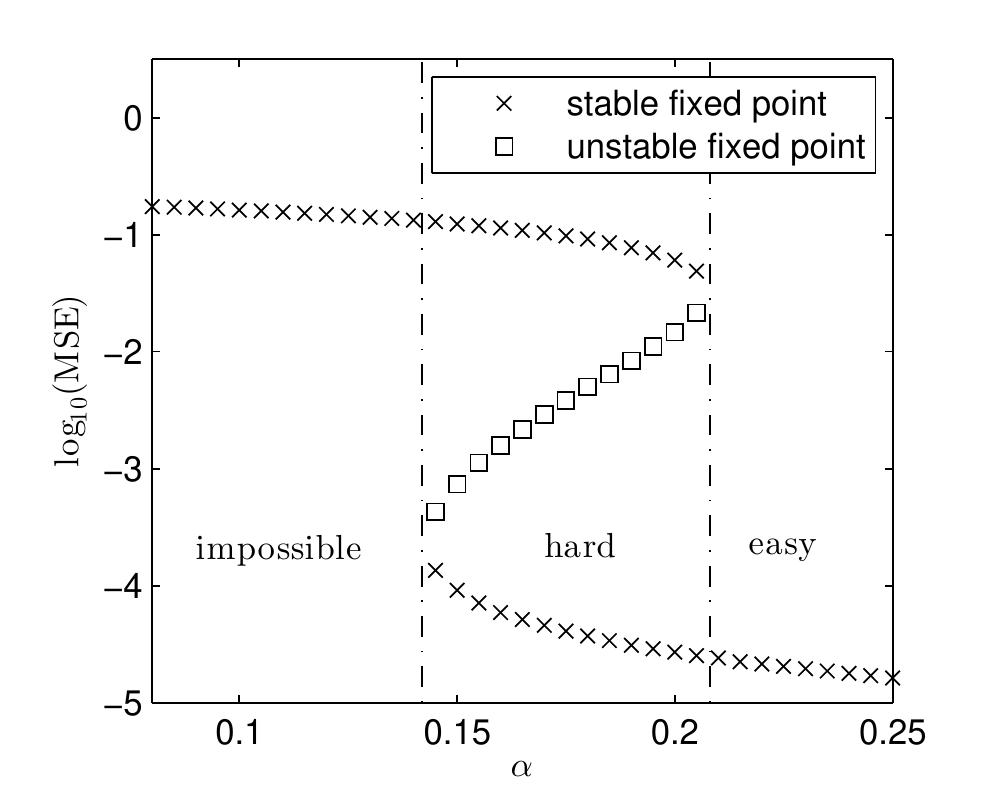}
 \caption{Fixed points of the state evolution.}
 \label{fig:3_cs_FP}
\end{subfigure}
\caption[Free entropy and fixed points in \cs]{(a) Free entropy as a function of the MSE for $\rho=0.1$, $\Delta=10^{-5}$. For high enough $\alpha$, there is only one maximum at low MSE. 
With decreasing $\alpha$, a local maximum appears for a higher value of the MSE. 
Lowering $\alpha$ further, that local maximum becomes the global maximum. Same plot as in~\cite{krzakaCS}.
(b) Fixed points of the state evolution equations ($\rho=0.1$, $\Delta=10^{-5}$).
The stable fixed points correspond to local maxima of the free entropy while the unstable fixed points are local minima.
In the hard phase, in which two stable fixed points exist, the algorithm will converge to the one with the high MSE if the intial MSE is bigger then the MSE of the unstable fixed point,
and to the low-MSE fixed point otherwise. 
For large-sized systems, the probability to initialize the algorithm close to the solution is vanishingly small and thus it is always the stable fixed point with the highest MSE 
that will be reached.}
\label{fig:3_csNRJ}
\end{figure}

The phase diagram in~\figref{fig:3_csPD_a} can be understood looking at the system's free entropy, just as for the Ising model in~\exref{ex:Ising}.
In Bayes optimal \cs, the free entropy~(\ref{eq:phi}) can be written as a function of only $\mx$:
\begin{align}
 \phi &= - \frac{1}{2}\mx \mhx  +\ix(\mhx) + \alpha \iz(\mx), \label{eq:phi_cs}
\end{align}
with
\begin{align}
 \mhx &= \frac{\alpha}{\Delta + \Qzx - \mx}, \\
 \iz(\mx) &= -\frac{1}{2} \left( \log\left( 2 \pi \left( \Delta + \Qzx - \mx \right) \right) + 1\right), \\
 \ix(\mx) &= \frac{1}{2} \left( \log\left(\frac{2 \pi}{\mhx}\right) + \mhx \Qzx \right) \nonumber \\
 &+ \frac{1}{\sqrt{\mhx}} \int \dd t f_0^{X}\left( \frac{t}{\sqrt{\mhx}}, \frac{1}{\mhx} \right) \log f_0^X \left( \frac{t}{\sqrt{\mhx}},\frac{1}{\mhx} \right).
\end{align}
The free entropy~(\ref{eq:phi_cs}) is plotted on~\figref{fig:3_cs_NRJ} as a function of the MSE (which is a function of $\mx$). 
It can have one or two local maxima depending on the values of $(\rho, \alpha)$.
At the phase transition, a second maximum appears at a location that is different from the first, which explains why the phase transition is of second order.
On~\figref{fig:3_cs_FP}, we show the fixed points of the state evolution equations, which correspond to the local extrema of the free entropy.

\subsubsection{Convergence issues}
\gamp can have convergence issues, for example when the measurement matrix $\Fv$ does not have zero mean or is low-rank.
In that case, some quantities in the algorithm diverge after a few iterations.
For non-zero mean matrices, the reason of this divergence has been explained and analyzed with state evolution equations in~\cite{caltaConvergence}.
Several approaches exist in order to solve such convergence issues, all requiring to change the algorithm.
\paragraph{Damping} of some of the variables in the algorithm can be used. 
While slowing down the algorithm, damping attenuates oscillations that can otherwise lead to diverging quantities.
A multitude of damping schemes can be implemented, most of them are not first-principled and it is not clear how to analyze the resulting algorithm.
For some special cases, damping was proven to allow convergence for any measurement matrix~\cite{ranganConvergence}. 
Other damping schemes use damping coefficients that vary from iteration to iteration depending on an energy function~\cite{vilaAdaptive}.
\paragraph{Sequential updates} of the algorithm's variables can replace the parrallel scheme of \gamp in which all estimators of each signal component are
update at each time step. The swept approximate message passing (SWAMP) was proposed in~\cite{swamp}. 
Like damping, sequential updating slows down the algorithm significantly, but greatly improves its convergence for certain measurement matrices.
\paragraph{UT-AMP} is a modified version of \amp that was proposed in~\cite{utamp}, and only makes minimal changes to the algorithm, which runs at the same speed.
In UT-AMP, the singular vector decomposition
\begin{equation}
 \Fv = \mathbf{U} \mathbf{\Lambda} \mathbf{V}
\end{equation}
is calculated and \gamp is applied to the system
\begin{align}
 \tilde{\yv} = \mathbf{U}^{\top} \yv = \mathbf{\Lambda} \mathbf{V} \xv.
\end{align}
While being very robust and not slowing down the algorithm as damping or sequential update, 
the limit of this trick is that it is restricted to \cs and cannot be applied to \gamp with a general sensing channel.


\section{Quantized sensing}
\label{sec:quantizedSensing}
Another example of a \glm is quantized sensing, studied with \gamp in~\cite{ulugbekOneBit}.
In quantized sensing, the measurements can only take a discrete set of values $\{ y_1, \dots, y_K\}$
that are assigned to $\y$ depending on the magnitude of a noisy version of $\z$:
\begin{align}
 \forall k\in \{1,\cdots,K\}, \quad \py(y_k|\z) &= \indic\left(\z+\xi \in [a_k,a_{k+1}] \right)  \quad \text{with } \xi \sim \NN(\xi;0,\Delta), \label{eq:quantizingChannel}
\end{align}
where $(a_1 < \dots < a_{K+1})$ are \textit{thresholds} that indicate how $z$ is discretized.
The corresponding function $f_0^Y$ used in \gamp is then
\begin{align}
  f_0^Y(y_k|\hat{z},\bar{z}) &= \frac{1}{2} \left( \erfc\left( -\frac{\hat{z} - a_k}{\sqrt{2 (\Delta + \bar{z})}} \right) - \erfc\left( -\frac{\hat{z} - a_{k+1}}{\sqrt{2 (\Delta + \bar{z})}} \right) \right),  \label{eq:quantizedCSf}
\end{align}
from which $f_1^Y$ and $f_2^Y$ can be easily obtained using relation~(\ref{eq:fDerivate}) and~(\ref{eq:erfcDerivative}).

\subsection{1 bit CS of continuous and binary signals}
In the case where $K=2$, each measurement takes 1 bit to be stored and we speak of 1-bit \cs.
This setting was studied with the replica method in~\cite{yingyingL1,yingying} and with \gamp in~\cite{ulugbekOneBit}.
The threshold $a_1$ is usually taken to be zero, but can be different. 
In~\cite{ulugbekAdaptive}, a setting is examined in which the threshold can be adapted to the measurements already taken.

\subsubsection{Continuous signals}
\begin{figure}[h]
 \centering
 \begin{subfigure}{0.42\textwidth}
 \includegraphics[width=\textwidth]{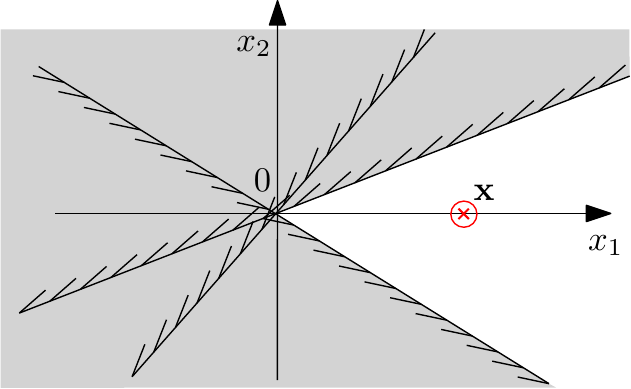}
 \caption{$1$-bit measurements in $2$ dimensions. }
 \label{fig:3_oneBit_geometry}
 \end{subfigure}
 \begin{subfigure}{0.48\textwidth}
 \includegraphics[width=\textwidth]{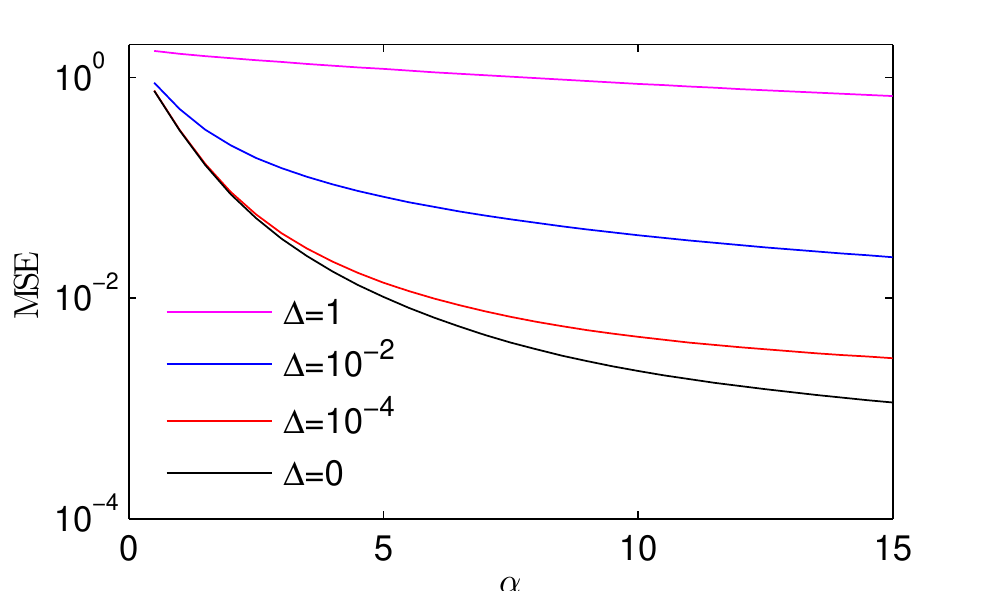}
 \caption{MSEs of $1$-bit \cs.}
 \label{fig:3_oneBit_continuous}
 \end{subfigure}
 \caption[Geometric insight into quantized sensing]{(a) Geometrical insight into 1-bit compressed sensing in dimension $N=2$.
 Each column of the measurement matrix $\Fv$ defines a hyperplane (a line in the present case).
 In the most favorable, noiseless $\Delta=0$ case, each measurement indicates on which side of the hyperplane the signal is located.
 White represents the region of signals that agree with the measurements.
  In the noisy setting, the frontiers defined by the hyperplanes would be fuzzy.
 If the threshold $a_1$ is $0$, it is easy to see that no matter how many measurements $M$ are available, $\xv$ can not be perfectly recovered,
 even if it is sparse: the white region will get narrower with increasing $M$, but will never allow to uniquely determine $\xv$.
 (b) MSE obtained with state evolution for $1$-bit \cs of a Gauss-Bernoulli signal with $\rho=0.25$ for different values of $\alpha$ and $\Delta$. 
 Unlike in \cs, there is no phase transition. }
 \label{fig:3_oneBit}
\end{figure}

It can be intuitively understood that a continuous signal cannot be perfectly reconstructed from 1-bit measurements.
A geometrical insight into this fact is given on~\figref{fig:3_oneBit_geometry}.
For this reason, there is no phase transition in $1$-bit \cs. 
Instead, the reconstruction performance improves continuously with the measurement rate:~\figref{fig:3_oneBit_continuous} shows 
the achievable MSE for binary sensing of Bernoulli-Gauss distributed signals.
Note that increasing the sparsity allows to obtain lower MSEs (not represented on the figure), but unlike in \cs, 
the sparsity constraint is not sufficient to allow perfect reconstruction.

\subsubsection{Quantized signals}
\begin{figure}[h]
  \begin{subfigure}{0.48\textwidth}
   \includegraphics[width=\textwidth]{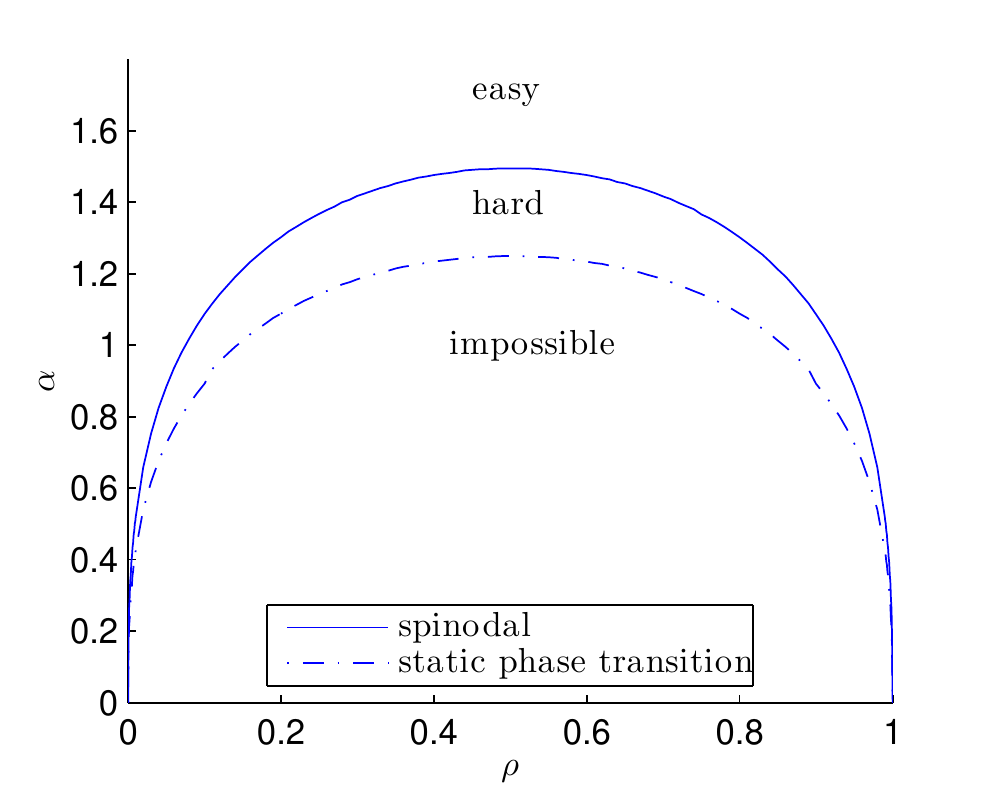}
   \caption{Binary signal with values $\{ -1,1\}$}
  \end{subfigure}
    \begin{subfigure}{0.48\textwidth}
   \includegraphics[width=\textwidth]{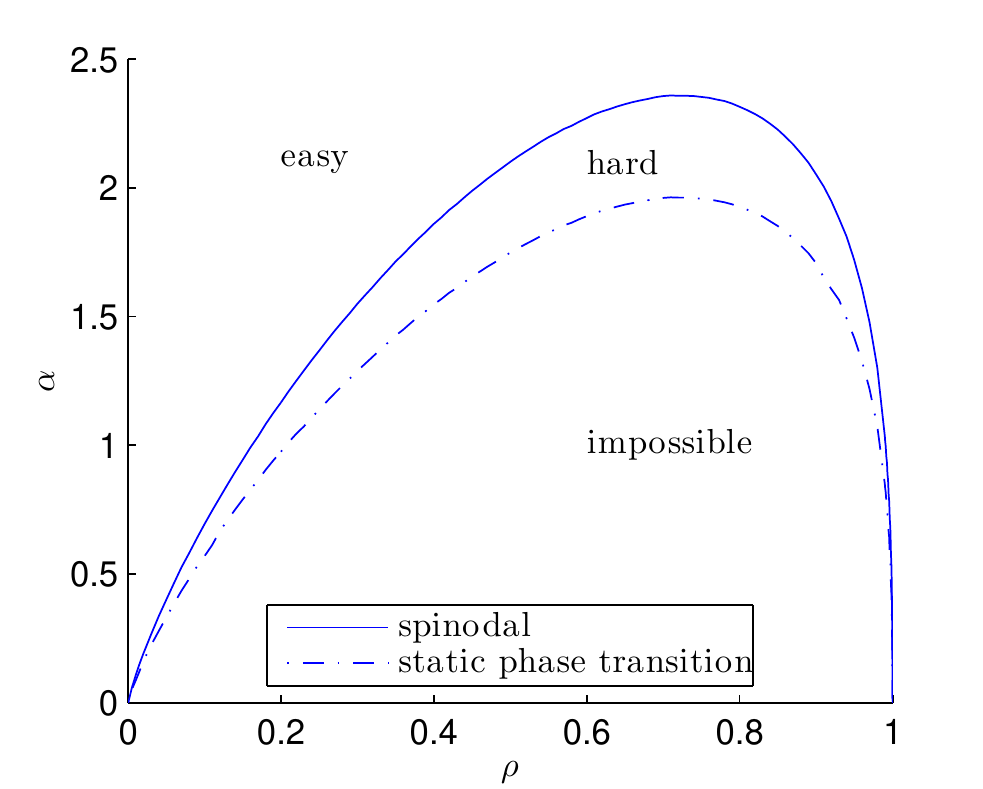}
   \caption{Binary signal with values $\{0,1\}$}
  \end{subfigure}
 \caption[Phase diagram of $1$-bit \cs of binary signals]{Phase diagrams obtained (by state evolution) for $1$-bit \cs of quantized signals.
 The signals follow the distribution $\px(x) = (1-\rho) \delta(x - x_-) + \rho \delta(x - x_+)$. As in \cs, the energy landscapes define 3 phases, in which inference is easy, hard or impossible.}
 \label{fig:3_binaryOneBit}
\end{figure}

On the other hand, a signal that is \textit{quantized} itself can be perfectly recovered from $1$-bit measurements.
Figure~\ref{fig:3_binaryOneBit} shows phase diagrams obtained for $1$-bit sensing of binary signals following the distribution \mynote{Binary prior}{0.65}
\begin{align}
 \px(x) &= (1-\rho) \delta(x-x_-) + \rho \delta(x - x_+), \label{eq:binaryPrior}
\end{align}
with $(x_-, x_+)$ equal to $(-1, 1)$ or $(0, 1)$.

The fact that perfect recovery is possible for quantized signals comes from the fact that the constraint imposed by the prior~(\ref{eq:binaryPrior}) is 
much stronger than the one imposed by the Bernoulli-Gaussian prior~(\ref{eq:gaussBernouilli}). In other words, the entropy of the binary prior is much smaller 
than the entropy of the Bernoulli-Gaussian prior. Therefore, quantized signals can also be perfectly recovered in settings with noisy measurements, which 
will be used in~\secref{sec:sparseSuperpositionCodes} in the context of sparse superposition codes.

Note that as in \cs, there is a hard phase in which inference is possible, but in which \gamp fails to perfectly recover the signal, as it gets trapped 
in a local free entropy maximum. In noiseless \cs, the position of the \textit{static} phase transition separating the impossible from the hard phase could 
be obtained by counting the number of equations and non-zero unknowns. This was related to the fact that \cs measurements are linear. 
As this is not the case in quantized sensing, there is \textit{a priori} no simple expression giving the position of the static phase transition.
The \textit{spinodal} separates the easy and hard phases.

\subsection{Applications: perceptron}
\label{sec:perceptron}
1-bit \cs is particularly interesting because in it, the measurement process is precisely the firing process of an idealized neuron (\exref{ex:neuralNetworks}). 
The sensing channel of 1-bit \cs is the equivalent of the neuron's \textit{activation function}, which is usually taken to be a sigmoid-shaped function such as 
the logistic function or the hyperbolic tangent. These are often approximated by the channel~(\ref{eq:quantizingChannel}) with $K=2$ and a well chosen noise level $\Delta$.

1-bit \cs can therefore be applied to the much older problem of the perceptron, presented in~\exref{ex:singleLayer}, from which we take over the notations.
The perceptron is a supervised learning problem in which a collection of $P$ known signals $\{ \xv_1 , \cdots, \xv_P\}$ (which are \textit{not} sparse) are given 
along with their correct classifications $\{ \mathbf{\sigma}_1, \cdots, \mathbf{\sigma}_P \}$ and the goal is to learn a matrix $\mathbf{J}$ such that:
\begin{align}
 \forall (\mu,p), \quad \sigma_{\mu p} = \Theta\left( \sum_i J_{\mu i} x_{i p} - \tau_{\mu} \right).
\end{align}
This problem can be restated as a \glm by considering that $\xv \in \RR^{P \times N}$ is a measurement matrix and that the vectors $\vec{J}_{\mu} = \{ J_{\mu 1}, \cdots, J_{\mu N} \}$ are 
signals of $\RR^{N}$ to recover:
\begin{align}
 \forall \mu, \quad \vec \sigma_{\mu} = \Theta\left( \xv \vec J_{\mu} - \tau_{\mu} \right).
\end{align}

\subsection{Neural networks inference} 
 Another application of 1-bit \cs is inference of neural connectivity matrices in recurrent neural networks~(\exref{ex:neuralNetworks}). 
The ultimate goal is to be able to infer the synaptic weights $\mathbf{J}$ of a real network of biological neurons 
whose activity is measured in an experiment.
Neural activities can be recorded with different experimental techniques such as direct measurements using arrays of electrodes that record the 
electrical potentials in individual neurons, or indirect measurements with fluorescence imaging. 
Fluorescence imaging can reveal neural activities by using fluorescent molecules that emit light in the presence of $Ca^+$ ions, which are 
released by firing neurons. A modified version of \gamp was proposed in~\cite{neuramp} in order to infer neural weights from fluorescence data.

More generally, there is a great interest in inference of neural weights~\cite{roudiIsing, PhysRevLett.110.210601}, due to the very rapidly increasing quantity and quality of experimental data.
In physics, the problem is known as the \textit{inverse Ising model}~\cite{PhysRevE.91.052136, mezardInverse, ricciInverse}.

\subsubsection{Generating model}
\begin{figure}
 \centering
 \begin{subfigure}{0.48\textwidth}
  \includegraphics[width=\textwidth]{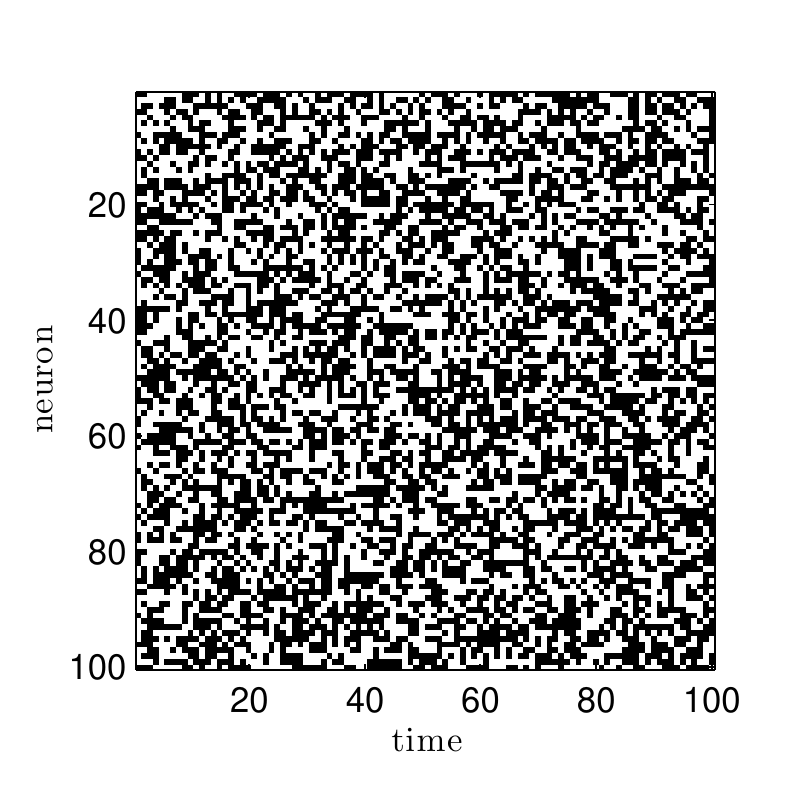}
  \caption{$\Delta = 100$}
 \end{subfigure}
 \begin{subfigure}{0.48\textwidth}
  \includegraphics[width=\textwidth]{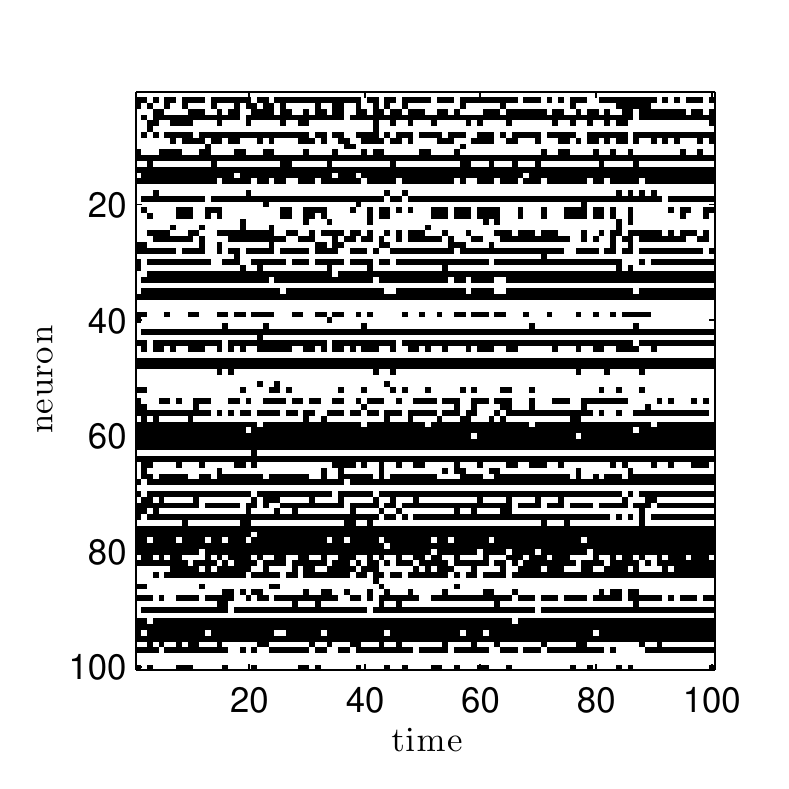}
  \caption{$\Delta = 0.01$}
 \end{subfigure}
 \caption[Neural firing patterns]{Firing patterns obtained with the model~(\ref{eq:generateActivity}) and different levels of noise $\Delta$.
 High noise produces random patterns, whereas patterns produced with a low noise are very correlated across time. 
 These correlations are problematic for inference using \gamp. Black pixels represent silent neurons ($\sigma=0$), white pixels represent firing neurons ($\sigma=1$). }
 \label{fig:activity1}
\end{figure}

In a preliminary study, I have tried to apply \gamp to inference of neural weights of a simulated network of neurons.
The setting was the following:
\begin{itemize}
 \item $N=100$ neurons are initialized in a random state $\sigv(0) \in \{0,1\}^N$. 
 \item A matrix $\Fv \in \RR^{N \times N}$ of synaptic weights is sampled from a distribution $p_{J}$.
 \item For $0<t<T$, the new state at time $t$ is determined from the state at time $t-1$:
 \begin{align}
  \sigv(t) = \Theta(\mathbf{J} \sigv(t-1) + \xiv(t)), \label{eq:generateActivity}
 \end{align}
 where $\xiv(t)$ is \iid \awgn of variance $\Delta$ and $\Theta$ is the Heaviside step function applied elementwise.
\end{itemize}
Figure~\ref{fig:activity1} shows an example of \textit{firing patterns} obtained by such a simulation.
These firing patterns have very different properties depending on the noise level $\Delta$ and the distribution of synaptic weights $p_J$.
For example, if $\Delta$ is very large, the firing of neurons at time $t$ is essentially random and nearly independent of the state at time $t-1$. 
On the contrary, if $\Delta=0$, the firing pattern is deterministic and entirely determined by the initial $\sigv(0)$ and by $\mathbf{J}$.
Depending on the parameters, the firing patterns can thus have very different aspects. In~\cite{brunel}, these firing patterns are classified in different phases, 
in which the firing of neurons is synchronous or asynchronous, regular or irregular, with fast or slow oscillations.

In~\figref{fig:activity1}, the firing patterns were produced using the following distribution of weights:
\begin{align}
 p_J(J) &= (1-\rho)\delta(J) + \rho \left[ \rho_+ \delta(J - J_+) + (1-\rho_+) \delta(J - J_-)\right], \label{eq:J_distribution} 
\end{align}
with $(\rho=0.2, \rho_+=0.8)$ and $(J_+=1/N, J_-=-4/N)$. The values of these parameters are used in other studies and are inspired by insights from biological experiments, in which the following was found. 
 (a) Neurons are connected to a small fraction of the other neurons. (b) Neurons have more excitatory than inhibitory synapses (hence $\rho_+>0.5$), but that the latter have a larger weight (hence $|J_-|>J_+$).
This model is very simplistic, but similar models are used as they produce firing patterns that have some of the properties observed in real firing patterns~\cite{destexhe2003high}.

\subsubsection{Inference} 
Once a firing pattern is produced, it can be used in order to perform inference of $\mathbf{J}$ with \gamp.
This can be done independently for each neuron $i$ by transposing~\eqref{eq:generateActivity}:
\begin{align}
 \left( \begin{tabular}{c}
  $\sigma_i(1)$ \\
  $\vdots$ \\
  $\sigma_i(T)$
 \end{tabular} \right) = \Theta \left[ \underbrace{\left( \begin{tabular}{c c c}
						$\sigma_1(0)$ & $\dots$ & $\sigma_N(0)$ \\
						$\vdots$ & & $\vdots$ \\
						$\sigma_1(T-1)$ & $\dots$ & $\sigma_N(T-1)$
						\end{tabular} \right) }_{\boldsymbol{\Sigma}}  
\left( \begin{tabular}{c}
  $J_{i1}$ \\
  $\vdots$ \\
  $J_{iN}$
 \end{tabular} \right) +  \left( \begin{tabular}{c}
  $\xi_i(0)$ \\
  $\vdots$ \\
  $\xi_i(T-1)$
 \end{tabular} \right) \right]. \label{eq:NNI}
\end{align}
The system~(\ref{eq:NNI}) is thus a simple 1-bit \cs problem. The measurement rate is $\alpha = \frac{T}{N}$ and the weights $\mathbf{J}$ are ternary.
Just like the binary signals in~\figref{fig:3_binaryOneBit}, ternary signals can be exactly reconstructed from 1-bit measurements provided $\alpha$ is big enough.

However, (\ref{eq:NNI}) presents a serious problem for applying \gamp: the measurement matrix $\boldsymbol{\Sigma}$ has a non-zero mean and more importantly, its elements can be strongly correlated, 
which can even be noted visually by looking at the firing patterns in~\figref{fig:activity1}.
To avoid divergences due to the non-zero mean, I used SWAMP~\cite{swamp}, which overcomes the problem and successfully performs 1-bit inference of ternary signals using 
a measurement matrix with \iid elements that are 0 or 1. It seems however that the correlations in $\boldsymbol{\Sigma}$ are usually too important to allow successful inference. 
Results were very dependent on the realization and and the reconstructions usually had very high MSEs. Taking very large measurement rates $\alpha$ did not seem to systematically improve results.

\subsubsection{Discussion}
The study was aborted due to difficulty to obtain reproducible results. 
The main reason for this seems to be the presence of strong correlations in the firing patterns, that act as measurement matrices.
Following future research directions could lead to interesting results:
\begin{itemize}
 \item Use different distributions $p_J$ that lead to firing patterns that present a greater variability.
  \item Make an analysis the noise level's influence. As $\Delta$ increases, $\boldsymbol{\Sigma}$ becomes more random and becomes a better measurement matrix. But at 
 the same time, a higher noise level generally makes inference harder.
 \item Introduce a fraction of neurons that fire at random, thus simulating external stimuli and allowing a greater variability of firing patterns.
 \item Build a model of correlated matrices and attempt to analyse the achievable inference performances with these matrices. In the pioneering work~\cite{kabaGAMP}, 
 the replica analysis was made with a more general model of random matrices than the \iid model used in this thesis. However, it does not seem to be directly useful for the present study.
\end{itemize}

\section{Conclusion}
Compressed sensing is a special case of a larger class of inference problems called 
generalized linear models.
Just like in community detection, inference of \glms can be performed using \bp.
The resulting algorithm is called \gamp and its performances can be compared to the theoretical performances 
of \glm inference, obtained by using the replica method.

In some special cases of \glms, such as compressed sensing or quantized sensing of binary signals, phase 
transitions separating \textit{easy}, \textit{hard} and \textit{impossible} inference exist, just as in community detection.
These phase transitions can be understood in terms of free entropy landscapes and of fixed points of the state evolution equations.

\gamp is highly successful for solving inference problems but also supervised learning problems that can be simply reformulated as inference problems.
The main limitation of \gamp is that it can encounter convergence issues for some measurement matrices or sensing channels.
Several techniques to overcome these issues exist
and can partially solve them, at the cost of speed.
Further examples of applications of \gamp are treated in~\chapref{chap:gampApplications}.
 
 \chapter{Generalized bilinear models}
\label{chap:generalizedBilinearModels}

In~\chapref{chap:generalizedLinearModels}, I have presented generalized linear models and
their theoretical analysis with the replica method.
Among the many applications of \glms, I presented the examples of compressed and quantized sensing;
further examples are treated in~\chapref{chap:gampApplications}.

Conceptually, \glms are easy to generalize to generalized \textit{bilinear} models, presented in this chapter.
As \glms, the \gbm setting can be analyzed with the replica method and message-passing algorithms for probabilistic inference can be derived.
However, the computational complexity of these problems is considerably higher than for \glms, which leads 
to much slower algorithms.
Furthermore, stability issues are much more present, such that message-passing algorithms often do not converge.

This chapter presents 3 bilinear problems of increasing complexity: blind gain calibration, generalized matrix factorization 
and generalized matrix compressed sensing.
My contributions to these problems will be presented in~\chapref{chap:blindSensorCal} and~\chapref{chap:matrixCS}.

\section{Gain calibration}
\label{sec:gainCal}

We focus on the noiseless setting in which measurements $\yv$ are generated as follows:
\begin{align}
\forall \mu \in [1,M] , \quad  \y_{\mu} = d_{\mu} \sum_{i=1}^N \F_{\mu i} \x_i,	\label{eq:gainCal}
\end{align}
where
\begin{align}
 \xv \in \RR^{N}, \quad \dv \in (\RR^{*+})^M, \quad \Fv \in \RR^{M \times N}. 
\end{align}
Equation~(\ref{eq:gainCal}) is linear both in $d$ and in $\x$ and is thus \textit{bilinear}.
As in \cs, the signal can be sparse and the measurement matrix $\Fv$ is supposed to have \iid random entries.
\begin{figure}
 \centering
 \begin{subfigure}{0.45\textwidth}
   \includegraphics[width=\textwidth]{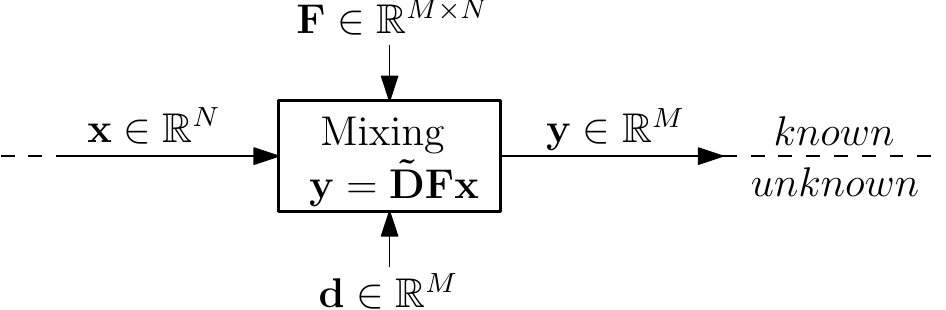}
   \caption{Supervised gain calibration.}
   \label{fig:4_gain_supervised}
 \end{subfigure}
 \hspace{1cm}
 \begin{subfigure}{0.45\textwidth}
   \includegraphics[width=\textwidth]{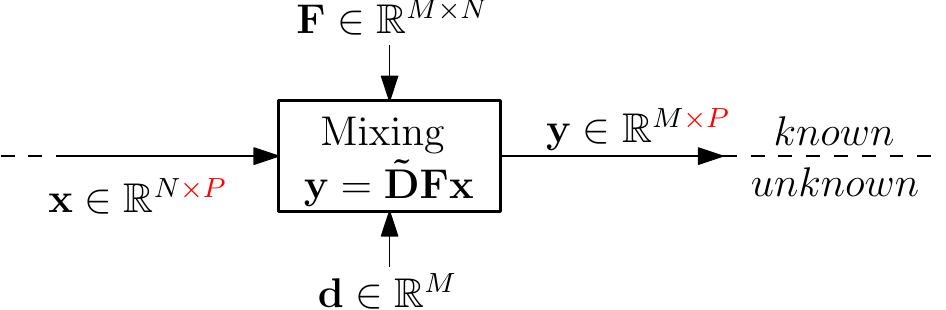}
   \caption{Blind gain calibration.}
   \label{fig:4_gain_blind}
 \end{subfigure}
 \caption[Blind gain calibration]{
 Noiseless gain calibration: The setting is similar to \cs, with $M$ multiplicative gains that are unknown and have to be inferred.
 The diagonal terms of the matrix $\tilde{\mathbf{D}}$ are the components of the vector $\dv$, its non-diagonal
 elements are zeros.
 (a) \textit{Supervised} gain calibration: measurements from a \textit{known} signal are taken and allow to obtain $\dv$.
 (b) \textit{Blind} gain calibration: only measurements from \textit{unknown} signals are available.
  $\xv$ is a set of $P$ independent signals, concatenated into a $N \times P$ matrix.
  Measuring $P>1$ independent signals is essential for inference to be possible.
  }
 \label{fig:4_gain}
\end{figure}

\subsubsection{Supervised vs blind calibration}
The difference between supervised and blind calibration is presented on~\figref{fig:4_gain}.
In a supervised setting, where it is possible to measure known signals, it is straightforward to recover $\dv$ by measuring a single signal, as
\begin{align}
 d_{\mu} &= \frac{\y_{\mu} }{\sum_{i=1}^N \F_{\mu i} \x_i}  \label{eq:supervised_gain_cal}.
\end{align}
In blind calibration, it is not possible to measure known signals. 
The problem is harder in the sense that there are more unknowns. 
As $\dv$ is fixed and independent of the signal measured, it introduces a correlation in the measurements of independent signals that can be exploited
in order to \textit{jointly} infer the signals and the gains.

A simple way to see this is to consider a simplified case, in which the elements of $\dv$ are either $1$ or $10$.
As the elements of $\Fv$ are iid, $\z_{\mu} = \sum_{i=1}^N \F_{\mu i} \x_i$ has the same variance (let's call it  $\sigma$) for all $\mu$s.
The variance of $\y_{\mu}$, however, is either $\sigma$ or $100 \sigma$. 
Measuring enough signals, the empirically calculated variances obtained for each $\mu$ approach the true variance, making it possible to determine 
if $d_{\mu}$ is $1$ or $10$.
If the gains are continuous variables, this naive method will not be efficient, as the convergence of the empirical variance to the true one is slow with increasing $P$.

\subsubsection{Scaling invariance}
An important remark about this problem is that it does not have a unique solution because of a fundamental scaling invariance:
For any scalar $\lambda$, the couple $(\lambda \dv, \frac{\xv}{\lambda})$ generates the same measurements as the couple $(\dv, \xv)$.
This ambiguity is lifted if the exact mean of $\dv$ or of $\xv$ is known. 
However, $\dv$ and $\xv$ are random variables, and even if their distribution is known, their empirical mean is not 
exactly equal to the mean of the distribution because of the finite size of $M$ and $N$.
Therefore, it is reasonable to consider this scaling invariance as unliftable.

For inference algorithms, this means that the MSE ceases to be a satisfying measure of success, as 
\begin{align}
 \mse(\xv,\lambda \xv) = (1-\lambda)^2 || \xv ||_2^2 ,
\end{align}
which does not reflect that the estimate $\lambda \xv$ cannot be improved without further informations.
Therefore a better measure of success is the normalized cross-correlation, which takes into account the scaling invariance: 
for $\xv \in \RR^N$ and $\xhv \in \RR^N$, \mynote{Normalized cross-correlation}{0.7}
\begin{align}
 \cc(\xv,\xhv) = \frac{| \xv^{\top} \xhv |}{||\xv||_2 ||\xhv||_2}.	\label{eq:crosscorr}
\end{align}
The normalized cross-correlation varies between $0$ (uncorrelated signals) and $1$ ($\xv$ and $\xhv$ are equal up to a multiplicative constant).
If the scaling invariance is restricted to positive scalars $\lambda$, a related measure of success is the normalized MSE: \mynote{Normalized MSE}{0.8}
\begin{align}
 \nmse(\xv,\xhv) = \frac{1}{2} \left| \left| \frac{\xv}{|| \xv ||_2} - \frac{\xhv}{|| \xhv ||_2} \right| \right|_2^2 = 1-\cc(\xv,\xhv), 	\label{eq:nmse}
 \end{align}
which varies from $1$ (uncorrelated signals) to $0$ ($\xv$ and $\xhv$ are equal up to a positive multiplicative constant).

\subsubsection{Bounds}
A first qualitative analysis of the problem can make it look hopeless: in fact, there are $M$ measurements for $N+M$ unknowns (the signal + the gains).
An obvious lower bound for $\alpha=M/N$ is the compressed sensing phase transition, as the problem reduces to \cs if the 
sensors are perfectly calibrated ($\dv$ is known):
\begin{equation}
 \alphacal(\rho) = \alphacs(\rho) .
\end{equation}

To overcome this problem, we consider a setting in which $P$ different real-valued signals are measured:
\begin{align}
 \forall \mu \in [1,M], l \in [1,P], \quad \y_{\mu l} = d_{\mu} \sum_{i=1}^N \F_{\mu i} \x_{i l}.
\end{align}
Considering only the non-zero components of the signal, there are $M P$ measurements for $\rho N P + M$ non-zero unknowns.
Inference should therefore be possible if $ MP \geq \rho N P + M$, or equivalently
\begin{align}
\alpha \geq  \alphagmin = \frac{P}{P-1} \rho.	\label{eq:gain_counting}
\end{align}
In fact, consider an oracle algorithm that knows the locations of the signal's zeros. 
For each of the $M$ sensors, the $P$ measurements can be combined into $P-1$ independent linear equations of the type
\begin{align}
 y_{\mu l} \sum_i F_{\mu i} x_{i m} - y_{\mu m} \sum_i F_{\mu i} x_{i l} =0.
\end{align}
In total, there are $M(P-1)$ such linear equations and $\rho N P$ unknowns, such that the 
system can be solved as soon as $\alpha \geq \alphagmin$, which confirms the bound~(\ref{eq:gain_counting}).

\subsection{Convex formulation}
\label{sec:convexGainFormulation}
Blind gain calibration was studied in~\cite{blindCalConvex} using convex optimization.
As in \cs, sparsity is enforced by an $L_1$ norm. 
For the problem to be convex, the change of variable $\dv \gets \dv^{\inv}$ is necessary, leading to the formulation
\begin{align}
 (\xhv, \dhv) &= \argmin_{\xv,\dv} || \xv ||_1 \quad \text{s.t.} \quad \tilde{\mathbf{D}} \yv = \Fv \xv, \, \Tr(\tilde{\mathbf{D}})=M	, \label{eq:convexGain}
\end{align}
where the diagonal terms of the matrix $\tilde{\mathbf{D}}$ are the components of the vector $\dv$ and its non-diagonal
 elements are zeros.
Note that the condition on the trace imposes the mean of $\dv$, thus lifting the scaling invariance. 

In~\cite{blindCalComplex}, the authors of~\cite{blindCalConvex} generalize the algorithm to the complex setting in which
\begin{align}
 \xv \in \CC^{N \times P}, \quad \Fv \in \CC^{M \times N},\quad \dv \in (\CC^*)^{M}.	\label{eq:complexGain}
\end{align}

\subsubsection{Complexity and convergence}
The results presented in~\cite{blindCalConvex} show that a small number $P$ of signals is usually sufficient for successful inference.
In the realized experiments, $P=3$ or $P=5$ signals allow perfect reconstruction in large parts of the $(\rho, \alpha)$ space.
As in \cs, a phase transitions is visible between regions of successful and unsuccessful inference.
For $P=21$, the experimental phase transition is nearly superposed with the Donoho-Tanner phase transition of \cs.

Being a convex minimization problem, the formulation of~\eqref{eq:convexGain} can be implemented with standard libraries such as
the CVX package for MATLAB~\cite{cvx,gb08}.
This ensures a fast and reliable convergence of the algorithm. 

In~\chapref{chap:blindSensorCal} I present a Bayesian algorithm for blind gain calibration, motivated by the 
fact that in \cs, the Bayesian approach outperforms the convex minimization approach.

\subsection{Applications}
The blind gain calibration problem can be encountered when signals are measured by physical devices (sensors) that introduce a multiplicative gain.
Ideally, sensor gains are known, either through a precise fabrication process, or experimentally determined after fabrication. 
The system can then be calibrated accordingly.
However, the gain of a sensor could vary over time (with the aging of the device or exterior conditions such as temperature, humidity...), thus
requiring regular calibration.
Supervised calibration might not always be possible (see~\appliref{appli:radioAstronomy}) or simply not desirable because it is not user-friendly.
Therefore blind calibration procedures can be necessary. Two applications of blind calibration are given below. 
Other applications include calibration of microphone arrays~\cite{microphones} or time-interleaved AD converters~\cite{adc1}. 

\begin{application}{Radio astronomy}
 \label{appli:radioAstronomy}
 In~\cite{radioInterferometryCalibration}, the authors present a modified version of the complex gain calibration algorithm proposed 
 in~\cite{blindCalComplex} and apply it to radio interferometry.
 
 Very long baseline interferometry~\cite{radioInterferometry,vlba} allows astronomical observations in the radio frequencies, for which conventional telescopes can not
 be used, as diffraction effects produced by lenses are far too important at radio frequencies.
 
 In radio interferometry, an array of antennae distributed on the ground over a surface of several square kilometers 
 \textit{emulates} a lens of the corresponding size, thus minimizing diffraction effects.
 In order to obtain good images, precise calibration procedures are necessary~\cite{RICalibration}.
 Different effects lead to decalibration. In~\cite{radioInterferometryCalibration}, their combined effect is 
 formulated as a complex gain calibration problem, in which all decalibration effects are treated with a single calibration parameter $\dv$, 
 the signal $\xv$ is the image of the sky and the measurement matrix $\Fv$ is known.
 
 In radio interferometry, calibration is necessary and can only performed blindly, as the only possible observation is the sky 
 itself: To the contrary of other applications, it is not possible to measure a known signal and use it for supervised calibration.
 The sparsity of images of the sky make it an ideal application of sparse inference~\cite{radioInterferometryCS}.
\end{application}

\begin{application}{Blind deconvolution}
 \label{appli:blindDeconvolution}
 Blind calibration is closely related to the problem of blind deconvolution.
 In imaging, deblurring images blurred by an imperfect measurement system or by camera movement 
 is a blind deconvolution task~\cite{blindDeconvolutionBook}. 
\end{application}
\begin{applicationSuite}
 The blurred image $\xv'$ is the convolution of the signal $\xv$ with a \textit{blurring kernel} $\kv$:
 \begin{align}
  \xv' = \kv \star \xv, \qquad \text{where} \quad [\kv \star \xv]_i = \sum_j k_{i-j} x_j. \label{eq:convolution}
 \end{align}
 Neglecting effects at the image borders, the convolution can be replaced by a circular convolution, which 
 has the property of being a multiplication in Fourier space. Therefore,
\begin{align}
 \FF(\xv') &= \FF(\kv) \odot \FF(\xv).
\end{align}
Noting $\yv \equiv \FF(\xv')$ and $\dv = \FF(\kv)$, the problem writes
\begin{align}
 y_{\mu} &= d_{\mu} \sum_i \FF_{\mu i} x_i,
\end{align}
(where $\FF_{\mu i}$ are the coefficients of the Fourier transform matrix),  which is nothing but complex gain calibration
if the kernel $\kv$ is unknown.
Usually one seeks to perform blind deconvolution from a single image, that is with $P=1$.
This can be possible provided one can make very strong assumptions on the kernel, in which case the counting bound~(\ref{eq:gain_counting}) 
can be beaten~\cite{BlindDeconvolutionIdentifiability}.
However, performing blind deconvolution using $P>1$ blurred images can be considered as well, provided the blurring kernel is the 
same for each of the pictures taken. This would for example be the case of pictures taken by the same camera whose objective introduces blurring.

Blind deconvolution algorithms can use image sparsity and convex approaches~\cite{blindDeconvolutionSparse,blindDeconvolutionConvex}.
\end{applicationSuite}

\section{Matrix factorization}
\begin{figure}[h]
 \centering
 \includegraphics[width=0.8\textwidth]{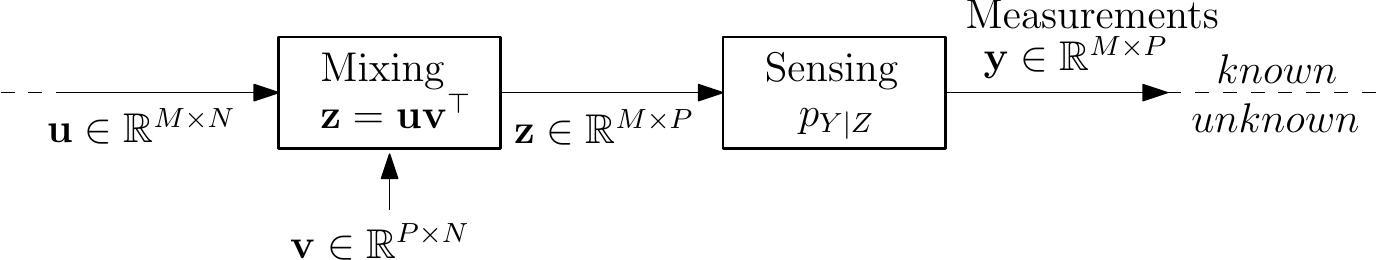}
 \caption[Generalized matrix factorization]{Generalized matrix factorization aims at solving inference problems with the generative model shown above.}
 \label{fig:4_gmf}
\end{figure}

As blind gain calibration, matrix factorization (\mf) is a bilinear inference problem.
In this section, I present a ``generalized'' version of it, illustrated by~\figref{fig:4_gmf}, as considered in~\cite{mfISIT,kabaMF}.

In matrix factorization, the number of unknowns is even higher than in blind calibration, 
as the measurement matrix itself is unknown.
For this reason I use the notations $(\uv,\vv)$ instead of $(\Fv,\xv)$ to clarify the equivalence of signal and matrix.
In a mixing step, $\uv \in \RR^{M\times N}$ and $\vv \in \RR^{P \times N}$ produce an intermediate variable $\zv = \uv \vv^{\top} \in \RR^{M \times P}$ whose components are given by
\begin{align}
 \z_{\mu l} &= \sum_{i} \u_{\mu i} \v_{l i}.	\label{eq:z_mf}
\end{align}
As~\eqref{eq:gainCal}, this equation is bilinear.
As in \glms, this mixing step is followed by a sensing step through a probabilitic channel $\py$ that produces the measurements,
hence the name \textit{generalized bilinear model}.
Specific settings of matrix factorization are discussed in~\secref{sec:mf_examples}, depending on 
the distributions of $\uv$ and $\vv$ and on the sensing channel.
We define the measurement rates
\begin{align}
 \alpha_U \equiv \frac{M}{N} \qquad \text{and} \qquad \alpha_V \equiv \frac{P}{N}.
\end{align}

\subsubsection{Invariances}
As in blind gain calibration, there is a scaling invariance in matrix factorization.
However, it is a much more general invariance: For every invertible matrix $\mathbf{C} \in GL(M,\RR)$,
the couples $(\uv,\vv)$ and $(\uv \mathbf{C} , \vv ( \mathbf{C}^{-1})^{\top})$ produce the same measurements.
As in blind gain calibration, this invariance might in theory be lifted if the distributions 
of $\uv$ and $\vv$ are known. 
However, for the same reason of finite system size, the invariance cannot be expected to be completely lifted.
Furthermore, it might introduce correlations between components of $\uv$ and $\vv$ that are supposed to be independent.

In some applications, the goal is to recover $\zv$: as $\zv$ is not affected by the invariances,
a good measure of success is the MSE on $\zv$.
In other applications, the goal is to recover $\uv$ and $\vv$, in which case 
the invariance might be a real issue.

\subsection{Algorithm and analysis}
As for \glms, a message-passing algorithm can be derived starting from the probabilitic approach to the inference problem.
This approach can be analyzed with the replica method, exactly as was done in~\chapref{chap:generalizedLinearModels}.
Derivations of the algorithm, called \bigamp, can be found in~\cite{bigamp1,kabaMF}, a derivation of the replica analysis in~\cite{kabaMF}.
As a very similar derivation is done in~\chapref{chap:matrixCS}, I will only present the results using notations that 
are coherent with those used in~\chapref{chap:generalizedLinearModels}.

\subsubsection{\bigamp}
\bigamp is presented in~\algoref{algo:bigamp}. It performs Bayesian inference of generalized matrix factorization in the same 
probabilistic framework as described for \gamp in~\secref{sec:additionalConditions}, starting from the posterior distribution
\begin{align}
 \p(\uv,\vv|\yv) \propto \prod_{\mu i} &\p_U(u_{\mu i})  \prod_{l i} \p_V(v_{l i})  
 \int \prod_{\mu l} \dd z_{\mu l} \py(y_{\mu l}|z_{\mu l}) \delta \left(z_{\mu l} -  \sum_{i=1}^{N} u_{ \mu i} v_{l i} \right).  \label{eq:proba_gmf}  
\end{align}
Its structure is very similar to \gamp: for each variable of the problem, there are two 
estimators along with their uncertainties, and the $(\ghv,\gbv)$ variables for the gradient descent. 
The update functions exactly correspond to those in \gamp (see~\appref{app:f}).
\begin{algorithm}
\caption{\bigamp}
\label{algo:bigamp}
 \textbf{Initialize} $\ghv_0=0$ and $(\uhv_0, \ubv_0, \vhv_0, \vbv_0)$ at random or according to $\pu$ and $\pv$. \\
  \textbf{Main loop:} while $t<t_{\rm max}$, calculate following quantities:
  \begin{align*}
  \Zbv_{t+1} &= \ubv_t \vbv_t^{\top} + \ubv_t ( \vhv_t^2 )^{\top} + \uhv_t^2 \vbv_t^{\top} \\
  \Zhv_{t+1} &= \uhv_t \vhv_t^{\top} - \ghv_t \odot \left( \ubv_t \left(  \vhv_t \odot  \vhv_{t-1} \right)^{\top} +  \left( \uhv_t \odot  \uhv_{t-1} \right) \vbv_t^{\top} \right) \\
  \gbv_{t+1} &= \gb^Y(\yv|\Zhv_{t+1}, \Zbv_{t+1}) \\
  \ghv_{t+1} &= \gh^Y(\yv|\Zhv_{t+1}, \Zbv_{t+1}) \\
  \Ubv_{t+1} &= - \left(  \gbv_{t+1} \left[ \vhv_t^2 +  \vbv_t \right] +   \ghv_{t+1}^2 \vbv_t\right)^{\inv} \\
  \Uhv_{t+1} &= \Ubv_{t+1} \odot \left( \ghv_{t+1} \vhv_t  - \uhv_t \odot \gbv_{t+1}\vhv_t^2  - \uhv_{t-1} \odot  (\ghv_{t+1} \odot \ghv_t) \vbv_{t-1} \right)  \\
  \ubv_{t+1} &= \fb^U(\Uhv_{t+1}, \Ubv_{t+1}) \\
  \uhv_{t+1} &= \fh^U(\Uhv_{t+1}, \Ubv_{t+1})  \\
  \Vbv_{t+1} &= - \left( \gbv_{t+1}^{\top} \left[  \uhv_t^2 +  \ubv_t \right]  +   (\ghv_{t+1}^2)^{\top} \ubv_t \right)^{\inv} \\
  \Vhv_{t+1} &= \Vbv_{t+1} \odot \left( \ghv_{t+1}^{\top} \uhv_t - \vhv_t \odot \gbv_{t+1}^{\top} \uhv_t^2  - \vhv_{t-1} \odot  (\ghv_{t+1} \odot \ghv_t)^{\top} \ubv_{t-1} \right)  \\
  \vbv_{t+1} &= \fb^V(\Vhv_{t+1}, \Vbv_{t+1}) \\
  \vhv_{t+1} &= \fh^V(\Vhv_{t+1}, \Vbv_{t+1})   
  \end{align*}
  \textbf{Stop} when $t=t_{\rm max}$, $|| \ubv_t||_2 + || \vbv_t||_2  < \epsilon$ or $|| \uhv_t - \uhv_{t-1} ||_2+|| \vhv_t - \vhv_{t-1} ||_2< \epsilon$.
\end{algorithm}

\subsubsection{Free entropy and state evolution equations}
The free entropy expressed as a saddle point reads \mynote{Generalized MF free entropy}{1.2}
\begin{align}
  \phi = \saddle   & \left\{ \alpha_U \left( -\mv \mhv - \frac{1}{2} \Qv \Qhv + \frac{1}{2} \qv \qhv +\iv(\Qmhv) \right) \right. \nonumber \\
  &\left. +\alpha_V \left( -\mymu \mhu - \frac{1}{2} \Qu \Qhu + \frac{1}{2} \qu \qhu +\iu(\Qmhu)  \right) +  \alpha_U \alpha_V \iz(\Qmz) \right\} ,	\label{eq:phiMF}
\end{align}
where $\iu, \iv$ are the exact equivalents of $\ix$~(\ref{eq:ix}), $\iz$ is identical to~(\ref{eq:iz}) and
\begin{align}
 \Qmz &= \Qmu \odot \Qmv,
\end{align}
such that the saddle point has to be performed over the 12 variables $(\mymu,\mhu,\qu,\qhu,\Qu,\Qhu)$ and $(\mv,\mhv,\qv,\qhv,\Qv,\Qhv)$.
The resulting state evolution equations are given by
\begin{align}
  \mymu &= \sqrt{\frac{\qhu}{\mhu ^2}} \int \dd t \, f_1^{U,0} \left(\frac{\sqrt{\qhu }}{\mhu }t , \frac{\qhu }{\mhu^2} \right)\fh^U \left( \frac{ \sqrt{\qhu } t }{\qhu  - \Qhu } , \frac{1}{\qhu  - \Qhu }  \right) ,   \label{eq:mu_DE_gen} \\
 \Qu-\qu &= \sqrt{\frac{\qhu}{\mhu^2}} \int \dd t \,  f_0^{U,0} \left(\frac{\sqrt{\qhu}}{\mhu}t , \frac{\qhu }{\mhu^2} \right)  \fb^U \left( \frac{ \sqrt{\qhu } t }{\qhu - \Qhu} , \frac{1}{\qhu - \Qhu} \right) ,  \label{eq:Qu_DE_gen}  \\
 \qu &= \sqrt{\frac{\qhu}{\mhu^2}} \int \dd t \,  f_0^{U,0} \left(\frac{\sqrt{\qhu}}{\mhu}t , \frac{\qhu}{\mhu^2} \right)  \left[ \fh^U \left( \frac{ \sqrt{\qhu} t }{\qhu - \Qhu} , \frac{1}{\qhu - \Qhu} \right) \right]^2   \label{eq:qu_DE_gen} ,
\end{align}
the same $3$ equations with $V \leftrightarrow U$ and \mynote{Generalized MF state evolution}{2.2} \mynote{Generalized MF state evolution}{-2.9}
\begin{align}
   \mhz &= \frac{1}{\mz}  \int \dd \y \int \DD t \frac{ \left[ \frac{\partial}{\partial t} f_0^{Y,0}\left(\y | \frac{\mz}{\sqrt{\qz}} t , \Qzz - \frac{\mz^2}{\qz} \right) \right] \left[ \frac{\partial}{\partial t} f_0^{Y}\left( \y |\sqrt{\qz} t , \Qz -\qz \right) \right]}{f_0^{Y}\left( \y |\sqrt{\qz} t , \Qz -\qz \right)} , \label{eq:mhz_DE_gen} \\
  \qhz &= \frac{1}{\qz}  \int \dd \y \int \DD t \left[ \frac{\frac{\partial}{\partial t} f_0^{Y}\left( \y |\sqrt{\qz} t , \Qz -\qz \right)}{f_0^{Y}\left( \y |\sqrt{\qz} t , \Qz -\qz \right)} \right]^2 f_0^{Y,0}\left(\y | \frac{\mz}{\sqrt{\qz}} t , \Qzz  - \frac{\mz^2}{\qz} \right),   \label{eq:qhz_DE_gen}\\
  \Qhz &= 2 \int \dd \y \int \DD t f_0^{Y,0}\left(\y | \frac{\mz}{\sqrt{\qz}} t , \Qzz  - \frac{\mz^2}{\qz} \right) \left( \frac{\frac{\partial}{\partial \Qz} f_0^{Y}\left( \y |\sqrt{\qz} t , \Qz -\qz \right)}{f_0^{Y}\left( \y | \sqrt{\qz} t , \Qz -\qz \right)} \right),  \label{eq:Qhz_DE_gen}
\end{align}
with 
\begin{align}
 \mhu &= \alphau \mv \mhz, & \qhu &= \alphau \qv \qhz, & \Qhu &= \alphau \Qv \Qhz, \\
 \mhv &= \alphav \mymu \mhz, & \qhv &= \alphav \qu \qhz, & \Qhv &= \alphav \Qu \Qhz.
\end{align}

As for \glms, the Bayes optimal setting greatly simplifies the state evolution to the following 3 equations: \mynote{Bayes optimal generalized MF state evolution}{2.5}
\begin{align}
  \mhz^{\iter+1} &= \frac{1}{\mz^{\iter}} \int \dd \y \int \DD t  \frac{\left[ \frac{\partial}{\partial t} f^Y_{0}\left(\y |\sqrt{\mz^{\iter}}t,\Qzz  -\mz^{\iter}\right)\right]^2}{f^Y_{0}\left(\y |\sqrt{\mz^{\iter}}t, \Qzz -\mz^{\iter}\right)} ,    \label{eq:bo_1}\\
 \mymu^{\iter+1} &= \frac{1}{\sqrt{\mhu^{\iter+1}}} \int \dd t \frac{\left[f^U_1\left(\frac{t}{\sqrt{\mhu^{\iter+1}}},\frac{1}{\mhu^{\iter+1}}\right) \right]^2}{f^U_0\left(\frac{t}{\sqrt{\mhu^{\iter+1}}},\frac{1}{\mhu^{\iter+1}}\right)} , \\
 \mv^{\iter+1} &= \frac{1}{\sqrt{\mhv^{\iter+1}}} \int \dd t \frac{\left[f^V_1\left(\frac{t}{\sqrt{\mhv^{\iter+1}}},\frac{1}{\mhv^{\iter+1}}\right) \right]^2}{f^V_0\left(\frac{t}{\sqrt{\mhv^{\iter+1}}},\frac{1}{\mhv^{\iter+1}}\right)} .	\label{eq:bo_3}
\end{align}

In~\cite{kabaMF}, several settings of generalized matrix factorization are examined with these Bayes optimal state evolution equations.
For reasons explained below, the theoretically predicted performances are unfortunately not always reached using~\algoref{algo:bigamp}.

\subsubsection{Complexity and convergence}
In most settings of matrix factorization, $M$, $N$ and $P$ are of the same order.
For this reason, matrix factorization algorithms are very slow and it is problematic to run them for $(M,N,P)$ bigger than a few hundreds.
In comparison, \gamp can perform inference on problems with $N=10000$ on a standard laptop.
An interesting special case of matrix factorization in which $N$ is of order one is examined in~\cite{tanaka,tibo1}.

Unlike \gamp for \cs or quantized sensing, \bigamp does not converge in the form in which it is presented in~\algoref{algo:bigamp}.
The convergence issue is similar to the one of \gamp when the matrix does not have zero mean or when it is low rank: after a few iterations, some of the estimators go to infinity.
To correct this behaviour, \bigamp can be modified to include damping. 
Such a damping scheme is presented in~\cite{bigamp1}, allowing to significantly improve convergence properties of \bigamp. 
Numerous experimental results are given in~\cite{bigamp2} for different applications of matrix factorization.
However, the performance predicted by the state evolution analysis can in general not be obtained, meaning that there is still room for improvements.
The reasons for the bad convergence properties of \bigamp and the necessity of damping are not well understood.
Therefore, damping schemes are mainly empirical and can probably be improved upon.

A very simple damping scheme that works quite well in some simple configuration is the following, that damps only one variable:
\begin{align}
 \Uhv_{t+1} \leftarrow \beta \Uhv_{t+1} + (1-\beta) \Uhv_{t},	\label{eq:damp_u}
\end{align}
with $\beta=0.3$, applied right before updating $\uhv^{t+1}$.
Furthermore, setting $\vbv=0$ and $\ubv=0$ in the update equations for $\Ubv$ and $\Vbv$ greatly reduces the risk of negative variances appearing, 
while emulating the Nishimori conditions~\cite{kabaMF}.

\subsection{Applications}
\label{sec:mf_examples}

\subsubsection{Dictionary learning}
\label{sec:DL}
Dictionary learning (\dl) can be considered as a more difficult version of \cs, in which the measurement matrix ($\uv$ or $\Fv$) itself is an unknown.
As in gain calibration, the \textit{blind} scenario is the most difficult, when only unknown signals can be measured.
The measurement matrix is then usually called the \textit{dictionary}. 
The sensing channel is AWGN and the signals ($\vv$ or $\xv$) are sparse, such that each measurement is a linear combination of a few entries of the dictionary.
A simple couting bound for \dl, obtained with the same reasoning as~\eqref{eq:gain_counting} is
\begin{align}
 P \geq \frac{\alpha}{\alpha - \rho} N, 	\label{eq:dl_counting}
\end{align}
showing that both $P$ and $M$ have to be of the same order as $N$ for successful inference.

It is possible to interpolate between \cs and \dl thanks to the ``blind matrix calibration'' setting~\cite{kabaMF,mfISIT}, 
in which a noisy estimate of $\uv$ is known:
\begin{align}
 \tilde{\uv} = \frac{\uv + \sqrt{\eta} \xiv}{\sqrt{1+\eta}},	\label{eq:blindMatrixCal}
\end{align}
where $\xiv$ is \awgn with of variance $1$.
Varying the parameter $\eta$ between $0$ and $\infty$ is interpolating between \cs and \dl and is 
a good setting to study in order to understand the problems arising in \dl.
In the state evolution, equations~(\ref{eq:mu_DE_gen}--\ref{eq:qu_DE_gen}) are replaced by equations~(\ref{eq:mx_DE_gen_add}--\ref{eq:qx_DE_gen_add}) for $\uv$,
as it is partially known.
In~\secref{sec:instability}, I will use the blind matrix calibration setting to examine the stability of the Nishimori line (\ref{eq:nishConds}) in bilinear inference problems.


\subsubsection{Low-rank matrix completion}
\label{sec:matrixComp}
Another interesting application of matrix factorization is low-rank matrix completion.
In that setting, the unknown to be infered is $\zv$, which is observed through a channel such that 
only a fraction $\epsilon$ of its components are measured.
The other elements of $\zv$ are unknown, but the matrix is known to be low-rank, therefore
\begin{align}
 \exists  \uv \in \RR^{M \times R}, \exists \vv \in \RR^{P \times R}, \quad \text{s.t.} \quad R<\min(M,P) \quad \text{and} \quad \zv=\uv \vv^{\top} .	\label{eq:lowRankedness}
\end{align}
The number of variables of the problem is therefore reduced to $R\times( M + P) < MP$, and a simple counting bound on $\epsilon$ is 
\begin{align}
 \epsilon \geq \frac{\alphau+\alphav}{\alphau \alphav}.	\label{eq:completion_counting}
\end{align}
A real-world application of low-rank matrix completion is the netflix prize~\cite{netflix}. 
In that challenge, the goal is to complete a matrix containing grades given by users to different movies
by observing only a fraction of the entries. The goal is to be able to \textit{predict} which movies a user likes
in order to make the best possible recommendation.

Low-rank matrix completion is an application for which \bigamp works quite well with a simple damping scheme.

\section{Matrix compressed sensing}
\label{sec:mcs}
\begin{figure}[h]
 \centering
 \includegraphics[width=1\textwidth]{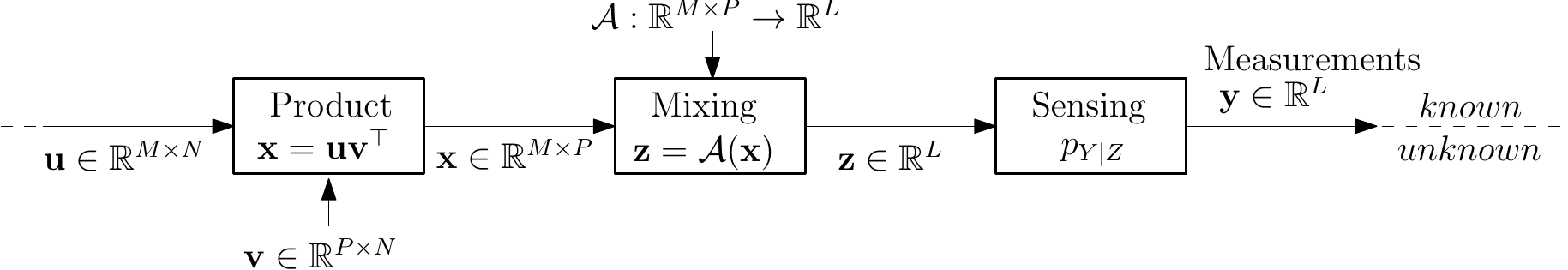}
 \caption[Matrix compressed sensing]{Setting for generalized matrix compressed sensing.}
 \label{fig:4_gmcs}
\end{figure}

A third example of bilinear inference problem is matrix compressed sensing (\mcs). 
\mcs is similar to \cs with the difference that the signal is a low-rank matrix instead of a sparse vector.
The measurements are made from linear combinations of the matrice's elements, with the usual probabilistic sensing channel $\py$ for the \textit{generalized} extension.
The setting is illustrated in~\figref{fig:4_gmcs}.

In~\chapref{chap:matrixCS}, I derive a Bayesian message-passing algorithm for \mcs and perform the theoretical analysis using the replica method.
\subsection{Setting}
As in matrix factorization, the matrices $\uv \in \RR^{M \times N}$ and $\vv \in \RR^{P \times N}$ are multiplied, but we call $\xv \in \RR^{M \times P}$ their product:
\begin{align}
 \xv &= \uv \vv^{\top}.	\label{eq:mcs_x}
\end{align}
The elements of $\xv$ are mixed with a linear operator $\A: \RR^{M \times P} \to \RR^{L}$ that can be represented by a matrix $\AM \in \RR^{L \times M \times P}$:
\begin{align}
 \zv &= \A(\xv) \in \RR^L, \label{eq:mcs_z}
\end{align}
or, written componentwise:
\begin{align}
 \z_l &= \left[ \A(\x) \right]_{l} = \sum_{\mu=1}^M \sum_{p=1}^P A^{l \mu p} \x_{\mu p}.	\label{eq:mcs_z_elementwise}
\end{align}
As for \cs, we generally consider that the matrix elements of $\AM$ are \iid random variables, which allows $\A$ to satisfy the generalization of the RIP 
for matrix compressed sensing with high probability~\cite{nuclearNormMin}.
As for matrix factorization, we consider the generalized setting, in which the measurement $\yv$ are taken through an element-wise, probabilistic sensing channel:
\begin{align}
 \yv \sim \py(\yv|\zv).
\end{align}

\subsection{Link to matrix factorization}
Matrix factorization can be seen as a special case of \mcs when $\A$ is the identity (in that case the \iid assumption on the elements of $\AM$ does obviously not hold).

But more generally, the setting is very similar to \mf if $L=MP$: if $\A$ has \iid elements, it is with high probability a bijection, and does not introduce any further difficulty into the problem.
If $L<MP$, and $\A$ has \iid elements, it is with high probability injective and the setting is very similar to the matrix completion scenario examined in~\secref{sec:matrixComp} with $\epsilon=\frac{L}{MP}$.

In that sense, matrix compressed sensing is the most general inference setting considered in \mythesis as it contains matrix factorization and thus generalized linear models,
which corres\-ponds to knowing $\uv$.

\subsection{Existing work for the \awgn channel}
Most previous work on matrix compressed sensing (which is known under different names, such as ``matrix sensing'', ``matrix completion'', ``affine rank minimization'') 
consider the \awgn sensing channel.
As  \cs, \mcs can be expressed as a minimization problem:
\begin{align}
 \xhv &= \argmin_{\xv} rank(\xv) \quad \text{ such that } \quad \A( \xv ) = \yv.	\label{eq:rankMin}
\end{align}
Note that the rank plays the role of the $L_0$ ``norm'' in \cs. Under this form, the minimization problem is hard to solve (it is in general NP-hard).
The algorithms developed in this context often rely on the minimization of the \textit{nuclear norm} of $\xv$, which is the sum of its singular values, 
while taking into account the measurements $\yv$~\cite{Donoho21052013,nuclearNormMin}. They therefore aim at solving the minimization problem
\begin{align}
 \xhv &= \argmin_{\xv} \frac{1}{2} || \yv - \A(\xv) ||_2^2 + \lambda || \xv ||_*,	\label{eq:nuclearNormMin}
\end{align}
which is a convex relaxation of~(\ref{eq:rankMin}), just as \lasso is a convex relaxation of the $L_0$ minimization problem in \cs.
As in \cs, there are conditions under which the convex relaxation~(\ref{eq:nuclearNormMin}) is equivalent to the rank minimization~(\ref{eq:rankMin}):
under these conditions, matrix compressed sensing can be solved using convex minimization techniques.

One limitation of this approach is that the equivalence of the problems does not hold in all ranges of parameters and that it performs the minimization in 
the high-dimensional space $\RR^{M \times P}$.

Other works rely on the decomposition of $\xv$ into the product $\uv \vv^{\top}$. 
This allows to reduce the search space and to take into account potential further requirements on $\uv$ and $\vv$ such as sparsity~\cite{lee2013near}, but has several disadvantages.
First, by using the $\uv \vv^{\top}$ decomposition the problem becomes non-convex. 
Second, decomposing $\xv$ into $\uv \vv^{\top}$ using singular values decomposition is computationally demanding and therefore has to be avoided if possible.

Such algorithms use alternating minimizations producing estimates of $\uv$ and $\vv$~\cite{hanLowRank, JainLowRank},
or focus on the case in which $\xv$ is positive semi-definite (therefore $\vv = \uv$)~\cite{laffertyLowRank} and use gradient descent.
\begin{algorithm}
\caption{Power factorization}
\label{algo:pf}
  \textbf{Initialization}: Initialize $\vhv_0$ at random. \\
  \textbf{Main loop:} while $t<t_{\rm max}$: \\
  Fix $\vhv_{t}$, update $\uhv_{t+1}$ such that
  \begin{align*}
    \uhv_{t+1} &= \argmin_{\uv} || \yv - \A(\uv \vhv_{t}^{\top})||_2^2.
  \end{align*}
  Fix $\uhv_{t+1}$, update $\vhv_{t+1}$ such that
  \begin{align*}
    \vhv_{t+1} &= \argmin_{\vv} || \yv - \A(\uv_{t+1} \vv^{\top})||_2^2.
  \end{align*}
  \textbf{Result:} $\xhv_t = \uhv_{t} \vhv_{t}^{\top}$.
\end{algorithm}
An example of a simple alternating minimization scheme is the ``Power Factorization'' algorithm proposed in~\cite{powerFactorization} and given in~\algoref{algo:pf}.
Note that both steps are simple and fast, that no singular value decomposition is needed and that results are claimed to be better than those obtained by nuclear norm minimization.
On the other hand, unlike nuclear norm minimization, power factorization is not convex and thus no theoretical bounds on its performance are known.



\section{Conclusion}
Generalized bilinear models are a vast class of inference problems with numerous applications.
One subclass is the generalized matrix factorization setting, that can be analyzed using the replica method, just as \glms.

Despite its similarity to \glms, inference of generalized matrix factorization using the Bayesian message-passing algorithm \bigamp is in general much more problematic 
for several reasons. First, unliftable invariances make the problem naturally ill-posed. 
Second, the complexity of inference is higher due to a higher number of variables to infer, such that 
inference is only practical for moderate signal sizes. 
Finally, the stability issues occasionally encountered in \gamp seem omnipresent in its counterpart \bigamp.
One reason for this instability is given in~\chapref{chap:matrixCS}.
As a result, convergence of \bigamp heavily relies on mainly heuristic damping and restarting strategies.
Improving those in order to obtain a more reliably algorithm is a major axis for future research.

Two other subclasses of generalized bilinear models are blind gain calibration and matrix compressed sensing, 
which I treat in more details in~\chapref{chap:blindSensorCal} and~\chapref{chap:matrixCS} respectively.
 
 \part{Main contributions}
 \label{part:3}
 
 \chapter{Vectorial GAMP and applications}
\label{chap:gampApplications}

Many physical signals are best represented by variables with both a real and an imaginary part.
This is the case for propagating waves such as sound and light, that have both an amplitude and a phase.
Two important applications in this case are magnetic resonance imaging and crystallography.

In this chapter, I give a derivation of \gamp for vectorial variables, from which \gamp for complex variables (\cgamp) can be obtained, and  I analyze the resulting 
complex \cs algorithm with the state evolution formalism.

I present two applications of vectorial \gamp. 
The first one is in the field of coding theory, and focusses on so-called \textit{superposition} codes.
The second one is \textit{phase retrieval}, for which I present both a theoretic study and the results of an optics experiment.
These two applications are presented in publications I have coauthored:~\cite{moiCodes,moiOptique}.


\section{Complex GAMP}
Consider the same setting as in \gamp except all variables are complex:
\begin{align}
 \xv \in \CC^{N} \quad \text{and} \quad \Fv \in \CC^{M \times N},	\label{eq:complexVariable1}
\end{align}
and $\zv$ and $\yv$ are generated as
\begin{align}
 \zv = \Fv \xv \in \CC^{M} \quad \text{and} \quad  \yv \sim \py(\yv | \zv) \in \CC^{M}. \label{eq:complexVariable3}
\end{align}
The posterior probability distribution still reads
\begin{align}
 \p(\xv | \yv, \Fv) &=  \frac{1}{\ZZ(\yv,\Fv)} \prod_{i=1}^N \px(\x_i) \prod_{\mu=1}^M \left( \int \dd \z_{\mu} \py(\y_{\mu}|\z_{\mu}) \delta(\z_{\mu} - \sum_{i=1}^N F_{\mu i} \x_i) \right),  \label{eq:complexGampPosterior}
\end{align}
with complex instead of real variables.
In general, the prior $\px$ and the channel $\py$ do not treat the real and imaginary parts of their respective variables independently.
For example, $\px$ can enforce the real and imaginary parts of $\xv$ to be zero at the \textit{same} locations.
This \textit{joint} sparsity assumption is stronger than the simpler assumption that both the real and the imaginary part of $\xv$ are sparse.
In order to take advantage of this, it is important to treat each component of $\xv$ (and of $\zv$ and $\yv$) as a single, complex variable, and not as two independent real variables.
Therefore the generalization of \gamp to the complex setting is not entirely trivial---although the resulting algorithms differ only very slightly, 
in a way that could be guessed. 
One possible systematic way to derive complex \gamp is presented below.

\subsection{Vectorial GAMP}
\label{sec:vectorial_gamp}
For the derivation, I consider a setting that is more general than the complex variables scenario of~(\ref{eq:complexVariable1}--\ref{eq:complexVariable3}).
Namely, let us consider the following scenario:
\begin{align}
 \forall i, \, \xv_i \in \RR^{\dun}, \quad  \forall (\mu, i), \, \Fv_{\mu i} \in \RR^{\ddeux \times \dun}, \quad  \forall \mu, \, \zv_{\mu} &= \sum_{i} \Fv_{\mu i} \xv_i \in \RR^{\ddeux},	\label{eq:vectorialVariable1}
\end{align}
with ($\dun, \ddeux$) of order one, and 
\begin{align}
  \forall \mu, \, \yv_{\mu} &\sim \py(\yv_{\mu} | \zv_{\mu} ) \in \RR^{\dtrois}, \label{eq:complexVariable4}
\end{align}
with $\dtrois$ of order one as well.

In order for the elements of $\z_{\mu}$ to be $O(1)$, we take the elements of $\Fv_{\mu i}$ to have variance $1/N$.
We take the matrices $\Fv_{\mu i}$ to be \iid random variables, but the elements of each $\Fv_{\mu i}$ are not necessarily independent.

The complex \gamp case corresponds to $\dun=\ddeux=\dtrois=2$ with
\begin{align}
 \xv_i &= \left( \begin{tabular}{c}
                $\Re(\x_i) $\\
                $\Im(\x_i)$
               \end{tabular}
               \right),
               \quad
 &  \zv_{\mu} &= \left( \begin{tabular}{c}
                $\Re(\z_{\mu}) $\\
                $\Im(\z_{\mu})$
               \end{tabular}
               \right),
               \quad
 &  \yv_{\mu} &= \left( \begin{tabular}{c}
                $\Re(\y_{\mu}) $\\
                $\Im(\y_{\mu})$
               \end{tabular}
               \right)
               \quad 	\label{eq:complex_vec1}
\end{align}
 and
 \begin{align}
 \Fv_{\mu i} &= \left( \begin{tabular}{cc}
                       $\Re(\F_{\mu i})$ & $- \Im(\F_{\mu i})$ \\
                       $\Im(\F_{\mu i})$ & $\Re(\F_{\mu i})$
                      \end{tabular}
                      \right). \label{eq:complex_vec2}
\end{align}

\subsubsection{Update functions}
Let us first introduce the equivalents of the $f$-functions used in real-valued \gamp.
The equivalent of means $\xh \in \RR$ and variances $\xb \in \RR^+$ are vectorial means $\xhv \in \RR^{\dun}$ and covariance matrices 
$\xbv \in \cov{\dun}$, where $\cov{\dun}$ is the ensemble of symmetric, positive-definite $\dun \times \dun$ matrices with real coefficients.
For any function $h: \RR^{\dun} \to \RR^+$, we define
\begin{align}
  f_0^h: \RR^{\dun} \times \cov{\dun}	&\rightarrow \RR  \nonumber \\
  (\xhv,\xbv) &\mapsto \int \dd \xv h(\xv) \NN\left( \xv; \xhv, \xbv \right),   \label{eq:f0_vec} \\
  f_1^h: \RR^{\dun} \times \cov{\dun}	&\rightarrow \RR^{\dun}  \nonumber 	\\
  (\xhv,\xbv) &\mapsto \int \dd \xv  \, \xv \, h(\xv) \NN\left( \xv; \xhv, \xbv \right), 	\label{eq:f1_vec} \\
 f_2^h:	\RR^{\dun} \times \cov{\dun}	&\rightarrow \cov{\dun}  \nonumber 	\\
 (\xhv,\xbv) &\mapsto \int \dd \xv \, \xv \xv^{\top} h(\xv) \NN\left( \xv; \xhv, \xbv \right),  \label{eq:f2_vec}
\end{align}
where $\NN\left( \xv; \xhv, \xbv \right)$ is a multivariate Gaussian (reminders are given in~\appref{sec:gaussians}).
From these, we define the update functions
\begin{align}
 \fh^h(\xhv,\xbv) &\equiv \frac{f_1^h(\xhv,\xbv)}{f_0^h(\xhv,\xbv)}, & &\in \RR^{\dun}   \label{eq:fh_vec} \\
 \fb^h(\xhv,\xbv) &\equiv \frac{f_2^h(\xhv,\xbv)}{f_0^h(\xhv,\xbv)} - \fh^h(\xhv,\xbv) \left( \fh^h(\xhv,\xbv) \right)^{\top} , & &\in \cov{\dun}  \label{eq:fb_vec}
\end{align}
and the auxiliary functions (using matrix inversions)
\begin{align}
 \gh^h(\xhv,\xbv) &\equiv \xbv^{-1} \left(  \fh^h(\xhv,\xbv) - \xhv  \right),  & &\in \RR^{\dun}  \label{eq:gh_vec}\\
 \gb^h(\xhv,\xbv) &\equiv \xbv^{-1} \left(  \fb^h(\xhv, \xbv) - \xbv  \right) \xbv^{-1}.	& & \in \sym{\dun} \label{eq:gb_vec}
\end{align}
As in \gamp, we will use these functions for $h=\px$ (denoted by '$X$') and $h=\py$ (denoted by '$\Y$') in order to calculate new estimators of $\xv$ and $\zv$ respectively.

\begin{SCfigure}[1][h]
 \centering
 \caption[\gamp factor graph]{Factor graph representing the posterior distribution~(\ref{eq:complexGampPosterior}).
 For clarity, $N=6, M=4$ and only the links from the first and last factor node are represented.
 As only one edge leads to the factor nodes representing the prior, it is not necessary to introduce messages on that edge.}
 \includegraphics[width=0.4\textwidth]{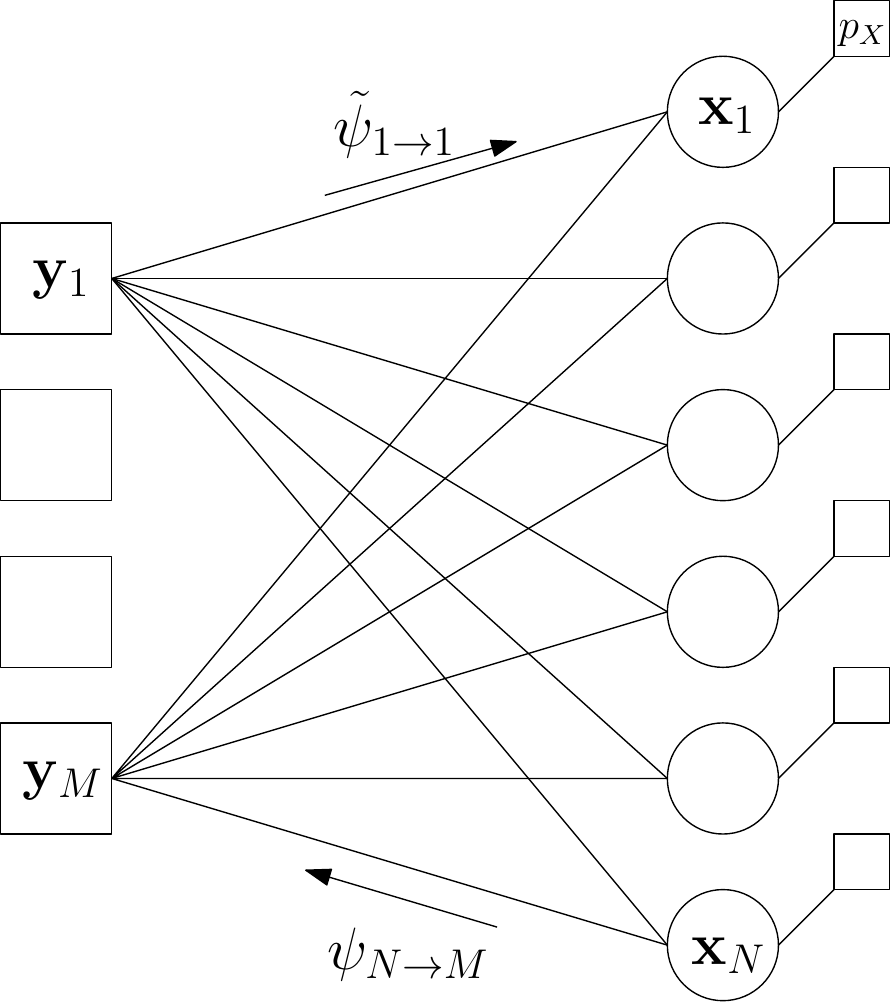}
 \label{fig:5_c_fg}
\end{SCfigure}

The factor graph representing the distribution~(\ref{eq:complexGampPosterior}) is shown on~\figref{fig:5_c_fg}.
With the help of it, I derive the \cgamp algorithm, following the exact same steps as for the derivation of \gamp, explained in~\secref{sec:bp_to_tap},
which the reader is suggested to re-read at this point.
The derivation of \gamp can be recovered by taking $\dun=\ddeux=\dtrois=1$.

\subsubsection{Step 1: BP}
We start by writing the \bp equations~(\ref{eq:bp_v}, \ref{eq:bp_f}) for the factor graph of~\figref{fig:5_c_fg}:
\begin{align}
 \messv_{i \to \mu}^{t}(\xv_i) &\propto \px(\xv_i) \prod_{\gamma \neq \mu} \messf_{\gamma \to i}^{t} (\xv_i) ,  	\label{eq:bp_v_gamp} \\
 \messf_{\mu \to i}^{t+1}(\xv_i) &\propto \int \prod_{j \neq i} \left( \dd \xv_j \messv_{j \to \mu}^{t}(\xv_j) \right) \int \dd \zv_{\mu} \py(\yv_{\mu} | \zv_{\mu}) \delta( \zv_{\mu} - \sum_i \Fv_{\mu i} \xv_i) . 	\label{eq:bp_f_gamp}
\end{align}
As each $\xv_i$ is a continuous, $\dun$-dimensional variable, these equations are problematic, as they would have to be calculated on a $d$-dimensional grid of points $\xv_i$, 
and as the right normalization constants could therefore only be approximated: Remember that the messages are probability distribution fuctions.  
Furthermore, $N-1$ $\dun$-dimensional integrations are required for each update of the $\messf$ messages: performing these numerically requires a lot of time and computational power.
In short, under this form, the \bp equations are close to useless.

\subsubsection{Step 2: AMP}
The second step consists in transforming equations~(\ref{eq:bp_v_gamp}, \ref{eq:bp_f_gamp}) such that they have a simple expression as a function of $\xv_i$, 
and reducing the number of integrals, such that they can easily be evaluated for any $\xv_i$.
For this, we use the (multidimensional) central limit theorem to approximate each component of $\zv_{\mu}$ by a Gaussian variable.
This is because 
\begin{align}
\zv_{\mu} &= \sum_{i=1}^N \Fv_{\mu i} \xv_i = \sum_{j \neq i} \Fv_{\mu j} \xv_j + \Fv_{\mu i} \xv_i 	\label{eq:zvmu}
\end{align}
is a sum of a large number $N-1$ of the random variables $\Fv_{\mu j} \xv_j$.
In~(\ref{eq:bp_v_gamp}), each $\xv_j$ (for $j\neq i$) is a random variable distributed with the \pdf $\messv_{j \to \mu}^{t}(\xv_j)$.
Let us call $\xhv_{j \to \mu}^t$ and $\xbv_{j \to \mu}^t$ the mean and covariance matrix of this pdf.
For the central limit theorem to hold, the random variables have to be independent, which actually is an assumption of \bp.
The mean of $\zv_{\mu}$ is then
\begin{align}
 \langle \zv_{\mu} \rangle &= \sum_{j \neq i} \Fv_{\mu j} \xhv_{j \to \mu}^t  + \Fv_{\mu i} \xv_i ,  \nonumber \\
 &\equiv \Zhv_{\mu \to i}^t + \Fv_{\mu i} \xv_i .
 \end{align}
 Note that $\xv_i$ is treated separately because in~\eqref{eq:bp_f_gamp} it is a constant.
 The covariance matrix of $\zv_{\mu}$ is 
 \begin{align}
  \Zbv_{\mu \to i}^t = \langle \zv_{\mu} \zv_{\mu}^{\top} \rangle - \langle \zv_{\mu} \rangle \langle \zv_{\mu} \rangle^{\top} &= \sum_{j \neq i} \Fv_{\mu j} \xbv_{j \to \mu}^t \Fv_{\mu j}^{\top}.
\end{align}
Note that as $\xv_i$ is an additive constant in~\eqref{eq:zvmu}, it does not contribute to this covariance matrix .
As a result,~\eqref{eq:bp_f_gamp} can be rewritten with \textit{one single} $\ddeux$-dimensional integral instead of $N$ of them:
\begin{align}
  \messf_{\mu \to i}^{t+1}(\xv_i) &\propto \int \dd \zv_{\mu} \py(\yv_{\mu} | \zv_{\mu}) \NN\left( \zv_{\mu}; \Zhv_{\mu \to i}^t + \Fv_{\mu i} \xv_i,  \Zbv_{\mu \to j}^t \right) . 	\label{eq:bp_f_gamp_gaussian}
\end{align}
We recognize the $f_0$-function defined in~\eqref{eq:f0_vec} (with $h=\py$, denoted by the exponent $Y$) 
and therefore we can write $\messf_{\mu \to i}$ in the simplified functional form
\begin{align}
  \messf_{\mu \to i}^{t+1}(\xv_i) \propto f_0^Y \left(\yv_{\mu}| \Zhv_{\mu \to i}^t + \Fv_{\mu i} \xv_i,  \Zbv_{\mu \to i}^t \right).
\end{align}

Now, let us express the messages $\messv_{i \to \mu}$ in a simpler form as well.
We start by noting that $|| \Fv_{\mu i} \xv_i ||_2 = O(\sqrt{\ddeux / N}) \ll || \Zhv_{\mu \to i}^t ||_2$ and approximate the function by the first terms of its Taylor series: 
\begin{align}
 \messf_{\mu \to i}^{t+1}(\xv_i) \propto  & f_0^Y \left(\yv_{\mu}| \Zhv_{\mu \to i}^t ,  \Zbv_{\mu \to i}^t \right)  \nonumber \\
					  &+ \nabla f_0^Y \left(\yv_{\mu}| \Zhv_{\mu \to i}^t ,  \Zbv_{\mu \to i}^t \right) \Fv_{\mu i} \xv_i  \nonumber \\
					  &+ \frac{1}{2} \left( \Fv_{\mu i} \xv_i \right)^{\top} \Jacobian\left( \nabla f_0^Y \right) \left(\yv_{\mu}| \Zhv_{\mu \to i}^t ,  \Zbv_{\mu \to i}^t \right)  \Fv_{\mu i} \xv_i   + o(1/N) , \label{eq:f_taylor}
\end{align}
where the gradient $\nabla$ is taken with respect to $\Zhv_{\mu \to i}^t$, and $\Jacobian(\nabla f)$ is the Hessian of $f$. 
Around $\xv_i =0$, we can approximate $\messf_{\mu \to i}$ by a multivariate Gaussian:
\begin{align}
 \messf_{\mu \to i}^{t+1} (\xv_i) \propto \NN\left( \xv_i; \hat{\mathbf{p}}_{\mu \to i}^t; \bar{\mathbf{p}}_{\mu \to i}^t \right) + o(1/N) 	\label{eq:messf_gaussian}
\end{align}
by identifying the coefficients of the Taylor expansion~\eqref{eq:f_taylor} with the Taylor expansion of a Gaussian with $\xv = O(\sqrt{1/N})$:
\begin{align}
 \NN\left( \xv; \mathbf{b}^{-1} \mathbf{a}, -\mathbf{b}^{-1} \right) \propto 1 - \mathbf{a}^{\top} \xv + \frac{1}{2} \xv^{\top} \left( \mathbf{b} + \mathbf{a} \mathbf{a}^{\top} \right) \xv + o(1/N).	\label{eq:gauss_taylor}
\end{align}
The expressions obtained by identification are
\begin{align}
  \bar{\mathbf{p}}_{\mu \to i}^t &= -\left( \Fv_{\mu i}^{\top} \, \gbv_{\mu \to i}^{t}  \, \Fv_{\mu i} \right)^{-1} ,	\\
  \hat{\mathbf{p}}_{\mu \to i}^t &= - \bar{\mathbf{p}}_{\mu \to i}^t \, \Fv_{\mu i}^{\top} \, \ghv_{\mu \to i}^t .
 \end{align}
Writing the messages $\messf$ under the form~(\ref{eq:messf_gaussian}) allows to find a simpler form for the $\messv$ messages as well.
The product of Gaussians in
\begin{align}
  \messv_{i \to \mu}^{t+1}(\xv_i) &\propto \px(\xv_i) \prod_{\gamma \neq \mu} \NN\left( \xv_i; \hat{\mathbf{p}}_{\gamma \to i}^{t+1}; \bar{\mathbf{p}}_{\gamma \to i}^{t+1} \right) 
\end{align}
is itself a Gaussian, whose mean and covariance matrix are given in the appendix by~\eqref{eq:productOfMultivariateGaussians}.
Using that formula, we obtain
\begin{align}
 \messv_{i \to \mu}^{t+1}(\xv_i) &\propto \px(\xv_i) \NN\left( \xv_i; \Xhv_{i \to \mu}^t, \Xbv_{i \to \mu}^t \right),
\end{align}
where
\begin{align}
 \Xbv_{i \to \mu}^t &= - \left( \sum_{\gamma \neq \mu} \Fv_{\gamma i}^{\top} \, \gbv_{\gamma \to i}^{t}  \, \Fv_{\gamma i} \right)^{-1}	, \\
 \Xhv_{i \to \mu}^t &= \Xbv_{i \to \mu}^t \sum_{\gamma \neq \mu} \Fv_{\gamma i}^{\top} \, \ghv_{\mu \to i}^t.
\end{align}
This allows to express  $\xhv_{j \to \mu}^t$ and $\xbv_{j \to \mu}^t$, the mean and covariance of $\messv_{i \to \mu}^t$, as
\begin{align}
 \xhv_{i \to \mu}^t &= \fh^X \left(\Xhv_{i \to \mu}^t, \Xbv_{i \to \mu}^t \right), \\
 \xbv_{i \to \mu}^t &= \fb^X \left(\Xhv_{i \to \mu}^t, \Xbv_{i \to \mu}^t \right).
\end{align}

These manipulations allow us to implement the \bp equations easily. 
Each message is entirely characterized by only two parameters: a mean and a variance, and the \amp algorithm 
iteratively updates these quantities at each time step. 
Algorithm~\ref{algo:vec_gamp_amp} gives a better overview of these update rules. Initialization and stopping conditions are as in~\algoref{algo:gamp}.

\begin{algorithm}
\caption{AMP version of vectorial GAMP}
\label{algo:vec_gamp_amp}
  \textbf{Main loop:} while $t<t_{\rm max}$,  calculate for all $(\mu, i)$:
  \begin{align*}
  \Zbv_{\mu \to i}^{t} &= \sum_{j \neq i} \Fv_{\mu j} \xbv_{j \to \mu}^{t-1} \Fv_{\mu j}^{\top} &   \Zhv_{\mu \to i}^{t} &= \sum_{j \neq i} \Fv_{\mu j} \xhv_{j \to \mu}^{t-1} \\
  \zbv_{\mu \to i}^{t} &= \fb^Y\left( \yv_{\mu}| \Zhv_{\mu \to i}^{t},  \Zbv_{\mu \to i}^{t} \right) &  \zhv_{\mu \to i}^{t} &= \fh^Y\left( \yv_{\mu}| \Zhv_{\mu \to i}^{t},  \Zbv_{\mu \to i}^{t}\right) \\
  \gbv_{\mu \to i}^{t} &= \left( \Zbv_{\mu \to i}^{t} \right)^{-1} \left(\zbv_{\mu \to i}^{t} - \Zbv_{\mu \to i}^{t} \right) \left( \Zbv_{\mu \to i}^{t} \right)^{-1} & \ghv_{\mu \to i}^{t} &= \left( \Zbv_{\mu \to i}^{t}\right)^{-1} \left(\zhv_{\mu \to i}^{t} -  \Zhv_{\mu \to i}^{t}  \right)  \\
  \Xbv_{i \to \mu}^{t} &=  - \left( \sum_{\gamma \neq \mu} \Fv_{\gamma i}^{\top} \, \gbv_{\gamma  \to i}^{t}  \, \Fv_{\gamma  i} \right)^{-1} &   \Xhv_{i \to \mu}^{t} &=   \Xbv_{i \to \mu}^t \sum_{\gamma \neq \mu} \Fv_{\gamma  i}^{\top} \, \ghv_{\gamma  \to i}^t  \\
  \xbv_{i \to \mu}^{t} &= \fb^X(\Xhv_{i \to \mu}^{t}, \Xbv_{i \to \mu}^{t}) &  \xhv_{i \to \mu}^{t} &= \fh^X(\Xhv_{i \to \mu}^{t}, \Xbv_{i \to \mu}^{t}) 
  \end{align*}
\end{algorithm}

\subsubsection{Step 3: TAP}
The \amp algorithm~\ref{algo:vec_gamp_amp} can be simplified by updating only \textit{local} quantities instead of messages.
This gives a so-called TAP version of the algorithm with lowered complexity.
Remembering the expression of the local belief~(\ref{eq:bp_marginals}), 
\begin{align}
  \messv_{i}^{t}(\xv_i) &\propto \px(\xv_i) \prod_{\mu} \messf_{\mu \to i}^{t} (\xv_i) , 
\end{align}
we introduce the local means and variances
\begin{align}
 \Xbv_{i}^t &= - \left( \sum_{\mu} \Fv_{\mu i}^{\top} \, \gbv_{\mu \to i}^{t}  \, \Fv_{\mu i} \right)^{-1}, &  \Xhv_{i}^t &= \Xbv_{i}^t \sum_{\mu} \Fv_{\mu i}^{\top} \, \ghv_{\mu \to i}^t,  \label{eq:localXh}\\
 \xbv_{i}^t &= \fb^X \left(\Xhv_{i}^t, \Xbv_{i}^t \right), &\xhv_{i}^t &= \fh^X \left(\Xhv_{i}^t, \Xbv_{i}^t \right) ,
\end{align}
and we do the same for the local estimates of $\Z$:
\begin{align}
 \Zbv_{\mu}^{t} &= \sum_{i} \Fv_{\mu i} \xbv_{i \to \mu}^{t-1} \Fv_{\mu i}^{\top}, &   \Zhv_{\mu}^{t} &= \sum_{i} \Fv_{\mu i} \xhv_{i \to \mu}^{t-1}, \\
 \zbv_{\mu}^{t} &= \fb^Y\left( \yv_{\mu}| \Zhv_{\mu}^{t},  \Zbv_{\mu}^{t} \right), &  \zhv_{\mu}^{t} &= \fh^Y\left( \yv_{\mu}| \Zhv_{\mu}^{t},  \Zbv_{\mu}^{t}\right), \\
 \gbv_{\mu}^{t} &= \left( \Zbv_{\mu}^{t} \right)^{-1} \left( \zbv_{\mu}^{t} -\Zbv_{\mu}^{t} \right) \left( \Zbv_{\mu}^{t} \right)^{-1}, &  \ghv_{\mu}^{t} &= \left( \Zbv_{\mu}^{t}\right)^{-1} \left( \zhv_{\mu}^{t} -\Zhv_{\mu}^{t}  \right) . 
\end{align}
We now see that the messages in~\algoref{algo:vec_gamp_amp} differ from the local quantities only by one single term of order $1/\sqrt{N}$ or $1/N$, which motivates us to write them as a function of the local quantities.
The tricky part for obtaining the TAP equations is to keep the necessary corrective terms by analyzing their order of magnitude as a function of $N$.
Let us start the analysis by noting that
\begin{align}
 \Xbv_{i \to \mu}^{t} = \Xbv_{i}^t + O(1/N),  \quad \text{and therefore} \quad \Xhv_{i \to \mu}^{t} = \Xhv_{i}^{t} -  \Xbv_{i}^{t}  \Fv_{\mu i}^{\top} \, \ghv_{\mu \to i}^t + O(1/N).
\end{align}
As a consequence,
\begin{align}
 \xhv_{i \to \mu}^{t} &= \fh^X(\Xhv_{i \to \mu}^{t}, \Xbv_{i \to \mu}^{t}) \nonumber \\
		       &= \fh^X\left(\Xhv_{i \to \mu}^{t},  \Xbv_{i}^{t} \right) +O(1/N) \nonumber \\
		       &= \fh^X\left(\Xhv_{i}^{t},  \Xbv_{i}^{t} \right) - \Jacobian(\fh^X)\left(\Xhv_{i}^{t},  \Xbv_{i}^{t} \right)  \Xbv_{i}^{t}  \Fv_{\mu i}^{\top} \, \ghv_{\mu \to i}^t +O(1/N) \nonumber \\
		       &= \xhv_{i}^t - \xbv_{i}^t \Fv_{\mu i}^{\top} \, \ghv_{\mu \to i}^t +O(1/N) ,
\end{align}
using the fact that $\Jacobian( \fh^X)(\Xhv,\Xbv) = \fb^X(\Xhv,\Xbv) \Xbv^{-1} $ (See~\appref{app:f}).
Let us now look at
\begin{align}
 \Zhv_{\mu \to i}^t &=  \Zhv_{\mu}^t - \Fv_{\mu i} \xhv_{i \to \mu}^{t-1} \nonumber \\
		    &= \Zhv_{\mu}^t - \Fv_{\mu i} \left( \xhv_{i}^{t-1} - \xbv_{i}^{t-1} \Fv_{\mu i}^{\top} \, \ghv_{\mu \to i}^{t-1} +O(1/N) \right) \nonumber\\
		    &= \Zhv_{\mu}^t - \Fv_{\mu i} \xhv_{i}^{t-1} + O(1/N), 
\end{align}
and therefore
\begin{align}
 \zhv_{\mu \to i}^t &= \fh^Y\left( \yv_{\mu}| \Zhv_{\mu \to i}^{t},  \Zbv_{\mu \to i}^{t}\right) \nonumber \\
		    &= \fh^Y\left( \yv_{\mu}| \Zhv_{\mu \to i}^{t},  \Zbv_{\mu}^{t}\right) + O(1/N) \nonumber \\
		    &= \fh^Y\left( \yv_{\mu}| \Zhv_{\mu}^{t},  \Zbv_{\mu}^{t}\right) - \Jacobian(\fh^Y)\left( \yv_{\mu}| \Zhv_{\mu}^{t},  \Zbv_{\mu}^{t}\right) \F_{\mu i} \xhv_{i}^{t-1} + O(1/N) \nonumber \\
		    &= \zhv_{\mu}^t - \zbv_{\mu}^t (\Zbv_{\mu}^t)^{-1} \F_{\mu i} \xhv_{i}^{t-1} + O(1/N),
\end{align}
from which we deduce
\begin{align}
 \ghv_{\mu \to i}^t &= \ghv_{\mu}^t - \gbv_{\mu}^t \Fv_{\mu i} \xhv_{i}^{t-1} + O(1/N).
\end{align}
From these expansions, let us try to obtain update equations of the beliefs as a function of only beliefs.
\begin{align}
 \Zhv_{\mu}^{t+1} &=  \sum_{i} \Fv_{\mu i} \xhv_{i \to \mu}^t \nonumber \\
	      &=  \sum_{i} \Fv_{\mu i} \left( \xhv_{i}^t - \xbv_{i}^t \Fv_{\mu i}^{\top} \, \ghv_{\mu \to i}^t\right) +O(1/N)   \nonumber \\
	      &= \sum_{i} \Fv_{\mu i} \xhv_{i}^t - \sum_{i} \Fv_{\mu i} \xbv_i^t \Fv_{\mu i}^{\top} \left( \ghv_{\mu}^t - \gbv_{\mu}^t \Fv_{\mu i} \xhv_{i}^{t-1} \right) + O(1/N) \nonumber \\
	      &= \sum_{i} \Fv_{\mu i} \xhv_{i}^t - \underbrace{\left( \sum_{i} \Fv_{\mu i} \xbv_i^t \Fv_{\mu i}^{\top} \right) \ghv_{\mu}^t}_{\text{Onsager term}} + O(1/N). 
\end{align}
Here, the important phenomenon is that the second term (called \textit{Onsager} term) is of order one. 
This is due to the fact that the matrix  $\Fv_{\mu i} \xbv_i^t \Fv_{\mu i}^{\top}$ has strictly positive diagonal terms. 
The parenthesis is therefore a sum of $N$ positive terms of order $1/N$ and is of order $1$. 
The manipulation with $\Xhv_{i}^t$ gives
\begin{align}
 \Xhv_{i}^{t+1} &=  \xhv_{i}^{t} - \Xbv_{i}^{t+1} \sum_{\mu} \Fv_{\mu i}^{\top} \, \ghv_{\mu}^{t+1} +O(1/N).
\end{align}
These two terms therefore differ from the ones one would have obtained by carelessly replacing all messages by their respective beliefs.
At this point, we neglect all terms that are $O(1/N)$ and we obtain the TAP version of vectorial \gamp, algorithm~(\ref{algo:vec_gamp_tap}).

\begin{algorithm}
\caption{TAP version of vectorial GAMP}
\label{algo:vec_gamp_tap}
  \textbf{Initialize} for all $(\mu, i)$, $\ghv_{\mu}^0=0$ and $(\xhv_i^0, \xbv_i^0)$ at random or according to $\px$. \\
  \textbf{Main loop:} while $t<t_{\rm max}$,  calculate for all $(\mu, i)$:
  \begin{align*}
  \Zbv_{\mu}^{t} &= \sum_{i} \Fv_{\mu i} \xbv_{i}^{t-1} \Fv_{\mu i}^{\top} &   \Zhv_{\mu}^{t} &= \sum_{i} \Fv_{\mu i} \xhv_{i}^{t-1} -  \Zbv_{\mu}^t \ghv_{\mu}^{t-1}  \\
  \gbv_{\mu}^{t} &=  \gb^Y\left( \yv_{\mu}| \Zhv_{\mu}^{t},\Zbv_{\mu}^{t} \right) &  \ghv_{\mu}^{t} &= \gh^Y\left( \yv_{\mu}| \Zhv_{\mu}^{t},\Zbv_{\mu}^{t} \right)\\
  \Xbv_{i}^{t} &=  - \left( \sum_{\mu} \Fv_{\mu i}^{\top} \, \gbv_{\mu}^{t}  \, \Fv_{\mu  i} \right)^{-1} &   \Xhv_{i}^{t} &=   \xhv_{i}^{t} - \Xbv_{i}^t \sum_{\mu} \Fv_{\mu i}^{\top} \, \ghv_{\mu}^t  \\
  \xbv_{i}^{t} &= \fb^X(\Xhv_{i}^{t}, \Xbv_{i}^{t}) &  \xhv_{i}^{t} &= \fh^X(\Xhv_{i}^{t}, \Xbv_{i}^{t}) 
  \end{align*}
  \textbf{Stop} when $t=t_{\rm max}$, $ \sum_i ||\xbv_i^t||_2 < \epsilon$ or $\sum_i || \xhv_i^t - \xhv_i^{t} ||_2< \epsilon$.
\end{algorithm}

The disadvantage of this algorithm is that, on the contrary of \gamp, it requires a matrix inversion (for the calculation of $\Xbv$),
 and that the update functions require to integrate over multi-dimensional variables.
Other ways of taking into account joint or structured sparsity exist, for example turbo-GAMP~\cite{turboGamp}.

\subsection{Complex \gamp}
Complex  \gamp is a special case of~\algoref{algo:vec_gamp_amp}, characterized by the relations~(\ref{eq:complex_vec1}, \ref{eq:complex_vec2}).
With a few additional assumptions, the algorithm can be rewritten with only scalars, such that no matrix inversion is necessary.
First, note that all covariance matrices are by definition of the form
\begin{align}
 \left( \begin{tabular}{cc}
         a & b\\
         b & c
        \end{tabular}
        \right).
\end{align}
In $\Zbv_{\mu}$ and $\Xbv_{i}$, covariance matrices are multiplied on the right and on the left by $\Fv_{\mu i}$, giving a matrix of the form
\begin{align}
 \footnotesize\left( \begin{tabular}{cc}
         $a_i \Re (F_{\mu i})^2 + 2b_i \Re(F_{\mu i}) \Im(F_{\mu i}) + c_i \Im(F_{\mu i})^2$ & $b_i \left( (\Re(F_{\mu i}))^2 - \Im(F_{\mu i})^2\right) - (a_i-c_i) \Re(F_{\mu i}) \Im(F_{\mu i})$ \\
         $b_i \left( \Re(F_{\mu i})^2 - \Im(F_{\mu i})^2\right) - (a_i+c_i) \Re(F_{\mu i}) \Im(F_{\mu i})$ &  $c_i \Re(F_{\mu i})^2 - 2b_i \Re(F_{\mu i}) \Im(F_{\mu i}) + a_i \Im(F_{\mu i})^2$
        \end{tabular}
        \right). \nonumber
\end{align}
Let us focus on the case in which $\Re(F_{\mu i})$ and $\Im(F_{\mu i})$ are independent of each other and distributed as $\NN(0,1/N)$.
We also assume that the values of $(a_i, b_i, c_i) \approx (a,b,c)$ for all $i$. 
Then, in the summation over $i$ of the matrices above,
\begin{itemize}
 \item $ \sum_i \Re(F_{\mu i}) \Im(F_{\mu i}) \approx 0$ because $\langle \Re(F_{\mu i}) \Im(F_{\mu i}) \rangle=0$,
 \item $ \sum_i \left( \Re(F_{\mu i})^2 - \Im(F_{\mu i})^2 \right) = O(1/\sqrt{N})$ because it is a sum of positive and negative terms,
 \item $ \sum_i (\Re(F_{\mu i}))^2 \approx \sum_i \Im(F_{\mu i})^2 \approx \sum_i \frac{|F_{\mu i}|^2}{2} = O(1)$ are the only terms with contributions to leading order.
\end{itemize}
As a result, the terms that do not contribute in the sum can be ignored and the summation made over diagonal matrices:
\begin{align}
   \Xbv_i&= - \left[ \sum_{\mu} \left( \begin{tabular}{cc}
         $ \frac{(a + c)}{2} |F_{\mu i}|^2$ & $0$ \\
         $0$ &  $ \frac{(a + c)}{2} |F_{\mu i}|^2$
        \end{tabular}
        \right) \right]^{-1}. \nonumber
\end{align}
Finally, as only the sum $a+c$ is necessary here, it is not necessary to compute the four terms of the covariance matrix, but only $a+c$.
Notice that $\Xbv_i$ is now a multiple of identity: the variance of both the real and the imaginary part of $\Xhv_i$ is $\frac{a+c}{2} |F_{\mu i}|^2$, therefore
the variance of the complex estimate $\Xh_i$ is $\Xb_i = (a+c) |F_{\mu i}|^2$.
As a result, let us redefine the $f$-functions (for $\xh\in \CC$ and $\xb \in \RR_+$):
\begin{align}
  f_0^h(\xh,\xb) &\equiv \int \dd x h(x) \CN\left( x; \xh, \xb \right),   & & \in \RR \label{eq:f0_complex} \\
  f_1^h(\xh,\xb) &\equiv \int \dd x  \, x \, h(x) \CN\left( x; \xh, \xb \right),  & &\in \CC	\label{eq:f1_complex} \\
 f_2^h(\xh,\xb) &\equiv  \int \dd x \, |x|^2 h(x) \CN\left( x; \xh, \xb \right), && \in \RR \label{eq:f2_complex}
\end{align}
as well as the update functions
\begin{align}
 \fh^h(\xh,\xb) &\equiv \frac{f_1^h(\xh,\xb)}{f_0^h(\xh,\xb)}, & &\in \CC   \label{eq:fh_complex} \\
 \fb^h(\xh,\xb) &\equiv  \frac{f_2^h(\xh,\xb)}{f_0^h(\xh,\xb)} - |\fh^h(\xh,\xb)|^2,  & &\in \RR  \label{eq:fb_complex}
\end{align}
where $\Cc$ is the circular-symmetric complex Gaussian
\begin{align}
 \Cc(x; \xh,\xb) &\equiv  \frac{1}{\pi \xb} e^{-\frac{|x - \xh|^2}{\xb}}.
\end{align}

Once this is done, all covariance matrices are multiples of the identity and can therefore be replaced by scalars.
Finally, noting that $\Fv_{\mu i}^{\top}$ is the matrix representation of the complex conjugate $F_{\mu i}^*$ and that 
$\Fv_{\mu i}^{\top} \Fv_{\mu i} = |F_{\mu i}|^2 \one$, the simplified, scalar version of complex \gamp is given in~\algoref{algo:cgamp}.

\begin{algorithm}
\caption{\cgamp}
\label{algo:cgamp}
 \textbf{Initialize} for all $(\mu, i)$, $\gh_{\mu}^0=0$ and $(\xh_i^0, \xb_i^0)$ at random or according to $\px$. \\
  \textbf{Main loop:} while $t<t_{\rm max}$, calculate for all $(\mu, i)$:
  \begin{align*}
  \Zb_{\mu}^{t+1} &= \sum_i |F_{\mu i}| ^2 \xb_i^{t}   & &\in \RR \\
  \Zh_{\mu}^{t+1} &= \sum_i F_{\mu i} \xh_i^{t} - \Zb_{\mu}^{t+1}  \gh_{\mu}^{t}  & &\in \CC  \\
  \gb_{\mu}^{t+1} &= \gb^Y(\y_{\mu}| \Zh_{\mu}^{t+1}, \Zb_{\mu}^{t+1})   & &\in \RR \\
  \gh_{\mu}^{t+1} &= \gh^Y(\y_{\mu}| \Zh_{\mu}^{t+1}, \Zb_{\mu}^{t+1})  & &\in \CC \\
  \Xb_{i}^{t+1} &= - \left( \sum_{\mu} |F_{\mu i}|^2 \gb_{\mu}^{t+1} \right)^{-1}  & &\in \RR   \\
  \Xh_{i}^{t+1} &= \xh_{i}^{t} + \Xb_{i}^{t+1} \sum_{\mu} F_{\mu i}^* \gh_{\mu}^{t+1}  & &\in \CC \\
  \xb_{i}^{t+1} &= \fb^X(\Xh_{i}^{t+1}, \Xb_{i}^{t+1})  & &\in \RR \\
  \xh_{i}^{t+1} &= \fh^X(\Xh_{i}^{t+1}, \Xb_{i}^{t+1})  & &\in \CC 
  \end{align*}
  \textbf{Stop} when $t=t_{\rm max}$, $  ||\xbv^t||_2 < \epsilon$ or $|| \xhv^t - \xhv^{t} ||_2< \epsilon$.
\end{algorithm}

This final version of \cgamp differs from \gamp only by the magnitudes $| \cdot |^2$ replacing the squares and by the complex conjugate $F_{\mu i}^*$.
For complex \cs, a shorter different derivation of \cgamp is provided in~\cite{conciseComplex} using expectation propagation.


\section{CS with fast operators and superpositions codes}
In this section, I describe the work published jointly with Jean Barbier and Florent Krzakala in~\cite{moiCodes},
that is in part based on my work on complex \cs.
\subsection{Complex \cs state evolution}
\label{sec:complexCsStateEvolution}
The replica analysis is only one way to derive the state evolution equations of \gamp and \cgamp.
Another way is to use the state evolution formalism, that analyzes the statistical fluctuations of the quantities iterated by the algorithm.
For the case of compressed sensing, this analysis is simpler to perform than the replica analysis, and here I derive it for complex \cs using~\algoref{algo:cgamp}.

First, let us explicitly write out the \awgn sensing channel in the complex case:
\begin{align}
 \py(y|z) &= \CN(y;z,\Delta),
\end{align}
which results in the update functions
\begin{align}
 \gh^Y(y|\Zh,\Zb) &= \frac{y - \Zh}{\Delta + \Zb},  \\
 \gb^Y(y|\Zh,\Zb) &= -\frac{1}{\Delta + \Zb},
\end{align}
which are the same as for real-valued \cs.
Next, we can make a further simplification, the so-called ``fully-TAP'' version of the algorithm, 
that is made by approximating every $|F_{\mu i}|^2$ by  $\frac{1}{N}$. 
This can be justified by noticing that the difference between the quantities calculated using $|F_{\mu i}|^2$ and the same quantities with $\frac{1}{N}$ instead is of order $O(1/N)$~\cite{krzakaCS}.
This leads to the variances $\Zb$, $\Xb$ to be index-independent, 
and so is $\gb$ as a consequence:

\begin{algorithm}
\caption{Complex \cs with scalar variances}
\label{algo:ccs}
  \textbf{Main loop:} while $t<t_{\rm max}$, calculate following quantities:
  \begin{align*}
  \Zb^{t+1} &= \frac{1}{N} \sum_i \xb_i^{t}    \\
  \Zh_{\mu}^{t+1} &= \sum_i F_{\mu i} \xh_i^{t} - \Zb^{t+1}  \frac{y_{\mu} - \Zh_{\mu}^{t}}{\Delta + \Zb^{t}}   \\
  \Xb^{t+1} &=    \frac{\Delta + \Zb^{t+1}}{\alpha} \\
  \Xh_{i}^{t+1} &= \xh_{i}^{t} +  \sum_{\mu} F_{\mu i}^* \frac{y_{\mu} - \Zh_{\mu}^{t+1}}{\alpha}   \\
  \xb_{i}^{t+1} &= \fb^X(\Xh_{i}^{t+1}, \Xb_{i}^{t+1})   \\
  \xh_{i}^{t+1} &= \fh^X(\Xh_{i}^{t+1}, \Xb_{i}^{t+1})   
  \end{align*}
\end{algorithm}

Algorithm~\ref{algo:ccs} can be analyzed by examining the statistical fluctuations of the updated quantities when the elements $F_{\mu i}$ 
have a circular Gaussian distribution with variance $1/N$. 
First, let us rewrite $\Xh_{i}$ starting from~\eqref{eq:localXh} by replacing $y_{\mu}$ by its expression $\sum_{i} F_{\mu i} \x_i + \xi_{\mu}$:
\begin{align}
 \Xh_{i}^{t+1} &= \frac{\Delta + \Zb^{t+1}}{\alpha} \sum_{\mu} F_{\mu i}^*  \frac{y_{\mu} - \Zh_{\mu \to i}^{t+1}}{\Delta + \Zb^{t+1}} \\
	      &= \frac{1}{\alpha} \sum_{\mu} F_{\mu i}^* \left( \sum_{j} F_{\mu j} x_j + \xi_{\mu} - \sum_{j \neq i} F_{\mu j} \xh_{i \to \mu}^t \right) \\
	      &= \frac{1}{\alpha} \sum_{\mu} F_{\mu i}^* F_{\mu i} x_i + \frac{1}{\alpha} \left( \sum_{\mu} F_{\mu i}^* \xi_{\mu} + \sum_{\mu} F_{\mu i}^* \sum_{j \neq i} F_{\mu j} (x_j - \xh_j^t) \right)
\end{align}
again, replacing $|F_{\mu i}|^2$ by $1/N$, we obtain the expression
\begin{align}
 \Xh_{i}^{t+1} &= x_i + \frac{1}{\alpha} \underbrace{\left(\sum_{\mu} F_{\mu i}^* \xi_{\mu} + \sum_{\mu} F_{\mu i}^* \sum_{j \neq i} F_{\mu j} (x_j - \xh_{j \to \mu}^t) \right)}_{r_i^t}
\end{align}
The term $r_i^t$ is a random variable of the $F_{\mu i}$ elements. 
It has zero mean and can be approximated as a complex Gaussian random variable, as it is the sum of a large number of independent random variables.
It can be verified that under the hypothesis that $F_{\mu i} \sim \CN(0,1/N)$, the real and imaginary parts of $r_i^t$ are independent.
Therefore $r_i^t$ is a circular Gaussian random variable, and its variance is $\alpha(\Delta +  E^t)$, where 
\begin{align}
 E^t &= \frac{1}{N}\sum_i |x_i - \xh_i^t|^2	\label{eq:complexMSE}
\end{align}
is the complex mean squared error at time $t$.  
This quantity can be evaluated at time $t+1$ by writing
\begin{align}
 E^t &= \int \dd x \px(x) \int \dd t \CN(t;0,1) \left| \fh^X\left( x + t \sqrt{\frac{\Delta+E^t}{\alpha}}, \frac{\Delta + \Zb^t}{\alpha} \right) - x \right|^2. \label{eq:E_DE}
\end{align}
The variance $\Zb^t$ can be expressed in a similar way by writing
\begin{align}
 \Zb^t &= \int \dd x \px(x) \int \dd t \CN(t;0,1) \fb^X\left( x + t \sqrt{\frac{\Delta+E^t}{\alpha}}, \frac{\Delta + \Zb^t}{\alpha} \right). \label{eq:V_DE}
\end{align}
Equations~(\ref{eq:E_DE}\ref{eq:V_DE}) are the state evolution equations for complex compressed sensing.
Note that in the Bayes-optimal case, the Nishimori conditions~(\ref{eq:nishConds}) impose that $E^t = \Zb^t$ and the state evolution can thus be written as a single equation, \mynote{Complex CS state evolution}{0.7}
\begin{align}
  E^t &= \int \dd x \px(x) \int \dd t \CN(t;0,1) \fb^X\left( x + t \sqrt{\frac{\Delta+E^t}{\alpha}}, \frac{\Delta + E^t}{\alpha} \right). \label{eq:complex_DE}
\end{align}

Preliminary results about the state evolution of the ``generalized'' setting are presented in \appref{app:PR_DE}. 

\subsubsection{Phase diagram for complex \cs}
\begin{SCfigure}[1][h]
 \includegraphics[width=0.6\textwidth]{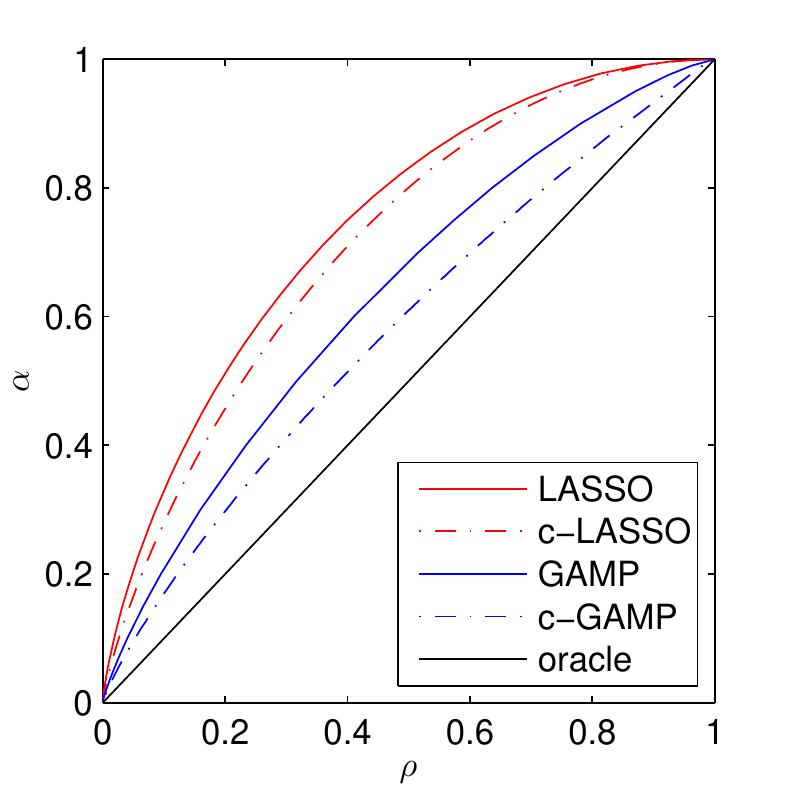}
  \caption[Phase diagram of complex \cs]{ Diagram comparing the positions of theoretical phase transitions for noiseless compressed sensing.
 The full lines are the phase transitions for real-valued \cs already shown in~\figref{fig:3_csPD_a}.
 They also apply to complex \cs if the real and imaginary part of the signal are treated independently.
 The dashed lines are the phase transitions when they are treated jointly and are jointly sparse.
 Both allow to improve on their respective full line phase transition, because joint sparsity is 
 a stronger constraint than sparsity of the real part and of the imaginary part. Empirical transitions are in excellent agreement with the theoretical ones.
 }
 \label{fig:5_complexDiag}
\end{SCfigure}

Equation~(\ref{eq:complex_DE}) allows to theoretically obtain the phase transitions of Bayes optimal \cgamp.
In~\figref{fig:5_complexDiag}, we consider two signal distributions:
\begin{align}
 \px^J(x) &= (1 - \rho)\delta(|x|) + \rho \CN(x;0,1), 	\label{eq:joint_sparse} \\
 \px^I(x) &= \left[ (1 - \rho)\delta(\Re(x)) + \rho \NN(\Re(x);0,1) \right]\left[ (1 - \rho)\delta(\Im(x)) + \rho \NN(\Im(x);0,1) \right], \label{eq:indep_sparse}
\end{align}
which we call the \textit{joint} and the \textit{independent} Gauss-Bernoulli distribution.
For $\px^I$, \cgamp treats the real and imaginary parts of the signal independently using the update functions already given in~\exref{ex:csGAMP}.
For $\px^J$, the update functions $f$ take a complex argument $\xh$ and a real argument $\xb$.
Their expressions are identical to the ones in~\exref{ex:csGAMP} replacing $\NN$ by $\CN$ and $(\cdot)^2$ by $|\cdot|^2$.
On~\figref{fig:5_complexDiag}, the theoretical ``\gamp'' and ``\cgamp'' transitions correspond to $\px^I$ and $\px^J$ respectively.
The diagram shows that taking into account the joint sparsity allows to increase the size of the region of possible recovery.

Another complex \cs algorithm exploiting joint sparsity has been proposed in~\cite{camp} under the name of CAMP.
Very similar to~\algoref{algo:vec_gamp_tap}, it solves the complex LASSO (c-LASSO) problem
\begin{align}
 \argmin_{\xv} \frac{1}{2} || \yv - \Fv \xv||_2^2 + \lambda || \xv ||_1
\end{align}
when $\lambda \to \infty$, where $||  \cdot ||_1$ is the complex $L_1$ norm: $|| \xv ||_1 = \sum_i |x_i|$.
It can be implemented with~\algoref{algo:ccs} using the update functions
\begin{align}
\fh^X_{\rm CAMP}(\xh,\xb) &= \xh \frac{\max\left( |\xh| - \xb,0 \right)}{\max\left( |\xh| - \xb,0 \right) + \xb} ,	\label{eq:complexSoftThresh} \\
 \fb^X_{\rm CAMP}(\xh,\xb) &= \left\{ \begin{array}{cc}
                                       \xb & \text{if } |\xh|>\xb, \\
                                       0 & \text{else.}
                                      \end{array}
                                      \right.  .
\end{align}
The corresponding state evolution is given in~\cite{camp} and leads to the ``c-LASSO'' transition on~\figref{fig:5_complexDiag}.
As it takes into account joint sparsity, it allows lowering the ``LASSO'' phase transition.

\subsection{Compressed sensing with fast operators}
\begin{figure}[h]
 \centering
 \begin{subfigure}{0.48\textwidth}
  \includegraphics[width=\textwidth]{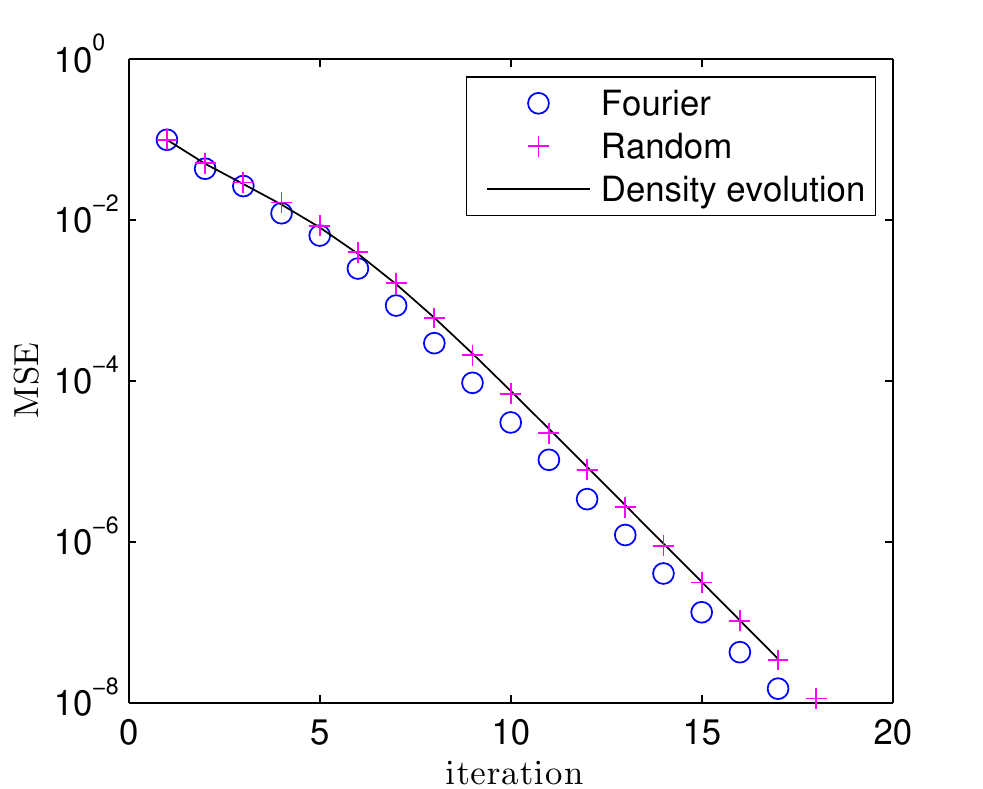}
  \caption{Experimental and theoretical convergence.}
  \label{fig:5_Fourier_convergence}
 \end{subfigure}
 \begin{subfigure}{0.48\textwidth}
   \includegraphics[width=\textwidth]{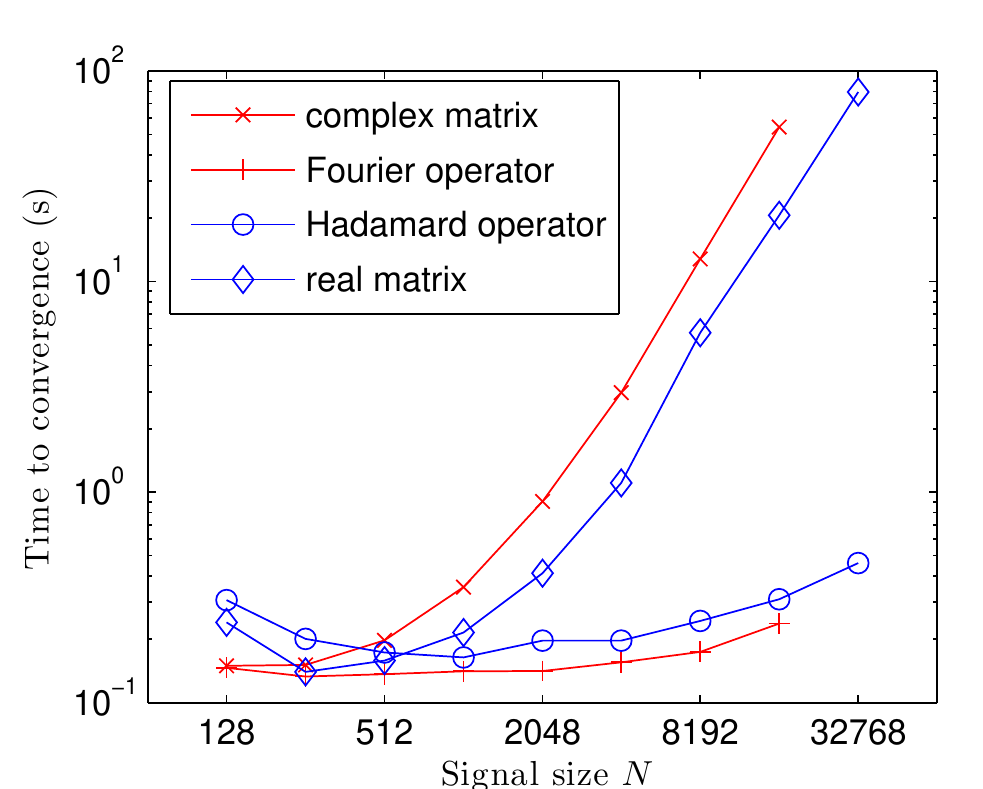}
  \caption{Times to convergence.}
  \label{fig:5_times} 
 \end{subfigure}
 \caption[Convergence of complex \cs]{ 
 (a) Evolution of the MSE of the signal estimate as a function of the iteration of the algorithm.
  The signal is sampled from the distribution with joint sparsity~(\ref{eq:joint_sparse}). 
  Experimental parameters are $N=8192$, $\rho=0.1$, $\alpha=0.3$ and $\Delta=0$.
  While the experiment with a random matrix closely matches the state evolution, the one 
  with a Fourier matrix clearly converges faster (in iterations).
  (b) Times required for convergence (defined as MSE$<10^{-6}$) of the algorithm 
  as a function of the signal size $N$ for complex \cs with random matrices and subsampled Fourier operators.
  The use of operators allows to be orders of magnitudes faster.
  }
 \label{fig:5_convergence}
\end{figure}

The main focus in~\cite{moiCodes} is the study of \cs with structured operators as measurement matrices, namely Hadamard operators for real-valued \cs
and Fourier operators for complex-valued \cs.
The characteristics of these operators are that
\begin{enumerate}
 \item They are not random but deterministic and in that sense \textit{structured}.
 Hadamard operators are represented by a matrix with whose entries are $\pm 1$.
 Fourier matrices have only matrix elements of the form $e^{\imath \phi_{\mu i}}$.
 Both are orthogonal matrices.
 \item They do not need to be stored in memory (as they are constructed on a very simple scheme)
 and can be applied to vectors of size $N$ in only $O(N \log N)$ operations---instead of $O(N^2)$ operations for matrix multiplication.
 These two combined properties allow to treat signals of much bigger size and in a significantly shorter time than with non-structured measurement matrices.
\end{enumerate}
The main finding is that empirically, despite violating the usual randomness assumption, both Hadamard and Fourier operators are good matrices for \cs and ensure convergence just
as well as random matrices. However, they do not follow the state evolution but instead converge \textit{faster} to the solution, as seen on~\figref{fig:5_Fourier_convergence}.
This is not astonishing as the derivation of state evolution is made with the hypothesis of matrices with \iid elements. 
The fact that convergence is faster is likely due to the fact that Fourier and Hadamard matrices are orthogonal, which is not exactly the case for random matrices.

\subsection{Superposition codes}
\label{sec:sparseSuperpositionCodes}
The second focus of~\cite{moiCodes} is the application of \gamp to superposition codes.
\subsubsection{The signal}
In superposition codes, the goal is to transmit a message $\xtv \in \XX^L$ through an \awgn channel with noise variance $\Delta$.
$\XX$ is a finite alphabet with $B$ elements and one way to represent its $b$-th element is to write it as a vector:
\begin{align}
 \xt \in \{0,1\}^B \quad \text{s.t.} \quad \xt_b=1 \quad \text{and} \quad \forall c\neq b, \, \, \xt_c=0
\end{align}
We call $N=BL$ and $\xv \in \{0,1\}^N$ the vector that is the concatenation of all $L$ $\xt_l$'s.
It has $L$ sections $\xv_l$ of size $B$ each.
\subsubsection{The code}
Coding $\xtv$ is made by multiplying $\xv$ with a matrix $\Fv \in \RR^{M \times N}$ that has as usual random \iid entries:
\begin{align}
 \forall \mu \in [1,M], \quad z_{\mu} = \sum_{i=1}^N F_{\mu i} x_i,	\label{eq:codes_z_1}
\end{align}
and the receptor gets a version of it corrupted by \awgn:
\begin{align}
 \forall \mu \in [1,M], \quad y_{\mu} = z_{\mu} + \xi_{\mu} \quad \quad \text{with} \quad \xi_{\mu} \sim \NN(\xi_{\mu};0,\Delta).
\end{align}
\subsubsection{Message passing decoder}
The setting is extremely similar to \cs, with the difference that each of the $L$ sections of the signal $\xv$ is sampled from the distribution
\begin{align}
 \px(\xv_l) &= \frac{1}{B} \sum_{b=1}^B \left[ \delta([\xv_l]_b-1) \prod_{c \neq b} \delta([\xv_l]_c) \right].	 \label{eq:codes_prior}
\end{align}
As in complex \cs, the components of $\xv$ are therefore not independent, which has to be taken into account using a vectorial version of \gamp.
In the notations of~\secref{sec:vectorial_gamp}, we have $\dun=B$ and $\ddeux=\dtrois=1$. 
Equation~(\ref{eq:codes_z_1}) can be written as
\begin{align}
 \forall \mu \in [1,M], \quad z_{\mu} = \sum_{l=L}^N \Fv_{\mu l}^{\top} \xv_l,
\end{align}
with $\Fv_{\mu l} \in \RR^{B}$.
In~\algoref{algo:vec_gamp_tap}, $\Zbv_{\mu}, \Zhv_{\mu}, \gbv_{\mu}$ and $\ghv_{\mu}$ are actually scalars, and it is 
easy to show that in the large $N$ limit, $\Xbv$ is diagonal.
As a result, the matrix formalism of~\algoref{algo:vec_gamp_tap} can be dropped for all quantities except for the update functions $\fh^X, \fb^X$.
Using the prior~(\ref{eq:codes_prior}), they result in the new updates
\begin{align}
 [\xhv_l]_b &= \frac{\exp\left(-\frac{1-2 [\Xhv_l]_b}{2 [\Xbv_l]_b} \right)}{\sum_{c} \exp\left(-\frac{1-2 [\Xhv_l]_c}{2 [\Xbv_l]_c} \right) }	\\
 [\xbv_l]_b &= [\xhv_l]_b \left( 1 - [\xhv_l]_b \right).
\end{align}

The constraint imposed by the prior~(\ref{eq:codes_prior}) is much stronger than the one in compressed sensing, as it enforces binary  values on the signal components.
For this reason, perfect signal reconstruction might be possible even in a noisy setting.

This message passing decoder was derived in~\cite{jeanCodes1} along with the corresponding replica analysis and state evolution equations.
As for real and complex variables, state evolution equations for vectorial variables can be demonstrated rigorously~\cite{Javanmard21102013}.
Both the replica theory and the experimental results, presented in~\cite{jeanCodes1, jeanCodes2, jeanThese} show that using a specially designed type of 
\textit{spatially coupled} measurement matrices, this decoder is \textit{capacity-achieving}: it allows transmission of information at the highest 
theoretically possible rate through an \awgn channel.

\section{Phase retrieval}
\label{sec:PR}

One interesting application of \gamp with complex signals is the problem of phase retrieval.
\subsubsection{Setting}
In phase retrieval, signal and matrix are complex, but measurements only provide the \textit{magnitude} of $\zv$:
\begin{align}
 \xv \in \CC^{N}, \quad \Fv \in \CC^{M \times N}, \quad \zv = \Fv \xv \in \CC^{M},
\end{align}
and considering a setting with complex noise before the measurements,
\begin{align}
\forall \mu,  \quad  y_{\mu} = |z_{\mu} + \xi_{\mu}| \quad \text{with} \quad \xi_{\mu} \sim \Cc(\xi_{\mu};0,\Delta).	\label{eq:pr_measurements}
\end{align}
\subsubsection{Invariances}
Just like the bilinear inference problems presented in~\chapref{chap:generalizedBilinearModels}, phase retrieval has 
an invariance that cannot be lifted: invariance up to a global phase.
As the phase is lost during measurements, it is easy to see that the signals $\xv$ and $e^{\imath \phi} \xv$ 
produce the same measurements $\yv$.
Therefore, the usual MSE is not an appropriate measure of success. 
Instead, one can use the normalized cross-correlation for complex signals:
\begin{align}
  \cc(\xv,\xhv) = \frac{| \xv^{\top} \xhv | }{||\xv||_2 ||\xhv||_2},	\label{eq:crosscorr_PR}
\end{align}
where $(\cdot)^{\top}$ indicates the complex transpose, or the normalized mean square error
\begin{align}
 \nmse(\xv,\xhv) = 1-\cc(\xv,\xhv).
\end{align}

\begin{figure}
\centering
\begin{subfigure}{0.45\textwidth}
 \includegraphics[width=2.5cm]{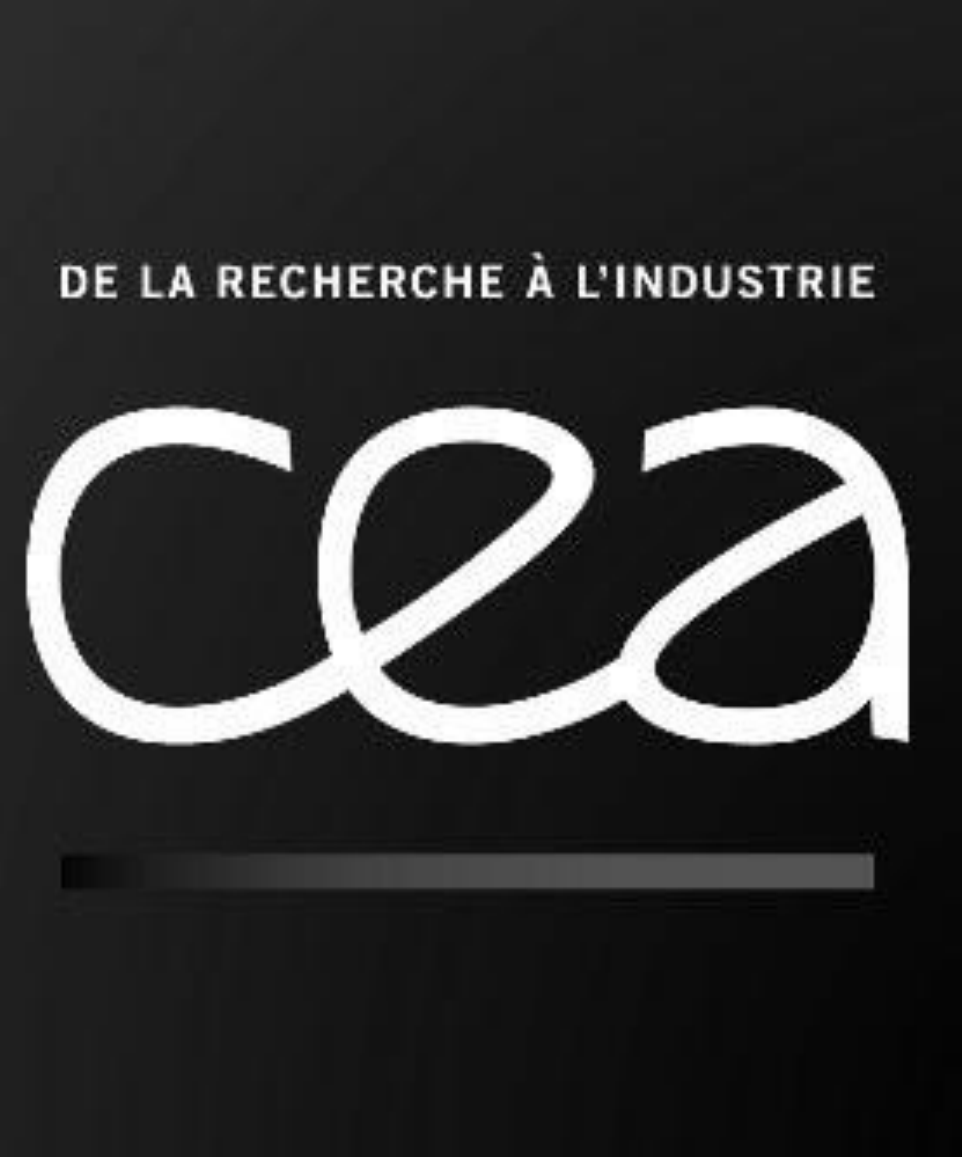} \hspace{1cm}
 \includegraphics[width=2.5cm]{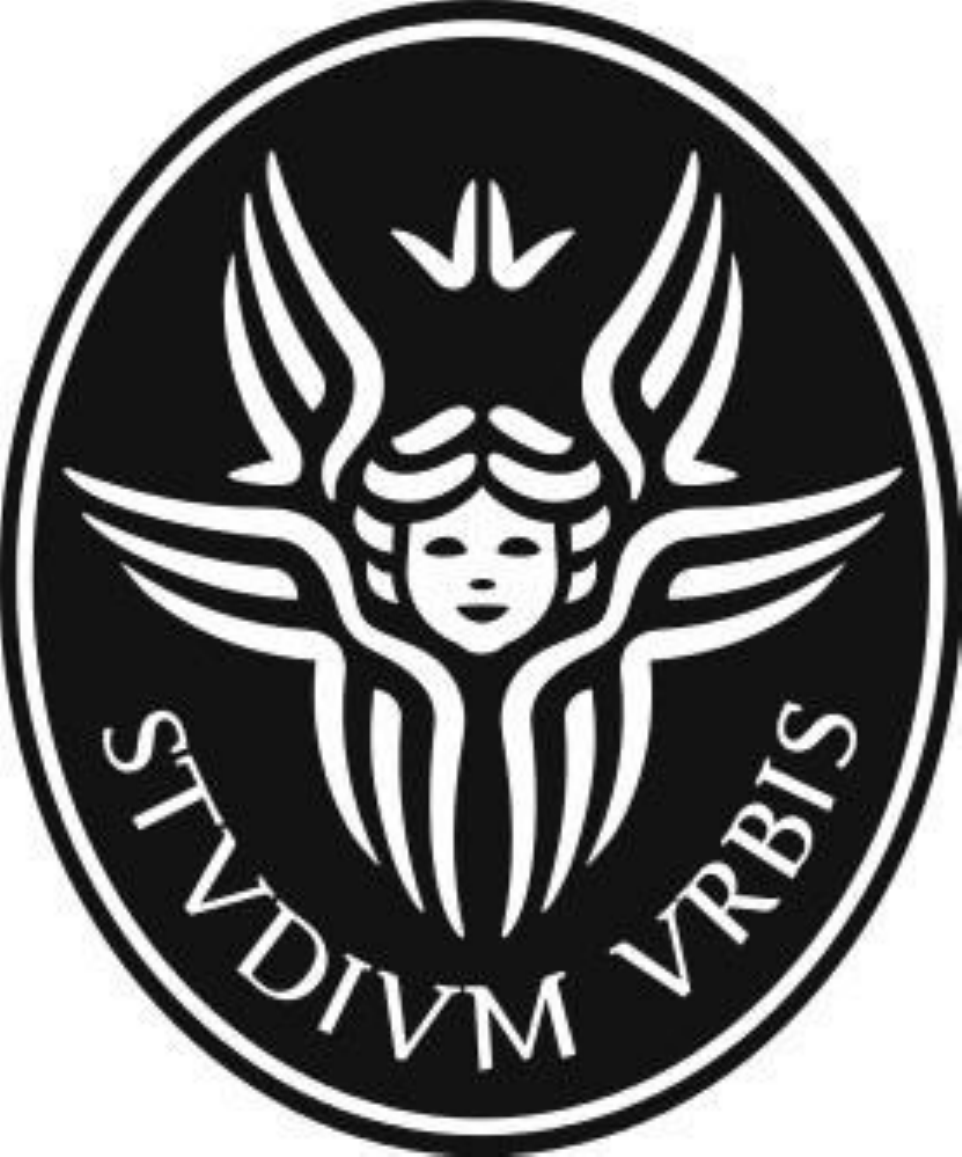}
 \caption{Original images.}
 \label{fig:5_orig}
 \end{subfigure}
 \hspace{1cm}
 \begin{subfigure}{0.45\textwidth}
  \includegraphics[width=2.5cm]{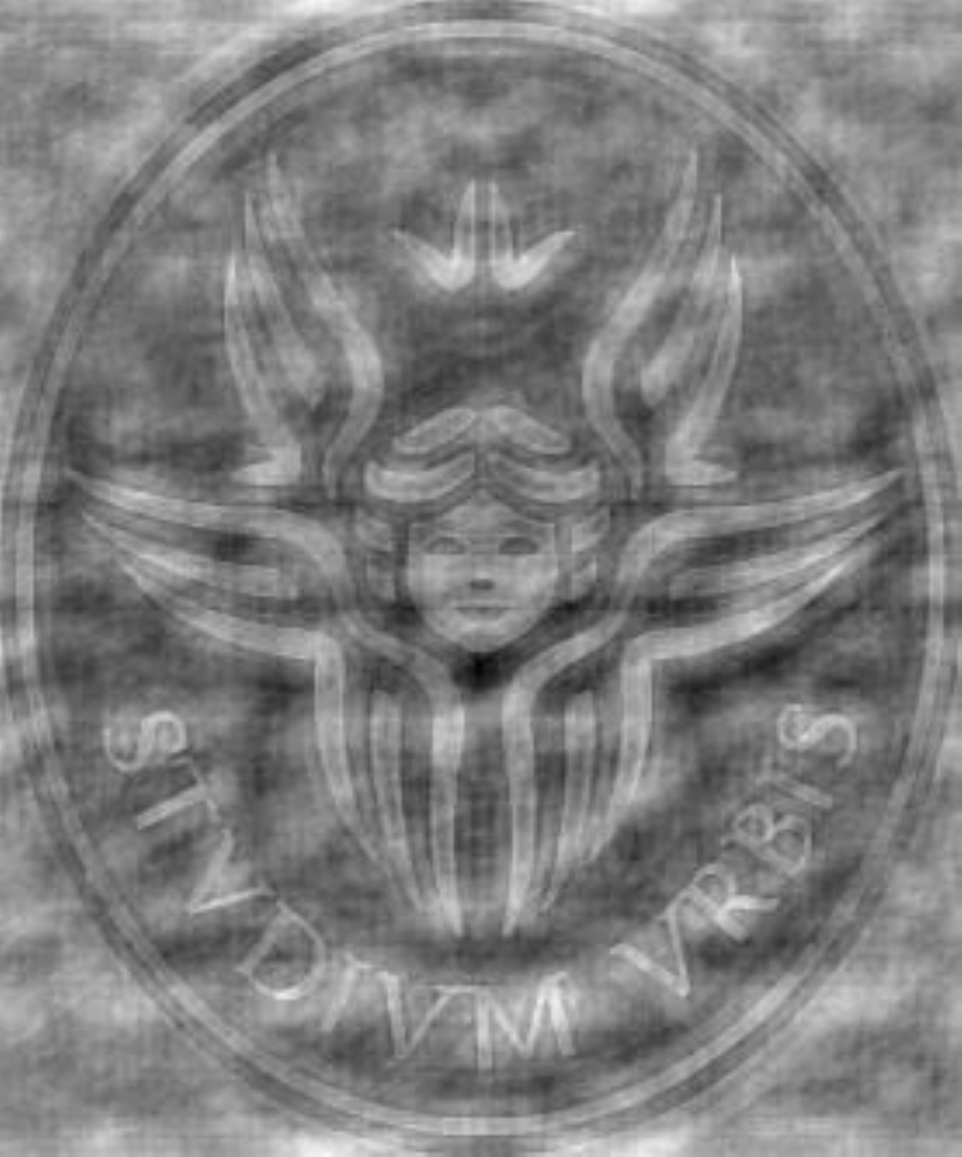}\hspace{1cm}
 \includegraphics[width=2.5cm]{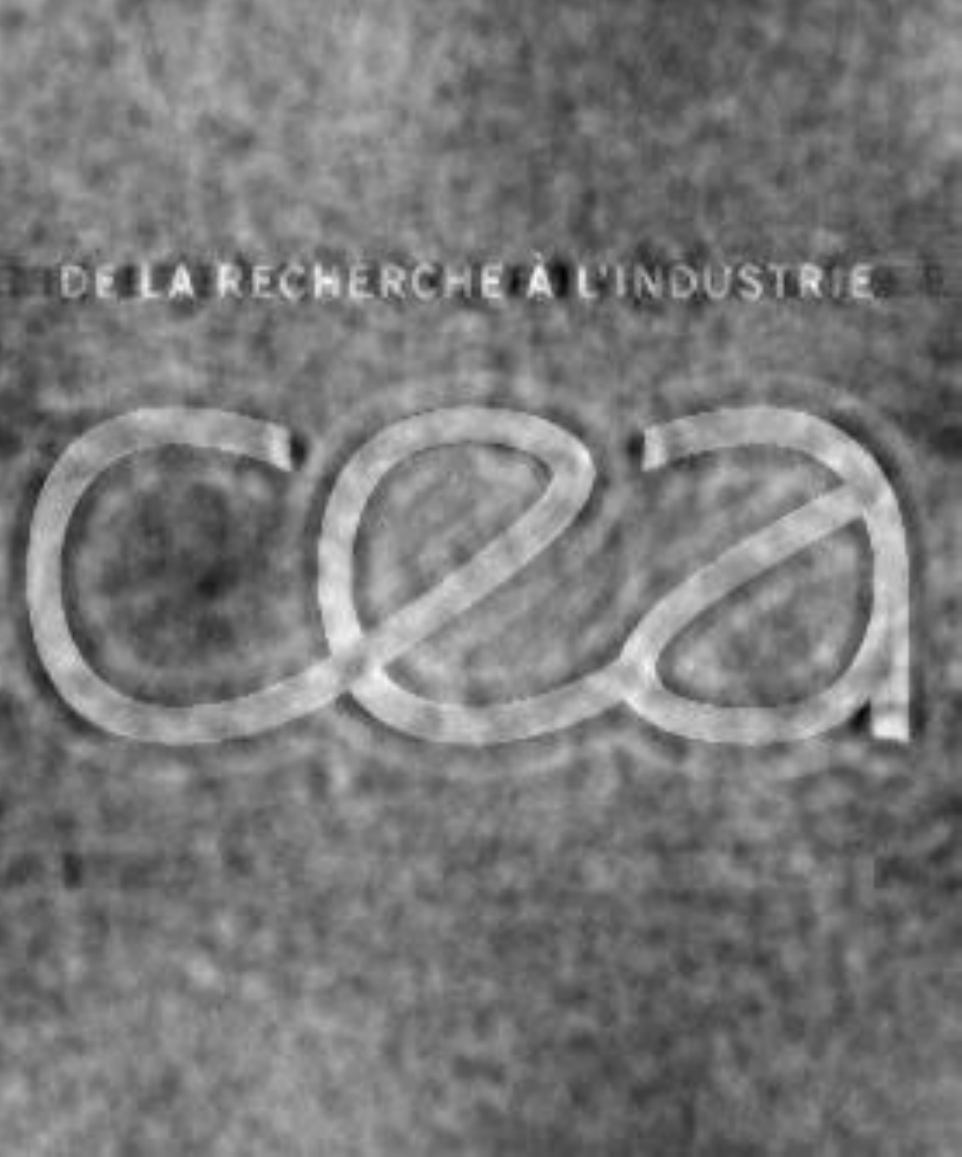}
 \caption{Reconstructions with switched phases.}
 \label{fig:5_recons}
 \end{subfigure}
 \caption[Importance of the phase in Fourier measurements]{Importance of the phase for reconstruction from Fourier measurements.
 (a) Two original, real-valued images.
 Their two-dimensional discrete Fourier transform is calculated and 
 only the magnitude is measured, as is the case in crystallography.
 (b) Reconstructions of the images made from the (right) magnitudes and 
 the \textit{switched} phases of the Fourier transforms.
 Knowledge of the correct phase is crucial for meaningful image reconstruction.}
 \label{fig:5_scrambled}
\end{figure}

\begin{application}{X-ray crystallography}
 \label{ex:crystallo}
 The goal of x-ray crystallography is to determine molecular structures, which is impossible with traditional 
 imaging techniques such as microscopy because of their intrinsic resolution limits.
 
 To this effect, pure crystals of the molecule are synthesized and exposed to high-energy x-rays, which produces diffraction 
 patterns. These diffraction patterns are measured under different angles: 
 They are the magnitudes of the Fourier-transform of the electronic densities of the molecule.
 To reconstruct the electronic densities via direct Fourier inversion, the phases are necessary as well~(see~\figref{fig:5_scrambled}).
 However, these cannot be measured. It is therefore essential to produce good estimates of these phases using phase retrieval.
\end{application}

\subsubsection{Phase retrieval \gamp}
In~\cite{prgamp}, the authors use \cgamp for the phase retrieval problem.
This is done by considering the probability distribution corresponding to~\eqref{eq:pr_measurements},
\begin{align}
 \py(y|z) &= \int \dd \xi \Cc(\xi; 0, \Delta) \delta\left( y - |z + \xi| \right) ,	\label{eq:pr_output}
\end{align}
and the corresponding functions
\begin{align}
 f_0^Y(y|\zh,\zb) &= \frac{2 y}{\Delta+ \zb} e^{-\frac{y^2}{\Delta+\zb}} e^{-\frac{|\zh|^2}{\Delta+\zb}} I_0\left( \frac{2 y |\zh|}{\Delta+\zb} \right)	,  \label{eq:f0PR} \\
 f_1^Y(y|\zh,\zb) &= f_0^Y(y|\zh,\zb) \frac{\zh}{|\zh|} \left( \frac{\zb}{\Delta+\zb} \frac{ I_1\left( \frac{2 y |\zh|}{\Delta+\zb} \right)}{ I_0\left( \frac{2 y |\zh|}{\Delta+\zb} \right)} + \frac{\Delta |\zh|}{\Delta+\zb} \right) , \\
 f_2^Y(y|\zh,\zb) &= f_0^Y(y|\zh,\zb) \left( \frac{y^2}{\left( 1 + \Delta/\zb \right)^2 } + \frac{|\zh|^2}{\left( 1 + \zb/\Delta \right)^2} + \frac{1 + \frac{2 y |\zh|}{\Delta+\zb} \frac{ I_1\left( \frac{2 y |\zh|}{\Delta+\zb} \right)}{ I_0\left( \frac{2 y |\zh|}{\Delta+\zb} \right)}}{1/\Delta + 1/\zb} \right) ,
\end{align}
where $I_{0,1}$ are the modified Bessel functions of the first kind,
\begin{align}
 I_0(x) &= \frac{1}{\pi} \int_0^{\pi} \dd \theta e^{x \cos(\theta)} , \\
 I_1(x) &= \frac{1}{\pi} \int_0^{\pi} \dd \theta e^{x \cos(\theta)} \cos(\theta).
\end{align}

\subsubsection{Convergence}

Unfortunately, phase-retrieval \gamp encounters convergence problems that are very similar to the ones of \gamp for \cs with non zero-mean matrices. 
In particular, note that the variances $\Xb_i$ in~\algoref{algo:cgamp} can take negative values if no damping is used, at which point the algorithm diverges.
An additional damping scheme is therefore required, described in~\cite{prgamp}. The resulting algorithm is called \prgamp.

\begin{figure}[h]
\centering
\begin{subfigure}{0.48\textwidth}
 \includegraphics[width=\textwidth]{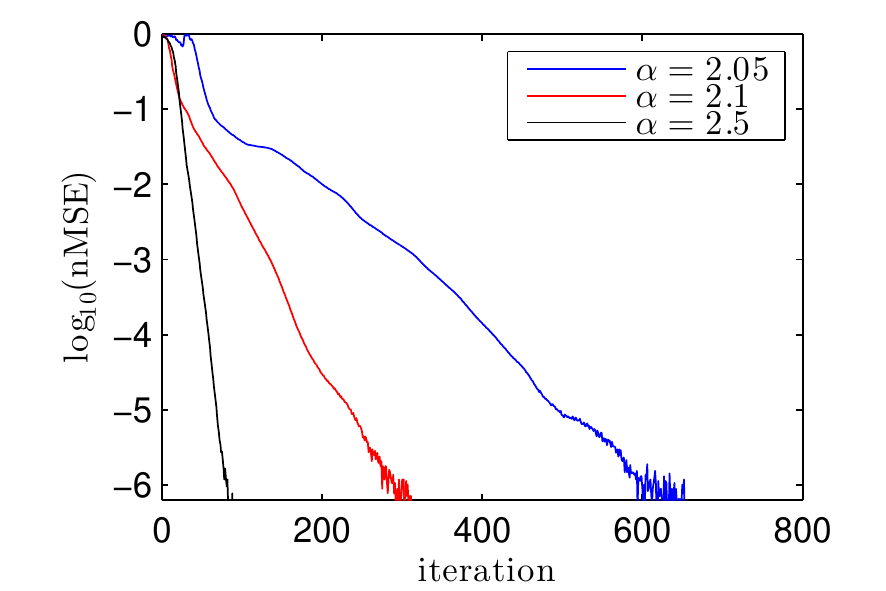}
 \caption{$\rho =1$}
 \label{fig:5_pr_evol_1}
\end{subfigure}
\begin{subfigure}{0.48\textwidth}
 \includegraphics[width=\textwidth]{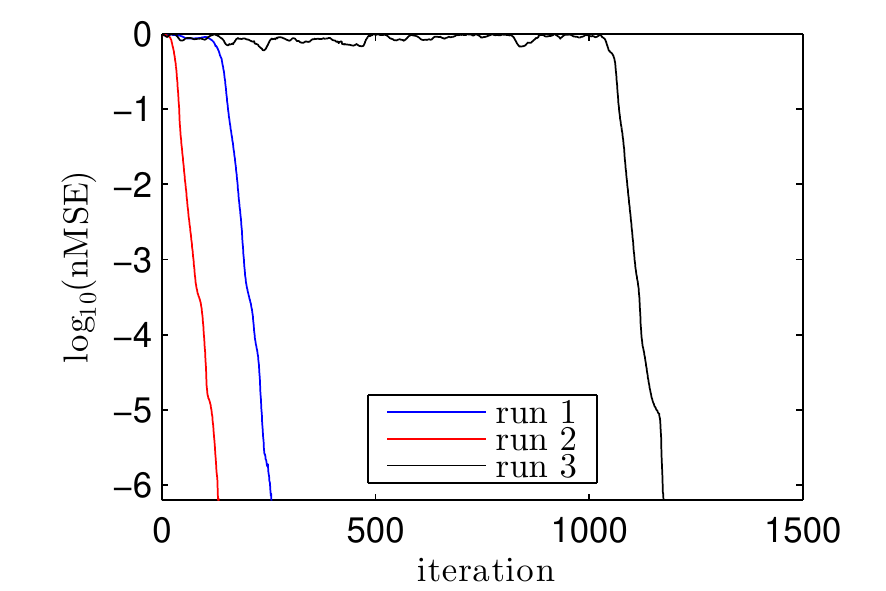}
 \caption{$\rho=0.1$}
 \label{fig:5_pr_evol_01}
\end{subfigure}
\caption[Experimental convergence of \prgamp]{Experimental convergence of \prgamp for non-sparse (a) and sparse signals (b) of size $N=500$ sampled from distribution~(\ref{eq:joint_sparse}).
(a) For non-sparse signals, a damping coefficient of $\approx 0.9$ is sufficient and results are reproducible.
The speed of convergence diminishes approaching $\alpha=2$, below which recovery is impossible: as in \cs, there is a phase transition.
(b) For sparse signals, more damping is needed ($\approx 0.1$) for \prgamp to converge. 
Convergence seems to depend crucially on initialization, but once a certain nMSE is reached, convergence is geometric as for \cs.
Here, $\alpha=1.1$.}
\label{fig:5_pr_evol}
\end{figure}

Additionally to damping, another way to improve convergence is learning the sparsity and the variance of the prior $\px$ with expectation-maximization, as done in~\cite{krzakaCS}. 
Even when they are known, learning them instead of imposing them increases the stability of the algorithm in many cases.

A general observation that can made is that \prgamp solves non-sparse problems more efficiently than sparse problems. 
This is illustrated by~\figref{fig:5_pr_evol}: for $\rho=1$, \prgamp works with little damping and the phase transition is at $\alpha=2$.
On the other hand, for $\rho=0.1$ and $\alpha=1.1$, a lot of damping is necessary and the time to converge varies a lot from instance to instance.
In many cases, the algorithm does not converge at all, such that it is not possible to define a clear phase transition.
The authors of~\cite{prgamp} propose to restart \prgamp from a different initialization if it does not converge and can in this way produce experimental phase diagrams.

\subsection{Imaging through scattering media}
In~\cite{moiOptique}, we propose to apply phase retrieval techniques to the challenge of imaging through scattering media.

\subsubsection{Scattering media and transmission matrix formalism}
\begin{figure}
 \centering
 \includegraphics[width=0.5\textwidth]{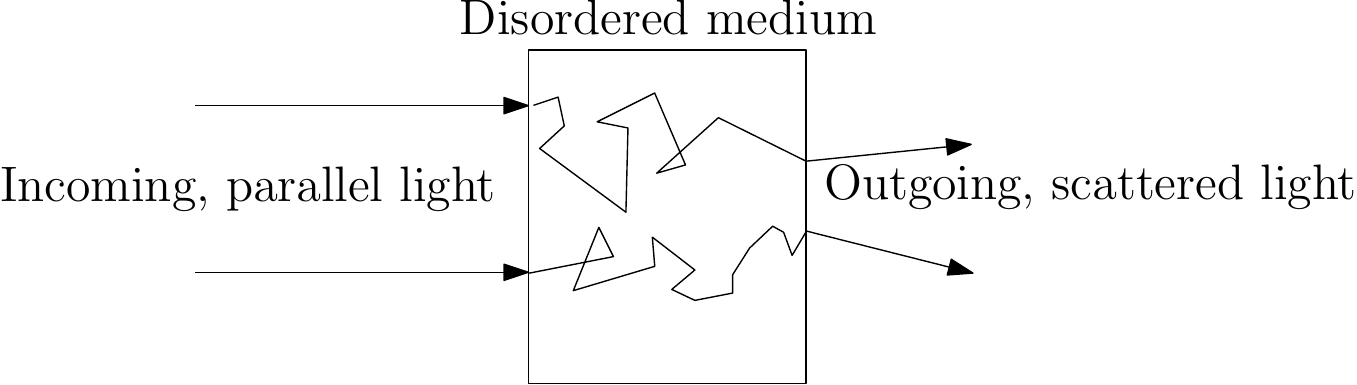}
 \caption[Scattering medium]{The trajectories of photons crossing a multiply scattering medium are deviated so radically 
 that seeing through such a medium is impossible.}
 \label{fig:5_scattering}
\end{figure}

In a scattering medium---such as fog or turbid water for example---transmitted light is not only attenuated: Photons 
crossing the medium are scattered multiple times by particles or impurities of the medium, thereby changing their trajectories (\figref{fig:5_scattering}).
As a result, the outgoing wavefront is radically different from the incoming wavefront: Seeing or imaging through such a medium is impossible.

Sending parallel coherent light on the medium produces a so called \textit{speckle} pattern which is the result of the interferences 
between the photons leaving the medium. 
Just as diffraction patterns in x-ray crystallography reflect crystalline structures,  speckles reflect the randomness of scattering media.
Despite its randomness, the system can be described in a simple way by its (complex-valued) \textit{transmission matrix} $\M$.
Discretizing a plane of incoming light and of outgoing light, the electromagnetic fields can be written as a (complex) incoming 
vector $\xv$ and an outgoing vector $\zv$, linked by the relation
\begin{align}
 \zv &= \M \xv + \xiv, \quad \text{with} \quad \xiv\sim \CN(\xiv;0,\Delta \indic). 	\label{eq:transmission_matrix}
\end{align}

With this relation, imaging through scattering media becomes a simple linear estimation problem.

\subsubsection{Challenges and solutions}
\begin{figure}[h]
 \centering
 \includegraphics[width=0.7\textwidth]{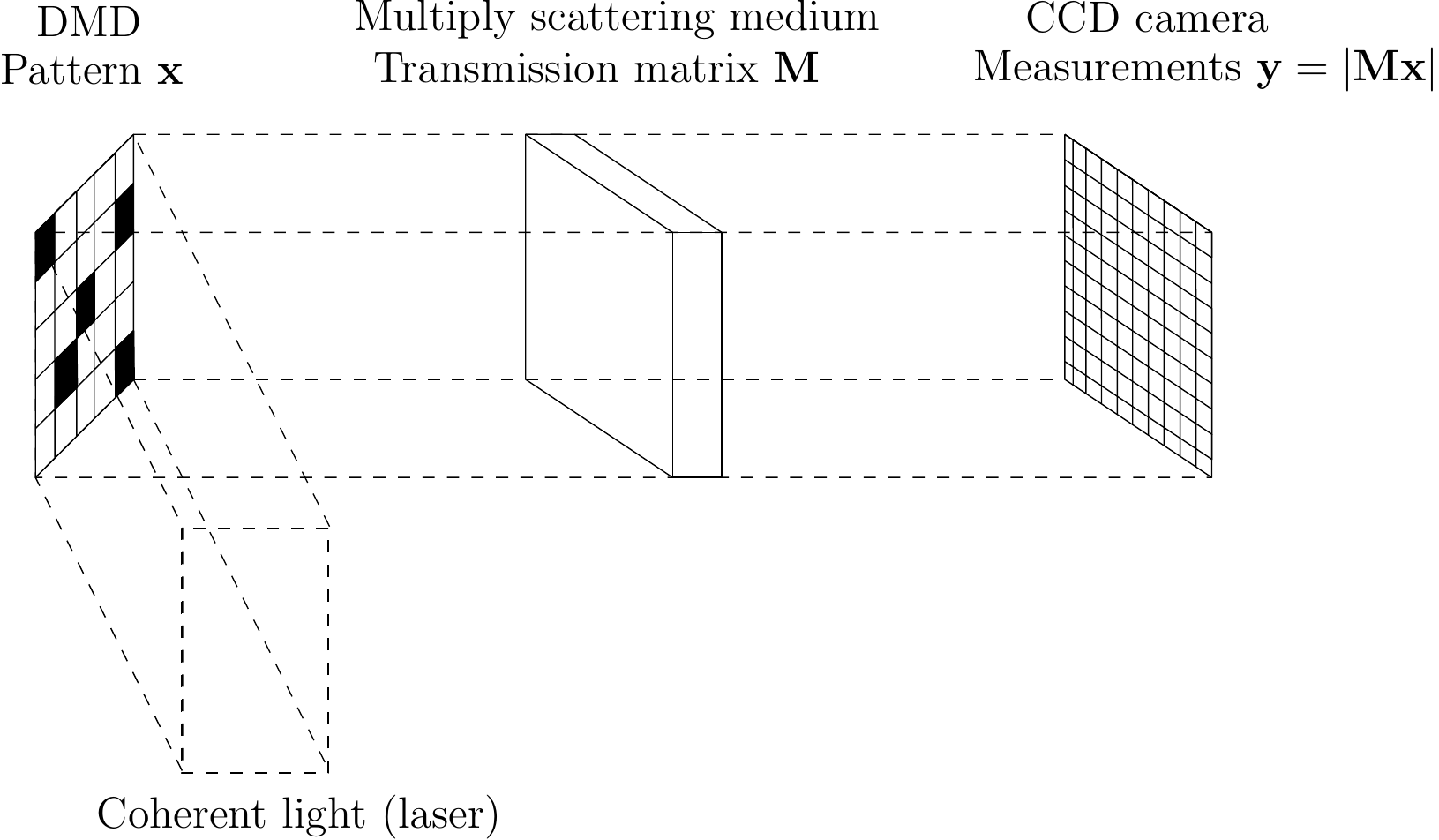}
 \caption[Experimental setup]{Experimental setup.
 A broad beam of parallel, coherent light is created with a laser and a telescope and is sent on the DMD.
 Depending on their current angles, the mirrors of the DMD send the incoming light on the layer of white paint (white pixels) 
 or away from it (black pixels). The incoming light is focused on the paint and the outgoing, scattered light on a CCD camera.
 }
 \label{fig:5_experiment}
\end{figure}

Despite the simple formula~(\ref{eq:transmission_matrix}), both theoretical and practical difficulties exist.
Taking these into account, the experimental setup of~\cite{moiOptique} is shown on~\figref{fig:5_experiment}.
\paragraph{Calibration through incoming light modulation.}
First of all, the transmission matrix $\M$ needs to be determined for each sample of scattering material.
This can be done in a supervised calibration step prior to imaging, by measuring the outputs of known input signals.
It is therefore necessary to modulate the incoming light in a controlled way. 
This can be done with \textit{spatial light modulators}, that can dephase light pixel by pixel~\cite{slm}.
The limit of this device is that it is relatively slow (few tens of Hertz).

In~\cite{moiOptique}, a digital micromirror device (\dmd) is used instead~\cite{dmd}. 
It is an array of $1920 \times 1080$ tilting micromirrors, that either project the light to the scattering medium or not, depending 
on their position. Unlike spatial light modulators, the phase of the light cannot be controlled, but only its 
intensity, in a binary way. This limitation is compensated by the higher functioning speed (over $20$kHz).

\paragraph{Medium stability.}
It is of crucial importance that the material is stable 
enough for $\M$ to vary on time scales large enough to allow both calibration and imaging.
For this reason, imaging through turbid liquids or gases---in which Brownian motion is present---seems for now impossible.
In biological tissues, the stability time is about a few milliseconds.
Therefore, in~\cite{moiOptique}, a $\sim 100$ microns thick layer of white paint is used: thick enough to mix the light and 
produce a complex interference pattern, it transmits sufficient intensities for imaging and is stable enough for $\M$ to vary only
weakly over a period of several minutes.

\paragraph{Phase problem.}
Equation~(\ref{eq:transmission_matrix}) is complex-valued. 
However, a CCD camera cannot capture the complex vector $\zv$, but only its intensity
\begin{align}
 \yv = | \zv |. \label{eq:magnitude}
\end{align}
Previous works have use reference beams in order to indirectly access the phase of $\zv$~\cite{interf1,interf2,interf3}.
This requires an interferometric setup, which is by nature very sensible to external perturbations.
In~\cite{moiOptique} instead, we use phase retrieval, allowing for a simpler experimental setup (see~\figref{fig:5_experiment}).

\subsubsection{Calibrating, imaging and focusing}
\paragraph{Calibration}
The determination of $\M$ in a supervised calibration step can be recast into an inference problem, just as the perceptron problem~(\secref{sec:perceptron}) 
was recast into a 1-bit \cs problem.
The supervised calibration consists in measuring the outputs $\yv =[ \yv_1, \cdots, \yv_P]$ of $P$ different, known inputs $\xv =[ \xv_1, \cdots, \xv_P]$.
Transposing the system, calibration is equivalent to solving the inference problem
\begin{align}
 \yv^{\top} = | \xv^{\top} \M^{\top} + \xiv |,	\label{eq:PR_calibration}
\end{align}
in which $\xv^{\top}$ is considered as a measurement matrix and $\M^{\top}$ as a signal to infer.
Note that:
\begin{itemize}
 \item The columns of $\M^{\top}$ correspond to the individual pixel of the CCD camera and can be treated independently. No sparsity assumption is made on $\M$, a Gaussian prior is used instead.
 \item The measurement matrix $\xv$ is composed of binary entries and does not have zero mean.
 This setting is not favorable to the use of \gamp, though such matrices can be handled using appropriate damping schemes~\cite{vilaAdaptive}.
 \item The measurement matrix $\xv$ being real, an additional invariance by complex conjugation prevents unique recovery of a signal.
 As for the global phase invariance, this has no effect on the algorithm's performances, but has to be kept in mind during tests on synthetic signals.
\end{itemize}

The results presented in~\cite{moiOptique} speak in favor of good reconstruction performances of the transmission matrix $\M$.
This is verified by looking at the nMSE between $\yv$ and $|\xv^{\top} \hat{\M}^{\top}|$, where $\hat{\M}$ is the estimation of $\M$ returned by the algorithm.
Additionally to \prgamp, several other algorithms were tested and the prVBEM algorithm~\cite{prvbem} was retained 
for its increased robustness compared to \prgamp.

The estimated transmission matrices $\M$  have entries that follow a circular Gaussian distribution, which is checked 
by looking at the distribution of their eigenvalues, that approximately follow a Marcenko-Pastur law.
This makes $\M$ an ideal matrix for imaging with \prgamp or other compressive phase retrieval algorithms.

\paragraph{Imaging}
Once the transmission matrix of a medium is determined, imaging experiments can be made, \ie experiments in which the incoming image $\xv$ is unknown and 
has to be recovered using $\yv$.
In the work leading to~\cite{moiOptique}, compressive imaging experiments were not successful. 
This is likely due to insufficient quality of the estimation of $\M$.
In a follow-up work~\cite{boshraImaging}, calibration of the matrix is performed using a modified version of \prgamp, 
allowing more precise estimation of $\M$ and successful imaging experiments.

\begin{figure}[h]
 \centering
 \includegraphics[width=0.5\textwidth]{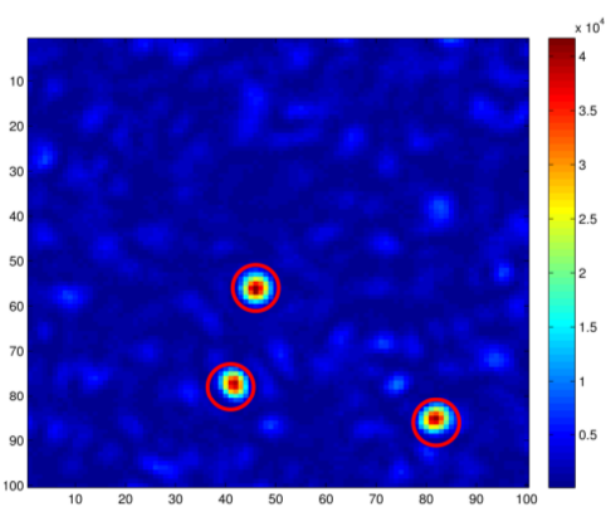}
 \caption[Focusing through a multiply scattering medium]{Focusing on three points through a multiply scattering medium.
 The circles indicate the target positions, in which the intensity is much higher than in the rest of the image. Figure from~\cite{moiOptique}.}
 \label{fig:5_focused}
\end{figure}

\paragraph{Focusing}
The knowledge of the transmission matrix $\M$ can as well be used for focusing through a multiply scattering medium.
This is achieved by creating an input pattern $\xv$ with the DMD that creates an output image with one or several points of high intensity.
Focusing is an optimization problem, but can also be treated a noisy inference problem: searching for an $\xv$ that produces an output 
as close as possible to the desired $\yv$.
Figure~\ref{fig:5_focused} shows the result of a focusing experiment presented in~\cite{moiOptique}.
The intensity reached on the three target points is about an order of magnitude higher than the background intensity.
The success of the focusing experiment is an indication of the quality of the estimation of $\M$.

\section{Conclusion}
In this chapter, I have proposed a version of \gamp for vectorial variables.
Among other possible uses, it allows a derivation of complex-valued \gamp.

Furthermore, I present two applications of \gamp to the concrete problems of coding and imaging through multiply scattering media.
The former application makes use of phase retrieval \gamp, which is an example in which converge of \gamp is not systematic and is not 
yet fully understood.

\vfill

 \chapter{Blind sensor calibration}
\label{chap:blindSensorCal}
In~\chapref{chap:generalizedBilinearModels}, I introduced the blind gain calibration problem as 
a bilinear inference problem, which can be treated using a convex formulation.
In~\cite{moiNips,moiCalibration},
I derive a Bayesian message passing algorithm for blind calibration, called \calamp and assess its performances experimentally.

Just like \gamp, \calamp allows to handle non-linearities in the measurement process and thus to consider a more general type of calibration.
Besides real and complex gain calibration, I have investigated two further examples of blind sensor calibration:
the faulty sensors problem and 1-bit threshold calibration.

\section{Setting}
In addition to the theoretical interest raised by the discovery of phase transitions in \cs, compressed sensing 
is already used both in experimental research and in real world applications such as medical imaging~\cite{sparseMRI}, 
in which it can lead to significant improvements.
One issue that can arise in real-world applications of \cs is a lack of knowledge or an uncertainty regarding the exact measurement process.
Even when the measurement matrix $\Fv$ of the mixing process is perfectly known, the sensing process might not exactly be known, as physical sensors 
are subject to failure, distortion or miscalibration.
As explained in~\chapref{chap:generalizedBilinearModels} in the context of gain calibration, supervised calibration of the sensors might not always 
be possible, in which case blind calibration procedures are necessary.
Several algorithms have been proposed for blind sensor calibration in the case of unknown multiplicative gains, relying on convex optimization~\cite{blindCalConvex,blindCalConvex2}.

\begin{figure}
 \centering
 \includegraphics[width=0.8\textwidth]{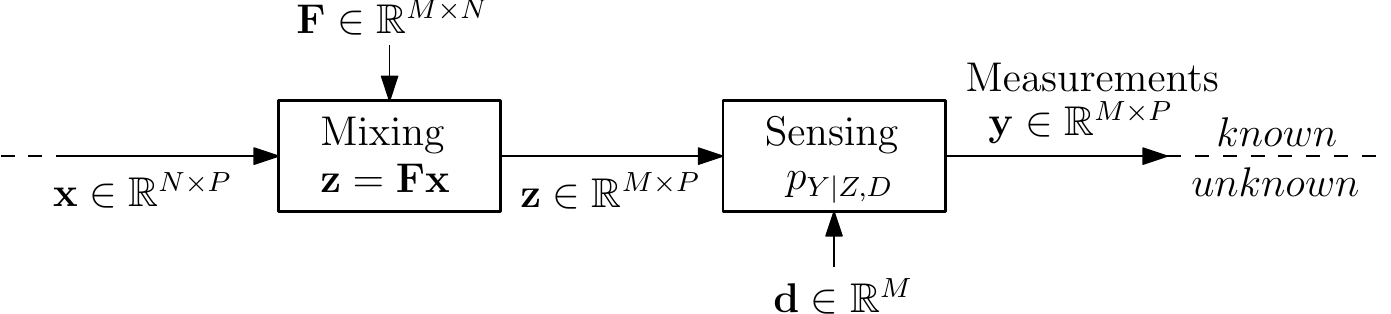}
 \caption[\calamp general setting]{General setting of blind sensor calibration: while the mixing matrix $\Fv$ is known,
 an unknown calibration parameter $d_{\mu}$ is associated to each sensor $\mu$.}
 \label{fig:6_setting}
\end{figure}

The \calamp algorithm proposed in~\cite{moiNips,moiCalibration} is based on \gamp and is therefore not restricted to gain calibration.
We consider the measurement process illustrated by~\figref{fig:6_setting}, in which each measurement $\y_{\mu}$ is generated from $\z_{\mu}$ \textit{and}
a sensor-dependent calibration parameter~$d_{\mu}$:
\begin{align}
 y_{\mu} \sim \pyd(\y_{\mu}|\z_{\mu},d_{\mu}).		\label{eq:generalCalibration}
\end{align}

\section{Cal-AMP}
A detailed derivation of \calamp is provided in~\cite{moiCalibration}.
As it is very similar to other derivations of message-passing algorithms (in particular the one provided for vectorial \gamp in~\chapref{chap:gampApplications}), 
I will only describe the differences to the latter.
\subsection{Derivation}
\begin{figure}
 \centering
  \includegraphics[width=0.5\textwidth]{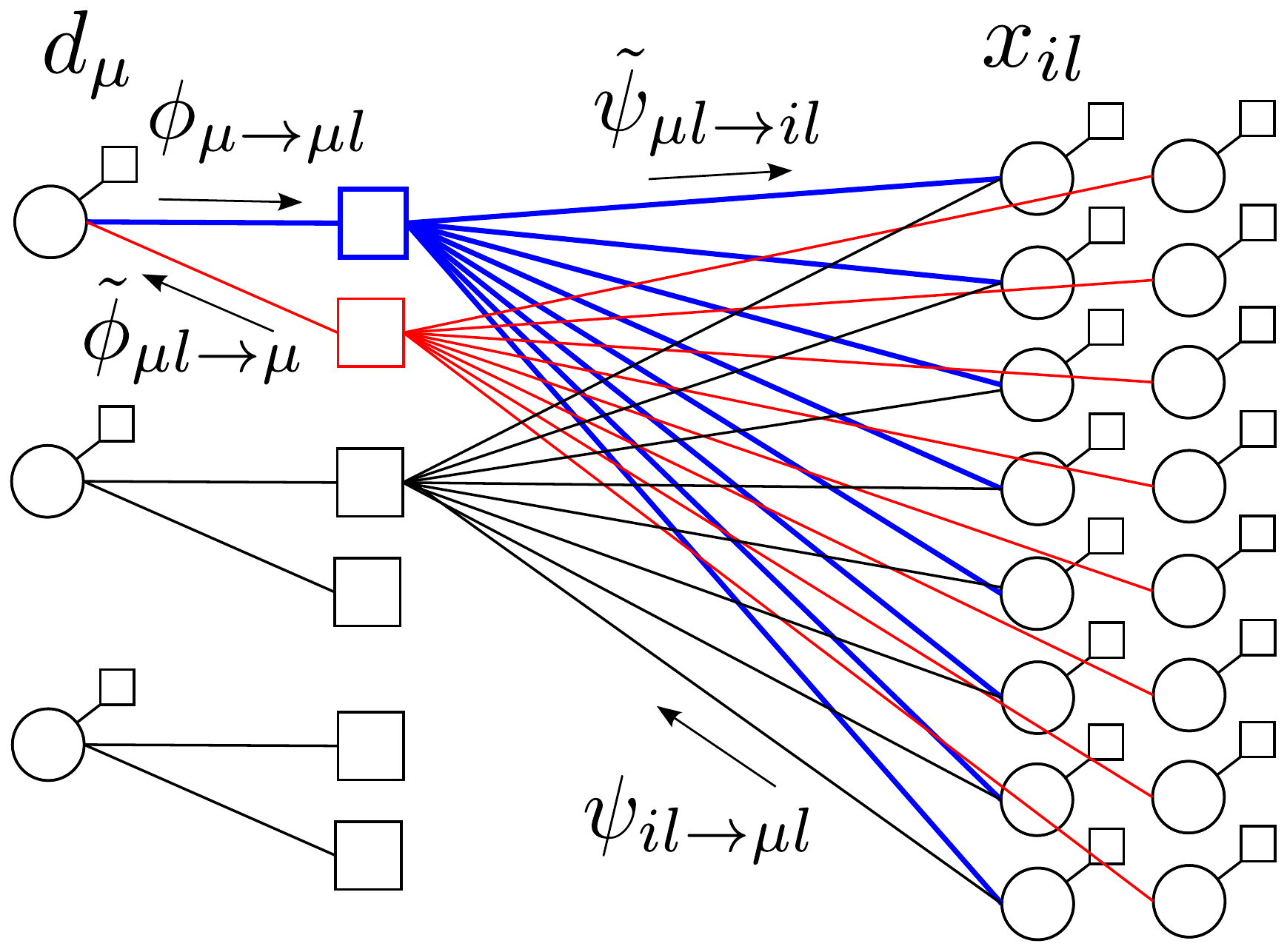}
 \caption[\calamp factor graph]{Factor graph representing the posterior distribution~(\ref{eq:posteriorCalibration}).
 For clarity, $N=8, M=3$ and $P=2$. Only the links from the three upper factor nodes are represented.}
 \label{fig:6_fg}
\end{figure}

First of all, the posterior distribution that is the starting point of \calamp is
\begin{align}
 p(\xv,\dv|\Fv,\yv) &= \frac{1}{\ZZ(\yv,\Fv)} \prod_{i,l=1}^{N,P} \px(x_{il}) \prod_{\mu=1}^M \pd(d_{\mu}) \prod_{l, \mu=1}^{P,M} \pyd(y_{\mu l}|z_{\mu l}, d_{\mu}).	\label{eq:posteriorCalibration} 
\end{align}
The factor graph representing this distribution is presented on~\figref{fig:6_fg}.
Compared to the factor graph on~\figref{fig:5_c_fg}, this factor graph has one additional type of messages that link the calibration parameters to the measurements.
Furthermore, a set of $P$ signals is considered, instead of just one signal, and as the measurements of all these signals are produced with the \textit{same} 
calibration parameters $d_{\mu}$, inference can be possible if $P$ is large enough.
The derivation of \calamp follows the same three steps as explained in Table~\ref{table:3_bpToTap} and detailed in~\chapref{chap:gampApplications}.

\subsubsection{Step 1: \bp}
We first write the \bp equations for the two types of messages.
For the signal variables, we have
\begin{align}
 \messv_{il \to \mu l}^t(x_{il}) &\propto \px(x_{il}) \prod_{\gamma \neq \mu} \messf_{\gamma l\to i l}^t(x_{il}), \\
 \messf_{\mu l \to i l}^{t+1}(x_{il}) &\propto \int \dd d_{\mu} \messvd_{\mu \to \mu l}^t(d_{\mu}) \int \left( \prod_{j \neq i} \dd x_{jl} \messv_{j l \to \mu l}^t(x_{jl}) \right) \pyd(y_{\mu}|\sum_i F_{\mu i} x_{il}, d_{\mu}),
\end{align}
and for the calibration variables,
\begin{align}
 \messvd_{\mu \to \mu l}^t(d_{\mu}) &\propto \pd(d_{\mu}) \prod_{m \neq l} \messfd_{\mu m \to \mu}(d_{\mu}) , \\
 \messfd_{\mu l \to \mu}^{t+1}(d_{\mu}) &\propto \int \left( \prod_i \dd x_{il} \messv_{il \to \mu}^t \right) \pyd(y_{\mu l}| \sum_{i} F_{\mu i}x_{il}, d_{\mu}) .
\end{align}

\subsubsection{Step 2: \amp}
In order to make the \bp equations tractable, we use the central limit theorem to approximate $\sum_i F_{\mu i} x_{il}$ by a Gaussian random variable. 
We obtain 
\begin{align}
 \messfd_{\mu l \to \mu}^{t+1}(d_{\mu}) &\propto f_0^Z(y_{\mu l} | \Zh_{\mu l}^{t+1}, \Zb_{\mu l}^{t+1}, d_{\mu}) \\
 \messvd_{\mu \to \mu l}^{t}(d_{\mu}) &\propto \pd(d_{\mu}) \prod_{m \neq l} f_0^Z(y_{\mu m} | \Zh_{\mu m}^{t}, \Zb_{\mu m}^{t}, d_{\mu})
\end{align}
with the usual $f_0^Z$ and
\begin{align}
 \Zh_{\mu l}^{t+1} &= \sum_i F_{\mu i} \xh_{il \to \mu l}^t, & \Zb_{\mu l}^{t+1} &= \sum_i F_{\mu i}^2 \xb_{il \to \mu l}^t,
\end{align}
in which $(\xh_{il \to \mu l}^t, \xb_{il \to \mu l}^t)$ are the mean and variance of the message $\messv_{il \to \mu}^t$.
These can be expressed as a function of estimators at an earlier time step, which allows to have iterative updates of means and variances only:
\begin{align}
 \xh_{il \to \mu l}^t &= \fh^X\left( \Xh_{il \to \mu l}^t, \Xb_{il \to \mu l}^t \right), & \xb_{il \to \mu l}^t &= \fb^X\left( \Xh_{il \to \mu l}^t, \Xb_{il \to \mu l}^t \right),
\end{align}
with
\begin{align}
 \Xb_{il \to \mu l}^{t+1} &= -\left( \sum_{\gamma \neq \mu} F_{\gamma i}^2 \gb_{\gamma l \to il}^{t+1} \right)^{-1} , & \Xh_{il \to \mu l}^{t+1} &= \Xb_{il \to \mu l}^{t+1} \sum_{\gamma \neq \mu} F_{\gamma i} \gh_{\gamma l \to il}^{t+1},  \\
 \gb_{\mu l \to il}^{t+1} &= \gb^C\left(y_{\mu l}| \Zhv_{\mu l \to il}^{t+1},\Zbv_{\mu l \to il}^{t+1} \right), &  \gh_{\mu l \to il}^{t+1} &= \gh^C\left( y_{\mu l}| \Zhv_{\mu l \to il}^{t+1},\Zbv_{\mu l \to il}^{t+1} \right).
\end{align}
The only difference with \gamp is that the functions $(\gh^Y, \gb^Y)$ are replaced by $(\gh^C, \gb^C)$ that are functions of vectors 
$(\Zhv_{\mu l \to il}^{t}, \Zbv_{\mu l \to il}^{t}) \in \RR^P$ , whose components are given by:
\begin{align}
 [\Zhv_{\mu l \to il}^{t}]_{l'} &=  \delta_{l',l} \Zh_{\mu l \to il}^{t} + (1-\delta_{l',l}) \Zh_{\mu l'}^{t}, \\
 [\Zbv_{\mu l \to il}^{t}]_{l'} &=  \delta_{l',l} \Zb_{\mu l \to il}^{t} + (1-\delta_{l',l}) \Zb_{\mu l'}^{t}.
\end{align}
Just as $(\gh^Y, \gb^Y)$ are defined from the functions $f_k^Z$ with~(\ref{eq:gh_vec}, \ref{eq:gb_vec}), $(\gh^C, \gb^C)$ are defined from
\begin{align}
 \left[ f_k^C\left( y_{\mu l}|  \Zhv_{\mu l}^{t},\Zbv_{\mu l}^{t} \right) \right]_{\mu l} &= \int \dd d_{\mu} \pd(d_{\mu})  f_k^Y\left(y_{\mu l}| \Zh_{\mu l}^{t},\Zb_{\mu l}^{t} \right)  \prod_{l' \neq l}f_0^Y\left(y_{\mu l}| \Zh_{\mu l'}^{t},\Zb_{\mu l}^{t} \right) ,	\label{eq:fkc}
\end{align}
for $k=\{0,1,2\}$. Note a slight difference in the time indices compared to~\cite{moiCalibration}, which turns out not to matter in practice.
\subsubsection{Step 3: \tap}
The resulting \amp algorithm can be brought into a \tap version by expressing messages as a function of beliefs and keeping the ``Onsager'' terms that appear.
The resulting \calamp is presented in~\algoref{algo:calamp}.

\begin{algorithm}
\caption{Cal-AMP}
\label{algo:calamp}
 \textbf{Initialize} $\ghv_0=0$ and $(\xhv_0, \xbv_0)$ at random or according to $\px$. \\
  \textbf{Main loop:} while $t<t_{\rm max}$, calculate following quantities:
  \begin{align*}
  \Zbv_{t+1} &= |\Fv|^2 \bar{\xv}_{t} \\
  \Zhv_{t+1} &= \Fv \xhv_{t} - \Zbv_{t+1} \odot \ghv_{t} \\
  \gbv_{t+1} &= \gb^C(\yv|\Zhv_{t+1}, \Zbv_{t+1}) \\
  \ghv_{t+1} &= \gh^C(\yv|\Zhv_{t+1}, \Zbv_{t+1}) \\
  \Xbv_{t+1} &= - \left(  |\Fv^2|^{\top} \gbv_{t+1} \right)^{\inv} \\
  \Xhv_{t+1} &= \xhv_{t} + \Xbv_{t+1} \odot \left( \Fv^{\top} \ghv_{t+1} \right)  \\
  \xbv_{t+1} &= \fb^X(\Xhv_{t+1}, \Xbv_{t+1}) \\
  \xhv_{t+1} &= \fh^X(\Xhv_{t+1}, \Xbv_{t+1}) 
  \end{align*}
   \textbf{Stop} when $t=t_{\rm max}$, $|| \xbv_t||_2 < \epsilon$ or $|| \xhv_t - \xhv_{t-1} ||_2< \epsilon$.
\end{algorithm}

\subsection{Comparison to \gamp}
The only difference between \gamp (\algoref{algo:gamp}) and \calamp are that
\begin{enumerate}
 \item in \gamp, the updated quantities are vectors, whereas in \calamp, they are matrices, which are the concatenation of $P$ vectors.
 \item the update functions $(\gh^Z, \gb^Z)$ are replaced by $(\gh^C, \gb^C)$.
\end{enumerate}
For $P=1$, \calamp is strictly identical to \gamp, with
\begin{align}
 \py(y|z) &= \int \dd d_{\mu} \pd(d_{\mu}) \pyd(y|z,d).
\end{align}
For $P>1$, the step involving $\gh^C$ and $\gb^C$ is the only one in which the $P$ signals are not treated independently.
If it is possible to perform perfect calibration of the sensors in a supervised way, the prior $\pd(d_{\mu})$ can be replaced 
by $\delta(d_{\mu} - d_{\mu}^{\mathrm{cal}})$, thus $(\gh^C,\gb^C)$ can be calculated independently for each of the $P$ signals 
and \calamp is identical to \gamp with perfectly calibrated sensors.

\subsubsection{Damping scheme}
As for \gamp, the stability of \calamp can be improved by damping some of the variables.
One can for example use the damping scheme proposed in~\cite{damping}, which corresponds to damping the variances $(\Zb,\Xb)$ and the means $(\Zh,\Xh)$ 
as follows:
\begin{align}
 \mathrm{var}^{t+1} &\equiv \left( \beta \frac{1}{\mathrm{var}_0^{t+1}} + \frac{1-\beta}{\beta} \frac{1}{\mathrm{var}^t}  \right)^{-1} , \\
 \mathrm{mean}^{t+1} &\equiv \beta' \mathrm{mean}_0^{t+1} + (1 - \beta') \mathrm{mean}^t ,
\end{align}
where $\beta \in (0,1]$, $\beta' \equiv \beta \mathrm{var}^{t+1} / \mathrm{var}_0^{t+1}$ and the quantities with index $0$ are before damping.

\section{Case studies}
In this section we numerically investigate several particular settings of blind sensor calibration.
\subsection{Gain calibration}
Gain calibration was introduced in~\secref{sec:gainCal}: each sensor multiplies the component $z_{\mu l}$ by an unknown gain $d_{\mu}^{-1}$.
In noisy, complex gain calibration, the measurement $y_{\mu l}$ is produced as follows:
\begin{align}
 y_{\mu l} &= \frac{z_{\mu l} + \xi_{\mu l}}{d_{\mu}}, \quad \text{with} \quad \xi_{\mu l} \sim \CN(\xi_{\mu l};0,\Delta), \quad d_{\mu} \in \CC^*.  \label{eq:blindGainCal}
\end{align}
The choice of \textit{dividing} instead of multiplying by the gain $d_{\mu}$ simply comes from the fact that this makes the calculation of $(\gh^C,\gb^C)$ easier.
In fact, 
\begin{align}
 f_0^Z(y|\Zh,\Zb,d) &= \frac{|d|}{|y|} \NN(d; \frac{\Zh}{y}, \frac{\Delta+\Zb}{y^2})
\end{align}
and therefore
\begin{align}
 \fh^C(y|\Zh,\Zb) &= \frac{\Delta \Zb}{\Delta + \Zb} \left( \frac{\Zh}{\Zb} + \frac{y \dh}{\Delta} \right), &  \fb^C(y|\Zh,\Zb) &= \frac{\Delta \Zb}{\Delta + \Zb} \left( 1 + \frac{\Delta \Zb}{\Delta + \Zb}\frac{y^2}{\Delta^2} \db \right) , 
\end{align}
where 
\begin{align}
 \dh_{\mu} &\equiv \fh^D(\Dh_{\mu},\Db_{\mu}), & \db_{\mu} &\equiv \fb^D(\Dh_{\mu},\Db_{\mu}), \\
 \Db_{\mu} &\equiv \left( \sum_{l} \frac{|y_{\mu l}|^2}{\Delta + \Zb_{\mu l}} \right)^{-1}, & \Dh_{\mu} &\equiv \Db_{\mu} \sum_{l} \frac{\Zh_{\mu l} y_{\mu l}^*}{\Delta + \Zb_{\mu l}} .
\end{align}

\subsubsection{Real gain calibration}
\begin{figure}
 \centering
 \includegraphics[width=1\textwidth]{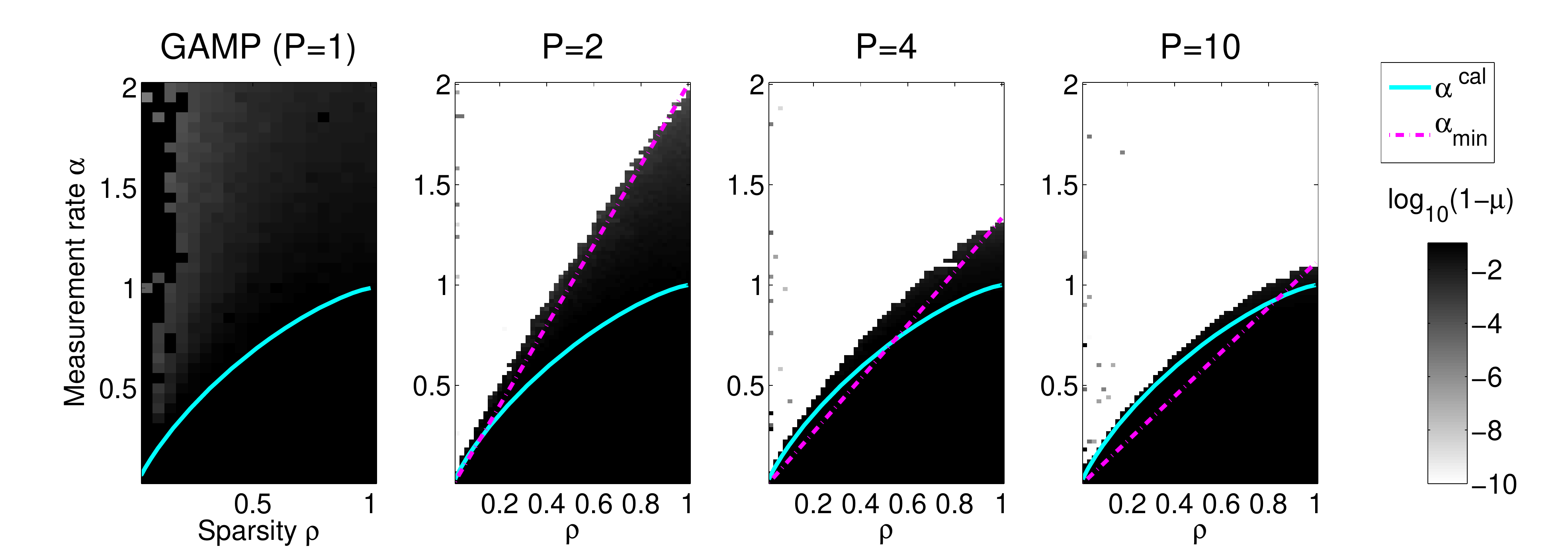}
 \caption[Phase diagram for real gain calibration]{Phase diagrams for real gain calibration.
 White indicates successful reconstruction, black indicates failure. Experiments were carried out for $N=1000$ and $w_d=1$.
 As the number of signals $P$ available for blind calibration increases, the lower bound $\alphagmin(\rho)$ from~\eqref{eq:gain_counting} tends to $\rho$, 
 and the observed phase transition gets closer to $\alpha^{\rm cal}=\alpha_{\rm CS}$, the transition of a perfectly calibrated algorithm.}
 \label{fig:6_real_gain}
\end{figure}
For real gain calibration,~\figref{fig:6_real_gain} shows the experimental phase diagrams obtained for a Gauss-Bernoulli distributed signal and gains uniformly distributed around $1$ with a width $w_d$:
\begin{align}
 \px(\xv) &= \prod_{i l} \left[ (1-\rho) \delta(x_{il}) + \rho \NN(x_{il};0,1) \right], & d_{\mu} &\sim \UU(d_{\mu};1,w_d) ,
\end{align}
with $w_d<2$. For such a distribution of gains, the update functions $(\fh^D,\fb^D)$ can be expressed analytically using 
\begin{align}
 \f_k^D(\Dh,\Db) &= J(P+k, \Dh,\Db, 1 - \frac{w_d}{2}, 1+\frac{w_d}{2})
\end{align}
where the function $J$ is defined in appendix~\ref{app:f}.
A damping coefficient of $\beta = 0.8$ was used, increasing the stability of the algorithm while not slowing it down significantly.

Note that the fact that this prior has a bounded support can lead to a bad behaviour of the algorithm. 
Using a slightly bigger $w_d$ in the prior than in the real distribution of gains (by a factor $1.1$ in our implementation) solves this issue.

\begin{figure}
\centering
 \begin{subfigure}{0.38\textwidth}
  \includegraphics[width=\textwidth]{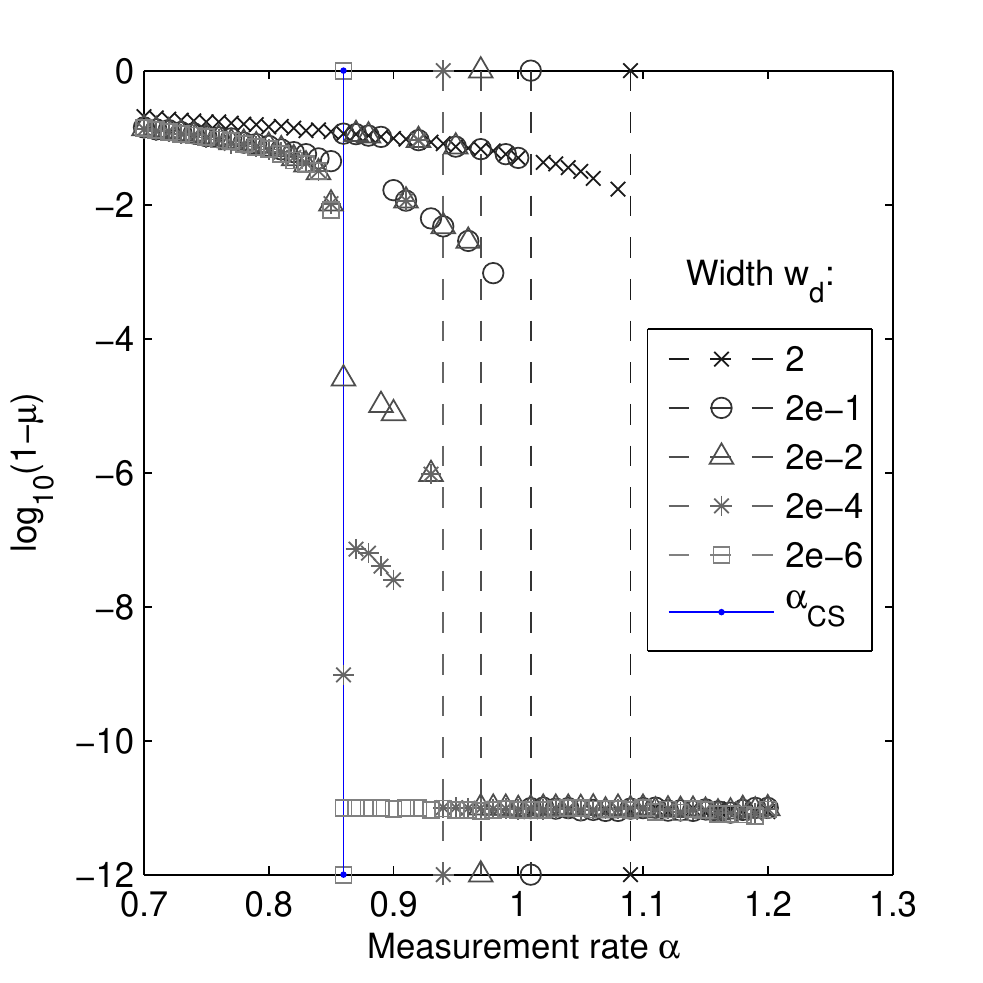}
  \caption{}
  \label{fig:6_gain_vary}
 \end{subfigure}
  \begin{subfigure}{0.38\textwidth}
  \includegraphics[width=\textwidth]{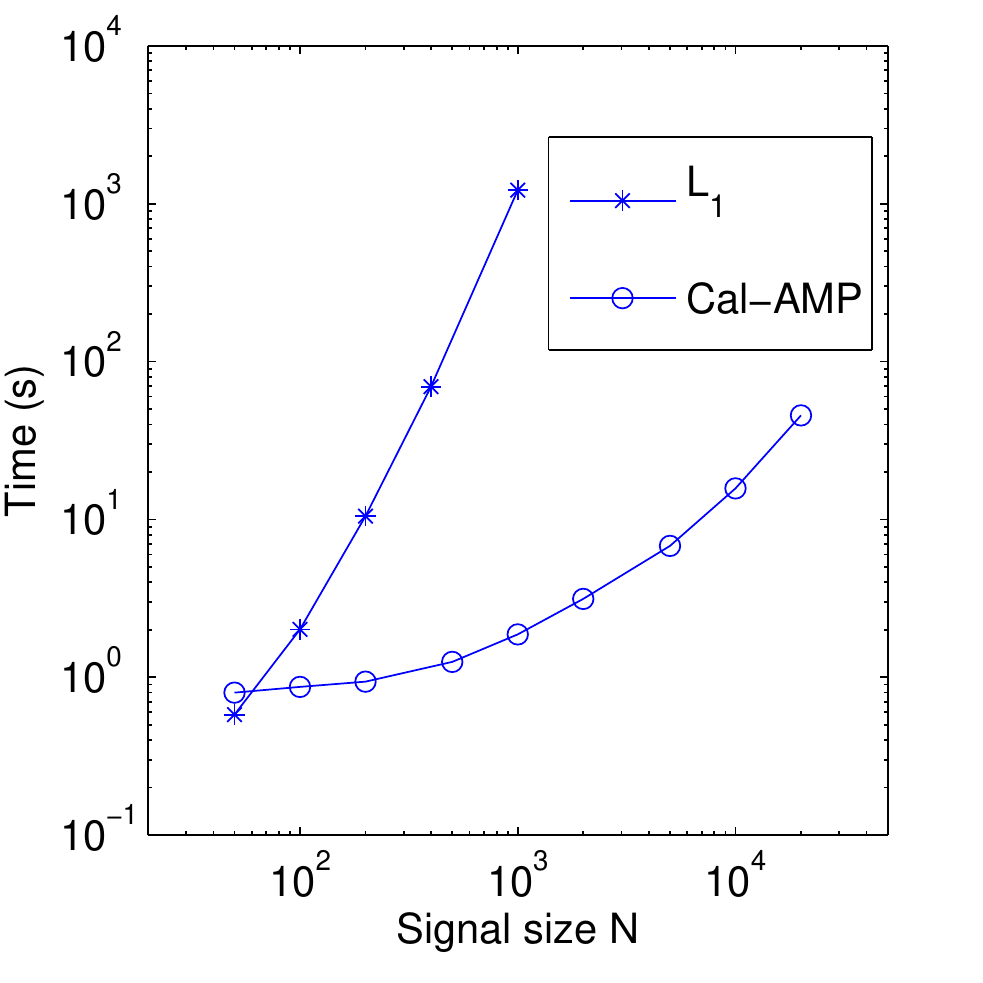}
  \caption{}
  \label{fig:6_compareTimes}
 \end{subfigure}
 \caption[Position of phase transition with decalibration amplitude]{(a) Displacement of the phase transition with varying decalibration amplitude $w_d$.
 Parameters are $\rho=0.7$, $P=4$ and $N=10 000$. 
 The vertical lines materialize the empirical positions of the phase transitions (all points to the 
 right of the line are perfectly reconstructed). 
 (b) Running times of \calamp compared to the $L_1$ minimizing algorithm of~\cite{blindCalConvex} using the CVX package~\cite{cvx}.
 Experiments were performed on a 2.4 GHz processor. Parameters were $\rho=0.2$, $\alpha=1$, $P=5$. 
 For big signal sizes, \calamp is orders of magnitudes faster.
 }
 \label{fig:6_wd_and_times}
\end{figure}

The exact position of the phase transition depends on the amplitude $w_d$ of the decalibration.
Figure~\ref{fig:6_gain_vary} shows how the empirical phase transition approaches $\alpha_{\rm CS}$ when $w_d \to 0$, \ie $d_{\mu} \to 1$ for all $\mu$.

\begin{figure}
 \centering
 \includegraphics[width=0.75\textwidth]{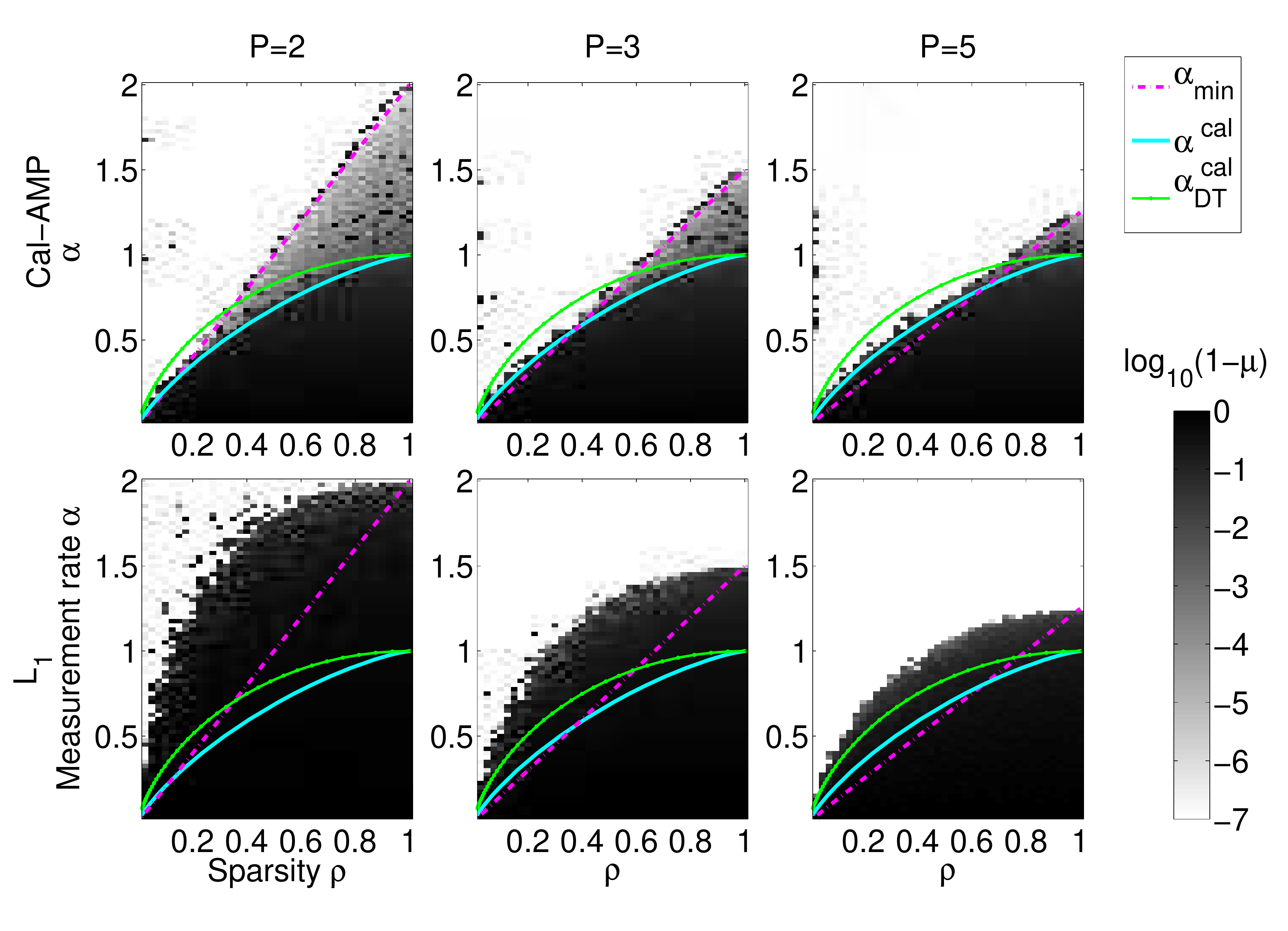}
 \caption[Comparison of gain calibration with \calamp and $L_1$ minimization]{
 Experimental phase diagrams for \calamp and $L_1$ minimization, for $N=100$ and $w_d=0.1$.
 While both algorithms show a similar qualitative behavior, the region of success (white) of \calamp is larger than the one of $L_1$ minimization.
 The line $\alphagmin$ is a lower bound from~\eqref{eq:gain_counting}, $\alpha^{\rm cal}$ is the phase transition of perfectly calibrated Bayesian \gamp and 
 $\alpha_{\rm DT}^{\rm cal}$ is the Donoho-Tanner phase transition of a perfectly calibrated $L_1$-based \cs algorithm.
 Just as the phase transition of \calamp approaches $\alpha^{\rm cal}$ with growing $P$, the one of the $L_1$ algorithm approaches $\alpha_{\rm DT}^{\rm cal}$.}
 \label{fig:6_gain_L1}
\end{figure}

In~\figref{fig:6_gain_L1} we compare the performances of \calamp and of the convex optimization approach of~\cite{blindCalConvex} (see~\secref{sec:convexGainFormulation}).
The convex algorithm can easily be implemented using the CVX package~\cite{cvx,gb08}. The figure shows that \calamp requires significantly less measurements for a successful reconstruction, 
especially for small $P$. This is similar to the improvement that Bayes optimal \gamp allows in \cs over LASSO.
Furthermore, as shown on~\figref{fig:6_compareTimes}, \calamp is significantly faster than the $L_1$ algorithm implemented with CVX.

\subsubsection{Complex gain calibration}
\calamp can be extended to the case of complex signals and gains in the same way as \cgamp was derived in~\chapref{chap:gampApplications}.
The only change in~\algoref{algo:calamp} is that $(\cdot)^{\top}$ indicates complex transposition.
Furthermore, the update functions are calculated with integrals over complex variables.

Complex gains are particularly useful: As often, the physical signal to measure is a propagating wave (sound or light), it is best represented by a complex number.
Complex gains allow to take into account both amplitude gains introduced by sensors and shifts of phases.
A particular case of complex gain calibration is therefore \textit{phase calibration}, in which the complex gain has known amplitude and unknown phase.
\begin{figure}
 \centering
 \includegraphics[width=0.8\textwidth]{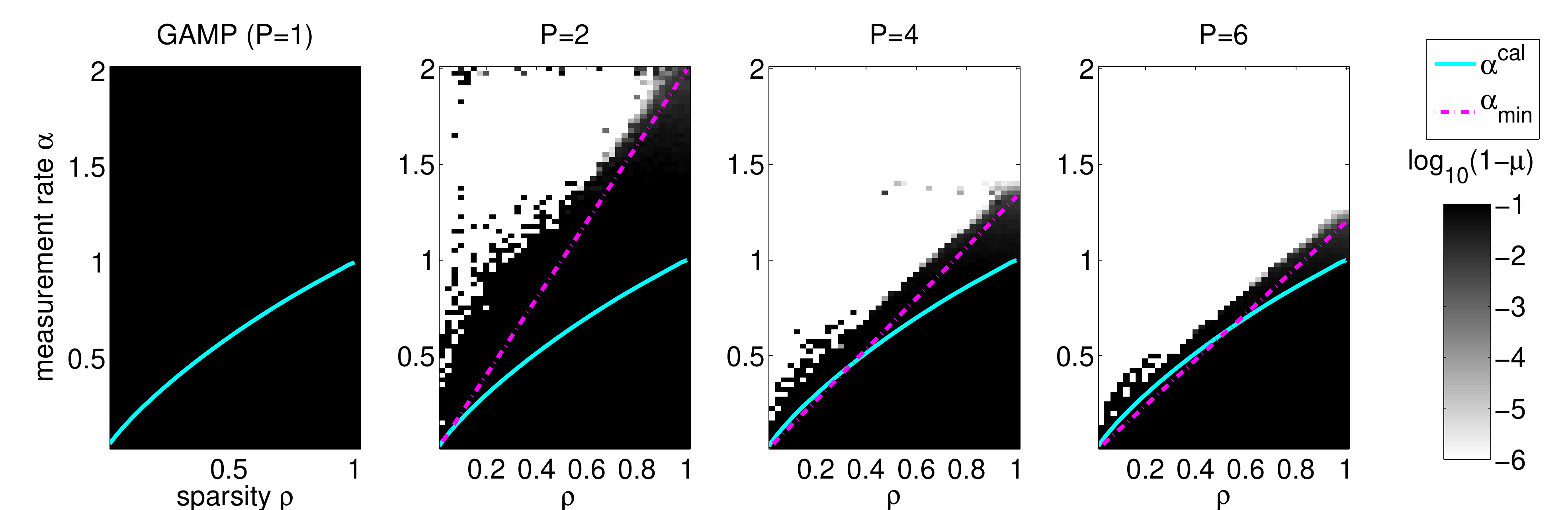}
 \caption[Phase diagram for complex gain calibration]{Phase diagram for complex gain calibration
  using \calamp with $N=500$. Here, $\alpha_{\rm cal}$ is the phase transition of a perfectly calibrated algorithm (c-GAMP phase transition of~\figref{fig:5_complexDiag}) 
  and $\alpha_{\rm min}$ is the lower bound of~\eqref{eq:gain_counting}.}
  \label{fig:6_complex_gains}
\end{figure}
For the experimental results presented on~\figref{fig:6_complex_gains}, the signal follows a complex Gauss-Bernoulli distribution and the complex gains a complex Gaussian distribution:
\begin{align}
  \px(\xv) &= \prod_{i l} \left[ (1-\rho) \delta(x_{il}) + \rho \CN(x_{il};0,1) \right], & d_{\mu} &\sim \CN(d_{\mu};0,10) .	\label{eq:complex_gains_distributions}
\end{align}
In the algorithm, the update functions used for the calibration parameters are
\begin{align}
 \fh^D(\Dh,\Db) &= \frac{\Dh}{|\Dh|} \frac{J(P+1,|\Dh|,\Db,0,\infty)}{J(P,|\Dh|,\Db,0,\infty))} , & \fb^D(\Dh,\Db) &= \Db.
\end{align}
Simpler than the Bayes optimal update functions, they lead to good results. Note that $\fh^D$ takes over the phase estimated by $\Dh$, as the prior on the phase is flat.
Just like for real gain calibration, the empirical phase transition is quite close to the lower bounds imposed by the counting bound $\alpha_{\rm min}$ 
and the perfectly calibrated algorithm.

\subsection{Faulty sensors}
A different example of blind sensor calibration which it might not be possible to recast into a convex minimization problem is the faulty sensors problem.
Without sparsity, this problem was treated in the context of wireless sensor networks, for example in~\cite{sensorFault1,sensorFault2}.
For a single signal, $P=1$, it was also treated using \gamp in~\cite{sensorFault3}.

We assume that a fraction $\fsf$ of sensors is faulty (characterized by $d_{\mu}=0$) and only records noise $\sim \NN(y_{\mu l}; m_f, \sigma_f)$, whereas the other sensors 
(characterized by $d_{\mu}=1$) are functional and record $z_{\mu l}$. We then have
\begin{align}
 \pyd(y|z,d) &= \delta(d-1) \delta(y-z) + \delta(d)\NN(y;m_f,\sigma_f), \\
 \pd(d) &= \fsf \delta(d) + (1-\fsf) \delta(d-1),
\end{align}
which leads to analytical expressions for the estimators $(\zhv,\zbv)$:
 \begin{align}
  \zh_{\mu l} &= \frac{\fsf \Zh_{\mu l} \pi^f_{\mu} + (1-\fsf) y_{\mu l} \pi^z_{\mu}}{\fsf \pi^f_{\mu} + (1-\fsf) \pi^z_{\mu}} ,  &  \zb_{\mu l} &= \frac{\fsf (\Zh_{\mu l}^2+\Zb_{\mu l}) \pi^f_{\mu} + (1-\fsf) |y_{\mu l}|^2 \pi^z_{\mu}}{\fsf \pi^f_{\mu} + (1-\fsf) \pi^z_{\mu}} - |\zh_{\mu l}|^2 ,  
 \end{align}
 with
 \begin{align}
 \pi^f_{\mu} &= \prod\limits_m \NN(y_{\mu m}; m_f, \sigma_f),  & \pi^z_{\mu} &= \prod\limits_m \NN(y_{\mu m}; \Zh_{\mu m}, \Zb_{\mu m}) .
 \end{align}
It is not known which sensors are functional and which ones are faulty.

If $m_f$ and $\sigma_f$ are sufficiently different from the mean and variance of the measurements taken by functional sensors, 
the problem can be expected to be easy. If not, nothing indicates \textit{a priori} which sensors are functional and which ones are faulty.
The algorithm thus has to solve a problem of combinatorial optimization consisting in finding which sensors are faulty.

\subsubsection{Perfect calibration}
If the sensors can be calibrated beforehand, \ie it is known which sensors are faulty, then the problem can be solved by a \cs algorithm by simply
discarding the faulty sensors.
This leads to an effective measurement rate $\alpha_{\rm eff}\equiv \alpha(1-\fsf)$, and the algorithm would succeed when $\alpha_{\rm eff} > \alpha_{\rm CS}$.
Therefore a perfectly calibrated algorithm would have a phase transition at 
\begin{align}
\alpha^{\rm cal}(\rho) \equiv \alpha_{\rm CS}(\rho)/(1-\fsf). \label{eq:alpha_cal_faulty}
\end{align}

\subsubsection{Experimental phase diagram}
\begin{figure}
 \centering
 \includegraphics[width=0.8\textwidth]{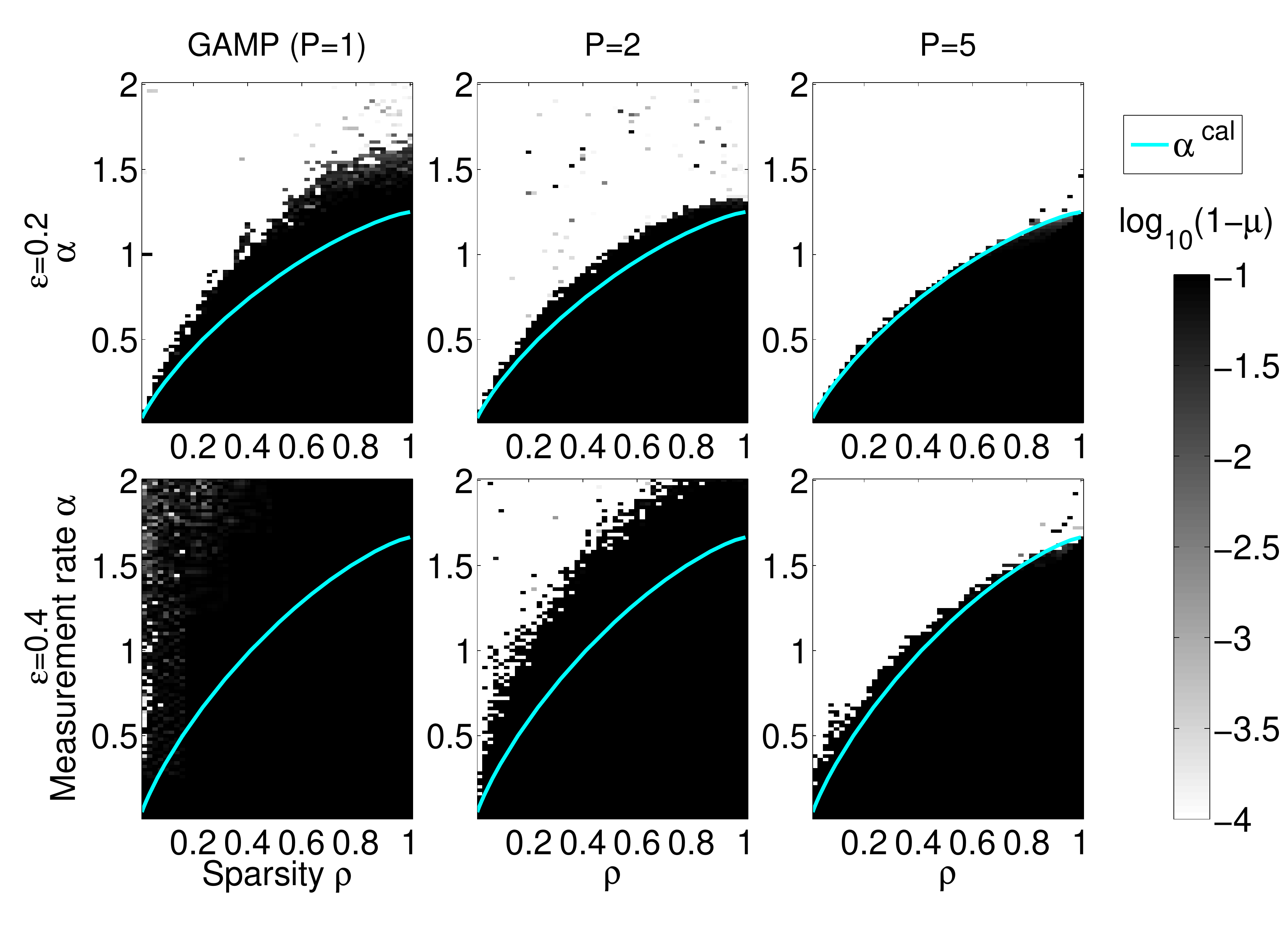}
 \caption[Phase diagram for the faulty sensors problem]{
 Experimental phase diagrams for the faulty sensors problem with $N=1000$.
 White indicates successful reconstruction, black indicates failure.
 The fraction of faulty sensors is of $0.2$ in the first row and $0.4$ in the second row.
 The line $\alpha_{\rm cal}$ from~\eqref{eq:alpha_cal_faulty} shows the performance of a perfectly calibrated algorithm.
 Increasing the number of samples $P$ allows to lower the phase transition down to $\alpha_{\rm cal}$.
 }
 \label{fig:6_faulty}
\end{figure}

Results of numerical experiments are shown on~\figref{fig:6_faulty}.
The signals have a Gauss-Bernoulli distribution, and we consider the hardest case in which $m_f=0$ and $\sigma_f=\rho$: the statistics 
of measurements taken by faulty and functional sensors are identical.
In some cases, $P=1$ signal is sufficient for correct reconstruction, in which case \gamp can be used. 
However, using \calamp and increasing $P$ allows to close the gap to the performances of a perfectly calibrated algorithm.

\subsection{1-bit threshold calibration}

A last example of application of \calamp is $1$-bit \cs with unknown thresholds.
The basic setting is the same as the one presented in~\secref{sec:quantizedSensing}, with the difference that 
the thresholds $d_{\mu}$ are unknown:
\begin{align}
 y_{\mu} &= \sign\left( z_{\mu} + d_{\mu} + \xi_{\mu} \right),		\label{eq:1bitCal}
\end{align}
where the thresholds are distributed following a distribution $\pd$ and the noise $\xi_{\mu} \sim \NN(\xi_{\mu};0,\Delta)$.

\subsubsection{Uncalibrated \gamp}
A first approach is to ignore the thresholds, that is to run \gamp considering them to be zero.
Then the thresholds can be incorporated into the noise, with a higher variance than $\Delta$ (and only Gaussian if $\pd$ is).
Figure~\ref{fig:3_oneBit_continuous} shows the degradation of reconstruction performances with increasing noise: 
if the variance of $\pd$ is bigger than $\Delta$, we can expect reconstruction performances to be significantly degraded.

\subsubsection{Perfect calibration}
If the thresholds are known, \gamp can be used to make inference using~\eqref{eq:quantizedCSf}.
The reconstruction performance depends on the distribution $\pd$ and is best when all thresholds are zero,
which comes from the fact that the mean of $z_{\mu}$ is zero. 

\subsubsection{Experimental setting using \calamp}
\begin{figure}
 \begin{subfigure}{0.48\textwidth}
  \includegraphics[width=\textwidth]{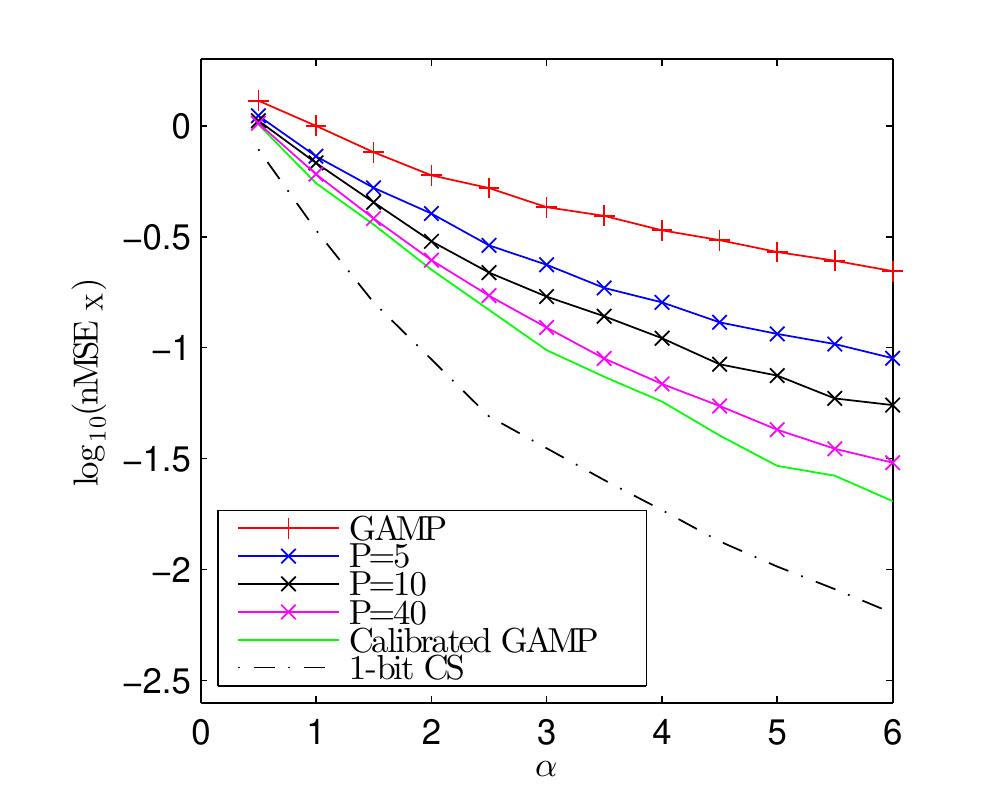}
  \caption{nMSE on the signal}
  \label{fig:6_1bitCal_signal}
 \end{subfigure}
 \begin{subfigure}{0.48\textwidth}
 \includegraphics[width=\textwidth]{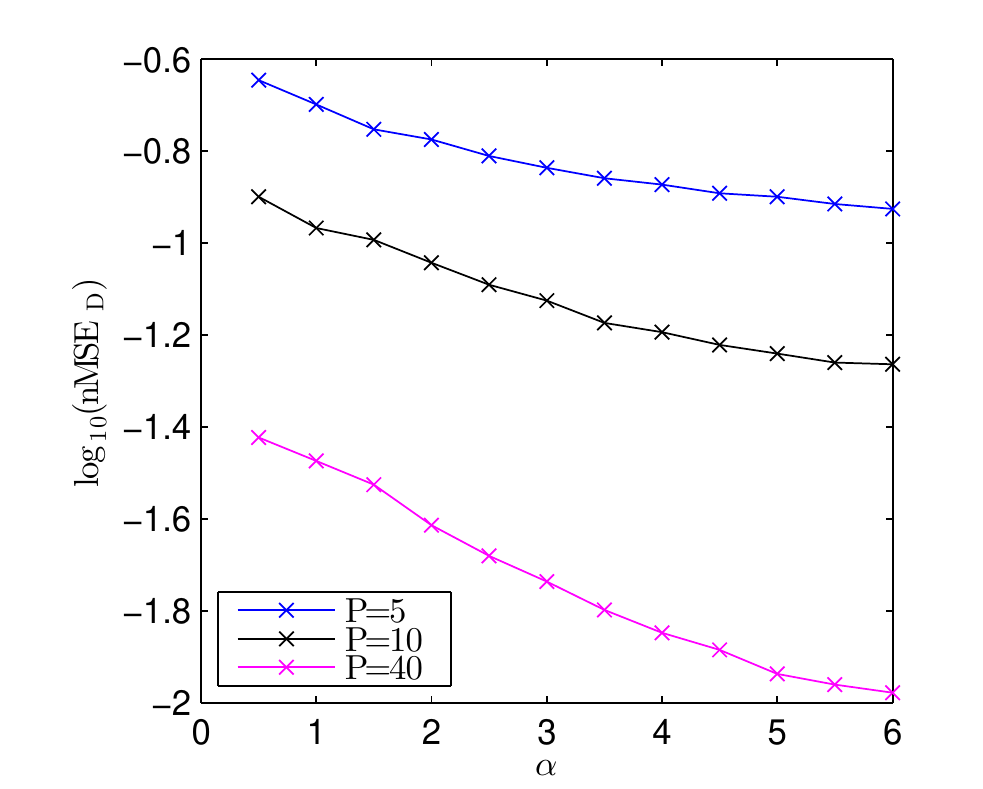}
 \caption{nMSE on the thresholds}
 \label{fig:6_1bitCal_thresholds}
\end{subfigure}
\caption[Bayes optimal $1$-bit threshold calibration]{Achieved normalized MSEs for Bayes optimal $1$-bit threshold calibration.
The thresholds are drawn from the discrete distribution~(\ref{eq:1bitCalDistribution}).
(a) nMSE on the signal for different settings. 
Additionally to \calamp with $P=5,10$ and $40$, we show usual \gamp ($P=1$, red) and calibrated \gamp (green).
Increasing $P$ allows to approach the performances of calibrated \gamp. The dashed ``$1$-bit \cs'' line corresponds 
to thresholds all equal to zero.
(b) nMSE on the thresholds for $P=5,10$ and $40$.
Experiments were made with $\rho=0.25$, $N=4000$.}
\label{fig:6_1bitCal}
\end{figure}

The update functions for \calamp are obtained using 
\begin{align}
 f_0^Y(y_{\mu l}|\Zh_{\mu l},\Zb_{\mu m}) &= \sqrt{\frac{\pi \z_{\mu l}}{2}} \erfc\left( y_{\mu l} \frac{d_{\mu} - \zh_{\mu l}}{\sqrt{2 \zb_{\mu l}}} \right), 
\end{align}
and~\eqref{eq:fkc}. As the integrals in~\eqref{eq:fkc} need to be numerically evaluated for most distributions $\pd$, we restrict our study to the simple case where
$\pd$ has multiple discrete values:
\begin{align}
 \pd(d) &= \sum_{i=1}^{n_d} p_i \delta(d - D_i), \label{eq:1bitCalDistribution}
\end{align}
for which the integrals reduce to finite sums.
Figure~\ref{fig:6_1bitCal} shows results using $n_d=20$, $p_i=1/n_d$ and equidistant $D_i$s in the interval $[-1 , 1]$.
As in the other blind calibration settings considered, \calamp allows to approach the performance of a perfectly calibrated algorithm
by increasing $P$.

\begin{figure}
 \begin{subfigure}{0.30\textwidth}
  \includegraphics[width=\textwidth]{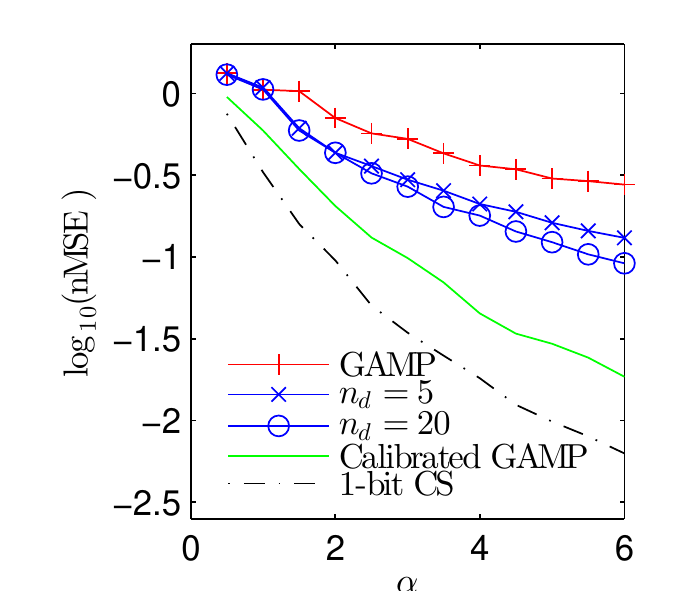}
  \caption{$P=5$}
  \label{fig:6_1bitCal_Uniform_5}
 \end{subfigure}
 \begin{subfigure}{0.30\textwidth}
 \includegraphics[width=\textwidth]{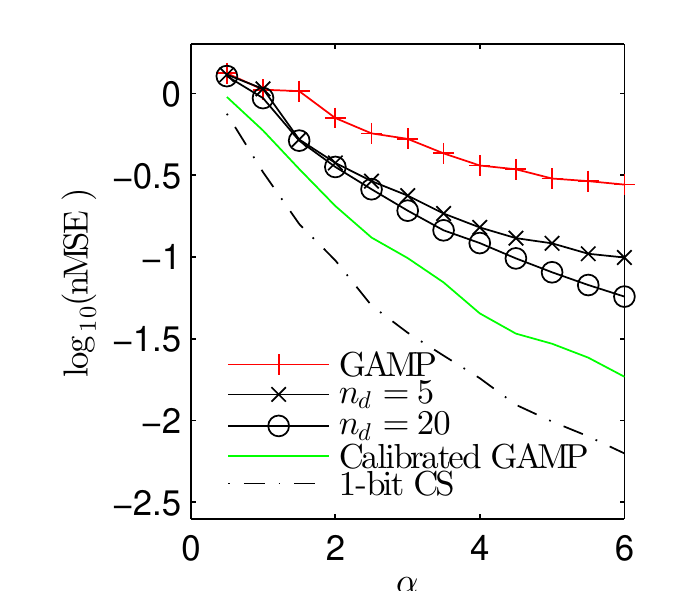}
 \caption{$P=10$}
 \label{fig:6_1bitCal_Uniform_10}
\end{subfigure}
 \begin{subfigure}{0.30\textwidth}
 \includegraphics[width=\textwidth]{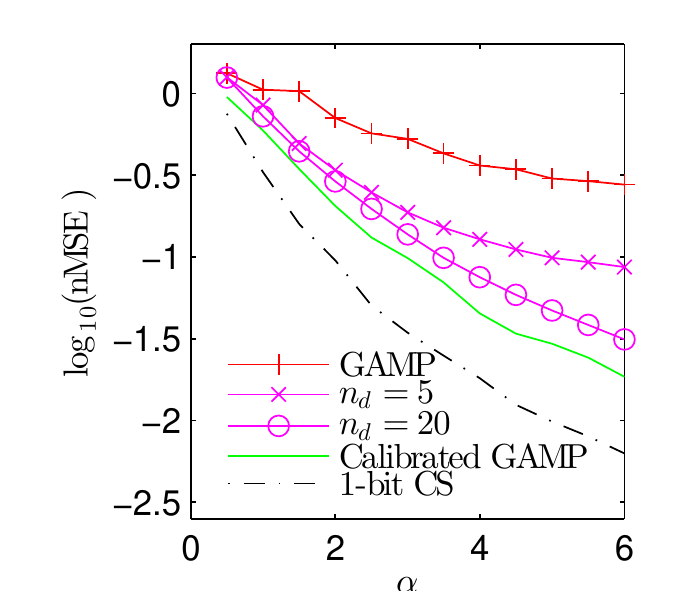}
 \caption{$P=40$}
 \label{fig:6_1bitCal_Uniform_40}
\end{subfigure}
\caption[$1$-bit threshold calibration with mismatching prior]{Achieved normalized MSEs on the signal for $1$-bit threshold calibration.
The thresholds are sampled from~(\ref{eq:uniform_thresholds}), and a mismatching prior is used 
for reconstruction using~(\ref{eq:1bitCalDistribution}).
Despite the gross error of taking a discrete prior for a continuous distribution, taking a high enough 
discretization $n_d$ of the interval allows approaching performances of a calibrated algorithm.
The plots are only separated for better legibility.
Experiments were made with $\rho=0.25$, $N=4000$.}
\label{fig:6_1bitCal_uniform}
\end{figure}

Figure~\ref{fig:6_1bitCal_uniform} shows results for a uniform distribution of thresholds
\begin{align}
 \pd^0(d) &= \UU(d,-1,1). \label{eq:uniform_thresholds}
\end{align}
Instead of numerically estimating the integrals necessary for the Bayes optimal update functions, we use 
the mismatching prior of~\eqref{eq:1bitCalDistribution} using $n_d=5$ and $n_d=20$, $p_i=1/n_d$ and equidistant $D_i$s in the interval $[-1 , 1]$.
Increasing $P$ and $n_d$ allows to approach the results of a perfectly calibrated algorithm.

\section{Encountered issues}
In addition to the results published in~\cite{moiNips, moiCalibration} and to the study of threshold calibration 
presented in this chapter, I have worked on applying \calamp to blind deconvolution and on deriving the state evolution of \calamp.
I have come across multiple issues in both of these tasks, which I describe in this section. 

\subsection{Blind deconvolution}
As shown in~\appliref{appli:blindDeconvolution}, the problem of blind deconvolution is very closely linked to the problem 
of complex blind gain calibration. 
Despite this fact, I could not successfully use \calamp for blind deconvolution.

The reason for this is the fact that if $\mathbf{k}$ is the convolution kernel, then
as formulated in~\eqref{eq:blindGainCal}, $\mathbf{d} \equiv \left( \mathcal{F}(\mathbf{k})\right)^{\inv}$, where $\mathcal{F}$ 
is the discrete Fourier transform. 
Although the coefficients of a typical convolution kernel $\mathbf{k}$  can be approximated by simple pdfs, 
the corresponding distributions for $\mathbf{d}$ do not have simple expressions. 

Furthermore, a typical convolution kernel has some coefficients close to zero, some other close to one, and its Fourier transform as well. 
This results in coefficients of $\mathbf{d}$ with a very big range of magnitudes. 
A prior taking this into account has such a big variance that blind calibration does not seem to work.
Another consequence of this big range of magnitudes is that the matrix $\tilde{\mathbf{D}} \Fv$ is ill-conditioned 
(as explained in~\exref{ex:linearMeasurements}), which is the source of additional problems.

Also, in a noisy setting, if $d_{\mu}$ has a very large amplitude, $y_{\mu}$ is basically noise, and these measurements carry no
information at all. The theoretically achievable bounds for blind deconvolution are therefore very different from those 
for blind complex gain calibration~\cite{BlindDeconvolutionIdentifiability}.

For these reasons, a more promising approach to blind deconvolution seems to be to treat it with a different factor graph in which the variables 
are the signals and the convolution kernel.
The associated $f_k^Y$ functions are similar to those for the blind gain calibration problem with $y_{\mu} = d_{\mu} z_{\mu}$: They 
do not have analytical expressions. A blind deconvolution algorithm similar to \calamp would require some more work to solve this issue.

\subsection{State evolution}
Just as for \gamp and \bigamp, it should be possible to describe the behaviour of \calamp with state evolution equations.
This would allow to predict the position of the phase transitions observed experimentally and to gain a better understanding of the specific settings studied.
I have used three different approaches:
\begin{itemize}
 \item Initially I have concentrated my efforts on the state evolution of the real gain calibration setting, starting from the algorithm (as done for complex
\cs in~\secref{sec:complexCsStateEvolution}). However, the statistical fluctuations of the quantities appearing in \calamp are more difficult 
to describe than those in \gamp for \cs. In the analysis done in~\secref{sec:complexCsStateEvolution}, the fluctuations can be described by a Gaussian distribution.
In real gain calibration, the fluctuations of the interesting quantities are more complex as they depend on a finite number $P$ of variables with different distributions.
Despite finding a few simplifications, I could not obtain a set of state evolution equations that matched the algorithm's behaviour. 
\item A second attempt was made using \textit{population dynamics}. Unlike state evolution, population dynamics does not give a simple closed set of equations 
that describes the algorithm's behaviour, but rather simulates its average behaviour. I could apply this approach successfully to \cs \gamp, but 
not to blind gain calibration.
\item Finally, I have used the replica method to derive the state evolution equations of the most general blind sensor calibration setting,
as in~\secref{sec:replicaGLM} for \gamp. Preliminary results are briefly presented in~\appref{app:calampStateEvolution}.
Just as the state evolution equations of \gamp~(\ref{eq:mh_DE_gen}--\ref{eq:Qh_DE_gen}) require integration over 2 variables, the state evolution equations for 
blind sensor calibration in general require integration over $2P$ variables. 
Finding an efficient and reliable way to perform these integrations numerically should in principle confirm the phase transitions observed empirically.
\end{itemize}
It can come as a surprise that the state evolution equations of generalized matrix factorization~(\ref{eq:bo_1}--\ref{eq:bo_3}) require integration over 2 variables only,
as the state evolution equations of \gamp, whereas those of \calamp require integration over $2P$ variables.
An explanation for this is that in generalized matrix factorization, we consider the limit $P\to \infty$, which allows to use the central limit theorem and replace $P$ integrals 
by a single one. On the other hand, in blind sensor calibration $P$ remains finite and therefore there is no general way of reducing the number of integrals.

\section{Conclusion}
In this chapter, the problem of blind gain calibration has been treated in a more general setting called blind \textit{sensor} calibration.
I derived a Bayesian message-passing algorithm called \calamp in~\cite{moiNips, moiCalibration}.
Experimental results of \calamp in several different settings of blind sensor calibration are presented in this chapter.
In the examples studied, \calamp converges very reliably to the solution, just as \gamp does for \cs: 
The convergence issues present in \bigamp for matrix factorization are not present in blind gain calibration.
The versatility of the blind sensor calibration setting could allow the use of \calamp in concrete applications such as astronomical imaging.

 \chapter{Analysis of matrix compressed sensing}
\label{chap:matrixCS}

In~\chapref{chap:generalizedBilinearModels}, I introduced matrix compressed sensing as a bilinear inference problem.
In this chapter, I perform the replica analysis of matrix compressed sensing in a probabilistic framework.
As for generalized linear models and matrix factorization, the analysis produces state evolution equations that
describe the asymptotic performance that can be reached in Bayesian inference of matrix compressed sensing.
These theoretical results are compared to the performance of
the recently introduced \pbig \cite{parker2015parametric} algorithm, that are in good agreement.
These results are presented in~\cite{moiMCS}. 
Furthermore, I analyse an instability of the Nishimori line in bilinear inference problems 
that explains the fragility of convergence in the \bigamp and \pbig algorithms.

Our analysis reveals a striking connection between the matrix
compressed sensing problem and the problem of matrix factorization as
studied in \cite{kabaMF}. These are two different
inference problems. In matrix compressed sensing we observe a set of
element-wise linear projections of the matrix, whereas in matrix
factorization we observe the elements of the matrix directly. Yet the
replica analysis of the two problems yields equivalent equations and
hence the asymptotic behaviour of the two problems, including the
phase transition, is closely linked. This analogy was already
remarked for the nuclear norm minimization in matrix compressed
sensing and matrix denoising in \cite{Donoho21052013}, or for matrix compressed sensing and matrix completion~\cite{Bolcskei}.

\section{Matrix compressed sensing}
\begin{figure}[h]
 \centering
 \includegraphics[width=1\textwidth]{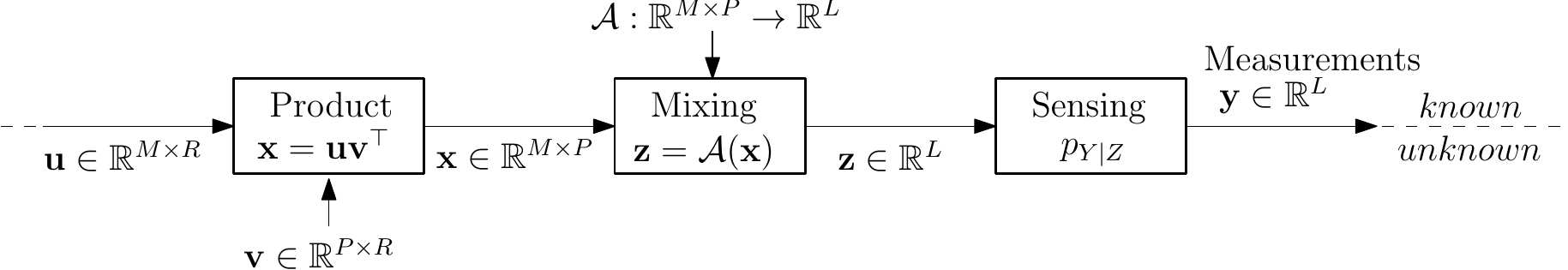}
 \caption[Generalized matrix compressed sensing]{Setting of generalized matrix compressed sensing (as in~\figref{fig:4_gmcs}).
 A low-rank matrix $\lr$ can be decomposed into a product of two smaller matrices $\uv$ and $\vv$.
 A linear operator $\A$ is applied to $\lr$, producing an intermediary variable $\zv$. 
 A measure $\yv$ of $\zv$ is obtained through a noisy channel. 
 The problem is closely linked to other inference problems: dropping the ``mixing'' block, one recovers 
 a generalized matrix factorization problem. Dropping the ``product'' block, one recovers a generalized linear model.}
 \label{fig:7_setting}
\end{figure}

We consider the setting described in~\secref{sec:mcs}, summarized by~\figref{fig:7_setting}.
Note that we replace the usual $N$ by $\rank$ in order to signify it is a rank.

\subsubsection{The probabilistic model and assumptions of our analysis.}
In order to enable the asymptotic analysis via the replica method we
introduce the following probabilistic model for matrix compressed
sensing.

\begin{itemize}
 \item We assume that elements of $\uv$ and $\vv$ are sampled
   independently at random such that 
 \begin{align}
\uv &\sim \prod_{\mu s} \p^0_U(u_{\mu s}), & \vv &\sim \prod_{ps} \p^0_V(v_{ps} ).    	\label{eq:priors0}
\end{align}
We assume the distributions $\p^0_U$ and $\p^0_V$ to have zero mean and respective variances $\Qzu$ and $\Qzv$ of order one.
 These distributions might not be known exactly: instead, we use zero-mean priors $\p_U$ and $\p_V$ believed to be close to $\p^0_U$ and $\p^0_V$.
 \item We assume the output distribution $\p^0_{Y|Z}$ to be separable: 
 \begin{align}
     \p^0_{Y|Z} = \prod_{l} p^0_{Y|Z}(y_l,z_l)\, .
\end{align}
In the inference we use a separable distribution $\py$ we believe to be close to it.
 \item We assume the matrix $\AM$ of the linear operator $\A$ to have normally distributed \iid elements 
 with zero mean and variance $1/(\rank \US \VS)$, such that the elements of $\zv$ have zero mean and variance one.
 This is the same assumption as is often made in compressed sensing, and differentiates the problem from matrix factorization, 
 in which $\A$ is the identity.
 \item We assume the dimensions $\US$, $\VS$ and $\YS$ to be large,
   but their following ratios to be of order one: 
 \begin{align}
   \alphau &= \frac{\YS}{\rank \US}, & \alphav &= \frac{\YS}{\rank \VS}, & \ppi &= \frac{M}{P}, & \alpha &= \frac{L}{\rank (M + P)}. \label{eq:alphas}
 \end{align}
On the other hand, $\rank$ can be small. $\alpha$ is a measurement ratio as in \cs: it is the ratio between the number of measurements and the number of unknowns.  

\end{itemize}

\subsubsection{Measures of recovery}
As in matrix factorization, there is an inherent ill-posedness when it comes to recovering the couple $(\uv,\vv)$. 
As a matter of fact, for any $\rank \times \rank$ invertible matrix $\mathbf{C}$, the couple $(\uv \mathbf{C} , \vv \left(\mathbf{C}^{-1} \right)^{\top})$ generates the same $\lr$ as $(\uv,\vv)$.
In some case, this ill-posedness can be lifted thanks to the distributions $p^0_U$ and $p^0_V$, but this is not always the case and might nevertheless be cause of trouble.
In that case, it is possible to have a very low $\msex$ but high $\mseu$ and $\msev$. 

A way to remove one degree of invariance is to consider the normalized mean squared errors
\begin{align}
 {\rm nMSE_U} &=  \frac{1}{2} \left\| \frac{\uv}{\| \uv \|_2} - \frac{\uhv}{\| \uhv \|_2} \right\|_2^2, & {\rm nMSE_V} &=  \frac{1}{2} \left\| \frac{\vv}{\| \vv \|_2} - \frac{\vhv}{\| \vhv \|_2} \right\|_2^2, \label{eq:nmses}
\end{align}
which absorbs the scaling invariance by positive multiplicative scalars, as in~\eqref{eq:nmse}.

\subsection{Notations}
If $\A$ is a linear operator and $\AM$ its matrix, we write $\A^2$ for the linear operator associated to $\AM^2$.
Using the matrix $\AM$, we can define two auxiliary linear operators $\Au: \RR^{\VS} \to \RR^{\YS \times \US}$ and 
$\Av: \RR^{\US} \to \RR^{\YS \times \VS}$ such that
\begin{align}
 [\Au(\vv)]_{\yil \ui} &\equiv \sum_{\vi} A_{\yil}^{\ui \vi} v_{\vi},   \label{eq:Au} \\
 [\Av(\uv)]_{\yil \vi} &\equiv \sum_{\ui} A_{\yil}^{\ui \vi} u_{\ui}.	\label{eq:Av}
\end{align}

\subsection{Message-passing algorithm}
\begin{figure}
 \centering
 \includegraphics[width=0.7\columnwidth]{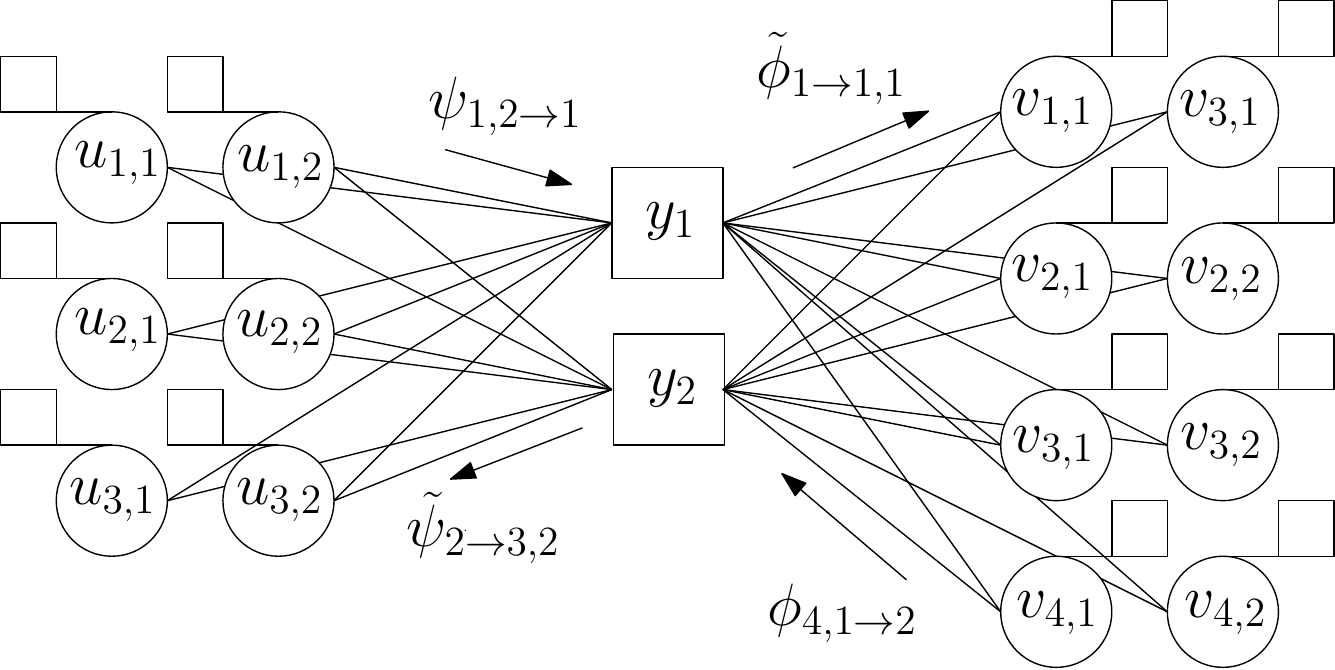}
 \caption[Factor graph of matrix compressed sensing]{Factor graph associated to the probability distribution~(\ref{eq:proba_expanded}). The two types of
 messages are strictly equivalent. Here, we used $\rank=2$, $\US=3$, $\VS=4$, $\YS=2$. Circle represent variables, squares represent constraints.
 The small squares represent the priors on the variables $\uv$ and $\vv$. Messages $(\messv, \messf)$ or $(\messvd, \messfd)$ are sent along each edge of the factor graph.}
 \label{fig:7_fg}
\end{figure}

As done in~\chapref{chap:gampApplications} for generalized linear models and in~\chapref{chap:blindSensorCal}, we
derive a Bayesian inference algorithm using belief propagation, starting from the posterior probability
\begin{align}
 \p(\uv,\vv|\yv,\A) \propto \prod_{\ui \ri} &\p_U(u_{\ui \ri})  \prod_{\vi \ri} \p_V(v_{\vi \ri})  \label{eq:proba_expanded}  \\
 &\int \prod_{\yil} \dd z_{\yil} \py(y_{\yil}|z_{\yil}) \delta \left(z_{\yil} - \sum_{\vi=1}^{\VS} \sum_{\ui=1}^{\US} A^{\vi \ui}_{\yil} \sum_{\ri=1}^{\rank} u_{ \vi \ri} v_{\ui \ri} \right),  \nonumber 
\end{align}
that is represented by the factor graph on~\figref{fig:7_fg}.
As the derivation is very similar to the one done in~\chapref{chap:gampApplications}, only the main steps are explained here. 
The full derivation can be found in~\cite{moiMCS}.

\subsubsection{Step 1: \bp}
As in blind sensor calibration, there are two types of message pairs $(\messv,\messf)$ and $(\messvd,\messfd)$, but here their role is completely symmetric.
Therefore, we will only treat explicitly the pair $(\messv,\messf)$: the result can be generalized straightforwardly to $(\messvd,\messfd)$.
The message-passing update equations read:
\begin{align}
 \messv_{\ui \ri \to \yil}^{t}(u_{\ui \ri}) &\propto \p_U(u_{\ui \ri}) \prod_{\yibis \neq \yil} \messf_{\yibis \to \ui \ri}^{t}(u_{\ui \ri})   \label{eq:mess_u} , \\
 \messf_{\yil \to \ui \ri}^{t+1}(u_{\ui \ri}) &\propto \int \left( \prod_{\vi \ribis} \dd v_{\vi \ribis} \messvd_{\vi \ribis \to \yil}^{t+1}(v_{\vi \ribis}) \prod_{(\ribis, \uibis) \neq (\ri, \ui)} \dd u_{\ui \ribis} \messv_{\ui \ribis \to \yil}^{t+1}(u_{\ui \ribis}) \right) \nonumber \\
 &\qquad \dd z \py(y_{\yil}|z) \delta(z - \A(\uv \vv^{\top})) .  \label{eq:mess_u_h}
\end{align}

\subsubsection{Step 2: \amp}
A first simplification can be made by replacing the $\rank(\US+\VS)$ integrals in~(\ref{eq:mess_u_h}) 
by a single one over the variable  $z$, which is the sum of $\rank(\US+\VS) -1$ random variables. 
We call $\hat{u}_{\ui \ri \to \yil}$ and $\bar{u}_{\ui \ri \to \yil}$ respectively the mean and variance of
the variable $u_{\ui \ri}$ distributed according to the distribution $\messv_{\ui \ri \to \yil}$ (and similarly 
for the variables $v_{\vi \ri}$). 
Note that the product of two independent random variables $u$ and $v$ has mean $\uh \vh$ and variance $\ub \vb + \uh^2 \vb + \ub \vh^2$.
By the central limit theorem, the variable $z_{\yil} = \sum_{\ui \vi} A_{\yil}^{\ui \vi} \sum_{\ri} u_{\ui \ri} v_{\vi \ri}$ is a Gaussian variable,
and its mean and variance are:
\begin{align}
 \hat{Z}_{\yil} &= \sum_{\ui \vi \ri} A_{\yil}^{\ui \vi} \hat{u}_{\ui \ri \to \yil} \hat{v}_{\vi \ri \to \yil} ,  \label{eq:zh} \\
 \bar{Z}_{\yil} &= \sum_{\ui \vi \ri} (A_{\yil}^{\ui \vi})^2 \left[  \bar{u}_{\ui \ri \to \yil} \bar{v}_{\vi \ri \to \yil} + (\hat{u}_{\ui \ri \to \yil})^2 \bar{v}_{\vi \ri \to \yil}  + \bar{u}_{\ui \ri \to \yil}(\hat{v}_{\vi \ri \to \yil})^2 \right] \nonumber   \\
		&+\sum_{\vi \ri} \sum_{\ui \neq \uibis} A_{\yil}^{\ui \vi} A_{\yil}^{\uibis \vi} \bar{v}_{\vi \ri \to \yil} \hat{u}_{\ui \ri \to \yil} \hat{u}_{\ui \ribis \to \yil}	    \nonumber \\
		&+\sum_{\ui \ri} \sum_{\vi \neq \vibis} A_{\yil}^{\ui \vi} A_{\yil}^{\ui \vibis} \bar{u}_{\ui \ri \to \yil} \hat{v}_{\vi \ri \to \yil} \hat{v}_{\vi \ribis \to \yil}	.      \label{eq:zb}
\end{align}
However, in eq.~(\ref{eq:mess_u_h}), $u_{\ui \ri}$ is fixed and thus $(\hat{u}_{\ui \ri \to \yil}, \bar{u}_{\ui \ri \to \yil})$ has to be replaced by $(u_{\ui \ri},0)$
in~(\ref{eq:zh},\ref{eq:zb}). 
Defining $(\hat{Z}_{\yil \to \ui \ri}, \bar{Z}_{\yil \to \ui \ri})$ to be $(\hat{Z}_{\yil},\bar{Z}_{\yil})$ with $(\hat{u}_{\ui \ri \to \yil}, \bar{u}_{\ui \ri \to \yil}) = (0,0)$ and
\begin{align}
 F_{\yil \ui \ri} &= \sum_{\vi} A_{\yil}^{\ui \vi} \hat{v}_{\vi \ri \to \yil} , \\
 H_{\yil \ui \ri} &=2 \sum_{\vi} \sum_{\uibis \neq \ui} A_{\yil}^{\ui \vi} A_{\yil}^{\uibis \vi} \hat{u}_{\ui \ribis \to \yil} \bar{v}_{\vi \ri \to \yil} ,\\
 G_{\yil \ui \ri} &= \sum_{\vi} (A_{\yil}^{\ui \vi})^2 \bar{v}_{\vi \ri \to \yil}  ,  
\end{align}
one can rewrite~(\ref{eq:mess_u_h}) with a single integral over a variable $z$ 
with a Gaussian distribution. 
Using the $f$-functions (see \appref{app:f}), the message~(\ref{eq:mess_u_h}) can be expressed as a simple function of the mean and variance of this Gaussian:
\begin{align}
 \messf_{\yil \to \ui \ri}(u_{\ui \ri}) \propto f_0^Y \left(y_{\yil} | \hat{Z}_{\yil \to \ui \ri} + F_{\yil \ui \ri} u_{\ui \ri} , \bar{Z}_{\yil \to \ui \ri} + H_{\yil \ui \ri} u_{\ui \ri} + G_{\yil \ui \ri} u_{\ui \ri}^2 \right) ,  \label{eq:messuhNice}
\end{align}
where we use the simplified notation $f_i^{Y} \equiv f_i^{\py}$.
Making a Taylor expansion of this equation, 
we can express the message~(\ref{eq:mess_u}) as
\begin{align}
 \messv_{\ui \ri \to \yil}(u_{\ui \ri}) &\propto p(u_{\ui \ri}) \NN\left( \hat{U}_{\ui \ri \to \yil}, \bar{U}_{\ui \ri \to \yil} \right),  \label{eq:messuNice}
\end{align}
with
\begin{align}
\bar{U}_{\ui \ri \to \yil} &=   -\left( \sum_{\yibis \neq \yil}  \left( F_{\yibis \ui \ri}^2 + G_{\yibis \ui \ri} \right) \bar{g}_{\yibis \to \ui \ri} + G_{\yibis \ui \ri} \hat{g}_{\yibis \to \ui \ri}^2  \right)^{-1},  \label{eq:Ub_final}  \\
\hat{U}_{\ui \ri \to \yil} &=  \bar{U}_{\ui \ri \to \yil} \sum_{\yibis \neq \yil} F_{\yibis \ui \ri}  \hat{g}_{\yibis \to \ui \ri}, \label{eq:Uh_final}					
\end{align}
where
\begin{align}
 \hat{g}_{\yibis \to \ui \ri} &= \gh^Y(y_{\yibis}| \hat{Z}_{\yibis \to \ui \ri}, \bar{Z}_{\yibis \to \ui \ri}),  &  \bar{g}_{\yibis \to \ui \ri} &= \gb^Y(y_{\yibis} | \hat{Z}_{\yibis \to \ui \ri}, \bar{Z}_{\yibis \to \ui \ri}),
\end{align}
and $(\hat{g}^Y(\cdot,\cdot),\bar{g}^Y(\cdot,\cdot))$ are simplified notations for the functions  $(\hat{g}^{\py}(\cdot,\cdot),\bar{g}^{\py}(\cdot,\cdot))$  defined in~\appref{app:f}.

This allows us to have a simple expression for the previously introduced mean and variance $\hat{u}_{\ui \ri \to \yil}$ and $\bar{u}_{\ui \ri \to \yil}$ of the message~(\ref{eq:messuNice}).
Using the notations~(\ref{eq:fh}, \ref{eq:fb}), 
\begin{align}
 \hat{u}_{\ui \ri \to \yil} &= \fh^U\left( \hat{U}_{\ui \ri \to \yil}, \bar{U}_{\ui \ri \to \yil} \right), & \bar{u}_{\ui \ri \to \yil} &= \fb^U\left(\hat{U}_{\ui \ri \to \yil}, \bar{U}_{\ui \ri \to \yil} \right),
\end{align}
where as before, we introduce the simplifying notation $f^U \equiv f^{p_U}$.
The exact same thing can be done for the messages $(\messvd,\messfd)$.
The result is a set of iterative equations on a set of means and variances
\begin{align}
 \left( \hat{Z}_{\cdot \to \cdot}^t,\bar{Z}_{\cdot \to \cdot}^t,\hat{g}_{\cdot \to \cdot}^t,\bar{g}_{\cdot \to \cdot}^t,\hat{U}_{\cdot \to \cdot}^t,\bar{U}_{\cdot \to \cdot}^t,\hat{u}_{\cdot \to \cdot}^t,\bar{u}_{\cdot \to \cdot}^t,\hat{V}_{\cdot \to \cdot}^t,\bar{V}_{\cdot \to \cdot}^t,\hat{v}_{\cdot \to \cdot}^t,\bar{v}_{\cdot \to \cdot}^t     \right)   \label{eq:all_bp_messages}
\end{align}
that constitutes the message-passing algorithm.

\subsubsection{Step 3: \tap}
This algorithm can be further simplified using the so-called Thouless-Andersen-Palmer (TAP) approximation introduced in the study of spin glasses~\cite{tap}.
The resulting algorithm~\ref{algo:pbigamp} was introduced as \pbig in~\cite{parker2015parametric}.

\subsubsection{Convergence}
As its counterparts for generalized linear models (\algoref{algo:gamp}) or generalized matrix factorization~(\algoref{algo:bigamp}), \algoref{algo:pbigamp} 
needs some adaptations that improve its convergence. 
One very simple empirical damping scheme that allows to improve convergence (though not guaranteeing it) consists in damping a single variable:
\begin{align}
 \Uhv_{t+1} \leftarrow \beta \Uhv_{t+1} + (1-\beta) \Uhv_{t},	\label{eq:simpleDamping_mcs}
\end{align}
with $\beta=0.3$, applied right after the calculation of $\Uhv_{t+1}$.
Furthermore, setting $\vbv=0$ and $\ubv=0$ in the update equations for $\Ubv$ and $\Vbv$ greatly reduces the risk of negative variances appearing, 
while emulating the Nishimori conditions~\cite{kabaMF}.
A more involved, adaptive damping strategy is presented in~\cite{vilaAdaptive}.
Notice that we defined the operators $\Au$ and $\Av$ used in algorithm~\ref{algo:pbigamp} as linear applications $\Au: \RR^{\VS} \to \RR^{\YS \times \US}$ and 
$\Av: \RR^{\US} \to \RR^{\YS \times \VS}$ in~(\ref{eq:Au},\ref{eq:Av}): In the algorithm, we apply them row-wise on the matrices they act on.

\begin{algorithm}
\caption{\pbig for matrix compressed sensing}
\label{algo:pbigamp}
  \textbf{Initialization:}  \\
  Initialize the means $(\hat{\uv}_0,\hat{\vv}_0)$ and the variances $(\bar{\uv}_0,\bar{\vv}_0)$ at random according to the distributions $\p_U^0$ and $\p_V^0$, and $\ghv_0=0$.

  \textbf{Main loop:} while $t<t_{\rm max}$, calculate following quantities:
  \begin{align*}
  \bar{\xv}_{t+1} &= \ubv_t \vbv_t^{\top} + \ubv_t ( \vhv_t^2 )^{\top} + \uhv_t^2 \vbv_t^{\top} \\
  \hat{\xv}_{t+1} &= \uhv_t \vhv_t^{\top} \\
  \bar{\zv}_{t+1} &= \A^2(\xbv_{t+1}) \\
  \hat{\zv}_{t+1} &= \A(\xhv_{t+1}) - \ghv_t \odot \left( \ubv_t \left( \Au(\vhv_t) \odot \Au(\vhv_{t-1}) \right)^{\top} +  \left( \Av(\uhv_t) \odot \Av(\uhv_{t-1}) \right) \vbv_t^{\top} \right) \\
  \gbv_{t+1} &= \gb^Y(\zhv_{t+1}, \zbv_{t+1}) \\
  \ghv_{t+1} &= \gh^Y(\zhv_{t+1}, \zbv_{t+1}) \\
  \Ubv_{t+1} &= - \left( \left[ \Au(\vhv_t)^2 + \Au^2(\vbv_t) \right] \gbv_{t+1} + \Au^2(\vbv_t) \ghv_{t+1}^2 \right)^{-1} \\
  \Uhv_{t+1} &= \Ubv_{t+1} \odot \left( \Au(\vhv_t) \ghv_{t+1} - \uhv_t \odot \Au(\vhv_t)^2 \gbv_{t+1} - \uhv_{t-1} \odot \Au^2(\vbv_{t-1}) \ghv_{t+1} \odot \ghv_t \right)  \\
  \ubv_{t+1} &= \fb^U(\Uhv_{t+1}, \Ubv_{t+1}) \\
  \uhv_{t+1} &= \fh^U(\Uhv_{t+1}, \Ubv_{t+1})  \\
  \Vbv_{t+1} &= - \left( \left[ \Av(\uhv_t)^2 + \Av^2(\ubv_t) \right] \gbv_{t+1} + \Av^2(\ubv_t) \ghv_{t+1}^2 \right)^{-1} \\
  \Vhv_{t+1} &= \Vbv_{t+1} \odot \left( \Av(\uhv_t) \ghv_{t+1} - \vhv_t \odot \Av(\uhv_t)^2 \gbv_{t+1} - \vhv_{t-1} \odot \Av^2(\ubv_{t-1}) \ghv_{t+1} \odot \ghv_t \right)  \\
  \vbv_{t+1} &= \fb^V(\Vhv_{t+1}, \Vbv_{t+1}) \\
  \vhv_{t+1} &= \fh^V(\Vhv_{t+1}, \Vbv_{t+1})   
  \end{align*}
  \textbf{Result : } $(\uhv,\vhv,\hat{\xv}, \hat{\zv})$ are the estimates for $(\uv,\vv,\xv,\zv)$ and $(\bar{\uv},\bar{\vv},\bar{\xv},\bar{\zv})$ 
  are uncertainties on these estimates.
\end{algorithm}

\section{Asymptotic analysis}
The problem of low-rank matrix compressed sensing can be analysed with statistical physics methods in the thermodynamic limit,
\ie when the dimensions of the signals
 $\US$ and $\VS$ and of the measurements $\YS$ go to infinity. $\rank$ can remain finite or go to infinity as well.
 On the other hand, the ratios defined in~(\ref{eq:alphas}) have to be fixed and finite. 
 As the analysis is very similar to the one in~\secref{sec:replicaGLM}, only the main steps are presented here.
 The complete derivation can be found in~\cite{moiMCS}.

\subsection{Replica analysis: free entropy}
The relevant partition function is:
\begin{equation}
 \ZZ(\yv,\AM) = \int \dd \uv\, \p_U(\uv) \int \dd \vv \, \p_V(\vv) \int \dd \zv \py\left(\yv | \zv \right) \delta \left[ \zv - \A(\uv \vv^{\top}) \right]. \label{eq:partition_mcs}
\end{equation}
Let us start by calculating
\begin{align}
\ZZ(\yv,\AM)^n = \int \prod_{a=1}^n \left\{ \dd \uv^a \, \p_U(\uv^a)  \dd \vv^a \, \p_V(\vv^a)  \dd \zv^a \py\left(\yv|\zv^a \right) \delta\left[ \zv^a - \A(\uv^a (\vv^a)^{\top}) \right] \right\} \label{eq:partitionReplicated_mcs} 
\end{align}
and its average with respect to the realizations of $\yv$, generated by $\uv^0$, $\vv^0$ and $\A$:
\begin{align}
 \langle \ZZ^n \rangle  = \int &\dd \uv^0 \, \p_U^0(\uv^0) \dd \vv^0 \, \p_V^0(\vv^0) \dd \mathbf{\AM} \, \p_{A}^0( \mathbf{\AM}) \dd \yv  \nonumber \\
    &\dd \zv^0 \py (\yv|\zv^0) \delta\left[ \zv^0 - \A(\uv^0 (\vv^0)^{\top}) \right] \ZZ(\yv, \AM)^n . \label{eq:partitionReplicatedAveraged_mcs}
\end{align}

We treat $z_{\yil}^a = [\A(\uv^a (\vv^a)^{\top})]_{\yil} $ as a random variable of $\A$ and look at the covariance between two of those variables:
\begin{align}
 \langle z_{\yil}^a z_{\yibis}^b \rangle &= \langle \left( \sum_{\ui \vi} A_{\yil}^{\ui \vi} \sum_{\ri} u_{\ui \ri}^a v_{\vi \ri}^a \right) \left( \sum_{\uibis \vibis} A_{\yibis}^{\uibis \vibis} \sum_{\ribis} u_{\uibis \ribis}^b v_{\vibis \ribis}^b \right) \rangle \nonumber \\
 &= \langle \sum_{\ui \uibis} \sum_{\vi \vibis} A_{\yil}^{\ui \vi} A_{\yibis}^{\uibis \vibis} \sum_{\ri \ribis} u_{\ui \ri}^a u_{\uibis \ribis}^b v_{\vi \ri}^a v_{\vibis \ribis}^b \rangle   \nonumber \\
 &= \sum_{\ui \uibis} \sum_{\vi \vibis} \langle A_{\yil}^{\ui \vi} A_{\yibis}^{\uibis \vibis} \rangle  \sum_{\ri \ribis} u_{\ui \ri}^a u_{\uibis \ribis}^b v_{\vi \ri}^a v_{\vibis \ribis}^b .
\end{align}
As the elements of $\AM$ are \iid with zero mean and variance $1/(\rank \US \VS)$, 
we have  
\begin{align}
 \langle A_{\yil}^{\ui \vi} A_{\yibis}^{\uibis \vibis} \rangle = \delta_{\yil, \yibis} \delta_{\ui, \uibis} \delta_{\vi, \vibis} \frac{1}{\rank \US \VS},
\end{align}
and thus
\begin{align}
  \langle z_{\yil}^a z_{\yibis}^b \rangle &= \delta_{\yil, \yibis} \frac{1}{\rank \US \VS} \sum_{\ri \ribis} \left( \left( \sum_{\ui} u_{\ui \ri}^a u_{\ui \ribis}^b \right) \left( \sum_{\vi} v_{\vi \ri}^a v_{\vi \ribis}^b \right) \right) \nonumber  \\
					&= \frac{\delta_{\yil, \yibis}}{\rank} \sum_{\ri \ribis} \left(  \left( \frac{1}{\US} \sum_{\ui} u_{\ui \ri}^a u_{\ui \ribis}^b \right) \left( \frac{1}{\VS} \sum_{\vi} v_{\vi \ri}^a v_{\vi \ribis}^b \right) \right).
\end{align}
We now make the following assumption:
\begin{align}
 \frac{1}{\US} \sum_{\ui} u_{\ui \ri}^a u_{\ui \ribis}^b &= \begin{cases}
                                                \Qu^{ab} = O(1) & {\rm if } \,  \ri=\ribis \\
                                                (\Qu^{ab})_{\ri \ribis} = O(\frac{1}{\sqrt{\US}}) & {\rm if } \, \ri \neq \ribis
                                               \end{cases}		\label{eq:QHyp}
\end{align}
This assumption corresponds to breaking the column-permutational symmetry and more generally the rotational symmetry between
different replicas. We thus assume that the $\ri$-th column of $\uv^a$ is correlated to the $\ri$-th column of $\uv^b$ and to none of the others.
We make the same assumption for $\vv$. Then,
\begin{align}
  \langle z_{\yil}^a z_{\yibis}^b \rangle &= \frac{\delta_{\yil, \yibis}}{\rank} \left(\sum_{\ri} \Qu^{ab} \Qv^{ab} + \sum_{\ri \neq \ribis} (\Qu^{ab})_{\ri \ribis} (\Qv^{ab})_{\ri \ribis} \right).
\end{align}
Due to the hypothesis~(\ref{eq:QHyp}), the second term vanishes, and
\begin{align}
   \langle z_{\yil}^a z_{\yibis}^b \rangle &= \delta_{\ri, \ribis} \Qu^{ab} \Qv^{ab}.
\end{align}
Note that by definition of $\Qu^{ab}$ in~(\ref{eq:QHyp}), $\Qu^{ab}=\Qu^{ba}$. 
$\za_{\yil} \equiv (z_{\yil}^0 \cdots z_{\yil}^n)$ is thus a multivariate Gaussian random  variable with mean $0$ and covariance matrix $\Qmz \equiv \Qmu \odot \Qmv$, where the elements of the matrices $\Qmu$ and $\Qmv$ are given by:
\begin{align}
 \Qu^{ab} &\equiv \frac{1}{\US} \sum_{\ui} u_{\ui}^a u_{\ui}^b  ,  & \Qv^{ab} &\equiv \frac{1}{\VS} \sum_{\vi} v_{\vi}^a v_{\vi}^b . \label{eq:defQ}
\end{align}
With this, and introducing the conjugate variables $\Qmhu$ and $\Qmhv$ we obtain
\begin{align}
  \langle \ZZ^n \rangle  &= \int \dd \Qmu \dd \Qmhu e^{- \frac{\US \rank}{2} \Tr(\Qmu \Qmhu)} \left[ \dd \ua \pu(\ua) e^{\frac{1}{2} \ua^{\top} \Qmhu \ua} \right]^{\rank \US}  \nonumber \\
  &\int \dd \Qmv \dd \Qhv e^{-\frac{\VS \rank}{2} \Tr(\Qmv \Qmhv)} \left[  \dd \va \pv(\va) e^{\frac{1}{2} \va^{\top} \Qmhv \va} \right]^{\rank \VS} \nonumber \\
  & \left[ \int \dd \za \NN(\za; 0,\Qmz) \int \dd y \py(y|\za) \right]^{\YS}.
\end{align}
We take $\US$, $\VS$ and $\YS$ going to infinity with constant ratios, and rewrite
\begin{align}
  \langle \ZZ^n \rangle  &= \int \dd \Qmu \Qmhu \Qmv \Qmhv e^{-\US \rank \left[ \ac_n(\Qmu,\Qmhu,\Qmv,\Qmhv) \right] }
\end{align}
and to use the saddle point method, according to which
\begin{align}
  \log \left( \langle \ZZ^n \rangle  \right) &= -\US \rank \min_{\Qmu,\Qmhu,\Qmv,\Qmhv} \ac_n(\Qmu,\Qmhu,\Qmv,\Qmhv) + O(1).  \label{eq:saddlePoint_mcs}
\end{align}
We are therefore left with a minimization problem over the space of the matrices $\Qmu,\Qmhu,\Qmv$ and $\Qmhv$, representing $ 2 (n+1) (n+2)$ parameters (as the matrices are symmetric).

\subsection{Replica symmetric assumption}
With the replica symmetric hypothesis, 
the extremization is only over 12 variables: $(\mymu, \mhu, \qu, \qhu, \Qu , \Qhu)$ and $(\mv, \mhv,\qv, \qhv, \Qv , \Qhv)$ .
The function $\ac_n$ to extremize is:
\begin{align}
 \ac_n(\Qmu,\Qmv,\Qmhu,\Qmhv) \equiv &\left[ \frac{1}{2} \Tr \Qmu \Qmhu - \log \left( \int \dd \ua \pu(\ua) e^{\frac{1}{2} \ua^{\top} \Qmhu \ua} \right) \right] \nonumber  \\
		    + \frac{\US}{\VS} &\left[ \frac{1}{2} \Tr \Qmv \Qmhv  - \log \left( \int \dd \va \pu(\va) e^{\frac{1}{2} \va^{\top} \Qmhv \va} \right) \right] \nonumber \\
		    - \frac{\YS}{\rank \VS} &\log \left( \int \dd \za \NN(\za;0,\Qmz) \int \dd y  \py(y|\za) \right)  . 		\label{eq:action_mcs}
\end{align}
Taking its derivative with respect to $n$ and to $n \to 0$ limit, we obtain the free entropy $\phi$ as an extremum
\begin{align}
 \phi =  {\rm SP}  & \left\{ \frac{1}{1 + \ppi} \left( -\mymu \mhu - \frac{1}{2} \Qu \Qhu + \frac{1}{2} \qu \qhu +\iu(\Qmhu) \right)  \right. \nonumber  \\
	      +  \frac{\ppi}{1 + \ppi} & \left.\left( -\mv \mhv - \frac{1}{2} \Qv \Qhv + \frac{1}{2} \qv \qhv  +\iv(\Qmhv) \right) + \alpha \iz(\Qmu \odot \Qmv)	\right\} 	\label{eq:phi_mcs}
\end{align}
over a set of $12$ variables. Note that the shift from a minimum in~(\ref{eq:saddlePoint_mcs}) to an extremum in the equation 
above is a consequence to the hazardous $n\to 0$ limit in the replica method.
The functions $\iu$ and $\iv$ are the equivalent of $\ix$ in~\eqref{eq:ix}, the function $\iz$ is the same as in~\eqref{eq:iz}. 


\subsubsection{Equivalence to generalized matrix factorization}

It is interesting to notice that if $\YS = \US \VS$ and $\rank = O(\US)$, this free entropy is the same as in generalized matrix factorization~(\ref{eq:phiMF})~\cite{kabaMF}.
This is not an entirely obvious fact, as the two problems are different and that they are identical only if $\A$ is the identity: in generalized matrix factorization,
\begin{align}
 \zv = \xv.
\end{align}
In order to perform the theoretical analysis of generalized matrix factorization as in~\cite{kabaMF}, it is important to take the limit $\rank \to \infty$. 
In fact, it is this limit that ensures that each entry of $\zv$ is the sum of a large number of random variables, which allows to consider that it has a Gaussian distribution.
This is a condition both in the derivation of the message-passing algorithm and in the replica analysis.
For that reason, generalized matrix factorization with finite $\rank$ leads to different algorithms and theoretical bounds~\cite{tanaka,tibo1}.
However, in matrix compressed sensing, the mixing of coefficients with $\A$ ensures that even if $\rank=1$, each element of $\zv$ can be considered to have 
a Gaussian distribution. Thanks to this, both the algorithm and the analysis are the same, independently of $\rank$.

Let us examine the case in which $\YS = \US \VS$ and $\rank = O(\US)$ and the two problems are strictly equivalent.
What differentiates the generalized matrix compressed sensing from the generalized matrix factorization case is that $\A$ is not the identity.
However, as $\A$'s coefficients are Gaussian \iid, it is with high probability a bijection when $\YS=\US \VS$, and in this 
sense the mixing step does not introduce any further difficulty into the problem compared to matrix factorization.
If $\YS > \US \VS$, matrix compressed sensing is not ``compressive'' and therefore
easier than the corresponding matrix factorization problem, because more 
measures are available. 
If $\YS < \US \VS$, matrix compressed sensing is ``compressive'' and equivalent to 
low rank matrix completion, \ie to the matrix factorization setting in which only a fraction of the matrix entries is observed.
Note that the two problems are treated jointly in~\cite{Bolcskei}.
 
\subsection{State evolution equations}
As done in~\secref{sec:stateEvol}, we can obtain state evolution equations from the free entropy by setting its derivatives to zero.
We obtain the set of equations
\begin{align}
 \mymu &= \sqrt{\frac{\qhu }{\mhu ^2}} \int \dd t \, f_1^{U,0} \left(\frac{\sqrt{\qhu }}{\mhu }t , \frac{\qhu }{\mhu^2} \right)\fh^U \left( \frac{ \sqrt{\qhu } t }{\qhu  - \Qhu } , \frac{1}{\qhu  - \Qhu }  \right) ,   \label{eq:mu_DE_gen_mcs} \\
 \Qu-\qu &= \sqrt{\frac{\qhu }{\mhu ^2}} \int \dd t \,  f_0^{U,0} \left(\frac{\sqrt{\qhu }}{\mhu }t , \frac{\qhu }{\mhu^2} \right)  \fb^U \left( \frac{ \sqrt{\qhu } t }{\qhu - \Qhu} , \frac{1}{\qhu - \Qhu} \right) ,  \label{eq:Qu_DE_gen_mcs}  \\
 \qu &= \sqrt{\frac{\qhu}{\mhu^2}} \int \dd t \,  f_0^{U,0} \left(\frac{\sqrt{\qhu}}{\mhu}t , \frac{\qhu}{\mhu^2} \right)  \left[ \fh^U \left( \frac{ \sqrt{\qhu} t }{\qhu - \Qhu} , \frac{1}{\qhu - \Qh_u} \right) \right]^2 ,  \label{eq:qu_DE_gen_mcs}
\end{align}
the same equations hold replacing $U$ by $V$, and
\begin{align}
  \mhz &= \frac{1}{\mz}  \int \dd y \int \DD t \frac{ \left[ \frac{\partial}{\partial t} f_0^{Y,0}(y| \frac{\mz}{\sqrt{\qz}} t , \Qzz - \frac{\mz^2}{\qz} ) \right] \left[ \frac{\partial}{\partial t} f_0^{Y}(y|  \sqrt{\qz} t , \Qz - \qz ) \right]}{f_0^{Y}(y|  \sqrt{\qz} t , \Qz -\qz )} ,  \label{eq:mh_DE_gen_mcs} \\
  \qhz &= \frac{1}{\qz}  \int \dd y \int \DD t \left[ \frac{\frac{\partial}{\partial t} f_0^{Y}(y|  \sqrt{\qz} t , \Qz -\qz )}{f_0^{Y}(y|  \sqrt{\qz} t , \Qz -\qz )} \right]^2 f_0^{Y,0}(y|  \frac{\mz}{\sqrt{\qz}} t , \Qzz  - \frac{\mz^2}{\qz} ) ,  \label{eq:qh_DE_gen_mcs}\\
  \Qhz &= 2 \int \dd y \int \DD t f_0^{Y,0}(y|  \frac{\mz}{\sqrt{\qz}} t , \Qzz  - \frac{\mz^2}{\qz} ) \left[ \frac{\frac{\partial}{\partial \Qz} f_0^{Y}(y|  \sqrt{\qz} t , \Qz -\qz )}{f_0^{Y}(y|  \sqrt{\qz} t , \Qz - \qz )} \right] , \label{eq:Qh_DE_gen_mcs}
\end{align}
and remembering that $\mz=\mymu \mv, \qz=\qu \qv$, $\Qz = \Qu \Qv$ and the definitions~(\ref{eq:alphas}):
\begin{align}
\mhu &= \alphau \mv \mhz,  &  \qhu &= \alphau \qv \qhz , &  \Qhu &= \alphau \Qv \Qhz ,  \label{eq:mhu_mh}\\
 \mhv &= \alphav \mymu \mhz,  & \qhv &= \alphav \qu \qhz , &  \Qhv &= \alphav \Qu \Qhz . \label{eq:mhv_mh}
\end{align}
 The equations~(\ref{eq:mu_DE_gen_mcs}, \ref{eq:Qu_DE_gen_mcs}, \ref{eq:qu_DE_gen_mcs}) along with their equivalents for $v$, 
 the equations~(\ref{eq:mh_DE_gen_mcs}, \ref{eq:qh_DE_gen_mcs}, \ref{eq:Qh_DE_gen_mcs}) and~(\ref{eq:mhu_mh}, \ref{eq:mhv_mh}) 
 constitute a closed set of equations that hold at the extrema of $\phi$~(\ref{eq:phi_mcs}).
 
 When they are iterated, they constitute the so-called state evolution equations. 
 These can also be obtained by the analysis of the BP algorithm and are known to accurately describe the 
 algorithm's behaviour when the replica symmetric hypothesis is indeed correct.
 
 As noted before, if $\YS = \US \VS$, these state evolution equations are identical to the ones in matrix factorization~\cite{kabaMF}.
 Therefore, they reduce to the state evolution of \gamp when $\uv$ is known, which corresponds to fixing $\mymu=\qu=\Qu=\Qzu$ in the equations.

\subsection{Bayes optimal analysis}
If we suppose exact knowledge of the true signal distributions and of the true measurement channel,
the state evolution equations greatly simplify because of the so-called Nishimori conditions~\cite{zdeboReview}.
In our case, these ensure that following equalities hold:
\begin{align}
 Q = Q^0 , \quad \hat{Q} = 0, \quad m = q, \quad \hat{m} &= \hat{q}	\label{eq:nishConds_mcs}
\end{align}
both for $u$ and $v$.
Then, we only need to keep track of the variables $(\mymu, \mhu, \mv, \mhv)$,
and  the state evolution is obtained by choosing initial values for $(\mymu^0,\mv^0)$ and iterating for $\iter \geq 0$ the equations
\begin{align}
 \mhz^{\iter+1} &= \frac{1}{\mymu^{\iter} \mv^{\iter}} \int \dd y \int \DD t  \frac{\left[ \frac{\partial}{\partial t} f^Y_{0}(y|\sqrt{\mymu^{\iter} \mv^{\iter}}t,\Qzu \Qzv  -\mymu^{\iter} \mv^{\iter})\right]^2}{f^Y_{0}(y|\sqrt{\mymu^{\iter} \mv^{\iter}}t, \Qzu \Qzv -\mymu^{\iter} \mv^{\iter})} ,   \label{eq:mh_DE_BO_mcs} \\
 \mymu^{\iter+1} &= \frac{1}{\sqrt{\alphau \mv^{\iter} \mhz^{\iter+1}}} \int \dd t \frac{\left[f^U_1(\frac{t}{\sqrt{\alphau \mv^{\iter} \mhz^{\iter+1}}},\frac{1}{\alphau \mv^{\iter} \mhz^{\iter+1}})\right]^2}{f^U_0(\frac{t}{\sqrt{\alphau \mv^{\iter} \mhz^{\iter+1}}},\frac{1}{\alphau \mv^{\iter} \mhz^{\iter+1}})} , \label{eq:mu_DE_BO} \\
 \mv^{\iter+1} &= \frac{1}{\sqrt{\alphav \mymu^{\iter} \mhz^{\iter+1}}} \int \dd t \frac{\left[f^V_1(\frac{t}{\sqrt{\alphav \mymu^{\iter} \mhz^{\iter+1}}},\frac{1}{\alphav \mymu^{\iter} \mhz^{\iter+1}})\right]^2}{f^V_0(\frac{t}{\sqrt{\alphav \mymu^{\iter} \mhz^{\iter+1}}},\frac{1}{\alphav \mymu^{\iter} \mhz^{\iter+1}})} , \label{eq:mv_DE_BO} 
\end{align}
until convergence. From $\mymu$ and $\mv$, one can simply deduce the normalized 
mean squared errors by the following relations:
\begin{align}
 {\rm nMSE_U} &=  1 - \frac{\mymu}{\Qzu}, & {\rm nMSE_V} &=  1 - \frac{\mv}{\Qzv} , & {\rm nMSE_X} &= 1- \frac{\mymu \mv}{\Qzu \Qzv}.  \label{eq:mses_BO}
\end{align}
The initialization values $(\mymu^0,  \mv^0)$ indicate how close to the solution the algorithm is at initialization. 
In case of a random initialization of the algorithm, the expected initial ${\rm nMSE_U}$ and ${\rm nMSE_V}$ are of order $1/\US$ and $1/\VS$ respectively,
and they should therefore be set to these values (or less) in the state evolution equations.

Note that state evolution run with matching priors without imposing the Nishimori conditions~(\ref{eq:nishConds_mcs}) should in principle 
give the exact same results as the Bayes optimal state evolution analysis presented above, and thus naturally follow the so-called ``Nishimori line'' defined by~(\ref{eq:nishConds_mcs}).
However, as shown in~\cite{caltaConvergence}, the Nishimori line can be unstable: In that case, numerical fluctuations around it will be amplified 
under iterations of state evolution that will thus give a different result than its counterpart with imposed Nishimori conditions.
We analyse this instability in the following section.

\section{Stability analysis}
\label{sec:instability}
In this section, we propose an explanation for the instability of both \bigamp and \pbig by 
using the state evolution.
In~\cite{caltaConvergence}, it is shown that the non-convergence of \gamp with non-zero mean matrices 
comes from an instability of the Nishimori line. 
Here, we make a similar analysis for \bigamp/\pbig.

\subsection{Blind matrix calibration state evolution}
We consider the setting of blind matrix calibration, presented in~\secref{sec:DL} and which allows to 
interpolate between \cs and dictionary learning.
The elements of the matrix $\uv$ follow a Gaussian distribution (no sparsity) and 
are partially known by direct noisy measurements $\tilde{\uv}$:
\begin{align}
 \tilde{\u}_{\mu i} = \frac{\u_{\mu i} + \sqrt{\eta} \xi_{\mu i}}{\sqrt{1+\eta}}	\quad \text{with} \quad \xi_{\mu i} \sim \NN(\xi_{\mu i};0,1), 
\end{align}
while the elements of $\vv$ follow a Gauss-Bernoulli distribution.
The state evolution equations~(\ref{eq:mu_DE_gen}--\ref{eq:Qhz_DE_gen}) can be implemented using~(\ref{eq:mx_DE_gen_add}--\ref{eq:qx_DE_gen_add}).
Some quantities have analytical expressions. For the \awgn channel,
\begin{align}
 \mhz &= \frac{1}{\Delta+\Qz-\qz}, & \qhz &= \frac{\Delta+ \Qzz + \qz - 2\mz}{(\Delta+\Qz-\qz)^2}, & \Qhz &= \frac{\Qzz - \Qz + 2(\qz - \mz)}{(\Delta+\Qz-\qz)^2}.
\end{align}
For the $\uv$ part, we define
\begin{align}
 \xhC &= \frac{\sqrt{\qh_u}}{\mh_u} , & \xbC &=\xhC^2, & \uhC &= \frac{\sqrt{\qh_u}}{\qh_u - \Qh_u} , & \ubC &= \frac{1}{\qh_u-\Qh_u}.
\end{align}
with the help of which we write
\begin{align}
 \mymu &= \frac{\xbC \ubC (\eta +1) + \eta ( \ubC + \frac{\uhC}{\xhC} \xbC) + \eta^2 (1 + \xbC)\frac{\uhC}{\xhC} }{ \left( \eta + \xbC (\eta+1) \right) \left( \eta + \ubC (\eta +1)\right)} , \\
 \qu &= \frac{\ubC^2 (\eta+1) + 2\eta \ubC \frac{\uhC}{\xhC} + \eta^2 \left( \frac{\uhC}{\xhC} \right)^2(1 + \xbC) }{\left( \eta + \ubC(\eta+1)\right)^2},  \\
 \Qu &=  \qu + \frac{\ubC \eta}{\eta+\ubC (\eta+1)}.
\end{align}

We focus on the case in which inference \textit{is} Bayes optimal. 
In this setting, we usually use the simplified state evolution equations~(\ref{eq:bo_1}-\ref{eq:bo_3}), 
by imposing the Nishimori conditions
\begin{align}
 \qhu&=\mhu, & \Qhu &=0, & \qhv&=\mhv, & \Qhv &= 0, \\
 \qu&=\mymu, & \Qu &= \Qu^0, & \qv&=\mv, & \Qv&=\Qv^0,
\end{align}
which reduces the number of state evolution equations to $3$.
However, in this study, we implement the $9$ state evolution equations in the Bayes optimal setting and compare the results 
to those obtained by imposing the Nishimori conditions.

\subsection{Instability of the Nishimori line}
\begin{figure}[h]
 \centering
 \includegraphics[width=1\textwidth]{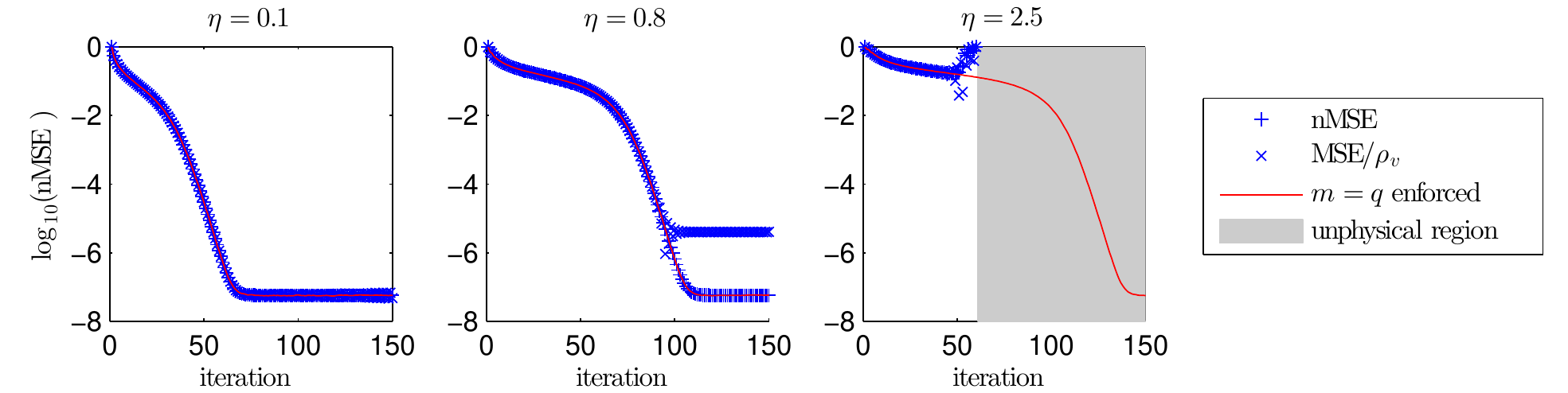}
 \caption[Nishimori line instability: MSE]{Instability of the Nishimori line in blind matrix calibration.
 The setting used is $\rho_V=0.2$, $\alphau=0.5$,$\alphav=4$, $\Delta=10^{-8}$ and different values of $\eta$.
 The state evolution equations were run enforcing or not the Nishimori condition $m=q$ and $Q=Q^0$ (both for $u$ and $v$).
 When $\eta=0.1$, enforcing or not the Nishimori conditions leads to the same result.
 For $\eta=0.8$ and $\eta=2.5$, imposing the Nishimori conditions still predicts success of the algorithm.
 However, when the Nishimori conditions are not imposed, oscillations develop. 
 While the nMSE still goes down to the value predicted by the Bayes optimal analysis, the MSE does not.
 For $\eta=2.5$, $Q_V$ becomes negative after a few iterations, leading to an unphysical region.
 The instability can also be observed by looking at the the deviation of $Q_V$ from  $\rho_V$ (\figref{fig:7_stab2}).
 }
 \label{fig:7_stab1}
\end{figure}

\begin{figure}[h]
 \centering
 \includegraphics[width=0.6\textwidth]{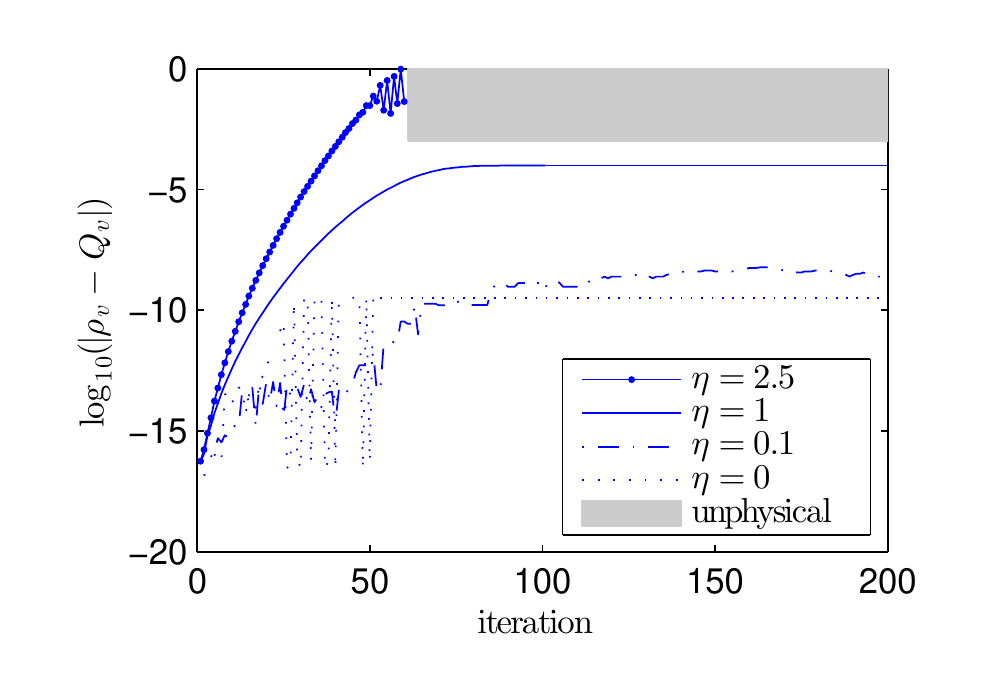}
 \caption[Nishimori line instability: $Q_v$]{Instability of the Nishimori line in blind matrix calibration.
 Same setting as in~\figref{fig:7_stab1} (Nishimori conditions are not imposed). 
 In all cases, the sign of $\rho_V-Q_V$ changes at each iteration: $Q_V$ oscillates around $\rho_V$.
 For $\eta=0$ and $\eta=0.1$, $Q_V$ stays very close to $\rho_v$: the Nishimori line is stable. 
 For $\eta=1$ and $\eta=2.5$, the amplitude of oscillations grows until a point at which $Q_V$ becomes negative,
 which is unphysical because it is a variance.
 This instability of the Nishimori line is likely to be one of the origins of the bad convergence of the algorithm.}
 \label{fig:7_stab2}
\end{figure}

Results are presented in~\figref{fig:7_stab1} and~\figref{fig:7_stab2}.
We observe that as by increasing $\eta$, the Nishimori line becomes unstable: when the Nishimori conditions are not imposed,
the system naturally drifts away from them. 
This is interesting because in the algorithm, the Nishimori conditions cannot be imposed. 
Therefore, the instability observed in the state evolution is likely to take place as well in the algorithm.

On~\figref{fig:7_stab1}, we show that this instability
first causes the MSE to not go as low as it should, while the nMSE still does. 
This is due to the scaling invariance of the problem.
Although one could suppose this invariance to be lifted by the use of priors $p_U$ and $p_V$ with a given variance,
it turns out that these priors are not sufficient to stabilize the variance of the estimates $\uhv$ and $\vhv$.
Note that when the Nishimori conditions are not imposed, the nMSE writes:
\begin{align}
 {\rm nMSE_V} &= 1 - \frac{\mv}{\sqrt{\Qzv Q_V}},
\end{align}
and is thus different from
\begin{align}
 \frac{{\rm MSE_V}}{\Qzv} &= 1 - \frac{\mv}{\Qzv}.
\end{align}

The instability causes oscillations, that lead to parameters to take ``unphysical'' values 
(some of the variances take \textit{negative} values), if the amplitude of the oscillations grows too much.
Such negative variances cause the algorithm to break or diverge, which is in fact the observed behaviour of \bigamp and \pbig
without damping.

\subsubsection{Discussion}
As mentioned before, the damping strategies used to make \bigamp and \pbig converge are heuristic, do not allow
systematic convergence and significantly slow down the algorithm.
Furthermore, it is difficult to analyse them using the state evolution formalism, as they correlate 
estimates of several previous time steps.

A possible axis of investigation to find a cure to the non-convergence of \bigamp and \pbig is
to study the state evolution equations (without imposing the Nishimori conditions) with additional parameter learning. 
Parameter learning empirically improves algorithm convergence, and can be taken into account in the state evolution equations~\cite{krzakaCS}.
Finding a parameter learning scheme that stabilizes the Nishimori line could allow more systematic convergence of the algorithms.

Another remark is that the state evolution equations for \textit{non} Bayes optimal inference diverge after a few iterations.
For that reason, performing an analysis of bilinear inference using mismatching priors is not straightforward.
In particular schemes using $L_1$ minimization as a sparsity 
promoter, as could be done for \lasso, would be interesting to analyse. 
Investigating into a damping scheme for the state evolution equations, allowing them to converge to a fixed point could solve that issue 
and reveal fundamental limits of dictionary learning using $L_1$ minimization.


\section{Case Study}
In this section, we focus on one specific setting of matrix compressed sensing for which the Bayes optimal state evolution equations are practical to implement. 
An analysis of their fixed points leads to an understanding of different phases and of the phase transitions between them.

We look at the setting in which both $\uv$ and $\vv$ follow a Bernoulli-Gauss distribution:
\begin{align}
 \p_U(u) &= (1-\rho_U) \delta(u) + \rho_U \, \NN(u;0,1), \\
 \p_V(v) &= (1-\rho_V) \delta(v) + \rho_V \, \NN(v;0,1),
\end{align}
and the measurements are taken through an \awgn channel:
\begin{align}
 \forall \yil \in [1, \YS], \quad y_{\yil} = [ \A(\uv \vv^{\top})]_{\yil} + \xi_{\yil}, \qquad \quad \text{with} \quad \xi_{\yil} \sim \NN(\xi_{\yil};0,\Delta).
\end{align}
Note that most previous works~\cite{lee2013near,nuclearNormMin,JainLowRank,laffertyLowRank} consider this channel.
For the \awgn channel,~\eqref{eq:mh_DE_BO_mcs} has a simple analytical expression:
\begin{align}
 \hat{m}^{\iter+1} &= \frac{1}{\Delta + \Qzu \Qzv - \mymu^{\iter} \mv^{\iter}}.
\end{align}
Further simplifying the setting to the special case $\US=\VS$ and $\rho_U=\rho_V=\rho$, the Bayes optimal state evolution equations~(\ref{eq:mh_DE_BO_mcs}--\ref{eq:mv_DE_BO}) can be written as one single equation
\begin{align}
 m &= \sqrt{\frac{\Delta+\rho^2-m^2}{\alphau m}} \int \dd t \frac{\left[ f_1^U(\sqrt{\frac{\Delta+\rho^2-m^2}{\alphau m}}t , \frac{\Delta+\rho^2-m^2}{\alphau m}) \right]^2}{f_0(\sqrt{\frac{\Delta+\rho^2-m^2}{\alphau m}}t , \frac{\Delta+\rho^2-m^2}{\alphau m})}, \label{eq:simple_DE}
\end{align}
in which the iteration-time indices of $m$, $\iter$ (left hand side) and $\iter-1$ (right hand side), are left out for better legibility. 
The global measurement rate is
\begin{align}
 \alpha \equiv \frac{\YS}{2 \US \rank} = \frac{\alphau}{2} ,	\label{eq:global_alpha}
\end{align}
and is the natural quantity to compare $\rho$ to.

\subsection{Phases and phase transitions}
As in compressed sensing or in matrix factorization, the analysis of the free entropy and state evolution equations 
reveals the existence of different phases in which the difficulty of the problem is different.
In our case study, the free entropy $\phi$ has the following expression:
\begin{align}
 \phi(m) &= - m \hat{m} - \frac{\alpha}{4} \log\left( 2 \pi \left( \Delta + \rho^2 - m^2\right)\right) \nonumber \\
 &+ \frac{2}{\sqrt{\hat{m}}} \int \dd t f_0^U \left( \frac{t}{\sqrt{\hat{m}}}, \frac{1}{\hat{m}} \right) \left[ \frac{t^2}{2} + \log\left( \sqrt{\frac{2 \pi}{\hat{m}}} f_0^U \left( \frac{t}{\sqrt{\hat{m}}}, \frac{1}{\hat{m}} \right) \right) \right],
\end{align}
with
\begin{align}
  \hat{m} &= \frac{1}{\Delta + \rho^2 - m^2}.
\end{align}
The integral can best be numerically evaluated replacing $\int$ by $2 \left( \int_{0}^{20} + \int_{20}^{20 \sqrt{1+\hat{m}}} \right) $, which allows a reliable 
numerical evaluation for all possible values of $\hat{m}$.

\begin{figure}[h]
 \centering
  \includegraphics[width=0.5\textwidth]{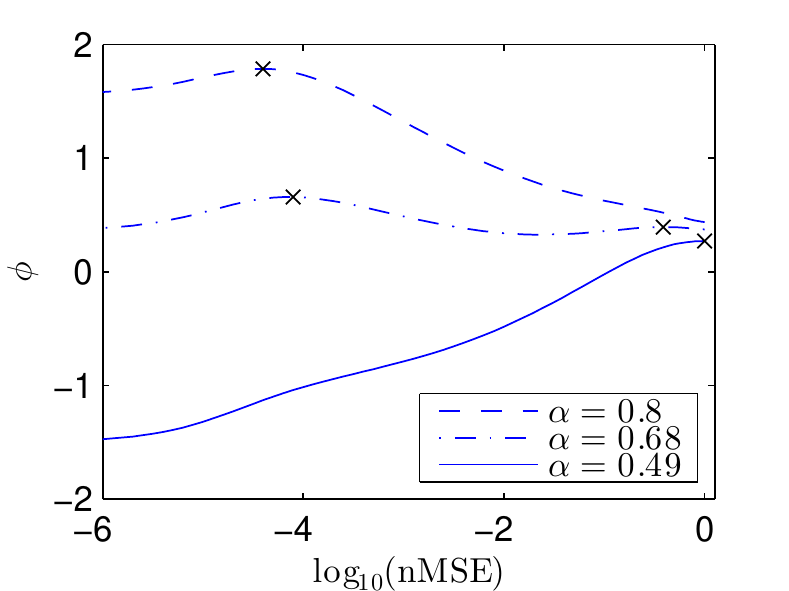}
  \caption[Matrix compressed sensing: Free entropy landscape]{Free entropy landscapes for $\rho=0.5$, $\Delta=10^{-5}$. 
  Crosses represent local maxima. There are three types of them: either at nMSE$=1$ (as for $\alpha=0.49$),
  or at ${\rm nMSE} \approx \Delta$, or in an intermediary region. In case there are several local maxima (as for $\alpha=0.68$), 
  the algorithm will perform sub-optimally, getting stuck in the local maximum of highest nMSE instead of converging to the global maximum (``hard but possible'' phase). }
\label{fig:phases}
\end{figure}

Figure~\ref{fig:phases} shows the free entropy landscapes for $\rho=0.1$ and different values of $\alpha$.
Instead of using $m$ as $x$-axes, we use the normalized mean squared error 
which is a more natural quantity to measure the quality of reconstruction.

We can define three different phases depending on the positions of the free entropy maxima.
In the noiseless setting, these are:
\begin{enumerate}
 \item An ``impossible'' phase, in which the global maximum of the free entropy is not at nMSE$=0$.
 In that phase, no algorithm can find the correct solution.
 \item A ``hard but possible'' phase, in which the free entropy has its global maximum at nMSE$=0$, but also a \textit{local} maximum at non-zero nMSE.
 In that phase, it is possible to find the correct solution, by correctly sampling from the posterior distribution~(\ref{eq:proba_expanded}).
 However, algorithms such as \pbig get stuck in the local free entropy maximum instead of finding the global maximum.
 \item An ``easy'' phase, in which the free entropy function has a single maximum at nMSE$=0$.
\end{enumerate}
In a noisy setting as in~\figref{fig:phases}, the lowest achievable nMSE is of the order of the AWGN $\Delta$ instead of $0$.

\subsubsection{State evolution fixed points}
\begin{figure}[h]
 \centering
 \begin{subfigure}{0.48\textwidth}
  \includegraphics[width=\textwidth]{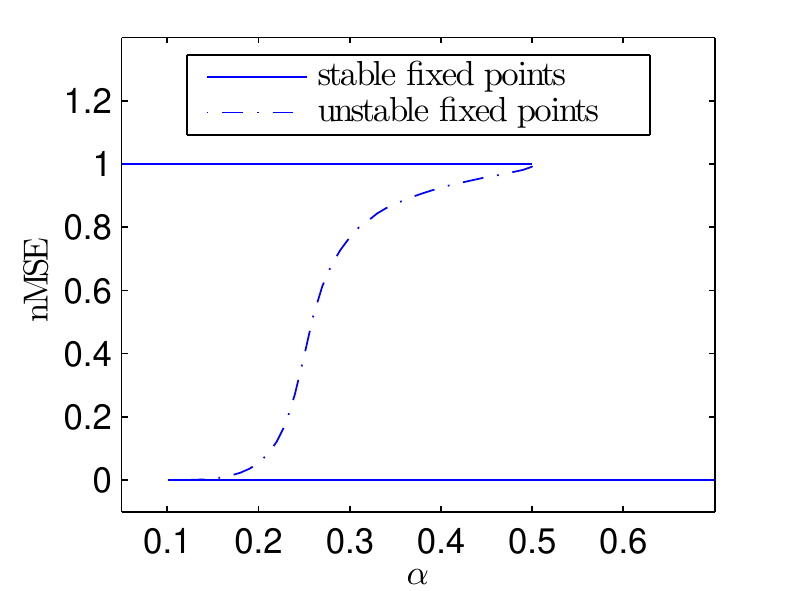}
  \caption{$\rho=0.1$}
  \label{fig:fp01}
 \end{subfigure}
 \begin{subfigure}{0.48\textwidth}
  \includegraphics[width=\textwidth]{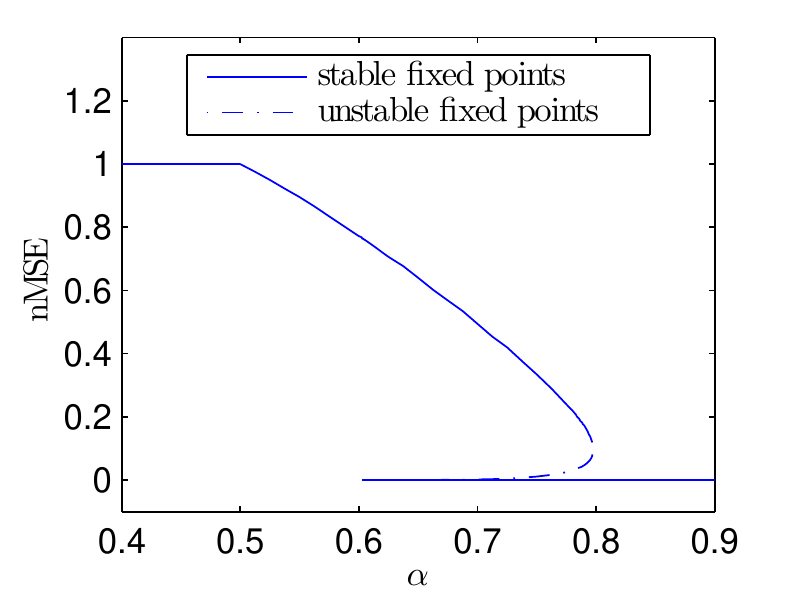}
  \caption{$\rho=0.6$}
  \label{fig:fp06}
 \end{subfigure}
\caption[State evolution fixed points]{Fixed points of the state evolution~(\ref{eq:simple_DE}) for two different sparsities $\rho$.
For values of $\alpha$ for which two stable fixed points exist, the iterated state evolution equation converges to the one of higher nMSE 
if the initial nMSE is higher than the unstable fixed point, and to the one of lower nMSE if not.}
\label{fig:fp}
\end{figure}

The state evolution equation~(\ref{eq:simple_DE}) can either be iterated or considered as a fixed point equation.
Figure~\ref{fig:fp} shows the fixed points of~(\ref{eq:simple_DE}), which are all local extrema of the free entropy $\phi$.
On the other hand, iterating the state evolution equations gives only one of the local maxima.

The plots allow to see more clearly the ``impossible'', ``hard but possible'' and ``easy'' phases.
They show that in the ``hard but possible'' phase, the state evolution has an unstable fixed point, which corresponds 
to a local minimum of the free entropy.
Two interesting facts can be noticed:
\begin{enumerate}
 \item In the noiseless setting, the impossible/possible phase
   transition (the apparition of the ${\rm nMSE}=0$ fixed point) takes place at $\alpha=\rho$.
 This can be expected because it is the critical $\alpha$ at which the number of available equations is equal to the total number of non-zero components of the unknowns, just as in compressed sensing.
 \item The fixed point at nMSE=1 always exists for $\alpha \in [0, 0.5]$. 
 This is more unexpected as it is not the case in compressed sensing. A consequence of this is the existence of a large ``hard but possible'' phase for small values of $\rho$. 
 Also, the measurement rate necessary for ``easy'' recovery is \textit{at least} $0.5$, even for very small $\rho$. 
 This radically differs from the low-$\rho$ regime in compressed sensing, in which a measurement rate $\alpha \propto \rho$ is sufficient for easy recovery.
\end{enumerate}

\begin{figure}[h]
 \centering
 \includegraphics[width=0.5\textwidth]{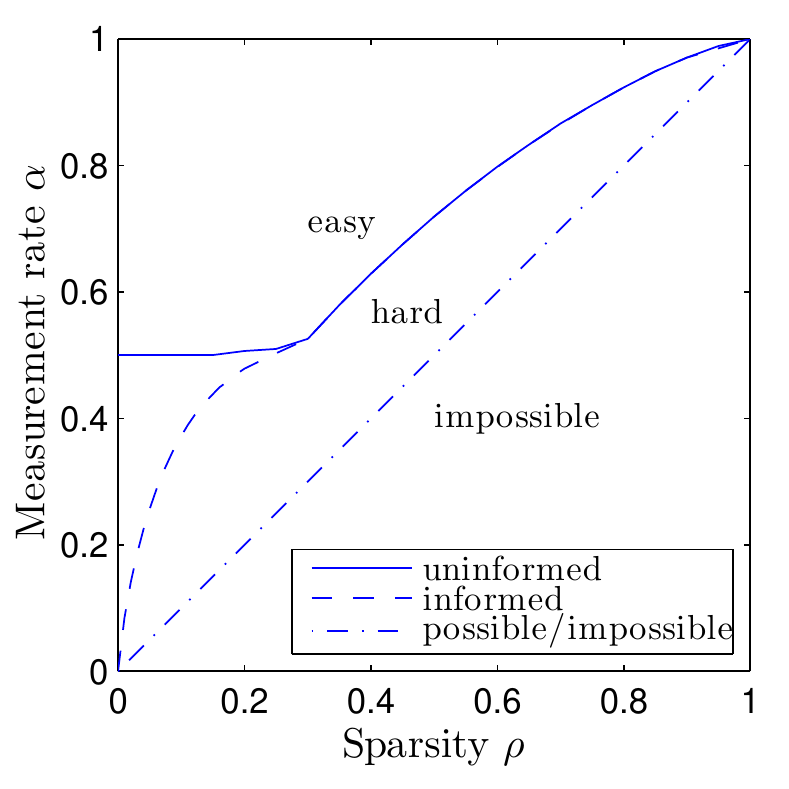}
 \caption[Phase diagram for matrix compressed sensing]{Phase diagram for the considered case-study obtained with (\ref{eq:simple_DE}).  $\Delta=10^{-12}$ and success is defined by a final nMSE$<10^{-10}$.
 The disappearing of the state evolution fixed point (or equivalently, of a free entropy maximum) with nMSE of order 1 marks the frontier between the ``hard'' and the ``easy'' phase.
 The possible/impossible frontier represented corresponds to the noiseless case.}
 \label{fig:full_phase_diagram}
\end{figure}

Figure~\ref{fig:full_phase_diagram} shows the full phase diagram for
the case-study problem, with the easy, hard and impossible phases.
The ``uninformed'' line is obtained by starting the state evolution
starting from nMSE$= 1 - \epsilon$, with an infinitesimally small $\epsilon$, and defines the transition between the ``easy'' and the ``hard'' phase.
 Interestingly, the entire region with $\alpha<0.5$ is in the hard
 phase, even at low values of $\rho$, due to the existence of the stable
 fixed point at ${\rm nMSE}=1$.
 In the ``hard'' phase, inference is possible provided a good estimation of the signal is already known. 
 The effect of such a partial knowledge can be simulated by running the state evolution equation (\ref{eq:simple_DE}) starting with nMSE$=0.9$, 
 leading to the ``informed'' line, for which $\alpha \to 0$ when $\rho \to 0$. The exact position of this line depends on the starting nMSE.

\subsection{Comparison with algorithmic performances}
\begin{figure}[h!]
 \centering
 \begin{subfigure}{0.48\textwidth}
  \includegraphics[width=\textwidth]{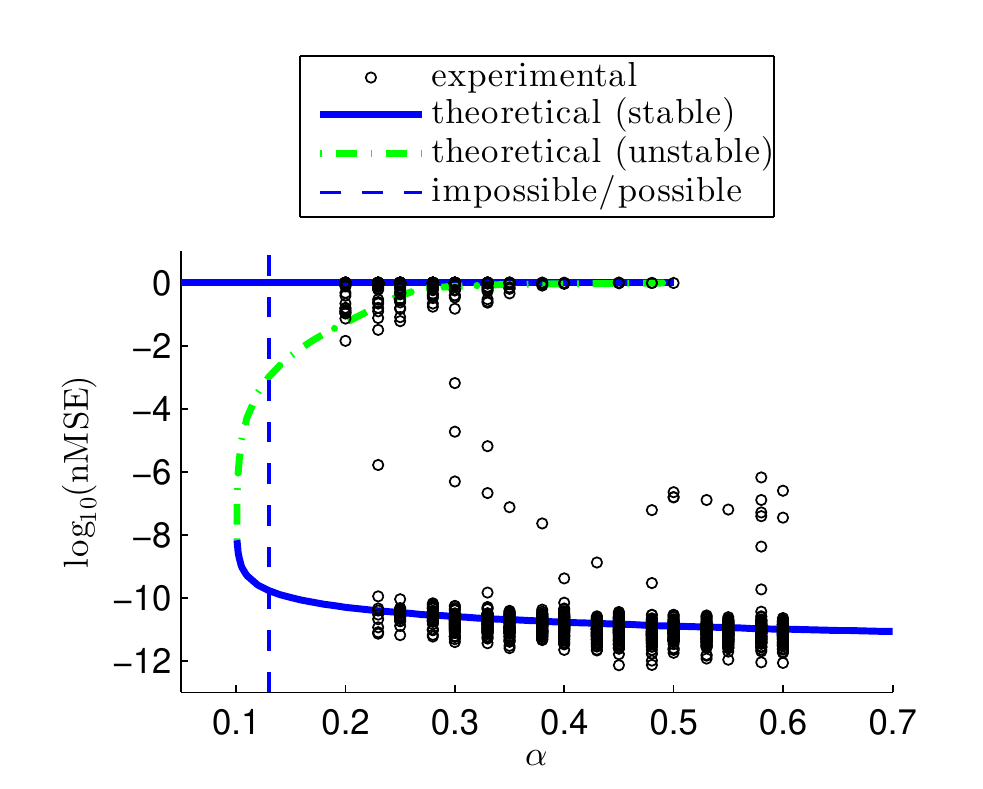}
  \caption{$\rho=0.1$, $M=50$}
  \label{fig:comp01N50}
 \end{subfigure}
 \begin{subfigure}{0.48\textwidth}
  \includegraphics[width=\textwidth]{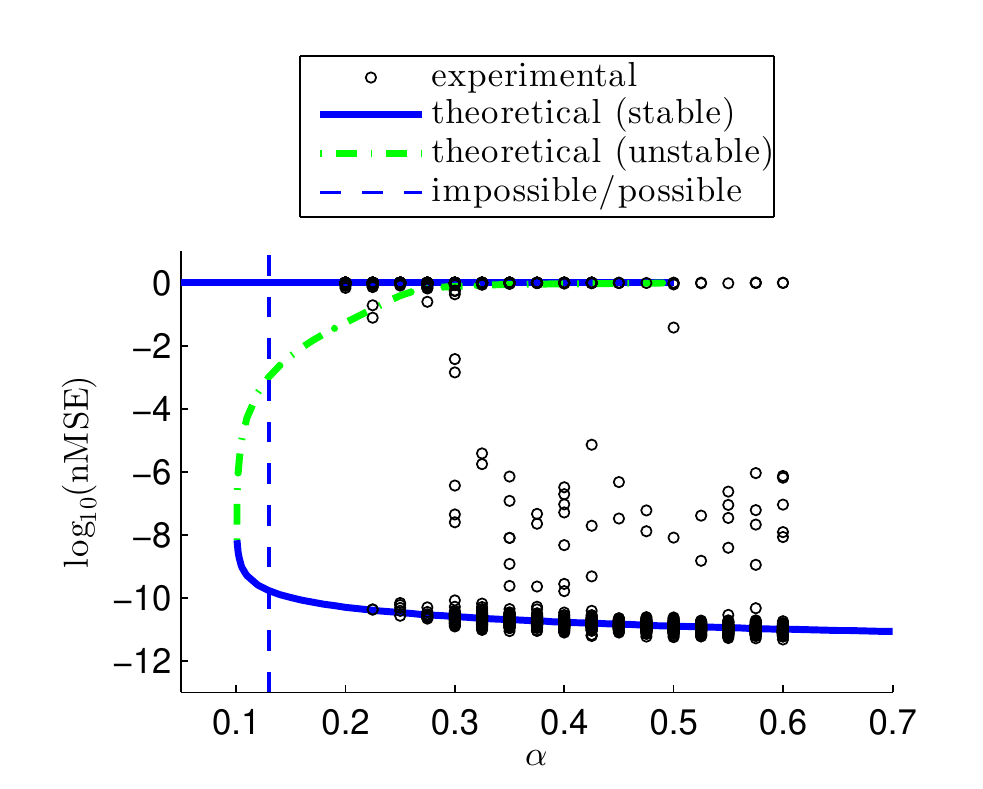}
  \caption{$\rho=0.1$,  $M=200$}
  \label{fig:comp01N200}
 \end{subfigure}
\caption[Experimental and theoretical fixed points 1]{Comparison of fixed points obtained by state evolution and experimentally reached estimates.
Parameters  are $\rho=0.1$, $\Delta=10^{-12}$ with (a) $M=50$, (b) $M=200$.
For each $\alpha$ there are $100$ experimental points.
The experimental fixed points are relatively close to the fixed points of the state evolution.
Note that the spreading around the theoretical line diminishes with growing $M$.
In the thermodynamic limit $M \to \infty$, all experimental points would be on the fixed point of \textit{highest} nMSE.
At finite $M$, the probability to initialize the algorithm \textit{below} the unstable fixed point allows some instances to 
converge to the low-nMSE fixed point.}
\label{fig:comp01}
\end{figure}

\begin{figure}[h!]
 \centering
 \begin{subfigure}{0.48\textwidth}
  \includegraphics[width=\textwidth]{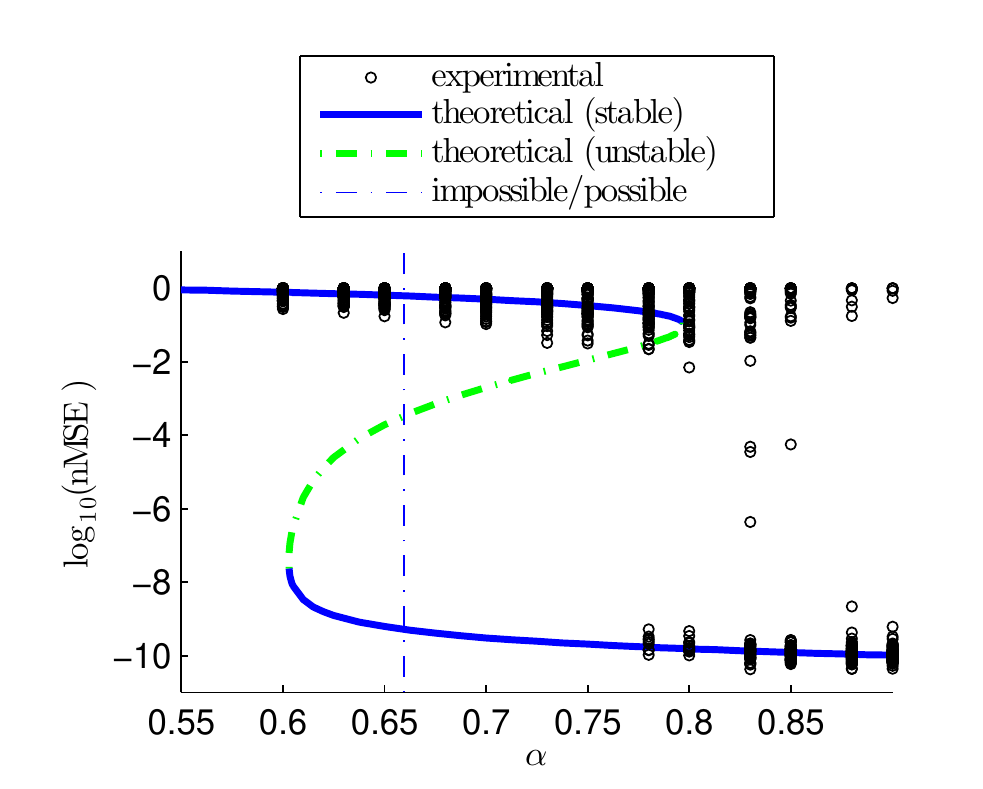}
  \caption{$\rho=0.6$, $M=50$}
  \label{fig:comp06N50}
 \end{subfigure}
 \begin{subfigure}{0.48\textwidth}
  \includegraphics[width=\textwidth]{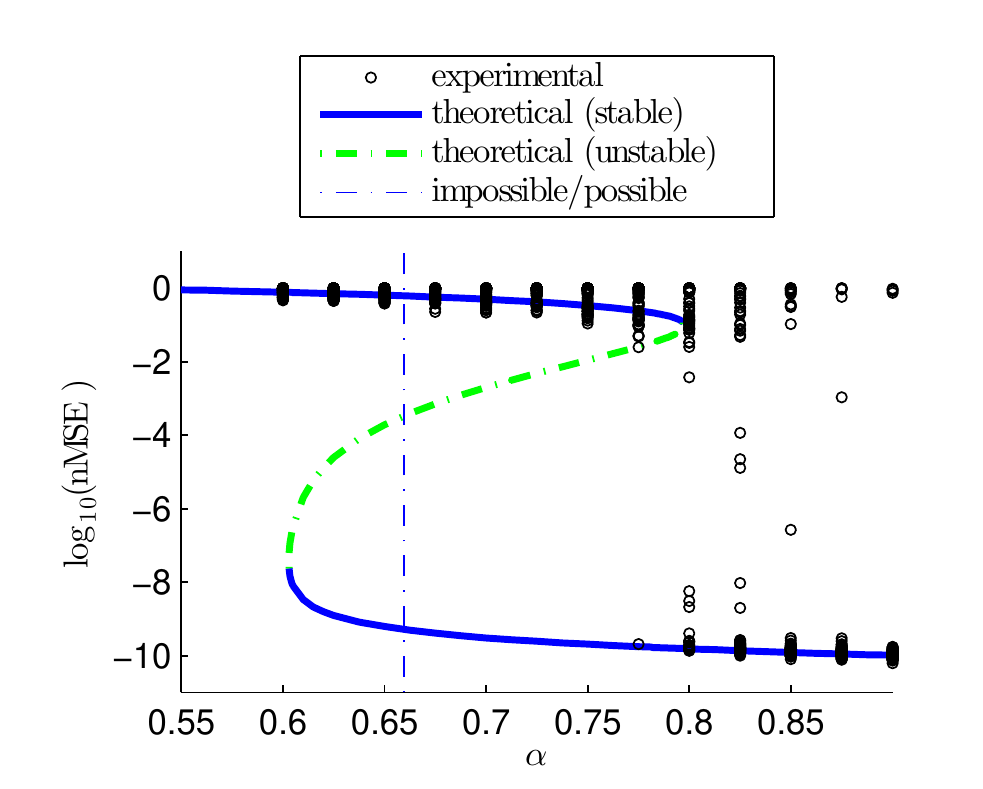}
  \caption{$\rho=0.6$,  $M=200$}
  \label{fig:comp06N200}
 \end{subfigure}
\caption[Experimental and theoretical fixed points 2]{Comparison of fixed points obtained by state evolution and experimentally reached estimates.
Parameters  are $\rho=0.6$, $\Delta=3.6\times 10^{-11}$ with (a) $M=50$, (b) $M=200$. 
For each $\alpha$ there are $100$ experimental points.
Unlike for the $\rho=0.1$ case on~\figref{fig:comp01}, the algorithm 
fails for an important fraction of instances in the ``easy'' phase.
This phenomenon is not explained by the state evolution analysis and 
might be a finite size effect. However, as $\alpha$ grows the probability of success goes to $1$ (see~\figref{fig:PT_06}).
Unlike for $\rho=0.1$, the probability of recovery inside the ``hard'' phase is much smaller, due to the lower nMSE of the unstable fixed point.}
\label{fig:comp06}
\end{figure}

Figures~\ref{fig:comp01} and~\ref{fig:comp06}  present a comparison of the theoretical fixed point analysis performed above with 
the actual performances of \pbig. Experiments were done by Philip Schniter.

For the experiments, rank $\rank=1$ was used. In this setting, the only invariance left is a scaling invariance: if $(\uv,\vv)$ is the true solution, then for every 
$\gamma \neq 0$, $(\gamma \uv, \frac{1}{\gamma} \vv)$ is a solution as well. 
The final nMSE returned by the algorithm takes this invariance into account and is the average of the error on $\uv$ and the error on $\vv$:
\begin{align}
 \rm{nMSE} &= \frac{1}{2} \left[ \left( 1 - \frac{|\uv^{\top} \uhv|}{||\uv||_2 || \uhv ||_2 } \right) +  \left( 1 - \frac{|\vv^{\top} \vhv|}{||\vv||_2 ||\vhv||_2} \right)  \right] ,	\label{eq:exp_nmse}
\end{align}
which will be compared to the theoretical expression~(\ref{eq:mses_BO}).
For each instance of the problem, the algorithm was allowed up to $20$ restarts  from different random initializations to reach a nMSE smaller than $10^{-6}$,
and the lowest of the reached nMSE was kept.


The results show that there is a good agreement between the theory and the performance of \pbig: most of the nMSEs reached by \pbig correspond to a stable fixed point of the state evolution.
The agreement with the theory becomes better with increasing system size. 
For smaller sizes, the experimental points are more spread around the theoretical fixed points. This can be well understood by analyzing the case of fixed points with nMSE=1.
The ``meaning'' of such fixed points is that the algorithm is unable to estimate the true signals better than at random. 
In the $\US \to \infty$ limit, the nMSE between the true signals and random signals is $1$ with probability $1$.
For finite values of $\US$ however, the nMSE between true and random signals follows a distribution on $[0,1]$ that gets more peaked on $1$ as $\US$ increases.
This explains the narrowing of the spread of experimental points around the fixed points as $\US$ increases.

\subsubsection{Succeeding in the hard phase: importance of the initialization}
An interesting consequence of this finite size effect is that for small $\US$, parts of the ``hard'' phase are quite easy.
The reason is that if the random initialization of the algorithm is such that the nMSE is \textit{smaller} than the nMSE of the \textit{unstable} fixed point, the algorithm naturally converges
 to the low-nMSE solution. Therefore, running the algorithm from a few different initializations can allow to converge to the correct solution even in the ``hard'' phase, 
 provided that $\US$ is small enough and that the unstable fixed point has a high enough nMSE.
 
\begin{figure}
 \centering
 \begin{subfigure}{0.48\textwidth}
  \includegraphics[width=0.8\textwidth]{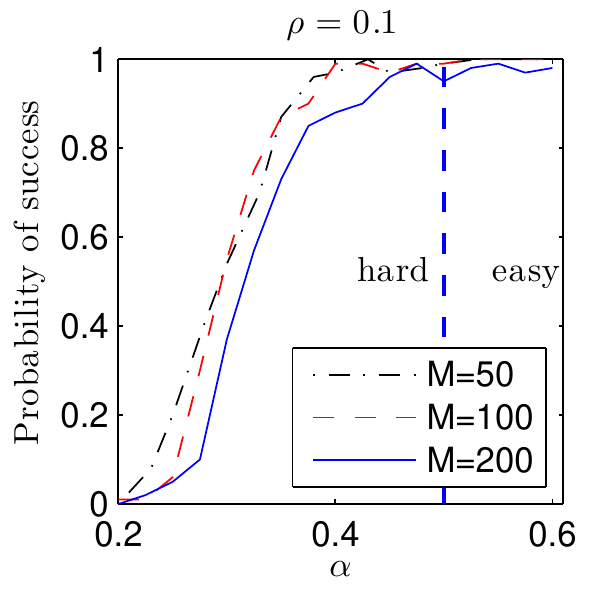}
  \caption{$\rho=0.1$}
  \label{fig:PT_01}
 \end{subfigure}
  \begin{subfigure}{0.48\textwidth}
  \includegraphics[width=0.8\textwidth]{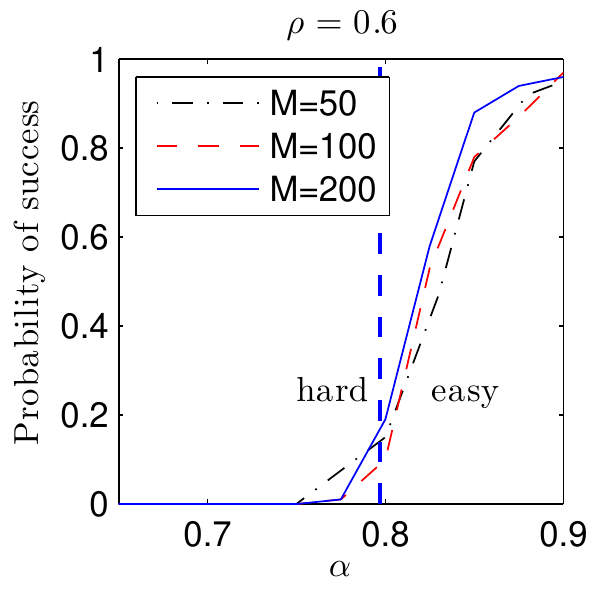}
  \caption{$\rho=0.6$}
  \label{fig:PT_06}
 \end{subfigure}
\caption[Experimental phase transition]{Empirical probability of success (defined by ${\rm nMSE}<10^{-6}$), for the experiments presented on figures~\ref{fig:comp01} and~\ref{fig:comp06}.
Due to the finite size, the position of the curves slightly vary for different values of $M$.
Finite size effects allow a fraction of successful instances inside the hard phase for $\rho=0.1$, but much less for $\rho=0.6$.}
\label{fig:PT}
\end{figure}

 Figure~\ref{fig:PT} shows that this effect is quite important for $\rho=0.1$, but nearly inexistent for $\rho=0.6$.
 The reason for this is the higher nMSE of the unstable fixed point for $\rho=0.1$ than for $\rho=0.6$.
 
Remember that in \pbig, the initial estimates of $\uv$ and $\vv$ are random. 
While in some regions of the phase diagram and with small signal sizes, running the algorithm from several of those random initial estimates might be sufficient, 
in general it would be preferable to have a procedure that systematically produces good initializations.
Previous works stress this fact as well~\cite{lee2013near,nuclearNormMin,JainLowRank,laffertyLowRank}.
 
Another difference between figures~\ref{fig:PT_01} and~\ref{fig:PT_06} is that in the latter, the algorithm fails for a significant fraction 
of instances inside the ``easy'' phase, which is not the case in the former. 
The fact that the fraction of such failed instances decreases with increasing signal size $\US$ seems to indicate that this is as well a finite size effect.
Unlike the previously examined finite size effect, this one cannot be explained from the state evolution, as it has a unique fixed point in the ``easy'' phase.

\section{Conclusion}

In this chapter, we provide an asymptotic analysis of Bayesian low-rank matrix compressed sensing.
We employ the replica method to obtain the so-called state evolution equations,
whose fixed points allow to determine if inference is easy, hard or
impossible. The state evolution equations describe the behaviour of the
associated message passing algorithm \pbig that was derived and studied
previously in \cite{parker2015parametric}, for whose derivation we provide the key steps.

An interesting point concerning the state evolution equations is that
they are the same as those for the matrix factorization problem
derived in \cite{kabaMF}. A related observation was made by \cite{Donoho21052013}.

We analyse in detail the phase
diagram for an \awgn sensing channel and Gauss-Bernoulli priors on
both the factors. We show  numerically that there is an excellent
agreement between the theoretical analysis and the performances of the
\pbig algorithm.
We observe that for the simulated system sizes, the algorithm performs better
than what could be expected from the asymptotic theoretical analysis. 
However, we explain this as a finite size effect in terms of state evolution fixed points and stress the importance of a good initial estimate 
in order to perform inference outside of the easy phase.

The stability analysis performed in~\secref{sec:instability} gives a theoretical explanation for why damping schemes are necessary both for \bigamp and 
\pbig to converge, and could be used to devise parameter learning schemes that stabilize them without the need to damp (and thus slow them down).

 \chapter*{Conclusion and perspectives}
\addcontentsline{toc}{chapter}{\protect\numberline{}Conclusion and perspectives}

The main focus in this thesis was the Bayesian treatment of generalized linear and bilinear inference problems,
using methods from statistical physics.
The replica method has proven to be very adapted to the theoretical analysis of these problems.
It reveals the existence of different \textit{phases}, in which inference is either possible or impossible, hard or easy.
Belief propagation allows to design fast algorithms that can often reach the performances predicted by the replica analysis.
However, in the case of bilinear inference or in the presence of certain non-linearities, belief propagation algorithms are limited by unreliable convergence properties.
Several concrete applications of generalized linear and bilinear inference problems were studied or simply mentioned,
showing how broad the applicability of these models is.

\section*{Open problems}
  Here are a few of problems I have come across, that are still open to my knowledge and that are interesting directions for future research.
\paragraph{Phase retrieval:} As explained in~\secref{sec:PR}, \gamp for phase retrieval does work, but
 is not very reliable, especially for sparse signals.
 As phase retrieval has many applications, it would be interesting to better understand where its difficulty stems from.
 The preliminary results in~\appref{app:PR_DE} could be extended to the non Bayes optimal case. 
 It seems plausible that as in bilinear inference, the convergence issues in phase retrieval come from an instability of the Nishimori line.
 
 \paragraph{Stabilizing the Nishimori line: } As seen in~\secref{sec:instability} (and possibly in phase retrieval), the state evolution equations 
 sometimes predict an instability in Bayes optimal message-passing algorithms. The empirical strategies that were proposed to make them converge 
  are effective in some cases, but not always. Furthermore, they slow down the algorithms significantly. 
 A more principled approach to stabilizing the algorithms could be tried by including parameter learning into the algorithm. 
 This has already proven to stabilize message-passing algorithm and can furthermore be analysed with state evolution equations. 
 By adding variables to the problems, it might be possible to stabilize the Nishimori line without slowing down the algorithm.
 More reliable bilinear inference algorithms could represent a breakthrough for many applications. 
 
 \paragraph{Overcoming invariances: } In several of the problems examined in this thesis, a fundamental invariance is present, 
 \eg the global phase invariance in phase retrieval. In some problems, this invariance is problematic. 
 In dictionary learning for example, the inferred signals are often not as sparse as they should, 
 which seems to be a consequence of the rotational invariance. It would be interesting to understand why 
 sparse priors do not seem able to enforce the right sparsity and whether this failure can be overcome. 
 

\section*{Beyond inference: deep learning}

The inference problems treated in this thesis mostly follow the scheme presented in~\secref{sec:inference}: 
Information about a signal is gathered in a measurement process and the goal is to reconstruct the initial signal.
In some other problems such as low-rank matrix completion, the underlying signal is not uniquely recoverable and 
is in most applications merely a useful low-dimensional representation of higher-dimensional data.
Representation learning~\cite{representationReview} is the key concept behind the power of deep learning~\cite{LeCun2015}, which has imposed itself 
as the state of the art technique in numerous machine learning~\cite{Hinton2006} and artificial intelligence tasks~\cite{masteringGo}.

Deep neural networks can be said to be \textit{unreasonably} effective considering the fact that their development is mainly heuristic 
and that little about them is understood on a theoretical level.
The work in this thesis contributes to understanding the basic building block of deep neural networks: The single, feedforward layer of neurons.
This understanding seems a prerequisite to being able to truly \textit{understand} deep neural networks.

 \appendix
 \part*{Appendices}
 
 \newpage
 
 \chapter{Useful functions}
\label{app:functions}

\section{Standard functions}
\subsection{Gaussians}
\label{sec:gaussians}
\subsubsection{Real Gaussians}
We note $\NN(x;\xh,\xb)$ the normalized Gaussian with mean $\xh \in \RR$ and variance $\xb \in \RR_+$:
\begin{equation}
 \NN(x;\xh,\xb) = \frac{1}{\sqrt{2 \pi \xb}} e^{-\frac{(x-\xh)^2}{2 \xb}}.	\label{eq:gaussian}
\end{equation}
We can note that:
\begin{equation}
 \NN(\alpha x ;\xh,\xb) = \frac{1}{\alpha} \NN(x; \frac{\xh}{\alpha}, \frac{\xb}{\alpha^2}).	\label{eq:gaussian_change}
\end{equation}
Its derivatives with respect to its mean and variance are:
\begin{align}
 \frac{\partial}{\partial \xh} \NN(x;\xh,\xb)& = \frac{x-\xh}{\xb} \NN(x;\xh,\xb),	\label{eq:gaussianDerivate} \\
 \frac{\partial}{\partial \xb} \NN(x;\xh, \xb) &= \frac{(x-\xh)^2 - \xb}{2 \xb^2} \NN(x;\xh,\xb).	\label{eq:gaussianDerivateVar}
\end{align}
The following formula for a product of Gaussians with the same argument but different means and variances is very useful:
\begin{align}
\prod_{i=1}^N \NN(x; \xh_i, \xb_i) &= \NN(x; \xh, \xb) \frac{\prod_{i=1}^N \NN(\xh_i;0,\xb_i)}{\NN(\xh;0,\xb)},		\label{eq:productOfGaussians} 
\end{align}
with
\begin{align}
 \xb^{-1} &= \sum_{i=1}^N \xb_i^{-1} , \quad  &\xh &= \xb \sum_{i=1}^N \frac{\xh_i}{\xb_i}.  \nonumber
\end{align}
and in the case of two Gaussians, (\ref{eq:productOfGaussians}) particularizes to:
\begin{align}
 \NN(x; \xh_1, \xb_1) \NN(x; \xh_2, \xb_2) &= \NN(x; \xh, \xb) \NN(\xh_1; \xh_2, \xb_1 + \xb_2).	\label{eq:productTwoGaussians}
\end{align}

\subsubsection{Multivariate Gaussians}
A multidimensional variable $\xv \in \RR^d$ follows a non-degenerate multivariate normal distribution if it has the \pdf
\begin{align}
  \NN(\xv;\xhv,\xbv) = \frac{1}{\sqrt{(2 \pi)^d \det(\xbv)}} e^{-\frac{1}{2}(\xv - \xhv)^{\top} \xbv^{-1} (\xv - \xhv) }.	\label{eq:multivariateGaussian}
\end{align}
where $\xhv \in \RR^d$ is the mean of the random variable $\xv$ and its covariance matrix $\xbv \in \cov{d}$ is symmetric and positive definite.
The relations verified by real Gaussians have very close equivalents for multivariate Gaussians:
\begin{align}
 \NN(\alpha \xv ;\xhv,\xbv) &= \frac{1}{\alpha^d} \NN(\xv; \frac{\xhv}{\alpha}, \frac{\xbv}{\alpha^2}),	\label{eq:multivariateGaussian_change} \\
 \nabla \NN(\xv;\xhv,\xbv)& = (\xv-\xhv)^{\top} \xbv^{-1} \NN(\xv;\xhv,\xbv),	\label{eq:multivariateGaussianDerivate}  \\
 \Jacobian(\nabla \NN)(\xv;\xhv,\xbv) &= \xbv^{-1} \left( -1 + (\xv-\xhv)(\xv-\xhv)^{\top} \xbv^{-1} \right) \NN(\xv;\xhv,\xbv). \label{eq:multivariateGaussianHessian}
\end{align}
The formula for the product reads
\begin{align}
\prod_{i=1}^N \NN(\xv; \xhv_i, \xbv_i) &= \NN(\xv; \xhv, \xbv) \frac{\prod_{i=1}^N \NN(\xhv_i;0,\xbv_i)}{\NN(\xhv;0,\xbv)},		\label{eq:productOfMultivariateGaussians} 
\end{align}
with
\begin{align}
 \xbv^{-1} &= \sum_{i=1}^N \xbv_i^{-1} , \quad  &\xhv &= \xbv \sum_{i=1}^N \xbv_i^{-1} \xhv_i . \nonumber
\end{align}

\subsection{Other useful functions and integrals}
\subsubsection{Complementary error function}
The complementary error function is defined by
\begin{align}
 \erfc(x) &= \frac{2}{\sqrt{\pi}}\int_x^{\infty} e^{-t^2} \dd t. 	\label{eq:erfc}
\end{align}
From this definition, we can obtain analytical expressions for the following integrals:
\begin{align}
 \int \dd x \NN(x;\xh,\xb) \indic(x>a) &= \frac{1}{2} \erfc \left(-\frac{\xh-a}{\sqrt{2 \xb}} \right), \\
 \int \dd x \NN(x;\xh,\xb) \indic(x\in [a,b]) &= \frac{1}{2} \left[ \erfc \left(-\frac{\xh-a}{\sqrt{2 \xb}} \right) -\erfc \left(-\frac{\xh-b}{\sqrt{2 \xb}} \right) \right].
\end{align}
The derivative of $\erfc$ is given by
\begin{align}
 \frac{\dd}{\dd x} \erfc(x) &= -\frac{2}{\sqrt{\pi}}e^{-x^2}. \label{eq:erfcDerivative}
\end{align}

\subsubsection{Gamma functions}
The gamma function is defined as
\begin{align}
 \Gamma(s) &= \int_0^{\infty} t^{s-1} e^{-t} \dd t	\label{eq:gammaFunction}
\end{align}
and the lower incomplete gamma function is defined as
\begin{align}
 \gamma(s,x) = \int_0^x t^{s-1}e^{-t} \dd t .	\label{eq:incompleteGamma}
\end{align}
Using them, we can obtain an analytical expression for the integral
\begin{align}
 J(N,\xh,\xb,a,b) &= \int x^N \NN(x; \xh,\xb) \indic(x\in [a,b]) \dd x  \\
      &= \frac{1}{2} \sum_{i=0}^N \left[ \binom{N}{i} \xh^{N-i} \sqrt{2 \xb}^{i+1} \Gamma\left(\frac{i+1}{2}\right) \right. \nonumber \\
      & \quad \left. \left( \sigma^i_b \gamma\left( \frac{i+1}{2}, \frac{(b-\xh)^2}{2 \xb} \right) -  \sigma^i_a \gamma\left( \frac{i+1}{2}, \frac{(a-\xh)^2}{2 \xb} \right)\right) \right]
\end{align}
where $\sigma_x^i$ is $1$ if $i$ is even and $\sign(x-\xh)$ if $i$ is uneven.

\section{Update functions}
\label{app:f}
For any non negative function $h$ and $i \in \{0,1,2\} $, we define the $i$-th moment of the product of $h$ multiplied by a Gaussian of mean $\xh \in \RR$ and variance $\xb \in \RR_+$:
\begin{align}
  f_0^h: \RR \times \RR_+	&\rightarrow \RR  \nonumber \\
  (\xh,\xb) &\mapsto \int \dd x h(x) \NN\left( x; \xh, \xb \right), \\
  f_1^h: \RR \times \RR_+	&\rightarrow \RR  \nonumber 	\\
  (\xh,\xb) &\mapsto \int \dd x  \, x \, h(x) \NN\left( x; \xh, \xb \right), 	 \\
 f_2^h:	\RR \times \RR_+	&\rightarrow \RR_+  \nonumber 	\\
 (\xh,\xb) &\mapsto \int \dd x \, x^2 h(x) \NN\left( x; \xh, \xb \right), 
\end{align}
From these, we define the update functions
\begin{align}
 \fh^h(\xh,\xb) &\equiv \frac{f_1^h(\xh,\xb)}{f_0^h(\xh,\xb)}, & &\in \RR   \label{eq:fh} \\
 \fb^h(\xh,\xb) &\equiv \frac{f_2^h(\xh,\xb)}{f_0^h(\xh,\xb)} - \left[ \fh^h(\xh,\xb) \right]^2 , & &\in \RR_+ \label{eq:fb}
\end{align}
which are the mean and variance of the distribution $h(x) \NN(x;\xh,\xb)$ and serve as update functions for the MMSE estimators and their related uncertainty.
We also define the auxiliary functions
\begin{align}
 \gh^h(\xh,\xb) &\equiv  \frac{ \fh^h(\xh,\xb) - \xh}{\xb},  & &\in \RR  \\
 \gb^h(\xh,\xb) &\equiv \frac{\fb^h(\xh, \xb) - \xb}{\xb^2}.	& & \in \RR. 
\end{align}
that allow notational compactness in the algorithms.

From~(\ref{eq:gaussianDerivate}) and~(\ref{eq:gaussianDerivateVar}) we obtain the relations:
\begin{align}
 \frac{\partial}{\partial \xh} \f_i^h(\xh,\xb) &= \frac{1}{\xb} \left( \f_{i+1}^h(\xh,\xb) - \xh \f_i^h(\xh,\xb) \right),	\label{eq:fDerivate} \\
 \frac{\partial}{\partial \xb} \f_i^h(\xh,\xb) &= \frac{1}{2 \xb^2} \left( \f_{i+2}^h(\xh,\xb) -2 \xh \f_{i+1}^h(\xh,\xb) - (\xb - \xh^2)\f_i^h(\xh,\xb) \right). 	\label{eq:fDerivateVar}
\end{align}
These are useful in the derivations of the message-passing algorithms and also allow to easily obtain $f_1^h, f_2^h$ from $f_0^h$.
The following relations are used as well:
\begin{align}
 \frac{\partial}{\partial \xh} \fh^h(\xh,\xb) &= \frac{1}{\xb} \fb^h(\xh,\xb),	\\
 \frac{\partial}{\partial s} f_i^h(\sqrt{s}t, \rho - s) &= -\frac{e^{\frac{t^2}{2}}}{2s} \frac{\partial}{\partial t} \left( e^{-\frac{t^2}{2}} \frac{\partial}{\partial t} f_i^h(\sqrt{s}t, \rho-s) \right).  \label{eq:fDerivateDifficult}
\end{align}

A useful feature of these update functions is the following: if $H(x) = h(x+\xi)$ with $\xi \sim \NN(\xi;0,\Delta)$, then
\begin{align}
 f_0^H(\xh,\xb) &= f_0^h(\xh,\xb+\Delta).
\end{align}
As a consequence, obtaining the update functions for a sensing channel with \awgn is straightforward once the functions for the noiseless version have been calculated. 
\subsubsection{Multivariate version}
These update functions can be extended to mutidimensional variables $\xv \in \RR^d$ and non negative functions $h: \RR^d \times \cov{d} \to \RR_+$:
\begin{align}
  f_0^h: \RR^{d} \times \cov{d}	&\rightarrow \RR  \nonumber \\
  (\xhv,\xbv) &\mapsto \int \dd \xv h(\xv) \NN\left( \xv; \xhv, \xbv \right), \\
  f_1^h: \RR^{d} \times \cov{d}	&\rightarrow \RR^{d}  \nonumber 	\\
  (\xhv,\xbv) &\mapsto \int \dd \xv  \, \xv \, h(\xv) \NN\left( \xv; \xhv, \xbv \right), 	 \\
 f_2^h:	\RR^{d} \times \cov{d}	&\rightarrow \cov{d}  \nonumber 	\\
 (\xhv,\xbv) &\mapsto \int \dd \xv \, \xv \xv^{\top} h(\xv) \NN\left( \xv; \xhv, \xbv \right), 
\end{align}
From these, we define the update functions
\begin{align}
 \fh^h(\xhv,\xbv) &\equiv \frac{f_1^h(\xhv,\xbv)}{f_0^h(\xhv,\xbv)}, & &\in \RR^{d}   \\
 \fb^h(\xhv,\xbv) &\equiv \frac{f_2^h(\xhv,\xbv)}{f_0^h(\xhv,\xbv)} - \fh^h(\xhv,\xbv) \left( \fh^h(\xhv,\xbv) \right)^{\top} , & &\in \cov{d}  
\end{align}
and the auxiliary functions
\begin{align}
 \gh^h(\xhv,\xbv) &\equiv \xbv^{-1} \left(  \fh^h(\xhv,\xbv) - \xhv  \right),  & &\in \RR^{d}  \\
 \gb^h(\xhv,\xbv) &\equiv \xbv^{-1} \left(  \fb^h(\xhv, \xbv) - \xbv  \right) \xbv^{-1}.	& & \in \sym{d} 
\end{align}
The gradient and Jacobian of $f_0^h$ with respect to its first argument read:
\begin{align}
 \nabla f_0^h(\xhv,\xbv) &= \xbv^{-1} \left( f_1^h(\xhv,\xbv) - \xhv f_0^h(\xhv, \xbv) \right), \\
 \Jacobian(\nabla f_0^h) (\xhv,\xbv) &= \xbv^{-1} \left[ \left( -1 + \xhv \xhv^{\top} \xbv^{-1} \right) f_0^h(\xhv,\xbv) + f_2^h(\xhv,\xbv) \xbv^{-1} - \left( f_1^h(\xhv,\xbv) \xhv^{\top} + \xhv f_1^h(\xhv,\xbv) \right) \xbv^{-1} \right] . \nonumber
\end{align}

 \chapter{Introducing the conjugate variable $\Qmh$}

\label{app:replicaDetails}
In~\eqref{eq:ZwithDeltas}, Dirac $\delta$ functions enforce the relations~(\ref{eq:defQx}).
We use the integral representation of these $\delta$ functions to carry on the calculation:
\begin{align}
 \delta(x) &= \frac{1}{2 \pi} \int \rm{d} \tilde{x} e^{\imath \tilde{x} x} = \delta \left( \imath x \right) = \frac{1}{2 \pi} \int \dd \tilde{x} e^{-\tilde{x} x}.
\end{align}
This leads to
\begin{align}
 \delta \left( N \Qx^{ab} - \sum_i x_i^a x_i^b \right) = \frac{1}{2 \pi} \int \dd \tilde{Q}_X^{ab} \, e^{-\tilde{Q}_X^{ab} \left( N \Qx^{ab} - \sum_i x_i^a x_i^b \right)},
\end{align}
and the product of all the $\delta$ functions thus gives
\begin{align}
 \prod_{a \leq b} \delta \left( N \Qx^{ab} - \sum_i x_i^a x_i^b \right) \propto \int \dd \tilde{\mathbf{Q}}_X \, \exp \left(-N \sum_{a \leq b} \tilde{Q}_X^{ab} Q_X^{ab} \right) \exp \left(\sum_{a \leq b} \sum_i \tilde{Q}_X^{ab} x_i^a x_i^b \right) . \label{eq:productOfDeltas}
\end{align}
Note that the summation is over $a \leq b$ because $Q_X^{ab} = Q_X^{ba}$.
Finally, we make the change of variables
\begin{align}
 \forall a, \hat{Q}_X^{a} &= 2 \tilde{Q}_X^{aa}, \nonumber \\
 \forall a, \forall b\neq a, \hat{Q}_X^{ab} &= 4 \tilde{Q}_X^{ab},
\end{align}
which allows us to write the sums in~\eqref{eq:productOfDeltas} more compactly:
\begin{align}
 \sum_{a\leq b} \tilde{Q}_X^{ab} Q_X^{ab} &= \frac{1}{2} \Tr( \Qmx \Qmhx) , \nonumber \\
 \sum_{a \leq b} \tilde{Q}_X^{ab} x_i^a x_i^b &= \frac{1}{2} \xa_i^{\top} \Qmhx \xa_i,
\end{align}
where we introduce the vector $\xa_i = ( x_i^0 , \cdots, x_i^n)^{\top}$. 
Changing the integration variables from $\tilde{\mathbf{Q}}_X$ to $\Qmhx$, we obtain~\eqref{eq:ZwithQhat}.


 \chapter{Blind sensor calibration state evolution}
\label{app:calampStateEvolution}

The blind sensor calibration setting presented in~\chapref{chap:blindSensorCal} can be analysed 
using the replica method in a way very similar to the analysis performed in~\secref{sec:replicaGLM} for generalized linear models.
These are preliminary results: no implementation of the state evolution equations presented below was made and therefore their correctness has not been 
verified by comparison with experimental results.

In the Bayes optimal case, the state evolution equations derived in~\chapref{chap:generalizedLinearModels} for \gamp are:
\begin{align}
 \mhx^{\iter+1} &= \frac{\alpha}{\mx^{\iter}} \int \dd \y \int \dd t \NN(t;0,1)  \frac{\left[ \frac{\partial}{\partial t} f^Y_{0}\left(\y |\sqrt{\mx^{\iter}}t,\Qzx  -\mx^{\iter}\right)\right]^2}{f^Y_{0}\left(\y |\sqrt{\mx^{\iter}}t, \Qzx -\mx^{\iter}\right)} ,   \\
 \mx^{\iter+1} &= \frac{1}{\sqrt{\mhx^{\iter+1}}} \int \dd t \frac{\left[f^X_1\left(\frac{t}{\sqrt{\mhx^{\iter+1}}},\frac{1}{\mhx^{\iter+1}}\right) \right]^2}{f^X_0\left(\frac{t}{\sqrt{\mhx^{\iter+1}}},\frac{1}{\mhx^{\iter+1}}\right)}.
 \end{align}
For Bayes optimal blind sensor calibration from $P$ independent signals, the equation for $\mx$ remain unchanged, but:
\begin{align}
 \mhx^{\iter+1} &=  \frac{\alpha}{\mx^{\iter}} \int \dd \yv \int \dd \tv \NN(\tv;0,\indic)  \frac{ \frac{1}{P} \sum_{l=1}^P \left[ \frac{\partial}{\partial t_l} f^C_{0}\left(\yv |\sqrt{\mx^{\iter}} \tv,(\Qzx  -\mx^{\iter}) \one \right)\right]^2}{f^C_{0}\left(\yv |\sqrt{\mx^{\iter}}\tv, (\Qzx -\mx^{\iter}) \one \right)} ,  \label{eq:mhxCal}
\end{align}
where $f^C_0$ is given by~\eqref{eq:fkc}. Both $\tv$ and $\yv$ are $P$-dimensional vectors.

In general, integration over $2P$ variables is therefore necessary in order to evaluate $\mhx$.
In particular settings, it might be possible to reduce the number of necessary integrations.
For instance, in the real gain calibration setting, the function to integrate depends only on the three scalars $\yv^{\top} \yv$, $\yv^{\top} \tv$ and $\tv^{\top} \tv$,
and the integration over $2P$ variables can therefore be reduce to an integration over $3$ variables.
Similarly, in the faulty sensors setting, it seems that a proper reformulation could reduce the number of integration variables to $4$.
No numerical results have been obtained yet but a careful analysis of the functions to integrate could allow to obtain the exact positions of the phase 
transitions observed experimentally.
 
 \chapter{Sparse phase retrieval state evolution}
\label{app:PR_DE}

The complex generalized model setting presented in~\chapref{chap:gampApplications} can be analysed 
using the replica method in a way very similar to the analysis performed in~\secref{sec:replicaGLM} for generalized linear models.
These are preliminary results: no implementation of the state evolution equations presented below was made and therefore their correctness has not been 
verified by comparison with experimental results.

In the Bayes optimal case, the state evolution equations derived in~\chapref{chap:generalizedLinearModels} for \gamp are:
\begin{align}
 \mhx^{\iter+1} &= \frac{\alpha}{\mx^{\iter}} \int \dd \y \int \dd t \NN(t;0,1)  \frac{\left[ \frac{\partial}{\partial t} f^Y_{0}\left(\y |\sqrt{\mx^{\iter}}t,\Qzx  -\mx^{\iter}\right)\right]^2}{f^Y_{0}\left(\y |\sqrt{\mx^{\iter}}t, \Qzx -\mx^{\iter}\right)} ,   \\
 \mx^{\iter+1} &= \frac{1}{\sqrt{\mhx^{\iter+1}}} \int \dd t \frac{\left[f^X_1\left(\frac{t}{\sqrt{\mhx^{\iter+1}}},\frac{1}{\mhx^{\iter+1}}\right) \right]^2}{f^X_0\left(\frac{t}{\sqrt{\mhx^{\iter+1}}},\frac{1}{\mhx^{\iter+1}}\right)}.
 \end{align}
In the complex case, these equations change and become:
\begin{align}
 \mhx^{\iter+1} &= \frac{\alpha}{\mx^{\iter}} \int \dd \y \int \dd t \CN(t;0,1)  \frac{\frac{1}{2} \sum_{i=1}^2 \left[ \frac{\partial}{\partial t_i} f^Y_{0}\left(\y |\sqrt{\mx^{\iter}}t,\Qzx  -\mx^{\iter}\right)\right]^2 }{f^Y_{0}\left(\y |\sqrt{\mx^{\iter}}t, \Qzx -\mx^{\iter}\right)} ,   \\
 \mx^{\iter+1} &= \frac{1}{\sqrt{\mhx^{\iter+1}}} \int \dd t \frac{\left| f^X_1\left(\frac{t}{\sqrt{\mhx^{\iter+1}}},\frac{1}{\mhx^{\iter+1}}\right) \right|^2}{f^X_0\left(\frac{t}{\sqrt{\mhx^{\iter+1}}},\frac{1}{\mhx^{\iter+1}}\right)}.
\end{align}
where $t= t_1 + \imath t_2$ is a complex integration variable and $y = y_1 + \imath y_2$ as well.
For Bayes optimal complex \cs, these equations give back the state evolution equation~(\ref{eq:complex_DE}).
In the case of a the complex joint Gauss-Bernoulli prior~(\ref{eq:joint_sparse}), $\mx$ can be reduced to an integral over a single, real variable $u$:
\begin{align}
 \mx = 2 \rho^2 \frac{\mhx}{\mhx + 1} \int_0^{+\infty} \dd u \frac{ u^3 e^{-u^2}}{(1-\rho)(\mhx + 1) e^{-2\mhx u^2} + \rho}.
\end{align}
In the case of phase retrieval, $\mhx$ can be calculated from~\eqref{eq:f0PR}. Using the scaled versions $(\tilde{I}_0, \tilde{I}_1)$ of $(I_0, I_1)$:
\begin{align}
 \mhx &= \frac{4 \alpha}{(\Delta + \rho - \mx)^3} \int_0^{+ \infty} \dd y \int_0^{+\infty} \dd r \, y r e^{-r^2} e^{-\frac{(y-\sqrt{\mx}r)^2}{\Delta+\rho-\mx}} \frac{ \left[ y \tilde{I}_1\left( \frac{2 ry\sqrt{\mx}}{\Delta+\rho-\mx}\right) - r \sqrt{\mx} \tilde{I}_0\left( \frac{2 ry\sqrt{\mx}}{\Delta+\rho-\mx}\right) \right]^2}{\tilde{I}_0\left( \frac{2 ry\sqrt{\mx}}{\Delta+\rho-\mx}\right)} . \nonumber
\end{align}

As in~\appref{app:calampStateEvolution}, these are preliminary analytical results that need to be verified by implementation and comparison to algorithmic performances.

 
 
 \bibliographystyle{abbrv}
 \bibliography{references/refs}
 
\end{document}